%% file: main.tex
\documentclass{aa}  

\usepackage{graphicx}
\usepackage{txfonts}
\usepackage{lipsum}
\usepackage{subcaption}
\usepackage{lscape}
\usepackage{placeins}
\usepackage[colorlinks,allcolors=blue]{hyperref}
\usepackage[normalem]{ulem}
\usepackage{orcidlink}
\usepackage{booktabs}
\newcommand{\Msun}{\hbox{M$_\odot$}}
\defcitealias{Miller2026ApJ...996...69M}{M26}
\begin{document}
   \title{The white dwarf population of open clusters and their tidal tails}

   \subtitle{Tracers of contamination and stellar interactions}

   \titlerunning{WDs in open clusters and tidal-tails}
\author{
   Vikrant V. Jadhav\orcidlink{0000-0002-8672-3300}\inst{1,2}
   \corrauth{\tiny{vikrant-vinayak.jadhav@matfyz.cuni.cz}}
    \and
    Pavel Kroupa\orcidlink{0000-0002-7301-3377}\inst{1,2}\email{pkroupa@uni-bonn.de}
    \and
    David R. Miller\orcidlink{0000-0002-4591-1903}\inst{3}
    \and 
    Snehalata Sahu\orcidlink{0000-0002-0801-8745}\inst{4}
    \and
    Dinnbier Frantisek\orcidlink{0000-0001-5532-4211}\inst{5}
    \and
    Ladislav Šubr\orcidlink{0000-0003-1924-8834}\inst{1}
}

\institute{
   Astronomical Institute, Faculty of Mathematics and Physics, Charles University, V Holešovičkách 2, CZ-180 00 Praha 8, Czech Republic
   \and
   Helmholtz-Institut für Strahlen- und Kernphysik, Universität Bonn, Nussallee 14-16, D-53115 Bonn, Germany
   \and
   Department of Physics and Astronomy, University of British Columbia, Vancouver, BC V6T 1Z1, Canada
   \and
   Department of Physics, University of Warwick, Coventry CV4 7AL, UK
   \and
   Physics Division, National Center for Theoretical Sciences, National Taiwan University, Taipei 106319, Taiwan
}
\date{Received May 30, 2026/ Accepted July 12, 2026}
  \abstract
   {}
   {
   Recent \textit{Gaia} studies have identified numerous open clusters (OCs) and tidal-tail catalogues, enabling systematic searches for white dwarfs (WDs) associated with clusters and their extended structures. We compile a literature-based sample of OC--WD pairs to validate WD membership in cluster cores and tidal-tails, investigate the initial--final mass relation (IFMR), identify WDs formed through non-canonical evolution, and interpret the observed WD populations using a grid of $N$-body simulations.
   } 
   {
    We combine \textit{Gaia} DR3 cluster and tidal-tail catalogues with ultraviolet-to-infrared photometry to analyse the OC--WD pairs. WD masses, cooling ages, radii, effective temperatures, and luminosities are estimated using colour--magnitude diagrams and spectral energy distributions. These observations are interpreted in the context of $N$-body simulations.
   }
   {
   We identify 235 OC--WD pairs in 80 clusters, including 99 WDs in tidal tails. More than 28\% of the pairs are likely spurious, with contamination substantially higher in the tails ($>$48\%) than in the cluster cores ($>$13\%), indicating significant field-star contamination in current \textit{Gaia}-based catalogues. The Pleiades tidal tails also show severe contamination by old WDs. Simulations predict that the fraction of core WDs increases with cluster age, reaching $\gtrsim$10\%, whereas the observed fractions remain systematically lower, consistent with the WD deficit problem. Despite the high contamination rate, most tail WDs ($\approx$83\%) are consistent with having been born inside the tidal radius. We additionally identify 63 candidate binary-origin WDs and 47 new IFMR candidates.
   }
   {
   WDs provide a powerful probe of contamination in cluster and tidal-tail catalogues and place important constraints on cluster detection methods and $N$-body simulations. Resolving the WD deficit and improving membership validation will require improved observations, membership methods, WD physics, and spectroscopic follow-up, ultimately enabling stronger constraints on dynamical cluster evolution and the WD IFMR.
   }

\maketitle
\nolinenumbers
\section{Introduction} \label{sec:introduction}

Star formation mainly occurs in embedded star clusters, but most clusters dissolve over time due to interactions with the Galactic potential \citep{Kroupa1995MNRAS.277.1491K, Kroupa1995MNRAS.277.1507K, Lada2003ARA&A..41...57L}. As clusters dissolve, their stars escape into the Galactic field through tidal tails \citep{Kupper2008MNRAS.387.1248K, Dinnbier2020A&A...640A..85D, Jerabkova2021AA...647A.137J}. Only the more massive clusters that retain a bound population for longer or young clusters can still be identified today. Cluster membership is therefore valuable, as stars share common properties such as age, distance, and metallicity, which can be measured more precisely than for individual field stars.

Over the course of stellar evolution and assuming a canonical stellar initial mass function, more than 90\% of stars end up as white dwarfs (WDs; \citealt{Kroupa2001MNRAS.322..231K, Ritossa1999ApJ...515..381R}). WDs cool down with time, unlike main-sequence (MS) stars, whose observable properties remain mostly unchanged during the H-burning phase \citep{Mestel1952MNRAS.112..583M}. This makes them useful as clocks to estimate the time since their birth \citep{Tremblay2014ApJ...791...92T, Camisassa2025AN....34640118C}.

WDs can be characterised by their core composition and envelopes. 
The thin WD envelopes, which remain the only observable part of the WD, mostly consist of He and some H \citep{Blouin2024arXiv240903941B}. Based on the absorption lines in the spectra, the WDs are generally classified into the following spectral types: DA (H-dominated atmosphere); DB (He\textsc{i} lines); DC (featureless continuous spectra, likely due to lower temperature); DQ (C lines); DZ (metal lines).
The majority of the current WDs are evolved from MS stars with masses above 0.6 \Msun\ and below the WD-NS boundary, roughly 8--10 \Msun, which form a CO-core containing nearly all of the WD mass. At the high-mass end, sufficiently massive WD progenitors can undergo carbon ignition, leading to the formation of ONe-core WDs. ONe-cores are generally expected for most WDs above $\gtrsim$1.05 \Msun\ \citep{Siess2007A&A...476..893S, Camisassa2022MNRAS.511.5198C}, with empirical work suggesting a corresponding initial-mass cut-off of $\approx$6 \Msun\ (\citealt{ElBadry2018ApJ...860L..17E, Cummings2018ApJ...866...21C, Cunningham2024MNRAS.527.3602C, Miller2026ApJ...996...69M}; \citetalias{Miller2026ApJ...996...69M} hereafter).

Stars less massive than 0.6 \Msun\ form a He-core during the red giant phase and fail to ignite He fusion, thus forming a He-core WD. However, single-stellar-evolution He-core WDs do not exist because their progenitors remain on the MS for longer than the Hubble time. Thus, the observed He-core WDs ($m_{wd}\lesssim0.45$ \Msun) are proposed to be products of binary interactions \citep{Althaus1997ApJ...477..313A}. A massive progenitor can produce a lower-mass WD through mass-loss during common-envelope evolution or mergers. In that case, the progenitor lifetime inferred from single-star models is overestimated, while the cooling age should remain largely unaffected. Likewise, a binary interaction can accelerate WD formation, again changing the inferred progenitor lifetime without strongly affecting the cooling age. Mergers during the WD phase can increase the WD mass, which makes the progenitor lifetime appear shorter than it really is, and can also reheat the remnant, effectively resetting the cooling clock \citep{Kawka2023MNRAS.520.6299K}. This can allow a WD to appear much younger than the host cluster while still being a plausible member.

Combining WDs with star clusters provides additional constraints. Associations between open clusters (OCs) and WDs are used to derive the initial--final mass relation (IFMR) by linking WD cooling ages with stellar evolution models. Since WDs act as clocks, they can also be used to estimate cluster \citep{Hansen2004ApJS..155..551H, Chen2023ApJ...950..155C} and Galactic disc \citep{Knox1999MNRAS.306..736K, Kilic2017ApJ...837..162K} ages.
Among the millions of WD candidates known today \citep[e.g.,][]{Gentile2021MNRAS.508.3877G}, only a few thousand WDs have been linked with open \citep[e.g.,][]{Yan2026arXiv260216550Y} and globular \citep[e.g.,][]{Richer2013ApJ...778..104R, Sahu2022MNRAS.514.1122S} clusters. WDs in globular clusters are difficult to study due to their distance, faintness, and crowding. In contrast, nearby OCs provide better constraints on progenitor masses (\citealt{Cummings2018ApJ...866...21C, Prisegen2021A&A...645A..13P}; \citetalias{Miller2026ApJ...996...69M}). However, only $\approx 400$ WDs are currently linked to OCs \citep{Yan2026arXiv260216550Y}.

In addition to clusters themselves, the \textit{Gaia} data have also been used to identify extended structures around star clusters, which also contain WDs \citep{Roser2019AA...621L...2R, Jerabkova2021AA...647A.137J, Risbud2025AA...694A.258R}. These catalogues contain $>100$ WDs outside the tidal radii; however, the catalogues are highly incomplete with unknown purity \citep{Jadhav2025A&A...704A..50J}.
These WDs can increase the sample of known OC-WD associations. Simultaneously, the WDs, with their chronological properties, can be used to validate the membership catalogues.

Simulations provide a complementary approach to interpret these observations. Matching the observed number of WDs in clusters has long been a challenge, known as the WD deficit problem \citep{Tinsley1974PASP...86..554T, Weidemann1992AJ....104.1876W}. Proposed explanations include dynamical ejection due to mass-loss \citep{Fellhauer2003ApJ...595L..53F}, observational biases, and unresolved binaries \citep{Williams2007AJ....133.1490W}.

In this work, we aim to increase the number of OC–WD pairs by using WDs in the tidal tails of nearby clusters. These associations are used to both test membership catalogues, measure WD properties, and compare them with $N$-body simulation. In Section~\ref{sec:simulations}, we present the expected WD properties from simulations, including their spatial distribution and numbers. Section~\ref{sec:data} describes the data and sample, Section~\ref{sec:obs_results} presents the photometric analysis of WDs, and Section~\ref{sec:discussion} discusses membership contamination, binary-origin WDs, and constraints on the IFMR. Section~\ref{sec:discussion} concludes and summarises the work. Appendix~\ref{sec:sim_supplementary} contains the definitions and methods related to simulation-related quantities. Appendix~\ref{sec:supplementary_figures} contains auxiliary figures and details of the online catalogues. The diagnostic figures for each cluster are available in the arXiv version.

\section{Simulations and Preliminary Analysis}
\label{sec:simulations}

\begin{figure*}
    \centering
    \includegraphics[width=1\linewidth]{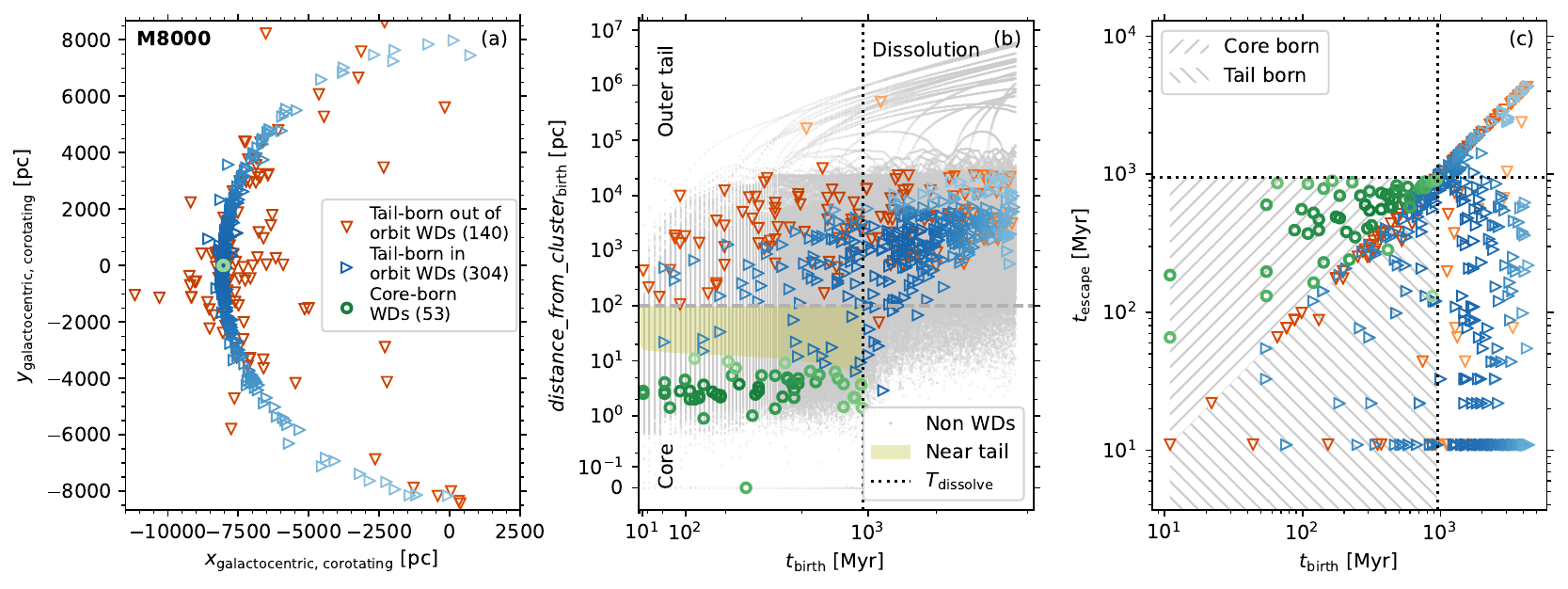}
    \caption{Birth properties of WDs in the simulated cluster M8000. (a) Birth location of WDs in the cluster-centric corotating frame. 
    Core-born WDs (green circles) and tail-born WDs with orbit similar to the cluster (blue triangles) and tail-born WDs with orbit different than the cluster (orange triangles) are shown separately.
    The markers are shaded darker in high-density regions of the respective phase space. The same colour scheme is used for all subplots.
    (b) Variation of the $distance\_from\_cluster$ with the birth epoch of the WDs. All stars in the simulation at all times are shown in the background (grey). The cluster dissolution time (black dotted line) and the near-tail region (olive patch) are marked for reference. Note that Asinh scaling is used for both axes.
    (c) Comparison of birth epoch to the escape epoch for all WDs within the model. The parameter spaces where WDs are formed in the core and in the tail are shaded.}
    \label{fig:sim_results}
\end{figure*}

\begin{figure}
    \centering
    \includegraphics[width=1\linewidth]{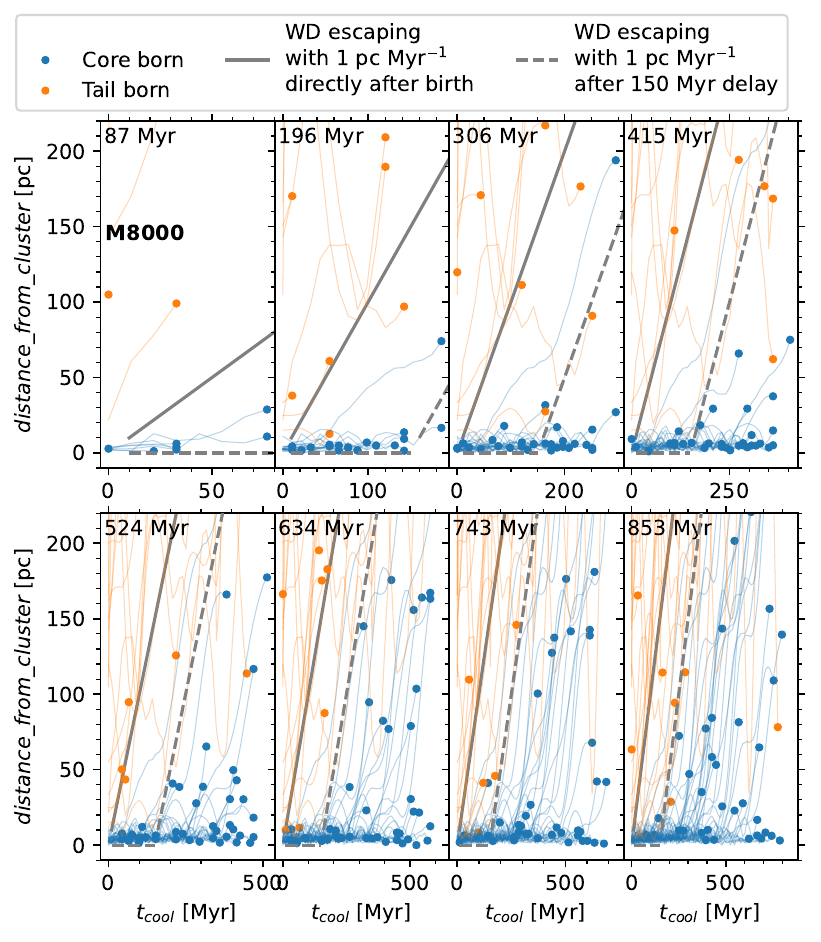}
    \caption{Comparison of the cooling time with the $distance\_from\_cluster$ for simulated M8000 snapshots from beginning to cluster dissolution time. Core-born (blue) and tail-born (orange) WDs and their past tracks (as thin lines) are plotted. The gray lines indicate the track of a WD moving away from the cluster at 1 pc Myr$^{-1}$, which escaped directly after its birth (solid line) and after a 150 Myr delay (dashed line).}
    \label{fig:sim_t_cool_distance}
\end{figure}

\begin{figure}
    \centering
    \includegraphics[width=1\linewidth]{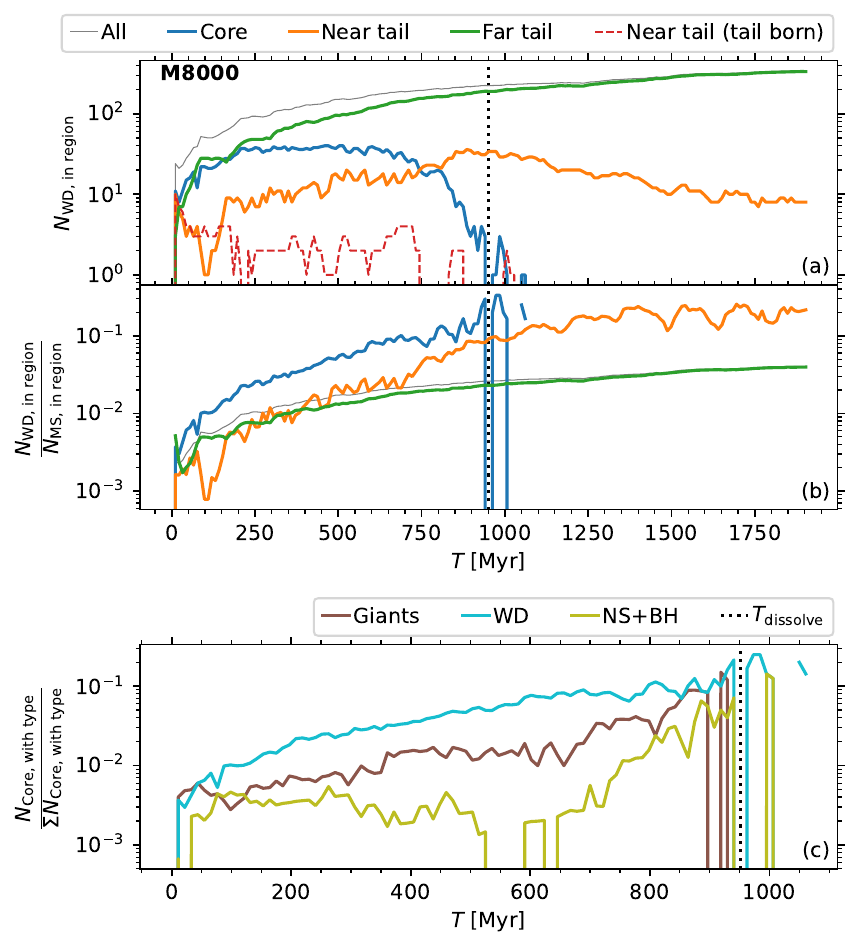}
    \caption{Population changes for various stellar types (MS, Giants, WD, and neutron stars+black holes; NS+BH) in the cluster, near-tail and far-tail for the simulated cluster M8000. 
    (a) Count of all (gray), core (blue), near-tail (orange), and far-tail (green) WDs as the cluster evolves. The dissolution time (vertical black dotted line) is shown for reference.
    The tail-born WDs in the near-tail are highlighted (red dashed curve).
    (b) The ratio of WDs to the MS stars within the above-mentioned regions. The colours are similar to those in panel (a).
    (c) Fraction giants (brown), WD (cyan), and NS+BH (olive) stars at any given time present in the core. The fraction of MS stars remains very close to 1 throughout the evolution.
    Expanded figures for models M8000 and M4100 are given in Figure~\ref{fig:bse_type_M8000} and \ref{fig:bse_type_M4100}, respectively.}
    \label{fig:sim_population_evolution_small}
\end{figure}

\subsection{\textsc{Nbody6} simulation setup}

We used a library of clusters with initial masses of 62--8000 \Msun\ originally calculated for studying the merger rate between B-type stars \citep{Dvorakova2024A&A...689A.234D}. The models used primordial binary fraction of 100\% \citep{Kroupa1995MNRAS.277.1491K, Kroupa1995MNRAS.277.1507K, Sana2012Sci...337..444S, Kobulnicky2014ApJS..213...34K, Belloni2017MNRAS.471.2812B}, canonical two-part power-law initial mass function \citep{Kroupa2001MNRAS.322..231K, Kroupa2026enap....2..173K} and including gas expulsion in the early evolution with star formation efficiency of 1/3 \citep{Lada1984ApJ...285..141L, Kroupa2001MNRAS.321..699K}. The clusters were modelled numerically using the code \textsc{Nbody6} \citep{Aarseth1999PASP..111.1333A, Aarseth2003gnbs.book.....A} with stellar and binary star evolution according to the description due to \citet{Hurley2000MNRAS.315..543H, Hurley2002MNRAS.329..897H} at the Solar metallicity ($Z = 0.014$). 
WDs are born with the velocity vector of the progenitor star unchanged; that is, no mass-loss kick is applied.

The simulations evolved for $\approx$1--14 Gyr. For reference, the clusters dissolve at $\approx 20$ Myr to $1$ Gyr depending on the initial cluster mass.
The models are named M62, ..., M2040, M4100, and M8000, corresponding to the approximate initial stellar mass of the cluster. The number of random realisations decreases with increasing cluster mass, ranging from 256 to 2 for the lowest- and highest-mass models, respectively. Figures generated using averaged or combined realisations are noted in the captions (e.g., Figure~\ref{fig:sim_WD_retention}b and Figure~\ref{fig:Hyades_missing_WDs}), while all other figures are based on a single realisation.
Section~\ref{sec:nbody_description} and \citet{Dvorakova2024A&A...689A.234D} provide for more details about the simulations.
Detailed explanations about the simulation-related terms (e.g., $distance\_along\_orbit$, $distance\_from\_orbit$, $distance\_from\_cluster$, and identification of the reference point for a dissolved cluster) are also given in Appendix~\ref{sec:sim_supplementary}.

\subsection{The spatial distribution of WDs in the cluster and the tail}

The tails of the clusters can span thousands of parsecs; however, most observational studies (particularly for OCs) are limited to the region near the cluster. We divided the stars into three ad-hoc regions: core (stars within the tidal radius, $R_\text{tidal}$), near-tail (region within $R_\text{tidal}$ to 100 pc), and far-tail (region outside 100 pc).
The far-tail region also includes most of the WDs whose orbits are significantly different than the cluster (Figure~\ref{fig:sim_results} b).
The `cluster' term refers to all stars (core and tail) associated with the cluster.
In this work, we primarily focus on the near-tail due to the observational significance. Similarly, the WDs born inside the tidal radius are referred to as core-born, while the WDs born outside the core are referred to as tail-born.

Figure~\ref{fig:sim_results} (a) shows the birth location of WDs in a simulation. The majority of the WDs are formed within the cluster's orbit, while a non-negligible ($\approx30$\%) fraction is born from stars leaving the cluster's orbit.
Figure~\ref{fig:sim_results} (b) shows the birth $distance\_from\_cluster$ for the WDs. The WDs are primarily born in the cluster core and the far-tail.
Figure~\ref{fig:sim_results} (c) compares the birth epoch with the time at which the star (either as WD or its progenitor) crossed the tidal radius.
Figure~\ref{fig:bse_type_M8000} and \ref{fig:bse_type_M4100} show the overall population changes for various stellar types in the core, near-tail, and far-tail.
Figure~\ref{fig:sim_t_cool_distance} shows the WD population in terms of observable quantities such as the cooling time and distance from the cluster. For comparison, we plotted the tracks for two WDs escaping the cluster with the approximate escape velocity of the cluster.

The WD births are concentrated within the core until the cluster dissolves, along with sporadic births all across the tidal tail. The birth rate of WDs (normalised using the total population) was found to be similar in the core and the tail. 
However, due to the large physical extent ($>100$ pc) of the tails, the WD linear\footnote{We measure linear number density, N pc$^{-1}$, instead of 3D number density, N pc$^{-3}$, due to the undefined 3D boundary of the tidal-tail region} number density in the core is roughly two orders of magnitudes larger than in the tails. 
Figure~\ref{fig:sim_t_cool_distance} shows that the majority of the near-tail-WDs have recently escaped from the cluster. In addition, the plots show that the core-born WDs stay within the cluster for a significant time and only begin to escape after a few hundred million years.
The near-tails of M8000 have tens of WDs, and many of them are young (and bright). The near-tails of lower-mass clusters harbour fewer WDs. However, they can host more WDs than the core region, depending on the two-body relaxation time of the cluster. This is caused by the slower escape velocities of the smaller cluster.
Overall, linking tail-WDs to the parent cluster will significantly boost the population of OC-WD pairs.

\subsection{Evolution of the WD population in the cluster and the tail}

Figure~\ref{fig:sim_population_evolution_small} (a) shows the evolution of the WD population with time. The number of WDs in the core creates a plateau ($\approx$30 WDs) during the majority of the cluster evolution. 
However, not all simulated clusters show such a plateau.
Near the dissolution time, the WDs (along with all members) start moving from the core to the near tail. As expected from Figure~\ref{fig:sim_t_cool_distance}, there are only a handful of tail-born WDs present in the near-tail (about a fifth). 
Figure~\ref{fig:sim_population_evolution_small} (b) shows that the ratio of WDs to MS stars is higher in the core region compared to the tails. 
Figure~\ref{fig:sim_population_evolution_small} (c) shows the evolution of stellar content within the core region. The fractional population of WD increases with time, and similar trends are seen for all cluster masses and also in the giant population.

\begin{figure}
    \centering
    \includegraphics[width=1\linewidth]{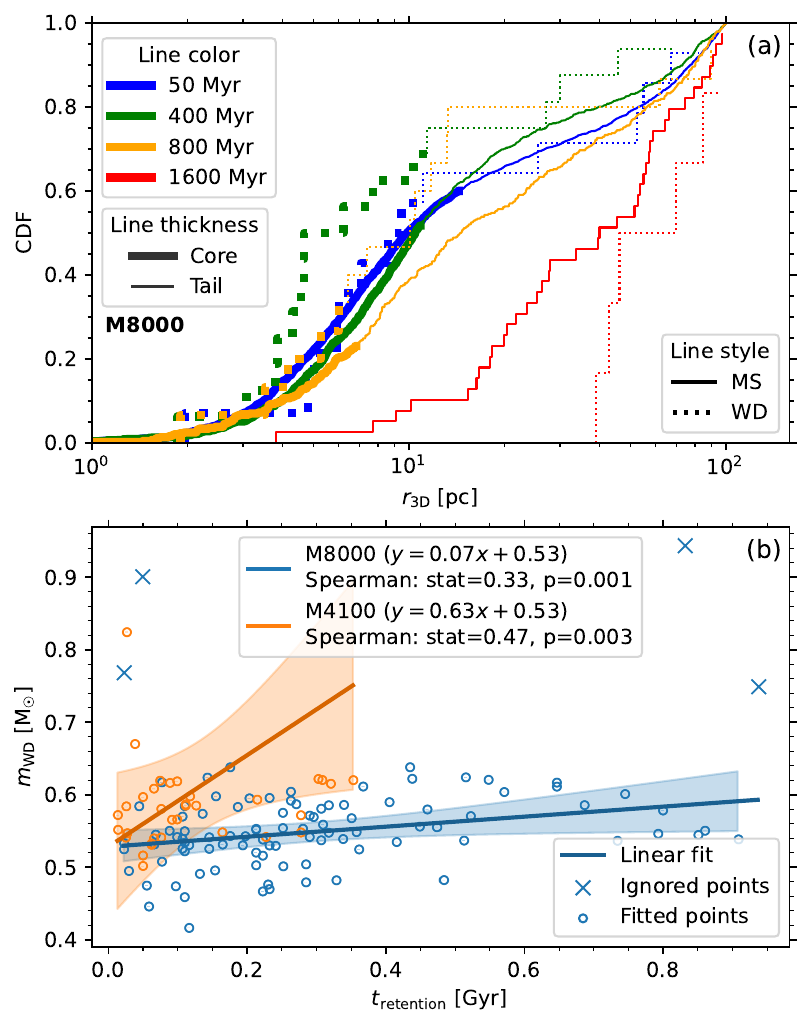}
    \caption{
    Radial segregation and WD retention time within the cluster core. 
    (a) The cumulative distribution of radii of WDs (dashed curves) and the corresponding MS populations (solid curves) in M8000. The curves are coloured according to the age of the cluster, and their thickness corresponds to the core (thicker) or tail (thinner) region. Note that the model at 1600 Myr is after the dissolution, thus only the tail population is shown.
    (b) Variation of the WD retention time within the cluster core compared to the WD mass in models M8000 (blue) and M4100 (orange). For better statistics during the fit, two and four random realisations of the M8000 and M4100 models are combined here, respectively.
    The linear fit and corresponding Spearman statistics are given in the legend. The crosses were not used for the fit.}
    \label{fig:sim_WD_retention}
\end{figure}

\subsection{Radial segregation and retention time of WDs}

The excess WD population in the core is a result of mass segregation. As the WD progenitors are among the most massive stars, they are preferentially retained in the cluster. The segregation increases with dynamical relaxation and is more prominent in more massive clusters. Figure~\ref{fig:sim_WD_retention} (a) shows the cumulative distribution profile of the WDs and MS stars. The WDs are more centrally segregated during the majority of the cluster evolution (100--800 Myr for M8000). Similar WD segregation has also been noted in simulations of globular clusters \citep{Baumgardt2003MNRAS.340..227B}.

Figure~\ref{fig:sim_WD_retention} (b) shows the variation of the retention time (defined as the difference between the time of birth and the time of escape) with WD mass. There is a moderate correlation between the $t_\text{retention}$ and the $m_\text{WD}$.
The velocities of the escaping WDs were comparable to those of non-WD escapees.
This is likely a result of the small mass range of the WD population, the stochastic nature of mass segregation, and the assumed WD physics (mass-loss and kicks).

\begin{figure*}[ht]
    \centering
    \includegraphics[width=1\linewidth]{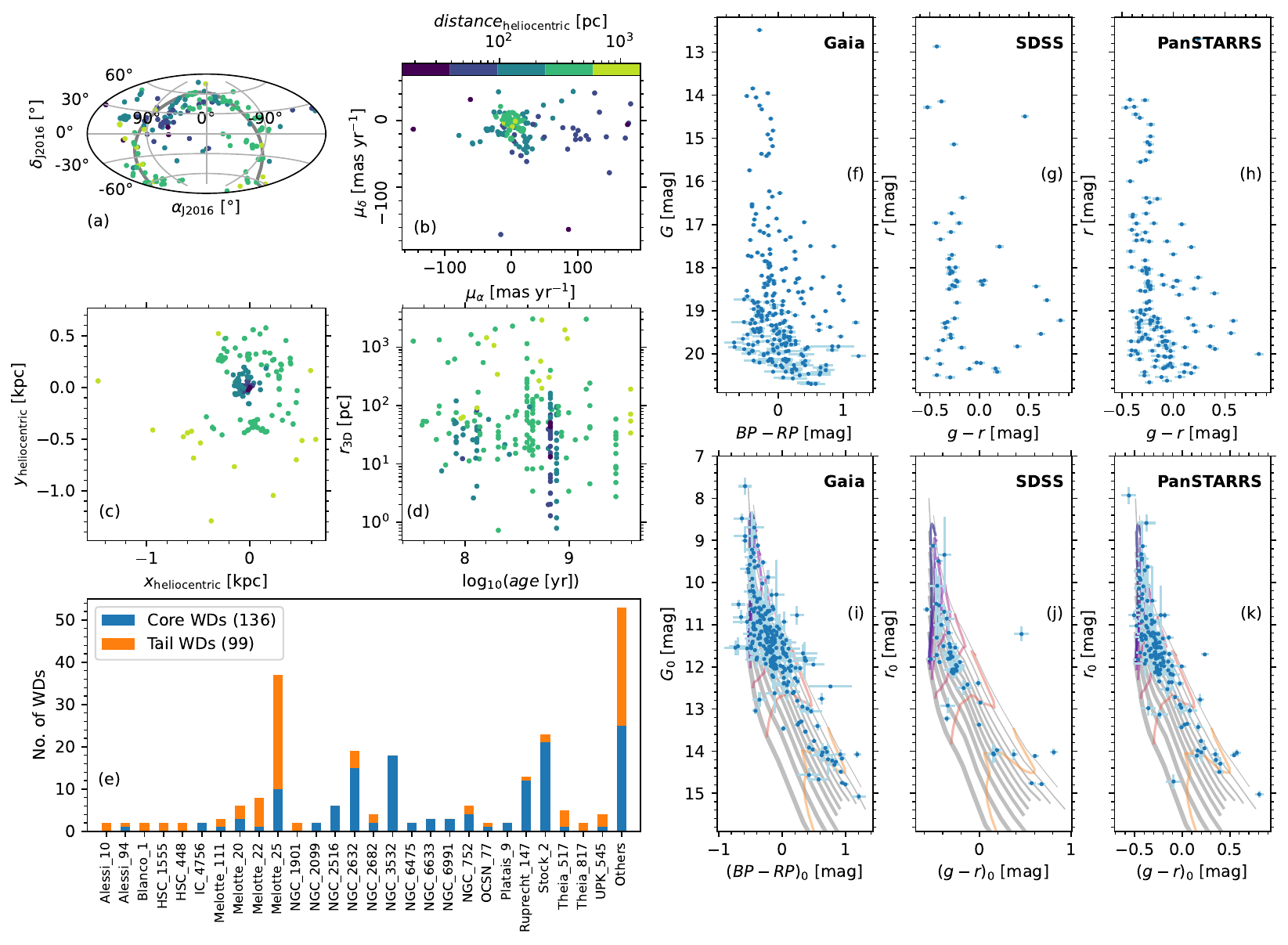}
    \caption{General properties of the 235 WD in the present sample. 
    (a) Spatial distribution on the sky in equatorial coordinates. The Galactic plane is highlighted by the grey curve. The WDs in panel (a)--(d) are coloured according to their heliocentric distances.
    (b) Vector point diagram showing the proper motions. 
    (c) Heliocentric position on the Galactic plane. The Galactic centre is at (8122, 0, -20.8) pc in this frame and the Galactic rotation is in the positive y direction (same as the \textsc{astropy} convention). 
    (d) Distribution of cluster age and 3D distance of the WD from the cluster centre.
    (e) Frequency of core (blue) and tail (orange) WDs in the 80 individual clusters. Clusters with only 1 WD are part of the `Others' column. 
    (f--h) Apparent (top row) CMDs of the WD sample using \textit{Gaia}, SDSS, and PanSTARRS data.
    (i--k) The corresponding absolute CMDs of the WD sample.
    The DA WD cooling curves are shown for reference (see Figure~\ref{fig:combo_Melotte_25} for details). 
    }
    \label{fig:sample_description}
\end{figure*}

\subsection{Caveats and scope for improvements in simulations}

The $N$-body simulations are at most an approximation of the real systems because of the assumptions about the physics, initial conditions, and the numerical precision. In these simulations, the analytical WD models (which are also not 100\% accurate) are not up to date. The prescriptions used to impart mass-loss driven recoil could also affect the evaporation of WDs into the tails \citep[as explored in][]{Fellhauer2003ApJ...595L..53F}.
However, such simulations provide the best currently available tools to understand the observations of tidal tails, which contain significant biases and incompleteness.

\section{Data and methods} \label{sec:data}

\subsection{Sample selection} \label{sec:sample_selection}

The OC-WD pair sample is compiled from four sources.
\citet{Prisegen2023A&A...678A..20P} compiled a list of 76 OC-WD pairs using \textit{Gaia} data among clusters with ages of 31--3100 Myr and heliocentric distances of 130--1400 pc.
\citet{Jadhav2025A&A...704A..50J} presented a list of high quality \textit{Gaia}-based catalogues of tidal-tails. There are 73 WDs associated with 9 clusters spanning ages of 80--1500 Myr and heliocentric distances of 47--440 pc.
\citet{Yan2026arXiv260216550Y} identified 439 WD candidates in 117 OCs.
Recently, \citetalias{Miller2026ApJ...996...69M} published a list of 166 OC-WD pairs from \textit{Gaia} data and literature search. \citetalias{Miller2026ApJ...996...69M} also includes a list of 57 WDs without \textit{Gaia} membership assessment, which are not included in the present sample. 
Note that 27 clusters were reclassified as moving groups in \citet{Hunt2024A&A...686A..42H}; however, we keep them in the sample due to their strong association and previous studies (see \citetalias{Miller2026ApJ...996...69M} for more details). The moving group classification is included in the Table~\ref{tab:cluster_params}.
We use the same `OC-WD pair' notation for these moving group WDs for simplicity.

We removed one WD (HSC 381) from the \citet{Prisegen2023A&A...678A..20P} sample and seven WDs (in six clusters: CWNU 515, HSC 2139, HSC 2156, HSC 2630, HSC 381, LISC-III 3668, NGC 2547) from the \citetalias{Miller2026ApJ...996...69M} sample because the parent cluster age is too young to form a WD ($\log_{10}(age) < 7.5$; $m_\text{progenitor}>10$ \Msun).
We additionally removed 328 WDs from the \citet{Yan2026arXiv260216550Y} sample that lacked 5D membership, and a further 11 WDs whose parent clusters were similarly too young. Although these objects are plausible cluster members, we excluded them to maintain consistency with the remainder of the compiled sample, for which 5D membership is available across the literature employing different membership determination methods.

Combining the four catalogs resulted in a total of 235 OC-WD pairs in 80 clusters and moving groups.
For each cluster, we investigated the spatial distribution, proper motions, parallaxes, colour-magnitude diagram (CMD), on-sky, and 3D separation from the cluster centre to identify tidal-tail members. The Jacobi radii mentioned in \citet{Hunt2024A&A...686A..42H} were used as guides to segregate the core and tail populations. This classification could be incorrect for clusters more distant than 300 pc due to positional uncertainty.
Overall, there are 99 tail-WDs and 136 core-WDs.
Only Hyades (Melotte 25), Stock 2, NGC 2632 (Praesepe), NGC 3532, Ruprecht 147, and Melotte 22 (Pleiades) have $>$10 associated WDs.
Figure~\ref{fig:sample_selection} shows the Venn diagram of the WDs taken from the catalogues and their overlap, and Table~\ref{tab:wd_params} contains the WD list with derived parameters.

Figure~\ref{fig:sample_description} shows the general properties of the WD sample. As expected, most WDs lie along the Galactic plane with heliocentric distances of 15 pc to 1.5 kpc and proper motions of up to 180 mas yr$^{-1}$. Panels (d) and (e) show that the majority of the WDs lie within the cluster core of each cluster.

\subsection{Archival photometry} \label{sec:archival_photometry}

We used pre-computed crossmatches \citep{Marrese2017A&A...607A.105M, Bianchi2020ApJS..250...36B} to collect archival photometry from other surveys: Galaxy Evolution Explorer (GALEX) GUVcat\_AIS GR6+7 \citep{Bianchi2017ApJS..230...24B}, Sloan Digital Sky Survey (SDSS) DR16 \citep{Ahumada2020ApJS..249....3A}, Panoramic Survey Telescope and Rapid Response System (PanSTARRS) DR1 \citep{Chambers2016arXiv161205560C}, Two Micron All-Sky Survey (2MASS;  \citealt{Cutri2003yCat.2246....0C}) and Wide-field Infrared Survey Explorer (WISE; only W1 and W2 filters; \citealt{Cutri2014yCat.2328....0C}). We also cross-matched the catalogue with UVIT DR1 \citep{Piridi2024ApJS..275...34P} using 1\arcsec\ radius.
These crossmatches were done using propagated proper motions, which are necessary due to the high proper motion targets in the sample.
Missing photometric errors were replaced by double the maximum available photometric error. Following \citet{Bergeron2019ApJ...876...67B}, we set a lower limit of 0.03 mag to all photometric errors. This accounts for calibration errors and ensures that bandpasses with small errors do not dominate the fits.
The zero distance errors were replaced with 0.1 pc uncertainty.

We visually checked the neighbourhoods of the WDs for possible contamination from nearby sources in the digital sky survey, PanStarrs, \textit{Gaia}, and 2MASS images\footnote{Using \textsc{Finder\_charts} \url{github.com/fkiwy/Finder_charts}}. We graded the stars based on the possible contamination as gold-crowding (no contamination within 5\arcsec), silver-crowding (other sources within 3\arcsec--5\arcsec), and bronze-crowding (other sources within 3\arcsec). There are 179, 33, and 26 WDs graded as gold, silver, and bronze-crowding, respectively.
For further analysis, we used only \textit{Gaia} data for the bronze-crowding WDs, \textit{Gaia} + SDSS + UVIT data for silver-crowding WDs, and all available data for gold-crowding WDs. Any remaining outlier magnitudes were identified visually in the spectral energy distributions (SEDs) and removed.

After these photometric quality checks, 
44, 3, 53, 139, 235, 21, and 25 WDs have
GALEX, UVIT, SDSS, PanStarrs, \textit{Gaia}, 2MASS, and WISE photometry, respectively.

\subsection{Cluster and Stellar parameters}

The cluster ages, metallicity, extinction, and distance were collected from the literature, giving preference in the following order: \citet{Netopil2016A&A...585A.150N}, \citetalias{Miller2026ApJ...996...69M}, \citet{Cummings2018ApJ...866...21C}, \citet{Bossini2019A&A...623A.108B}, \citet{Dias2021MNRAS.504..356D}, \citet{Cantat2020A&A...640A...1C}, and \citet{Hunt2024A&A...686A..42H}.
Missing metallicity was assumed to be solar, and missing errors in $\log_{10}(age)$ and metallicity were assumed to be 0.2 dex.

We collected radial velocities and spectral types of the WDs from Simbad and other literature \citep[e.g.,][]{Casewell2009MNRAS.395.1795C, Pasquini2019A&A...627L...8P, Pasquini2023MNRAS.522.3710P, Miller2026ApJ...996...69M}.
The age and metallicity of the WDs were assumed to be the same as those of the associated cluster. 
For WD distance estimates, we used either the cluster distance or the parallax-based distance estimate (\texttt{r\_med\_geo}; \citealt{Bailer2021AJ....161..147B}) depending on uncertainties and tail membership.
The WD extinction was assumed to be the same as the cluster for core-WDs. For tail-WDs, the 3D reddening map-based extinction was taken from \citet{Gentile2021MNRAS.508.3877G}.
Table~\ref{tab:cluster_params} lists the cluster parameters, and Figure~\ref{fig:cluster_parameters} presents the selection of the adopted stellar and cluster parameters from the literature.

\subsection{Progenitor masses and cluster cooling curve}

We calculate the progenitor mass similar to \citetalias{Miller2026ApJ...996...69M} using the Parsec isochrones. 
For a given cluster metallicity (assuming [M/H] $\approx$ [Fe/H]), we identify the least massive star on the asymptotic giant branch at different isochronal ages, producing a relation between the progenitor mass and the progenitor age. For each WD, the progenitor age is determined as $age_\text{cluster}-t_\text{cool}$. The progenitor mass is then calculated by interpolating the progenitor mass and age relation for the cluster metallicity.

\begin{figure*}
    \centering
    \includegraphics[width=1\linewidth]{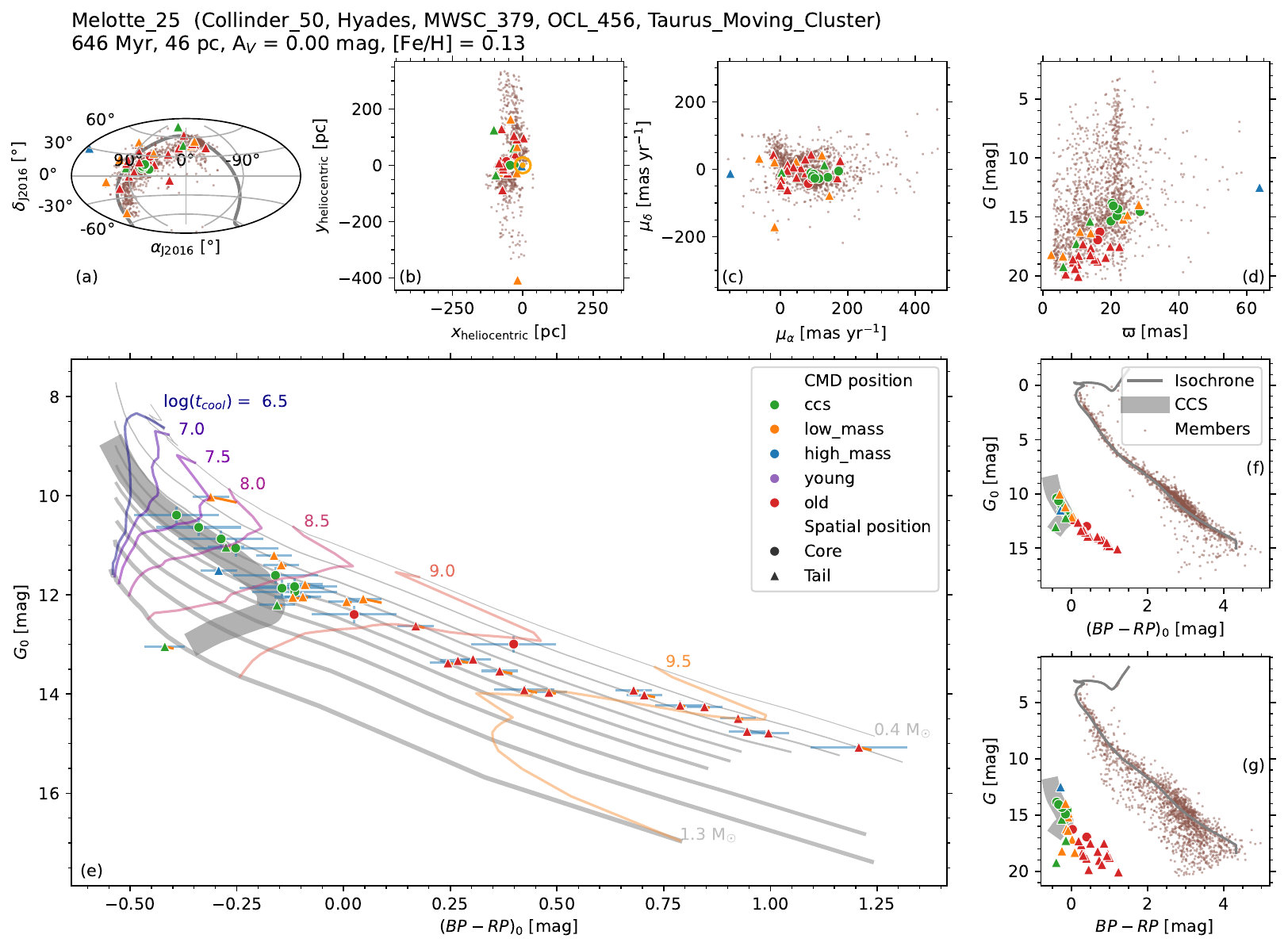}
    \caption{Diagnostic plots for Hyades. 
    (a) Spatial distribution of the core (circles) and tail (triangles) WDs coloured according to their CMD position (see Section~\ref{sec:cmd_diagnostics}). Non-WD cluster members (brown; cumulative sample from \citealt{Hunt2024A&A...686A..42H} and \citealt{Jadhav2025A&A...704A..50J}) and the Galactic plane (grey) are shown for reference.
    (b) Heliocentric position of WDs and other cluster members in the Galactic plane. The orange circle denotes the position of the Sun; when the Sun lies outside the plot boundaries, an orange wedge marks its direction beyond the plotted region.
    (c) Vector point diagram showing the proper motion of the WDs and other members.
    (d) Parallax and G band magnitude distribution for the WDs and other members.
    (e) Absolute \textit{Gaia} CMD of WDs. The \citet{Camisassa2025AN....34640118C} cooling curves are shown in the background, indicating the tracks for various WD masses (0.4 to 1.3 \Msun) and the cooling ages ($6<\log_{10}(t_\text{cool})<10$). The CCS is shown as the wide grey band. The movement in the CMD due to extinction is indicated by the orange lines.
    (f) Absolute \textit{Gaia} CMD showing the WDs and other members (brown). The Parsec isochrone (grey curve) and the CCS (grey band) are shown for reference.
    (g) Apparent \textit{Gaia} CMD of the WDs and other members. Reddened isochrone and CCS calculated for the cluster's extinction and distance modulus are shown for reference.
    }
    \label{fig:combo_Melotte_25}
\end{figure*}

As a cluster evolves, the most massive WDs are formed first, with progressively lower-mass WDs forming at later evolutionary times. Since WDs of different masses cool at different rates, the expected distribution of single-star evolution DA-type (H-atmosphere) WDs in a cluster CMD forms a characteristic sequence, which we define as the cluster cooling sequence (CCS), also sometimes referred to as the WD isochrone.
We use the \citetalias{Miller2026ApJ...996...69M} IFMR to estimate the WD masses and the DA-type WD cooling models from \citet{Camisassa2025AN....34640118C} to obtain the corresponding photometric magnitudes. We generated a grid of WDs which evolved between 30 Myr and the cluster age with a step size of $\Delta\log(age)=0.02$.
The stellar mass function was not considered in these calculations. Figure~\ref{fig:combo_Melotte_25} (e) shows the example of the Hyades CCS and the expected CMD location of the WDs.
The CCS is thus useful in visualising the expected properties of the WDs in the cluster, assuming they are formed by canonical single evolution and have an H atmosphere. The absolute and apparent faint-end limiting G-band magnitudes of the CCS for individual clusters are listed in Table~\ref{tab:cluster_params}.

\section{Observational results} \label{sec:obs_results}

\subsection{CMD-based WD parameters} \label{sec:cmd_based_params}

Figure~\ref{fig:sample_description} (h--k) show the apparent and absolute CMDs using \textit{Gaia}, SDSS, and PanSTARRS photometry.
The $((u-g),g)$ and $((u-r),r)$ CMDs (not shown in the figure) have overlapping models; hence, these CMDs are not suitable for interpolation-based estimations.

We used the WD cooling models from \citet{Camisassa2025AN....34640118C}\footnote{\url{https://evolgroup.fcaglp.unlp.edu.ar/modelos.html}} (also see \citealt{Koester2010MmSAI..81..921K, Althaus2013A&A...557A..19A, Camisassa2016ApJ...823..158C, Camisassa2017ApJ...839...11C, Camisassa2019A&A...625A..87C}) for doing the photometric analysis. The cooling models were interpolated in the CMD to convert the observables into WD parameters such as temperature, mass, cooling time, and $\log_{10} g$.
Figure~\ref{fig:wd_cmd} shows the detailed CMDs of an individual WD and the corresponding mass and cooling age estimation.
Linear interpolation was used to estimate WD mass, cooling age, luminosity, temperature, and radius. We used a Monte-Carlo approach to estimate the errors in the WD parameters by creating 100 realisations of the data by adding Gaussian noise to each point scaled according to the observational errors. Only the parameters with $>80\%$ valid estimates were retained. We use the \textit{Gaia}-CMD-based estimates for the rest of the analysis unless stated otherwise.
The measurements from the \textit{Gaia}, SDSS, and PanSTARRS CMDs are generally consistent with each other, but not always (Figure~\ref{fig:cmd_based_params}). Some WDs fell outside the cooling models and were not suitable for the comparison-based analysis. Even though 235, 139, and 53 WDs had \textit{Gaia}, PanSTARRS, and SDSS photometry, only 210, 132, and 50 WDs have CMD-based parameters, respectively.
The CMD-based parameters are given in Table~\ref{tab:wd_params}.

\begin{figure*}
    \centering
    \includegraphics[width=1\linewidth]{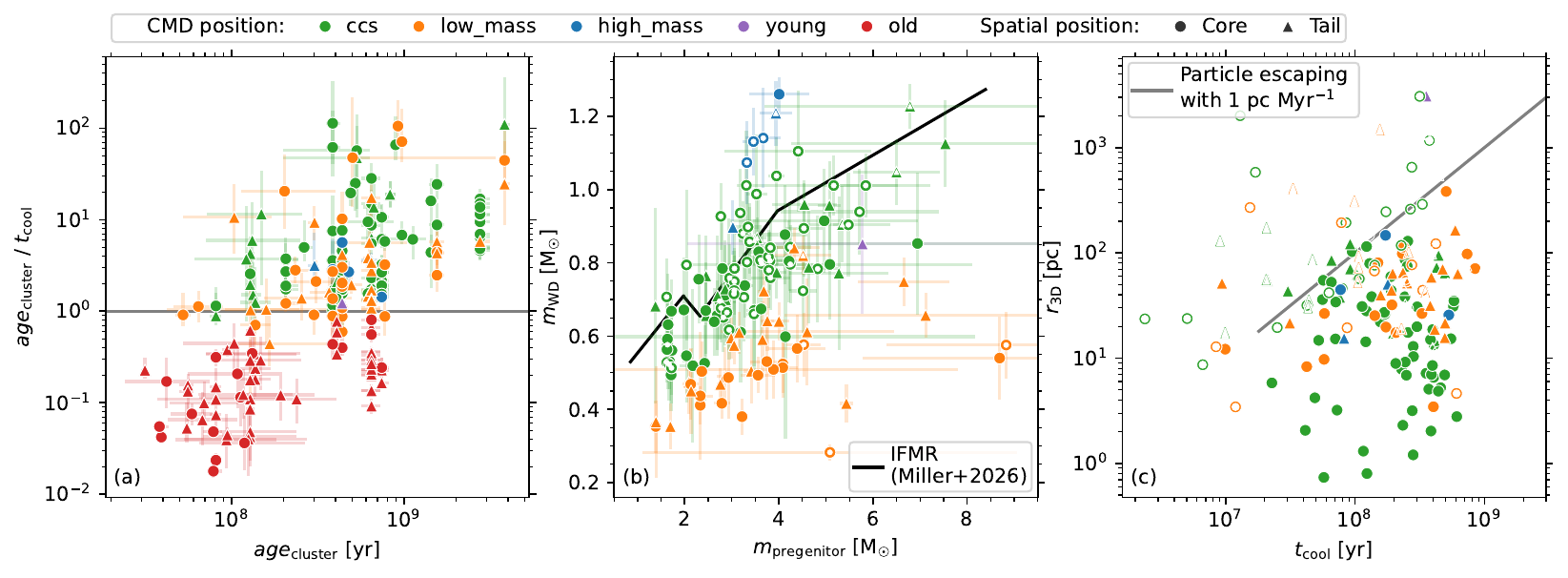}
    \caption{Comparison of WD parameters with cluster parameters and literature IFMR.
    (a) Comparison of the cluster age and the WD cooling time. The WDs are coloured and shaped similarly to Figure~\ref{fig:combo_Melotte_25}.
    (b) The IFMR of the WDs using photometric masses from this work. The hollow markers indicate the WDs that have spectroscopic mass estimates in \citetalias{Miller2026ApJ...996...69M}.
    The \citetalias{Miller2026ApJ...996...69M} IFMR is shown for reference.
    (c) Distribution of the WD cooling age with the distance from the cluster. Filled and hollow markers indicate whether the WD is nearer or farther than 400 pc from the Sun. The trajectory of a WD escaping at 1 pc Myr$^{-1}$ just after its birth is shown in grey.
    The \texttt{old} WDs are not shown in panels (b) and (c).}
    \label{fig:ifmr}
\end{figure*}

\subsection{SED-based WD parameters}

The SEDs were created using available photometry mentioned in Section~\ref{sec:archival_photometry}. 
They were fitted with Koester model spectra \citep{Koester2010MmSAI..81..921K, Tremblay2009ApJ...696.1755T} using \textsc{vosa}\footnote{\url{http://svo2.cab.inta-csic.es/theory/vosa/index.php}} to obtain the WD temperature, radius, and luminosity. Only 146 out of 235 WDs were successfully fitted with SEDs due to the unavailability of enough uncontaminated photometry.
The SED parameters are given in Table~\ref{tab:wd_params}.

\subsection{CMD and astro-photometric diagnostics} \label{sec:cmd_diagnostics}

Figure~\ref{fig:combo_Melotte_25} shows the CMD and other diagnostic plots for the Hyades. The spatial and velocity-space distributions reveal significant deviations from the mean cluster properties for the tail-WDs, as expected \citep{Risbud2025AA...694A.258R}. Figure~\ref{fig:combo_Melotte_25} (e)--(g) show the absolute and apparent \textit{Gaia} CMDs of the cluster. The absolute CMD still retains unexpected dispersion in lower MS indicative of binaries and contamination.
The WD cooling curves and CCS are shown for reference. Similar diagnostic plots for all clusters are available in Appendix~\ref{sec:arXiv_only}\footnote{only available in the arXiv version}.

The example CMD of Hyades WDs in Figure~\ref{fig:combo_Melotte_25} (e) also shows that many WDs significantly deviate from the CCS. A few WDs appear to have too low-mass compared to the CCS, while all the fainter ones are too old. 
We thus classified all WDs in to five categories based on the visual inspection of the CMD position, cooling age (Figure~\ref{fig:ifmr} a) and the WD mass (Figure~\ref{fig:ifmr} b): \texttt{CCS} (consistent with the CCS); \texttt{old} ($[age_\text{cluster}/t_\text{cool}]_\text{84th quantile}>1$); \texttt{young} (hotter and brighter than the WD models); \texttt{high\_mass} ($t_\text{cool}\lesssim age_\text{cluster}$ but more massive than the similar age CCS); \texttt{low\_mass} ($t_\text{cool}\lesssim age_\text{cluster}$ but less massive than the similar age CCS).

Figure~\ref{fig:ifmr} (a) shows the variation of the cooling time with the cluster age for all WDs. The relative scarcity of \texttt{old} WDs in older clusters is primarily driven by the larger heliocentric distances of those systems in combination with the \textit{Gaia} limiting magnitude. Future proper-motion membership determination with LSST is expected to substantially expand the WD census, while also inevitably introducing a healthy population of implausibly old contaminants into the oldest clusters.
Across all clusters, we find 101 \texttt{CCS}, 5 \texttt{young}, 10 \texttt{high\_mass}, 53 \texttt{low\_mass} and 66 \texttt{old} WDs.

\section{Discussion} \label{sec:discussion}

In this section, we discuss the parameters obtained from the CMD and SED-based analysis and their links to the membership and the IFMR.

\subsection{Robustness of the WD membership} \label{sec:robustness_of_wd_membership}

We compiled the list of OC-WD pairs from the literature, which is based on \textit{Gaia} astrometric data. However, all catalogues suffer from incompleteness and false positives. Contamination from a low-mass MS star is difficult to detect, as age, metallicity, or radial velocity are not trivial to obtain. WDs can be a useful tool to understand the false positives due to their cooling tracks.

A typical cluster member WD should lie on the CCS. The deviations from the CCS could be due to the following reasons: i) WD evolution deviating from simple single stellar evolution models due to interactions or other phenomenon, i.e. non-canonical stellar evolution, leading to high- or low-mass WDs \citep{Marsh1995MNRAS.275..828M, Kawka2023MNRAS.520.6299K}. ii) Incorrect photometric or distance priors. E.g., \textit{Gaia} BP and RP magnitudes are known to be corrupted for the fainter stars \citep{Riello2021A&A...649A...3R}. iii) The WD spectral type is not DA. The DA and DB cooling curves differ by $>$0.4 mag in G-band and $>$0.2 mag in the (BP-RP) colour. Thus, incorrect presumptions about the spectral class can lead to incorrect WD parameters and deviations from the CCS.
iv) False member. 
As mentioned in Section~\ref{sec:cmd_diagnostics}, only 43\% of WDs are consistent with the CCS. And 72\% WDs have $t_\text{cool}<age_\text{cluster}$. This means that at least 28\% of OC-WD pairs are erroneous. The 18 WDs disregarded during the sample selection (Section~\ref{sec:sample_selection}) associated with very young ($<31$ Myr) clusters also point towards contamination in the parent catalogues.

Out of the 66 \texttt{old} WDs, the majority (48) are classified as tail-WDs. The higher false member rate for tidal-tails (48\%) compared to the core members (13\%) is expected because the techniques used for tidal-tail detection are less robust than those used for core cluster membership.

The 28\% false-member rate from \texttt{old} (and typically faint) WDs suggests that contamination may also be present among WDs with $t_\text{cool}<age_\text{cluster}$. Because younger WDs are brighter and have higher-precision astrometry, their false-member rate is expected to be lower than 28\%. A dedicated, homogeneous search for OC-WD pairs using forthcoming high-precision data, coupled with simulations to quantify incompleteness and false-positive rates, is therefore required.

Figure~\ref{fig:WD_RVs} shows the radial velocities of the 38 WDs with literature radial velocities compared to the clusters' velocity. The sample is heterogeneous: some WD velocities have been corrected for gravitational redshift (e.g., most WDs in NGC 2682 and Hyades), while many others remain uncorrected (e.g., in Melotte 22).
For reference, the gravitational redshift arises due to the strong gravitational well of the compact WDs, leading to an increase in wavelength of the emitted light. This leads to the observed radial velocities being larger than the actual radial velocity.
The typical gravitational redshifts are $\approx$0.6 km s$^{-1}$ for Sun-like stars and $\gtrsim$30 km s$^{-1}$ for WDs \citep{Lopresto1980SoPh...66..245L, Williams2015AJ....150..194W}.
Without consistent gravitational redshift corrections, these radial velocities cannot be used to confirm or reject cluster membership. This highlights the need for homogeneous and robust radial velocity determinations when assessing OC-WD pairs.

\subsection{Possible contamination in cluster catalogues}

The false-membership of WDs also raises questions about the false-membership of the MS population of typical \textit{Gaia}-based OC catalogues (e.g., \citealt{Hunt2024A&A...686A..42H, Risbud2025AA...694A.258R}). The lower MS with magnitudes (and thus astrometric precision) similar to WDs might suffer from a similar level ($\approx13$\%) of contamination. Quantifying and reducing the contamination in the cluster members is thus an essential next step in improving the star cluster catalogues.

\subsection{Credibility of Pleiades tidal-tails}

One notable example of membership contamination is the Pleiades. \citet{Risbud2025AA...694A.258R} identified the tidal tails of Pleiades; however, they also cautioned that the tails are likely to be spurious due to the abnormal tail morphology. Among the 7 tail-WDs in Pleiades, 5 have $t_\text{cool}>age_\text{cluster}$. The contamination signatures inferred from the WDs are therefore consistent with earlier conjectures that the detected features do not represent genuine tidal tails. Recently, \citet{Boyle2025ApJ...994...24B} found that there is a large (bigger than 300 pc) comoving stellar complex with an age comparable to that of the Pleiades. The substantial spatial and kinematic overlap between the \citet{Risbud2025AA...694A.258R} Pleiades tidal-tails and this comoving complex suggests that the latter is being detected instead of the true Pleiades tidal-tails. Whether the genuine tidal tails remain undetected due to low contrast in phase space or have been disrupted by dynamical interactions remains an open question.

\subsection{The WD are rarely born in the tails}

Figure~\ref{fig:ifmr} (c) shows the 3D distance of each WD from its cluster compared with its cooling age for the WDs with valid $t_\text{cool}$ estimates.
Here, we ignored the WDs with distances greater than 400 pc due to the unreliability of the individual \textit{Gaia} parallaxes and the \texttt{old} WDs.
The figure shows that most of the WDs are nearer to the cluster than an escapee moving at 1 pc Myr$^{-1}$. The WD positions are similar to what is expected based on the simulations (Figure~\ref{fig:sim_t_cool_distance}).
This suggests that most of the WDs were formed within the cluster. Only 5 tail-WDs (out of the 29 non-\texttt{old} tail-WDs) formed in situ, while the rest likely escaped the cluster after their formation. 

\subsection{Comparison with literature}

Figure~\ref{fig:comparison_with_literature} shows the comparison of the WD parameters derived in this work with values from the literature. The masses ($m_\text{WD, Gaia CMD}$) and cooling ages ($t_\text{cool, Gaia CMD}$) derived from the \textit{Gaia} CMDs are generally consistent with those reported by \citet{Gentile2021MNRAS.508.3877G} and the photometric masses from \citetalias{Miller2026ApJ...996...69M}, which is expected given the similar methodologies and datasets used. In contrast, comparison with the spectroscopic masses from \citetalias{Miller2026ApJ...996...69M}, derived from Balmer-line fitting, shows larger deviations for several targets relative to both their photometric estimates and ours. These differences likely arise because photometric samples are more affected by contamination from non-DA WDs, unresolved binaries \citep{2023MNRAS.526.5800S}, spurious members, and less well-characterized clusters, whereas the spectroscopic sample includes only single DA WDs.

We also compared $m_\text{WD, Gaia CMD}$ with the masses reported by \citet{Prisegen2023A&A...678A..20P}, derived using \textit{Gaia}, \textit{GALEX}, SDSS, and Pan-STARRS photometry. Significant deviations are found for targets with masses $\gtrsim$0.6 \Msun, where $m_\text{WD, Gaia CMD}$ is systematically lower. This discrepancy is likely due to different methodologies, the adopted cluster priors and issues in the photometric zero point calibration \citep{Maiz2007ASPC..364..227M}. In addition, while current atmospheric models reproduce \textit{Gaia} optical colours well, combining \textit{Gaia} with other photometric datasets can introduce additional systematic uncertainties related to survey calibration. Moreover, ultraviolet and infrared photometry can introduce further uncertainties because model predictions outside the optical wavelength range are more sensitive to atmospheric composition, including the H/He ratio and the presence of metals \citep{Gentile2021MNRAS.508.3877G}.

The effective temperatures derived from the SED fits ($T_\text{eff, SED}$) are generally consistent with those from \citet{Gentile2021MNRAS.508.3877G} at lower temperatures ($T_{\rm eff}\lesssim15,000$ K), but show larger deviations at higher temperatures ($T_{\rm eff}\gtrsim20,000$ K), as shown in Figure~\ref{fig:comparison_with_literature} (d). Similarly, the $T_\text{eff, SED}$ values are systematically lower than the spectroscopic temperatures shown in Figure~\ref{fig:comparison_with_literature} (e), with the offsets more clearly visible in the temperature-difference plots in Figure~\ref{fig:comparison_with_literature} (f). Comparable discrepancies of $\approx5$\% were also reported by \citet[][their Figure~14]{Gentile2021MNRAS.508.3877G}, where they were attributed to issues in the \textit{Gaia} colour calibration. At higher temperatures ($T_{\rm eff}>20,000$ K), the deviations are likely amplified because \textit{Gaia} optical colours become progressively less sensitive to temperature as the peak emission shifts towards shorter wavelengths beyond the \textit{Gaia} passbands and the sampled SED \citep[see Figure~\ref{fig:comparison_with_literature} g and ][]{Jadhav2025JApA...46...20J}.

The SED-based radii are similar to the spectroscopic radii from \citetalias{Miller2026ApJ...996...69M}. 
The progenitor masses ($m_\text{progenitor, Gaia CMD}$) were found to be comparable to the spectroscopic estimates from \citetalias{Miller2026ApJ...996...69M}.  
The $m_\text{progenitor, Gaia CMD}$ estimates are also similar to the lower-mass photometric estimates from \citet{Prisegen2023A&A...678A..20P}. However, larger differences appear in the high-mass regime, mainly due to the larger differences in the inferred WD masses (Figure~\ref{fig:comparison_with_literature} b) and the differences in cluster parameter priors.

\subsection{WD initial--final mass relation}

Figure~\ref{fig:ifmr} (b) shows the initial and final mass of WDs using photometric estimations. As expected, the \texttt{CCS} WDs follow the \citetalias{Miller2026ApJ...996...69M} IFMR. In contrast, the \texttt{low\_mass} WDs lie well below the IFMR, and the \texttt{high\_mass} WDs remain above the IFMR, indicative of their likely binary origin. This analysis additionally identifies 47 \texttt{CCS} WDs that may further constrain the IFMR and were not included in the \citetalias{Miller2026ApJ...996...69M} sample. Particularly, \citetalias{Miller2026ApJ...996...69M} noted the absence of WDs in the $m_\text{progenitor}$ range of 2 to 2.7 \Msun. The piecewise linear \citetalias{Miller2026ApJ...996...69M} IFMR of lower-mass WDs ($\leq$0.7 \Msun) does not align with the higher-mass WDs ($\geq$0.9 \Msun). Thus, the additional candidates within this mass range will help refine the corresponding IFMR.

Figure~\ref{fig:cmd_based_params} shows the \textit{Gaia}, PanSTARRS, and SDSS CMD-based WD parameters obtained for the \texttt{CCS} WDs. The large differences in the mass estimates from the three CMDs highlight the need for spectroscopic follow-up and better calibration of the WD models and the photometry.

\begin{figure}
    \centering
    \includegraphics[height=\dimexpr\textheight-277.0pt\relax]{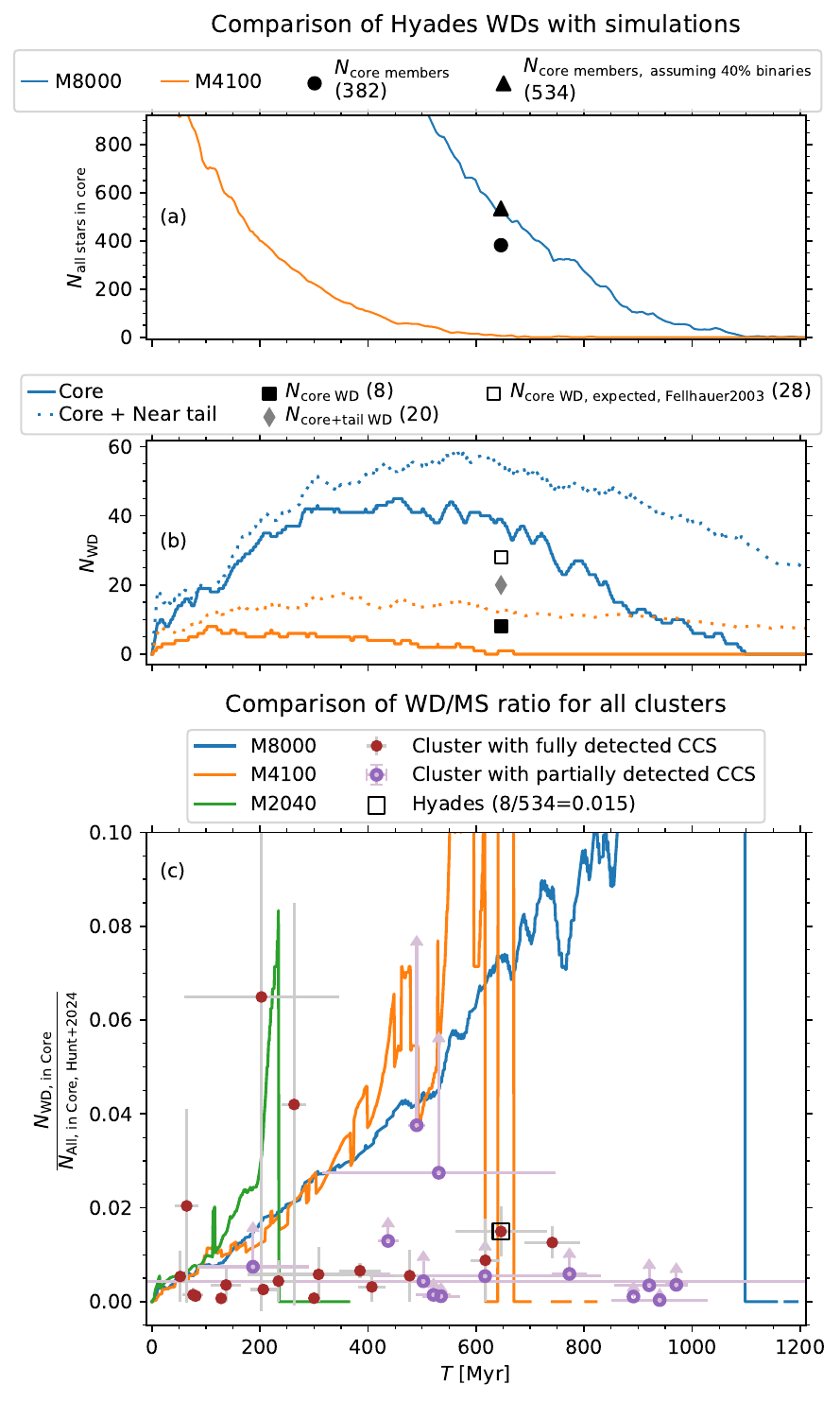}
    \caption{Demonstration of the missing WD problem. 
    (a) Evolution of the number of stars in the cluster core for models M8000 (blue) and M4100 (orange). The number of observed Hyades members (circle; from \citealt{Lodieu2019A&A...623A..35L}) and the number corrected for unresolved binaries (triangle) are also shown.
    (b) Evolution of the number of WDs in the cluster core (solid lines) and core plus tidal-tails (dotted lines) for the models M8000 and M4100. The number of observed WDs in the Hyades core (filled square), core+tail (filled grey diamond) and theoretically expected number of Hyades WDs from \citet{Fellhauer2003ApJ...595L..53F} (hollow square) are indicated.
    (c) Ratio of WDs to all non-degenerate stars in the core region of the $N$-body models (M8000 in blue; M4100 in orange; M2040 in green) and observed cluster cores (brown dots for clusters with all CCS brighter than 21 Gmag and purple circles for clusters with at least some CCS fainter than 21 Gmag). $N_\text{All, in Core, Hunt+2024}$ is the number of all core members in the \citet{Hunt2024A&A...686A..42H} catalogue. The Hyades are highlighted (black square).
    A 40\% binary fraction was assumed for all clusters, and the \texttt{old} WDs were removed.
    Note that $N_\text{All, in Core, Hunt+2024}$ is likely an underestimate due to faint star incompleteness, and ${N_\text{WD, in Core}}$ has various biases as mentioned in Section~\ref{sec:discussion}.
    The models used in this figure are averages of two, four, and eight random realisations of the M8000, M4100, and M2040 models, respectively.
    }
    \label{fig:Hyades_missing_WDs}
\end{figure}

\subsection{Expected number of WDs} \label{sec:expecter_number_of_wd}

A deficit of WDs has been reported in the Hyades and several OCs \citep{Tinsley1974PASP...86..554T, Kalirai2001AJ....122.3239K}, while some clusters show numbers consistent with expectations \citep{von1998AJ....115.1536V, Williams2018ApJ...867...62W}. 
Our simulations (similar to the previous expected WD count from \citealt{Fellhauer2003ApJ...595L..53F}) also indicate a WD deficit in Hyades. Figure~\ref{fig:Hyades_missing_WDs} (a) and (b) show the comparison of Hyades WDs with simulations. The number of all core members in the M8000 model matches with Hyades (accounting for unresolved binaries at 40\% binary fraction; \citealt{Torres2026arXiv260300229T}). 
Figure~\ref{fig:combo_Melotte_25} (g) shows that the whole CCS is brighter than 16 Gmag, thus all Hyades core-WDs should have good \textit{Gaia} astrometry.
Ignoring the \texttt{old} WDs (see Figure~\ref{fig:combo_Melotte_25} e), there are 8 WDs in the Hyades core and 12 in the tidal tails. In contrast, the M8000 simulation hosts $\approx$40 WDs in the core and 15 WDs in the near-tail at the Hyades age.

Figure~\ref{fig:Hyades_missing_WDs} (c) shows that the fractional population of WDs is consistent across the $N$-body models with initial masses of 2000 to 8000 \Msun. The WD fraction reaches $\gtrsim$10\% near dissolution.
In comparison, the majority of observed clusters with CCS brighter than 21 Gmag fall below the expected WD fraction, indicating the WD deficit.
Some of the deficit can be compensated for by the WDs present in binaries \citep{Williams2004ApJ...601.1067w, Jadhav2024A&A...688A.152J}, WDs which are too faint for \textit{Gaia} (e.g., as seen in NGC 2682 and all clusters with partially detected CCS), and WDs misclassified as non-members due to unreliable \textit{Gaia} astrometry. 
Similarly, our simulations assume zero kick velocity during the WD birth, which leads to overestimation of the expected WD numbers. 
A more rigorous examination of the incompleteness of WD surveys and modelling assumptions (e.g., kick velocities and WD physics in simulations) is needed to resolve this discrepancy.

\subsection{Non-canonical evolution of the WDs}

As given in Section~\ref{sec:cmd_diagnostics}, there are 53 \texttt{low\_mass} and 10 \texttt{high\_mass} WDs. These WDs are either false members or products of binary interactions. WDs from blue stragglers can be more massive than the CCS, while WDs formed by the donors in case A or case B mass transfer can have a mass lower than the CCS mass.

The frequency of stellar interactions in clusters is still theoretically uncertain, and the discussion in Section~\ref{sec:expecter_number_of_wd} highlights the limitations of current simulations. Nevertheless, alternative empirical indicators can be used to approximate the number of past mass-transfer events. For example, NGC 2682 (M67) hosts 21 blue stragglers, 5 yellow stragglers, 2 red stragglers, and 1 sub-subgiant candidate \citep{Mathieu2025ARA&A..63..467M}. These 29 objects are potential products of binary mass transfer or mergers and are either past or near the end of the H-burning phase. If extrapolated to their eventual WD progeny, this would imply an upper limit of 58 young non-CCS WDs in NGC 2682.

However, this estimate is clearly an upper bound. Some interactions may result in mergers, some systems may form through case C mass transfer (which does not significantly alter the WD mass of the donor), and a significant fraction of WDs may escape the cluster after formation. In addition, earlier interaction products will contribute to the population of older, cooled non-CCS WDs. Taking both effects into account, the true number of non-CCS WDs is expected to be significantly lower than 62 in NGC 2682. Observationally, 23 WD candidates have been identified as members of NGC 2682 \citep{Williams2018ApJ...867...62W, Canton2021AJ....161..169C}\footnote{Note that our WD sample includes only 4 of these 23 objects due to \textit{Gaia} detectability limits and selection criteria.}. Of these, two are classified as high-mass and two as low-mass relative to the CCS.

In the broader sample, 63 out of 235 WDs are identified as potential binary-origin candidates, while in NGC 2682, the corresponding fraction is 4 out of 23. These numbers indicate that binary-origin WDs constitute a non-negligible fraction of the WD population. A dedicated follow-up study of these 63 candidates is therefore required to determine whether they are contaminants or genuine products of binary interactions, and to better constrain the contribution of binary evolution to the WD population.

\subsection{Caveats and the scope for improvements}

Apart from false membership, several additional caveats apply to the present analysis. The parameters derived in this study depend sensitively on the adopted astrometric priors. In particular, individual distance estimates remain the least precise quantities and may propagate significant uncertainties into the inferred stellar parameters. Spectroscopic observations could lessen the impact of distance uncertainties. Furthermore, the ages of clusters with poorly defined MS turnoffs are intrinsically uncertain and must be determined more robustly to reliably constrain progenitor properties.

Similarly, SED-based analyses are strongly dependent on the number of available photometric bands and on the assumed extinction. The CMD-based analysis and the adopted IFMR relations are valid only for DA-type WDs; non-DA WDs or the presence of magnetic fields can introduce systematic deviations from these models \citepalias{Miller2026ApJ...996...69M}. In addition, WD cooling tracks exhibit discontinuities due to discrete sampling of the WD masses and the core compositions (e.g., He, CO, ONe). This can lead to incorrect interpolation and consequently bias the derived stellar parameters.

\section{Conclusions and summary} \label{sec:conclusions}

We analysed the WD population in OCs and their tidal tails using both $N$-body simulations and searches in \textit{Gaia} data. Based on the simulations and observational analysis, we derive the following conclusions.

\begin{itemize}
    \item We identify 235 OC–WD pairs in 80 clusters by combining cluster and tidal-tail membership catalogues based on the \textit{Gaia} DR3 data. We derive WD masses, cooling ages, radii, and effective temperatures using CMD- and SED-based analyses.

    \item More than 28\% of the identified OC–WD pairs are spurious ($t_\text{cool}>age_\text{cluster}$), with contamination being significantly higher in the tidal-tails ($>48$\%) than in the cluster cores ($>13$\%). Unfortunately, the WD sample is not big enough to comment on the purity of individual tail and cluster membership methods.
    \item The WD contamination in the core indicates that most \textit{Gaia} DR3 based cluster catalogues have $\gtrsim$13\% field-star contamination.
    Improved techniques are required to robustly detect clusters and tidal tails in the presence of contaminating comoving stars and increase the sample of known OC-WD pairs.

    \item Previous detections of the Pleiades tidal-tails suffer from significant contamination by very \texttt{old} WDs, reinforcing earlier suspicions that the Pleiades tidal-tails reported in \citet{Risbud2025AA...694A.258R} are not genuine.
    
    \item According to the simulations, the fraction of WDs within the core increases with time, reaching $\gtrsim$10\% near the cluster dissolution. 
    \item In contrast, the observed fraction of core-WDs is significantly lower for most clusters, consistent with the long-standing `WD deficit' problem.
    This open problem needs to be solved using improved simulations (e.g., improved kick velocity prescriptions) and understanding of the observational biases (deeper observations covering the complete CCS, improved membership determination, and more accurate cluster mass and radius estimates).
    
    \item The simulations suggest that WDs originate from three main populations: WDs retained in the cluster core, WDs escaping into tidal tails on orbits similar to that of the parent cluster, and fast-moving escapees on significantly different Galactic orbits. The near-tail regions generally contain fewer WDs than the cluster core, with most of these objects being dynamically escaped core-born WDs.
    \item Despite the high contamination rate, the majority ($\approx$83\%) of the tail-WDs analysed here are consistent with having formed in their host clusters and subsequently escaped through dynamical evolution. Thus, the tail-WDs are good candidates for IMFR analysis.

    \item We also provide a method to calculate the reference point (equivalent to the cluster centre) of a dissolved cluster (Section~\ref{sed:center_of_dissolved_cluster}).
    \item We identify 63 binary origin candidates and 47 new IFMR candidates. These objects require precise spectroscopic measurements for detailed follow-up studies.
\end{itemize}

The WD population provides important constraints for $N$-body simulations. Some improvements involve the direct modelling of WD physics, including kick velocity prescriptions and WD formation and evolution processes. At the same time, the observed WD dynamics can be used to test the assumed initial cluster conditions, such as primordial mass segregation and the initial binary population.

Observationally, future progress will require systematic high-resolution spectroscopic follow-up of WDs with $t_\text{cool}<age_\text{cluster}$, enabling precise mass determinations and radial-velocity (with gravitational redshift corrections) measurements to kinematically validate cluster membership. In parallel, independent constraints on cluster extinction, metallicity, and distances (derived from spectroscopy and higher-precision parallaxes) are necessary to reduce degeneracies in isochrone fitting. Improved membership determination will require more advanced methods applied to \textit{Gaia} DR4 data, together with the inclusion of fainter stars from the Legacy Survey of Space and Time (LSST).
Together, these efforts will enable more reliable identification of genuine cluster WDs and provide a firmer empirical foundation for studies of cluster dynamical evolution and WD progenitors.

\section*{Data availability}

The full versions of Table~\ref{tab:cluster_params} and Table~\ref{tab:wd_params} with additional columns are available at the CDS via \url{http://cdsweb.u-strasbg.fr/cgi-bin/qcat?J/A+A/}.

\begin{acknowledgements}
    We thank the anonymous referee for timely and constructive comments. 
    We acknowledge support through the DAAD-Eastern Europe Exchange grant at Bonn University and corresponding support from Charles University. We also acknowledge support through grant 26-21774S from the Czech Grant Agency.
    VJ thanks the Alexander von Humboldt Foundation for its support.
    DRM acknowledges support from the Natural Sciences and Engineering Research Council of Canada Discovery grants Nos. DG-RGPIN-2022-03051 and DG-RGPIN-2023-04486.
    FD appreciates support due to the National Science and Technology Council, the Ministry of Education (Higher Education Sprout Project NTU-114L104022-1), grant numbers 114-2124-M-002-003 and 115-2124-M-002-014, and the National Center for Theoretical Sciences of Taiwan.
    The work used the following tools for the analysis: 
    \textsc{Astropy} \citep{astropy:2022}; 
    \textsc{Astroquery} \citep{Ginsburg2019AJ....157...98G};
    \textsc{ClusterTools} \citep{Webb2023JOSS....8.4483W}; 
    \textsc{Galpy} \citep{Bovy2015ApJS..216...29B}; 
    \textsc{NumPy} \citep{2020Natur.585..357Harris}; 
    \textsc{Simbad} \citep{Wenger2000A&AS..143....9W};
    \textsc{SciPy} \citep{2020SciPy-NMeth}; 
    \textsc{topcat} \citep{2005ASPC..347...29TOPCAT}:
    \textsc{VizieR} \citep{Ochsenbein2000A&AS..143...23O}.
    This research has made use of the Unified Cluster Catalogue \citep[\url{https://ucc.ar};][]{Perren2023MNRAS.526.4107P}.
\end{acknowledgements}
\bibliographystyle{aa_url}
\bibliography{references}
\begin{appendix}
\onecolumn
\section{Simulation related information} \label{sec:sim_supplementary}

\subsection{Detailed description of the $N$-body modelling} \label{sec:nbody_description}

The $N$-body simulations were initialised with mass segregated clusters of mass segregation parameter $S$ = 0.5 obtained by a modification of a Plummer model by the recipe due to   \citet{Subr2008MNRAS.385.1673S, Subr2012ascl.soft06007S}. The initial half-mass radius $r_\mathrm{h}$ ranges from 0.17 to 0.31 pc depending on the cluster mass. The initial gas is represented by a Plummer model with the same half-mass radius and twice the mass as the stellar component. The gas mass was kept constant till 0.6 Myr, after which it exponentially decayed due to stellar feedback on a time-scale of $r_\mathrm{h}/(10\ \text{ km s}^{-1}) \approx 0.02$ to $0.03\ \mathrm{Myr}$ \citep{Kroupa2001MNRAS.321..699K}. The clusters are placed on circular orbits around the Galaxy at a radius of 8 kpc. The Galactic potential was modelled with three components based on \citet{Allen1991RMxAA..22..255A}.

All stars were initialised as binaries. While no higher order systems (e.g., triples) were initialised, dynamically formed higher order systems were treated self-consistently following the regularisation methods included in the code \textsc{nbody6} \citep{Aarseth1974A&A....37..183A}. The binary population was initialised differently for binaries with primary mass above and below 5 \Msun. For lower mass primaries, the mass ratio, period, and eccentricity distributions were taken from \citet{Kroupa1995MNRAS.277.1491K} and processed with the pre-MS eigenevolution \citep{Kroupa1995MNRAS.277.1507K}. For higher mass primaries, the orbital parameter distributions were taken from \citet[based on \citealt{Sana2012Sci...337..444S, Kobulnicky2014ApJS..213...34K}]{Belloni2017MNRAS.471.2812B}.
The binary fraction decreases rapidly from 100\% to $\approx$40--60\% within the first million years, depending on the initial cluster properties. And the survival rates are lower for binaries with large periods ($>10^4$ day) and low mass primaries (primary mass $<$ 0.5 \Msun; \citealt{Wu2026A&A...707A.304W}).
The single and binary star evolution (including mergers, common envelope evolution, wind mass accretion, Roche lobe overflow, tidal evolution, and WD cooling) are performed using the rapid analytical algorithms of \citet{Hurley2000MNRAS.315..543H} and \citet{Hurley2002MNRAS.329..897H}.

\subsection{Definitions}

The quantities used in this text are defined as follows (see \citealt{Jadhav2025A&A...704A..50J} for more details):

\begin{itemize}
    \item $distance\_along\_orbit$: This is a curved distance measured along the Galactic orbit, from the cluster centre to the point closest to the star on the cluster's orbit.
    \item $distance\_from\_orbit$: Closest straight line distance between the source and the cluster's orbit.
    \item $distance\_from\_cluster = \sqrt{distance\_along\_orbit^2 + distance\_from\_orbit^2}$
    \item Similar/different orbit stars: Majority of the tidal-tail stars slowly evaporate from the cluster, thus their orbits are quite similar to the parent cluster. 
    A star is considered to be in a similar orbit if its specific angular momentum ($sL_z$) and specific energy ($sE_{total}$) are within three standard deviations of the overall cluster median values at the given time. Sigma clipping was used while measuring the median and standard deviations.
    \item Cluster dissolution time, $T_\text{dissolve}$: The epoch at which the number of stars within the tidal radius fell below 10 ($N_\text{tidal}<10$).
\end{itemize}

\subsection{Reference point (equivalent to the cluster centre) of a dissolved cluster} \label{sed:center_of_dissolved_cluster}

A cluster centre is typically calculated by finding the centre of mass or density within an $N$-body simulation. 
We use the iterative centre of density algorithm by \citet{Harfst2007NewA...12..357H} for undissolved clusters.
A dissolved cluster, however, lacks an overdensity at the expected centre. Thus, we propose a new method to identify the centre of a dissolved cluster, to be used as a reference point for further calculations (e.g., cluster-centric radius or $distance\_along\_orbit$). The method is rooted in the conservation of energy and angular momentum. The method requires at least one snapshot (ideally more than 10) where the cluster centre can be identified using classical methods (e.g., centre of mass or density algorithms; \citealt{Casertano1985ApJ...298...80C, Harfst2007NewA...12..357H}). The details are as follows:

\begin{enumerate}
    \item Find the cluster centre using classical methods for all available snapshots. This will generally result in wrong centres for poor clusters ($N_\text{tidal}<10$).
    \item Measure the initial specific energy ($sE_\text{total,0}=sE_\text{potential,0}+sE_\text{kinetic,0}$) and specific angular momentum ($sL_\text{z,0}$) using the robust centre estimates of the early and undissolved clusters. Simultaneously, measure the standard deviation in energy and angular momentum ($\sigma_{sE_\text{total,0}},\ \sigma_{sL_\text{z,0}}$). This step can be done with a single snapshot or by combining all snapshots before the cluster dissolves, and the classical centre calculations start becoming incorrect.
    \item A way to test if the centres are correct is to check the time evolution of the cluster-core's $sE_\text{total}(t)$ and $sL_\text{z}(t)$. Any large deviations are likely results of incorrect centre estimation and should not be used while calculating the initial energy and angular momentum.
    \item Find those stars with the most similar energy and angular momentum as the initial setup.
    \begin{equation}
    \begin{split}
        |sE_{\text{total,}i}-sE_{\text{total,}0}|&=\Delta sE_\text{total} \\
        |sL_{\text{z,}i}-sL_{z,0}| &= \Delta sL_{z} \\
        \sqrt{\left(\dfrac{\Delta sE_\text{total}}{\sigma_{sE_\text{total,0}}}\right)^2 
        + \left(\dfrac{\Delta sL_z}{\sigma_{sL_{z,0}}}\right)^2} &<0.5
    \end{split}
    \end{equation}
    \item The median position and velocity of these stars represent the new reference point (i.e., the new cluster centre). Extreme outliers should be removed before determining the centre.
    \item Typically, the reference point lies in the region of lowest stellar density in the dissolved cluster compared to the neighbouring tail region. And the stellar distribution should be symmetric with respect to this point.
\end{enumerate}

We could correctly identify the reference point of M8000 up to $\approx$5 Gyr ($>$15 Galactic orbits) past the cluster dissolution. After this long time, the stars with similar energy span thousands of parsecs, and thus their median or central position becomes an unreliable reference point. 

\section{Supplementary table and figures} \label{sec:supplementary_figures}

\begin{table}[h]
    \caption{Cluster parameters used in this work. The extended version of this table with column descriptions, references, and additional parameters for all 80 clusters is available on CDS.}
    \centering

\begin{tabular}{lrr lll l}
\toprule
Cluster & $N_\text{WD,\, core}$ & $N_\text{WD,\, tail}$ & age [Myr] & A$_V$ [mag] & distance [pc] & [Fe/H] \\
\midrule
ABDMG & 0 & 1 & $133_{-20}^{+15}$ & $0.00_{-0.00}^{+0.10}$ & $19_{-3}^{+3}$ & $0.00\pm0.20$ \\
ASCC\_101 & 0 & 1 & $494_{-100}^{+77}$ & $0.05_{-0.02}^{+0.03}$ & $376_{-6}^{+6}$ & $0.00\pm0.20$ \\
ASCC\_113 & 0 & 1 & $282_{-9}^{+24}$ & $0.13_{--0.07}^{+0.07}$ & $528_{-5}^{+5}$ & $0.05\pm0.20$ \\
ASCC\_47 & 0 & 1 & $90_{-20}^{+20}$ & $0.39_{-0.10}^{+0.10}$ & $780_{-80}^{+120}$ & $0.00\pm0.20$ \\
ASCC\_99 & 1 & 0 & $308_{-105}^{+159}$ & $0.76_{-0.09}^{+0.09}$ & $294_{-0}^{+0}$ & $-0.04\pm0.08$ \\
... & ... & ... & ... & ... & ... & ... \\
\bottomrule
\end{tabular}
\tablefoot{The list of all columns: Cluster, \texttt{NWD\_total, NWD\_core, NWD\_tail, RAdeg, DEdeg, pmRA, pmDE, RV, dist, b\_dist, B\_dist, RefDist, OCType, age, b\_age, B\_age, RefAge, Fe\_H, e\_Fe\_H, RefFe\_H, AV, b\_AV, B\_AV, RefAV, CCSLimitApp, CCSLimitAbs}.
References for the cluster parameters: \citet{Netopil2016A&A...585A.150N, Cummings2018ApJ...866...21C, Bossini2019A&A...623A.108B, Cantat2020A&A...640A...1C, Dias2021MNRAS.504..356D, Hunt2024A&A...686A..42H, Miller2026ApJ...996...69M}.}
    \label{tab:cluster_params}
\end{table}

\begin{table}[h]
    \centering
    \caption{WD parameters from the CMD and SED fits. The extended version of the table with column descriptions, astrometry, photometry, and comments of all 235 WDs is available on CDS.}
\begin{tabular}{rlllllrl}
\toprule
GaiaDR3 & Cluster & $T_\text{eff, sed}$ [K] & $r_\text{sed}$ [km] & $t_\text{cool}$ [Myr] & $m_\text{WD, Gaia CMD}$ [M$_{\odot}$] & In tail & CMD position \\
\midrule
3251244858154433536 & ABDMG & $38000 \pm 1952$ & $3626 \pm 6$ & $76_{-63}^{+35}$ & $1.23_{-0.07}^{+0.06}$ & True & ccs \\
2098988107112755712 & ASCC\_101 &  &  & $253_{-25}^{+23}$ & $0.64_{-0.07}^{+0.08}$ & True & low\_mass \\
1871306874227157376 & ASCC\_113 & $32000 \pm 2975$ & $5931 \pm 57$ &  &  & True & young \\
5529347562661865088 & ASCC\_47 &  &  &  &  & True & young \\
4099156472803484416 & ASCC\_99 &  &  & $150_{-48}^{+53}$ & $0.57_{-0.08}^{+0.11}$ & False & low\_mass \\
... & ... & ... & ... & ... & ... & ... & ...\\
\bottomrule
\end{tabular}
\tablefoot{The list of all columns: \texttt{GaiaDR3, Cluster, RefID, RAdeg, DEdeg, pmRA, pmDE, rSky, r3D, b\_dist, dist, B\_dist, RefDist, b\_AV, AV, B\_AV, RefAV, spectralType, in\_tail, crowding\_grade, cmd\_position, \_RV, Ref\_RV, b\_MWD\_Gaia, MWD\_Gaia, B\_MWD\_Gaia, b\_MWD\_PS1, MWD\_PS1, B\_MWD\_PS1, b\_MWD\_SDSS, MWD\_SDSS, B\_MWD\_SDSS, b\_Teff\_Gaia, Teff\_Gaia, B\_Teff\_Gaia, b\_Teff\_PS1, Teff\_PS1, B\_Teff\_PS1, b\_Teff\_SDSS, Teff\_SDSS, B\_Teff\_SDSS, b\_T\_cool\_Gaia, T\_cool\_Gaia, B\_T\_cool\_Gaia, b\_T\_cool\_PS1, T\_cool\_PS1, B\_T\_cool\_PS1, b\_T\_cool\_SDSS, T\_cool\_SDSS, B\_T\_cool\_SDSS, b\_Mprog\_Gaia, Mprog\_Gaia, B\_Mprog\_Gaia, Teff\_SED, e\_Teff\_SED, logg\_SED, e\_logg\_SED, Lbol\_SED, e\_Lbol\_SED, Rad\_SED, e\_Rad\_SED, Chi2\_SED, logSF\_SED, Gmag, e\_Gmag, BPmag, e\_BPmag, RPmag, e\_RPmag, F148Wmag, e\_F148Wmag, F154Wmag, e\_F154Wmag, F169Mmag, e\_F169Mmag, F172Mmag, e\_F172Mmag, N219Mmag, e\_N219Mmag, N245Mmag, e\_N245Mmag, N263Mmag, e\_N263Mmag, N279Nmag, e\_N279Nmag, GALEX\_ID, FUVmag, e\_FUVmag, NUVmag, e\_NUVmag, PS1\_objID, PS1\_gmag, e\_PS1\_gmag, PS1\_rmag, e\_PS1\_rmag, PS1\_imag, e\_PS1\_imag, PS1\_zmag, e\_PS1\_zmag, PS1\_ymag, e\_PS1\_ymag, SDSS\_objID, SDSS\_umag, e\_SDSS\_umag, SDSS\_gmag, e\_SDSS\_gmag, SDSS\_rmag, e\_SDSS\_rmag, SDSS\_imag, e\_SDSS\_imag, SDSS\_zmag, e\_SDSS\_zmag, 2MASS\_ID, Jmag, e\_Jmag, Hmag, e\_Hmag, Kmag, e\_Kmag, AllWISE\_ID, W1mag, e\_W1mag, W2mag, e\_W2mag, W3mag, e\_W3mag, W4mag, e\_W4mag, GMag, e\_GMag, BPMag, e\_BPMag, RPMag, e\_RPMag, BPRPMag, e\_BPRPMag}.
References for the WD identification and membership: \citet{Roser2019A&A...621L...2R, Furnkranz2019A&A...624L..11F, Zhang2020ApJ...889...99Z, Oh2020MNRAS.498.1920O, Meingast2021A&A...645A..84M, Boffin2022MNRAS.514.3579B, Bhattacharya2022MNRAS.517.3525B, Prisegen2023A&A...678A..20P, Risbud2025AA...694A.258R, Miller2026ApJ...996...69M, Yan2026arXiv260216550Y, Jerabkova2021AA...647A.137J}.
References for the stellar parameters: \citet{Cummings2018ApJ...866...21C, Bossini2019A&A...623A.108B, Bailer2021AJ....161..147B, Dias2021MNRAS.504..356D, Gentile2021MNRAS.508.3877G,Hunt2024A&A...686A..42H,Miller2026ApJ...996...69M}.}
    \label{tab:wd_params}
\end{table}

\begin{figure}
    \centering
    \includegraphics[width=1\linewidth]{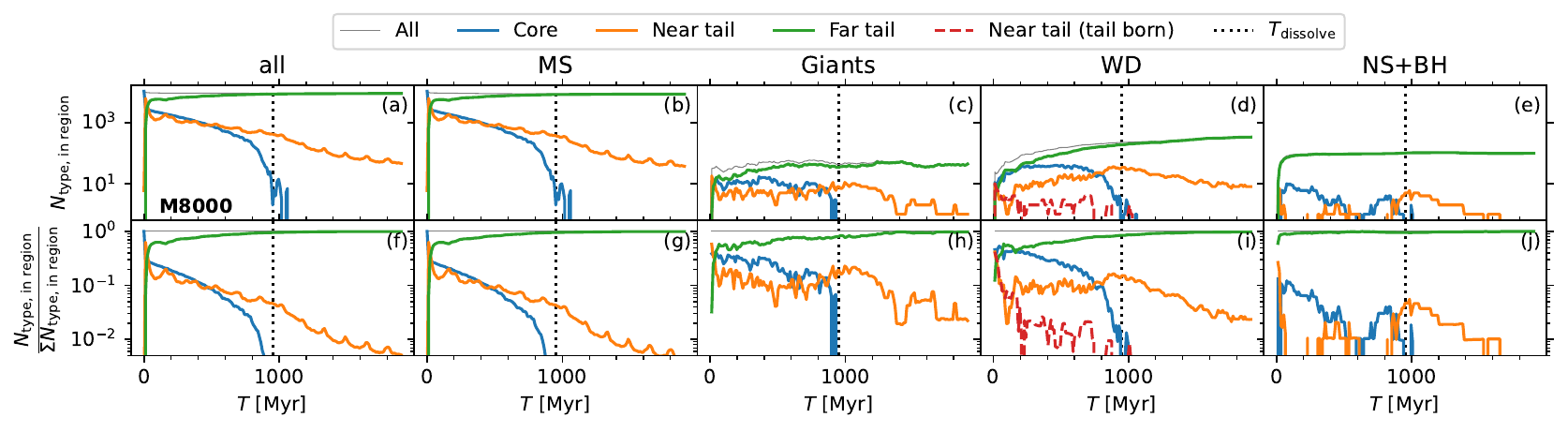}
    \includegraphics[width=1\linewidth]{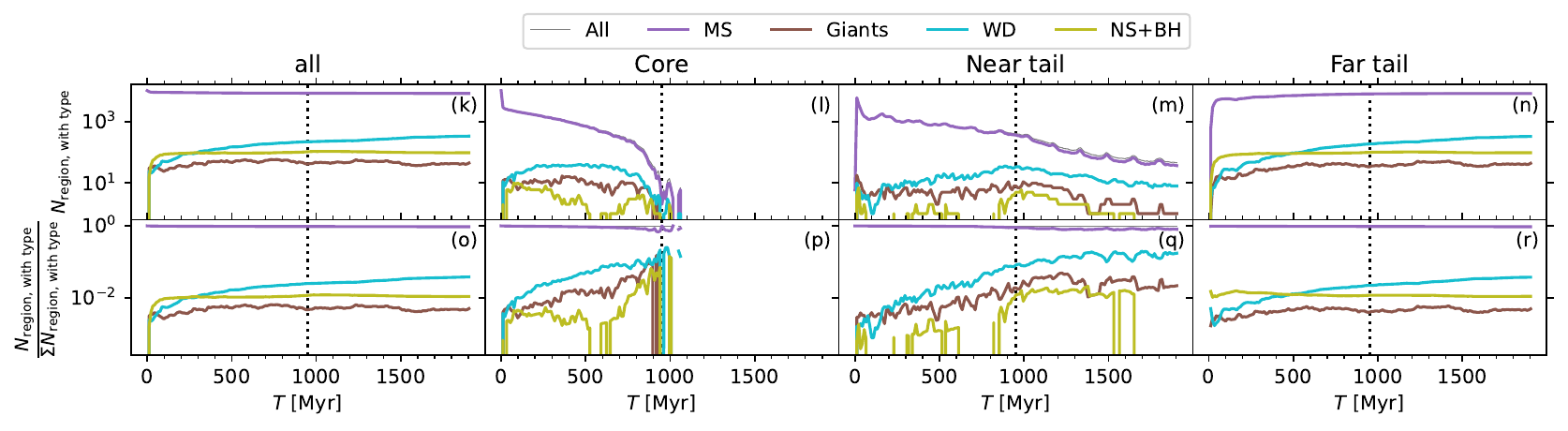}
    \caption{An expanded version of Figure~\ref{fig:sim_population_evolution_small} for one of the realisation of model M8000. 
    (a) Count of all (gray), core (blue), near-tail (orange), and far-tail (green) stars as the cluster evolves. The dissolution time (black dotted line) is shown for reference.
    (b)--(e) Stellar counts for MS, Giants, WDs, and NS+BHs.
    (f) Fraction of stars in the core, near-tail, and far-tail.
    (g)--(j) Fractional populations for MS, Giants, WDs, and NS+BHs.
    (k) Count of all (gray), MS (purple), giants (brown), WD (cyan), and NS+BH (olive) stars as the cluster evolves.
    (l)--(n) Stellar counts in the core, near-tail, and far-tail regions.
    (o) Fraction of stars in the above-mentioned evolutionary stages at any given time.
    (p)--(r) Fractional populations in the core, near-tail, and far-tail regions.
    The tail-born WDs in the near-tail are highlighted (red dashed curve) in (d) and (i).}
    \label{fig:bse_type_M8000}
\end{figure}
\begin{figure}
    \centering
    \includegraphics[width=1\linewidth]{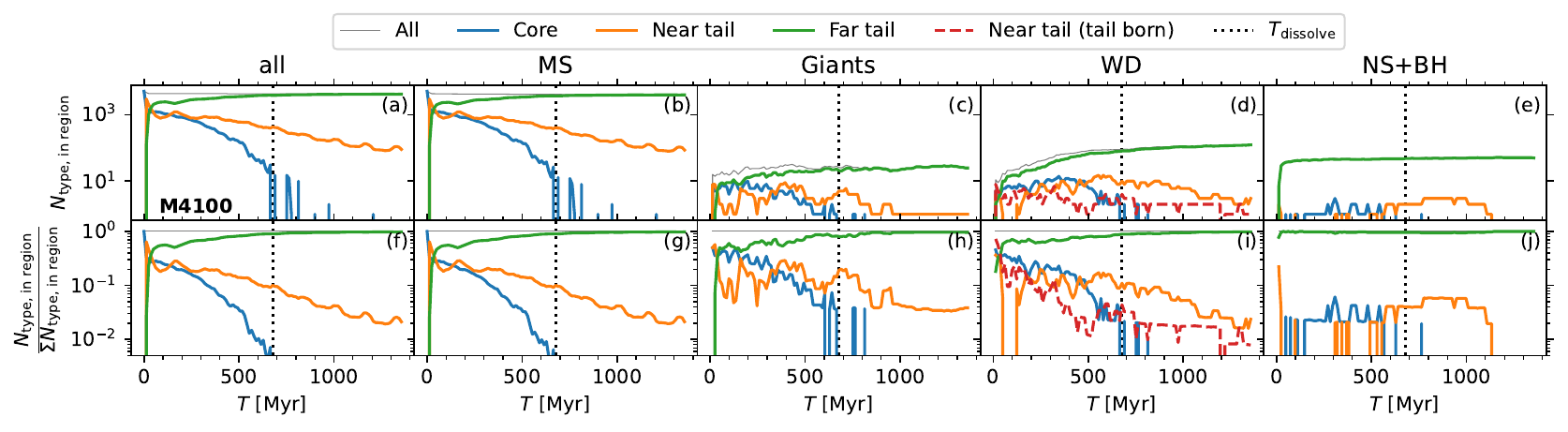}
    \includegraphics[width=1\linewidth]{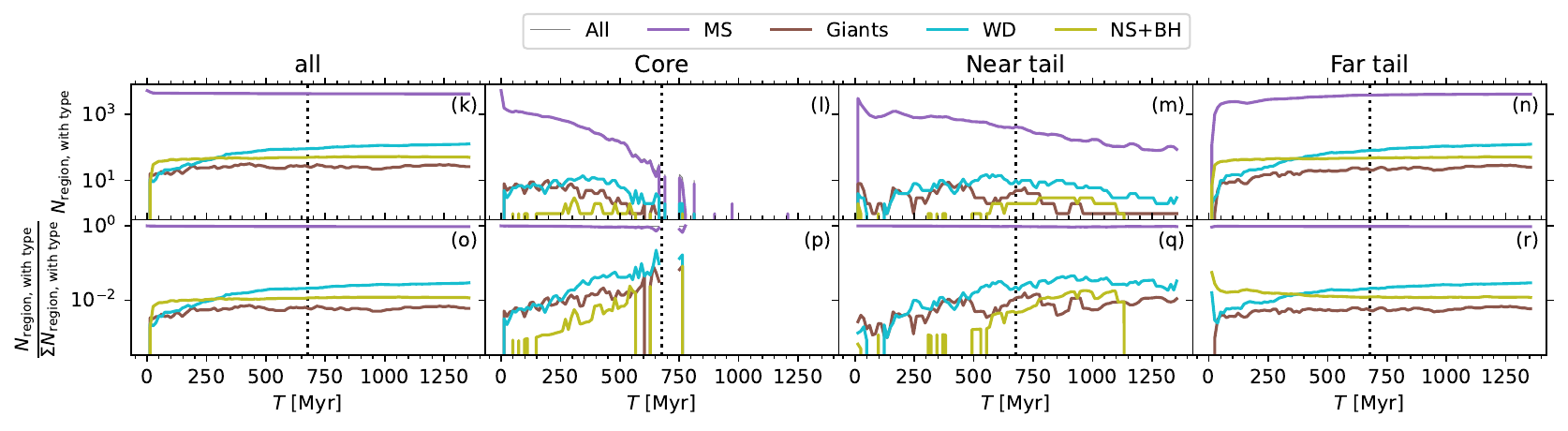}
    \caption{Same as Figure~\ref{fig:bse_type_M8000} for one of the realisation of model M4100.}
    \label{fig:bse_type_M4100}
\end{figure}

\begin{figure}
    \centering
    \includegraphics[width=0.4\linewidth]{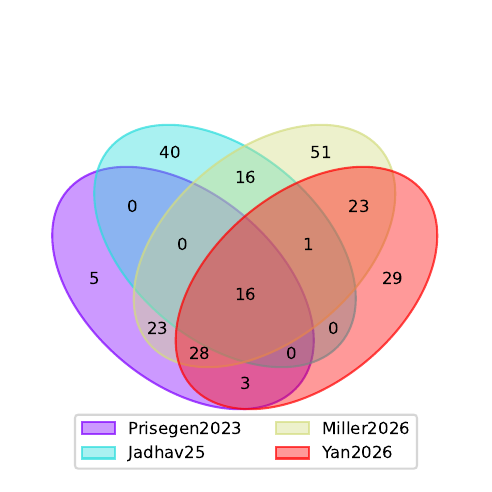}
    \caption{Sample selection and the overlap in the parent catalogs.}
    \label{fig:sample_selection}
\end{figure}

\begin{figure}
    \centering
    \includegraphics[width=0.99\linewidth]{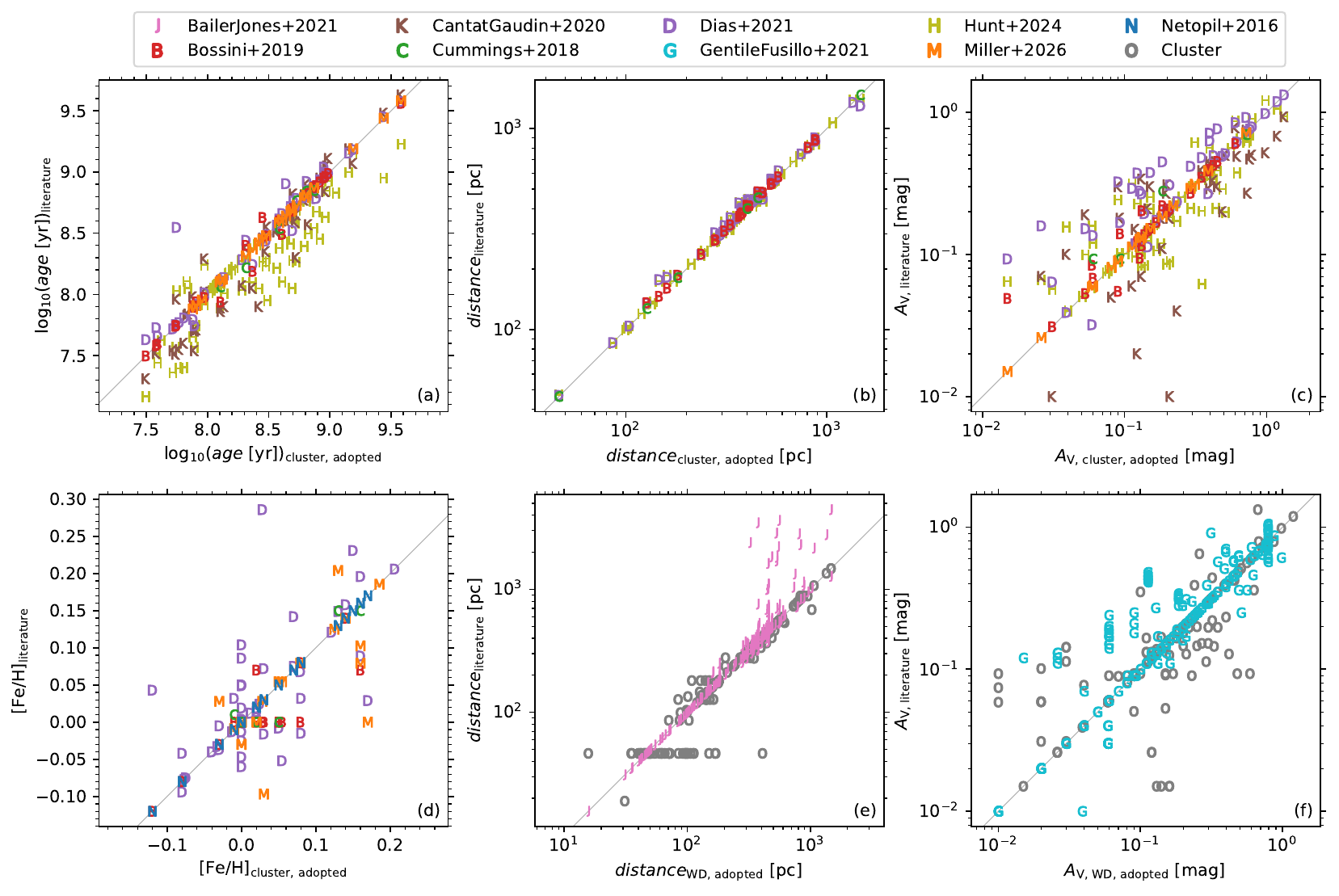}
    \caption{Comparison of parameters from the literature with those adopted in this work. 
    Panels (a)-(d) present comparisons between the adopted cluster parameters and literature values, while panels (e) and (f) compare the adopted individual stellar parameters (distance and extinction) with those of their associated clusters.
    Each reference is shown by a unique alphabet marker. As an example, for the cluster Ruprecht 147, we adopt $\log_{10}(age)$ of 9.44 from \citetalias{Miller2026ApJ...996...69M}.
    Other published age estimates for this cluster include $\log_{10}(age)$ = 9.45 \citep{Dias2021MNRAS.504..356D}, 9.48 \citep{Cantat2020A&A...640A...1C} and 8.95 \citep{Hunt2024A&A...686A..42H}, which are not used in this analysis.
    }
    \label{fig:cluster_parameters}
\end{figure}

\begin{figure}
    \centering
    \includegraphics[width=14cm]{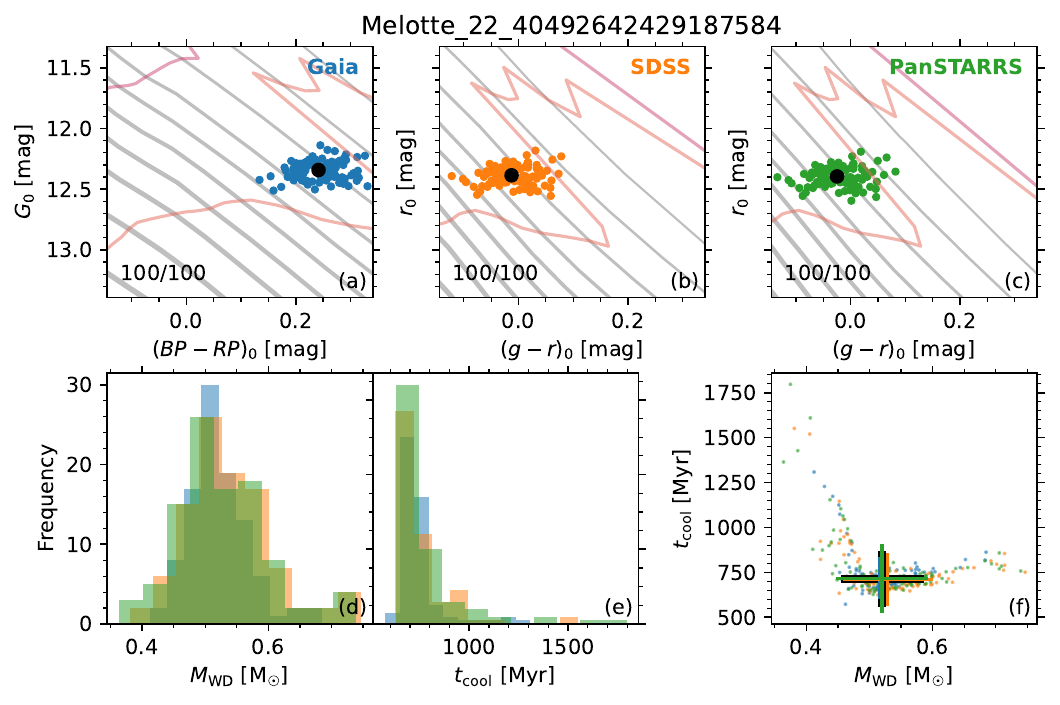}
    \caption{Diagrams showing the measurement of mass and cooling age of a WD (Gaia DR3 40492642429187584). The absolute \textit{Gaia} (blue), SDSS (orange), and PanSTARRS (green) CMDs are shown in (a)-(c) along with the corresponding cooling models. The noisy iterations are given as coloured dots while the observed value is shown in black. The number of valid solutions and total iterations are given in the bottom left corner. (d) Distribution of masses from the three CMDs. (e) Distribution of cooling ages. (f) Distribution of masses and cooling ages (as dots) and their medians with the standard deviation (crosses).}
    \label{fig:wd_cmd}
\end{figure}

\begin{figure}
    \centering
    \includegraphics[width=1\linewidth]{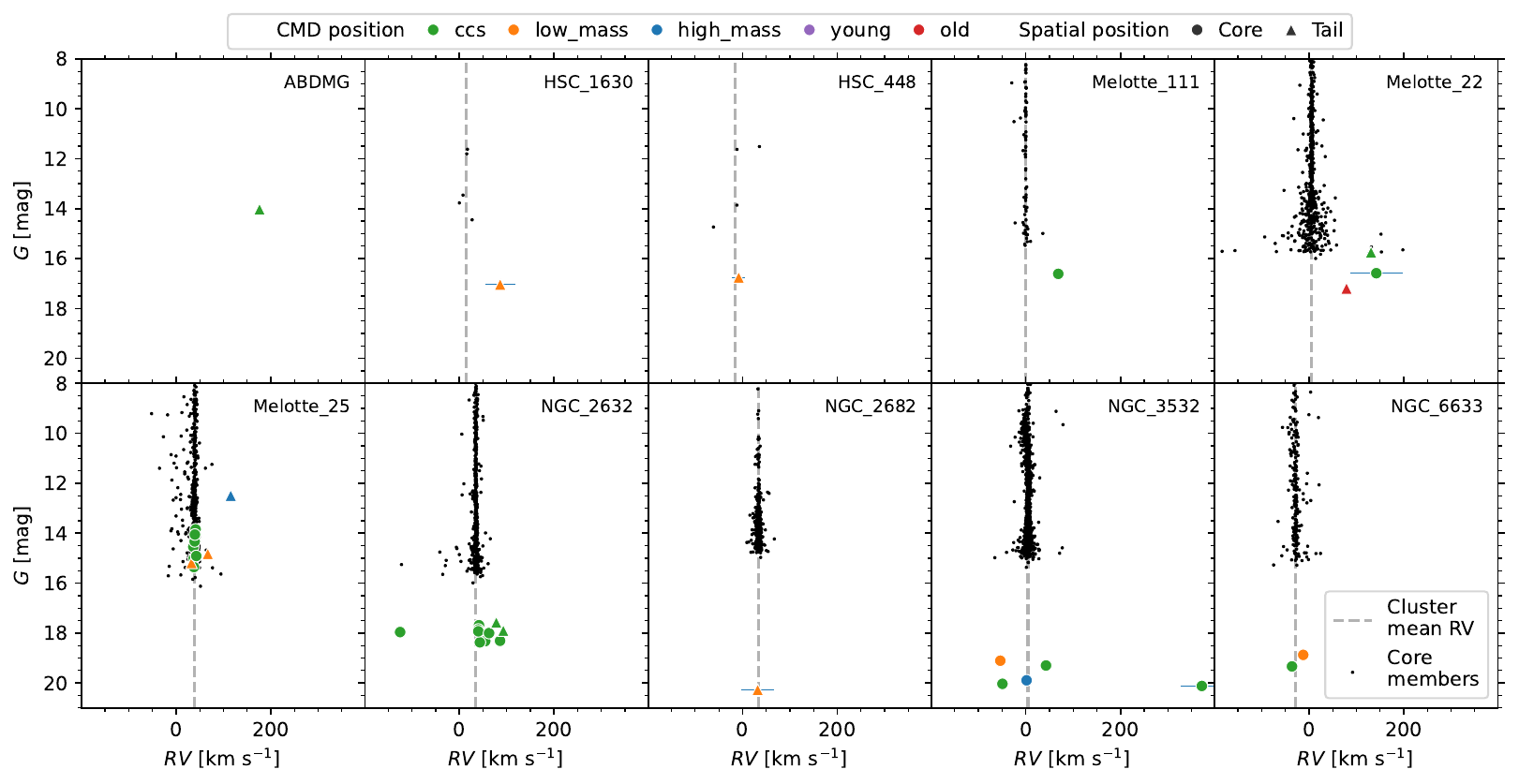}
    \caption{Variation of literature WD radial velocities with apparent \textit{Gaia} G-band magnitude. The WDs are shown in a scheme similar to Figure~\ref{fig:combo_Melotte_25} (e). The \textit{Gaia} DR3 radial velocities of the core members (black) and their mean (grey dashed line) are shown for reference. Note that some of the WD radial velocities are corrected for gravitational redshift while some are not (see Section~\ref{sec:robustness_of_wd_membership}).
    }
    \label{fig:WD_RVs}
\end{figure}

\begin{figure}
    \centering
    \includegraphics[width=1\linewidth]{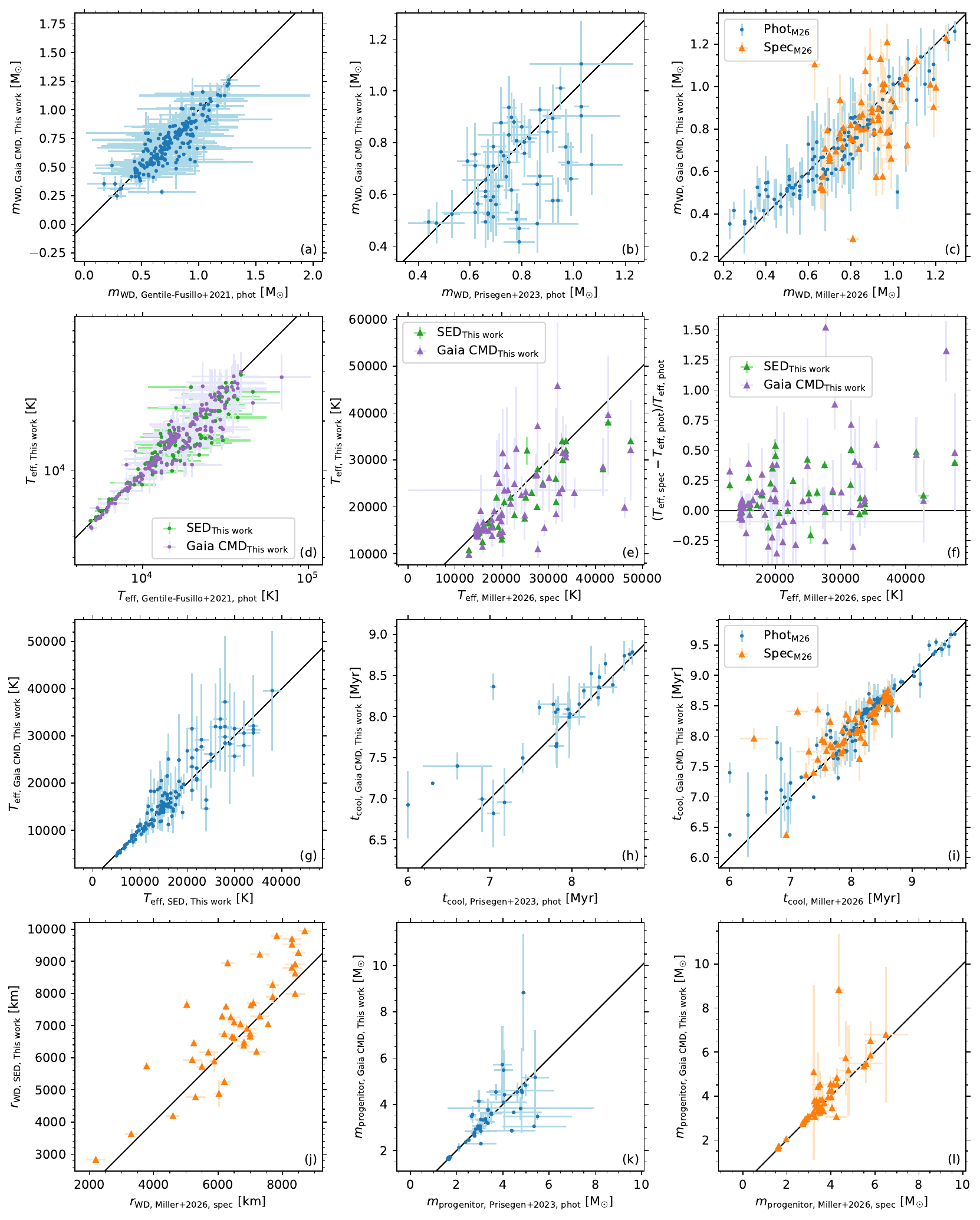}
    \caption{Comparison of WD parameters (mass, temperature, cooling age, radius, and progenitor mass) obtained in this work with literature. The line of equality is shown in grey. Refer to their individual legends for panels with two simultaneous comparisons. We use the H-atmosphere parameters from \citet{Gentile2021MNRAS.508.3877G} for the comparison in panels (a) and (d).
    }
    \label{fig:comparison_with_literature}
\end{figure}

\begin{figure*}
    \centering
    \includegraphics[width=1\linewidth]{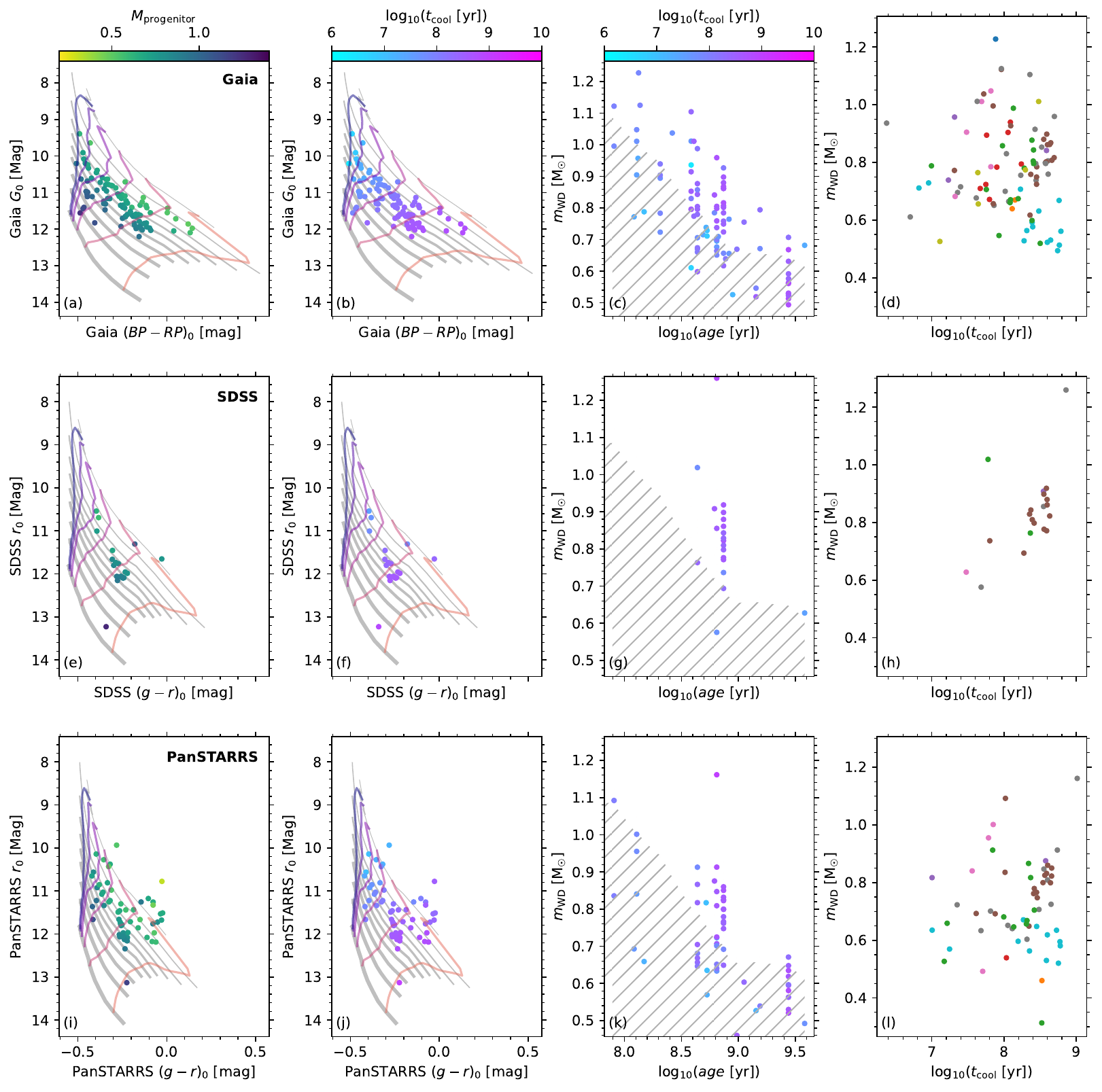}
    \caption{WD mass and cooling age estimates using \textit{Gaia} (first row), SDSS (second row), and PanSTARRS (third row) CMDs. The shaded region indicates the typically forbidden phase space for the WDs based on the cluster age. The last column shows every cluster in a separate colour. Only the \texttt{CCS} WDs are shown.}
    \label{fig:cmd_based_params}
\end{figure*}

\clearpage
\newpage
\section{arXiv only figures} \label{sec:arXiv_only}
\input{online_only}

\end{appendix}
\end{document}

%% file: online_only.tex
\begin{figure}[!hb]
\centering
\includegraphics[width=0.86\linewidth]{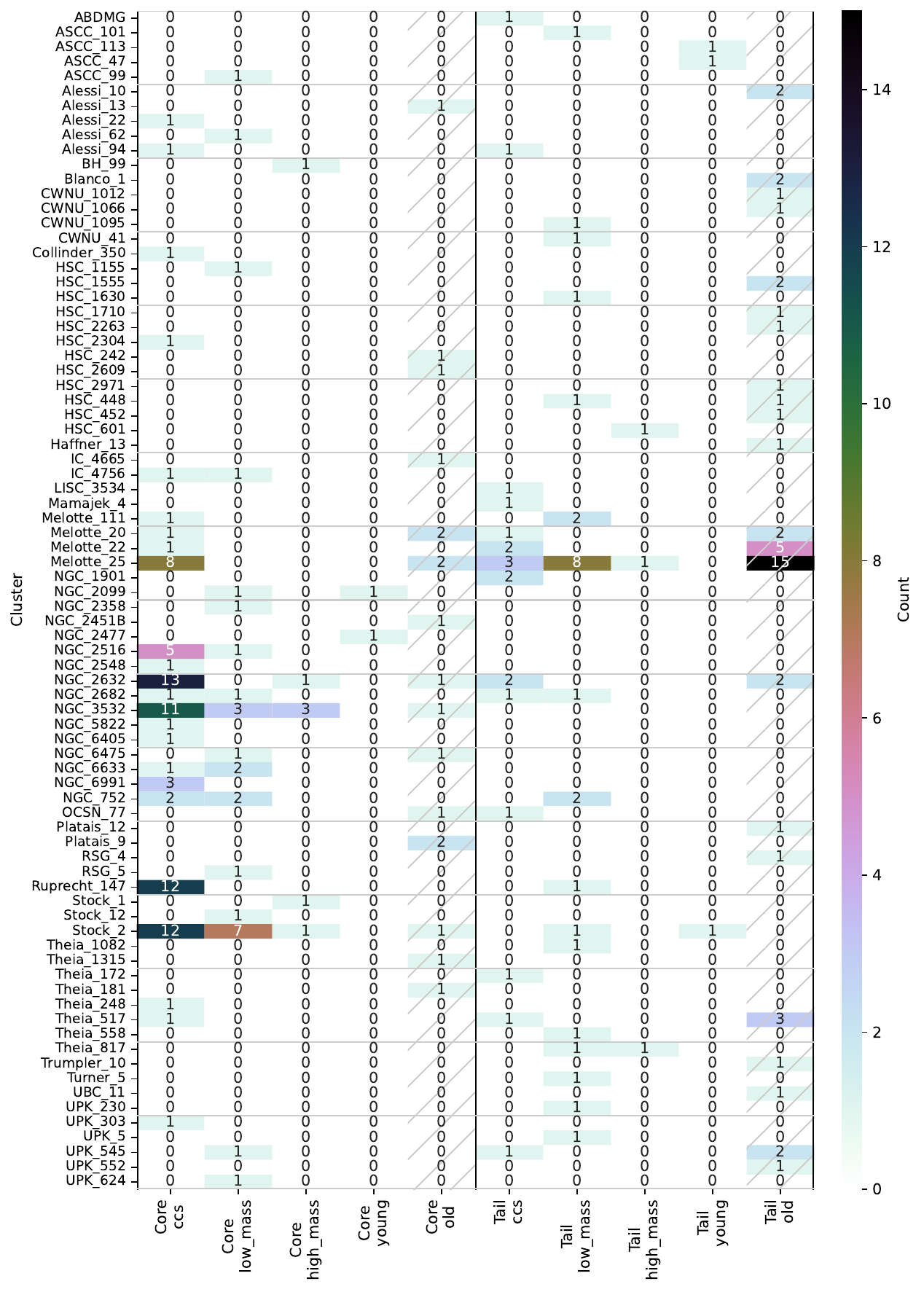}

\caption{Population counts of the WDs in individual cluster according to their spatial (core and tail) and CMD (\texttt{ccs, low\_mass, high\_mass, young,} and \texttt{old}) positions. The \texttt{old} WDs are hatched to highlight the likely contaminants.}
\label{fig:cmd_spatial_position_stats}
\end{figure}

\begin{figure}
\centering
\includegraphics[width=0.85\linewidth]{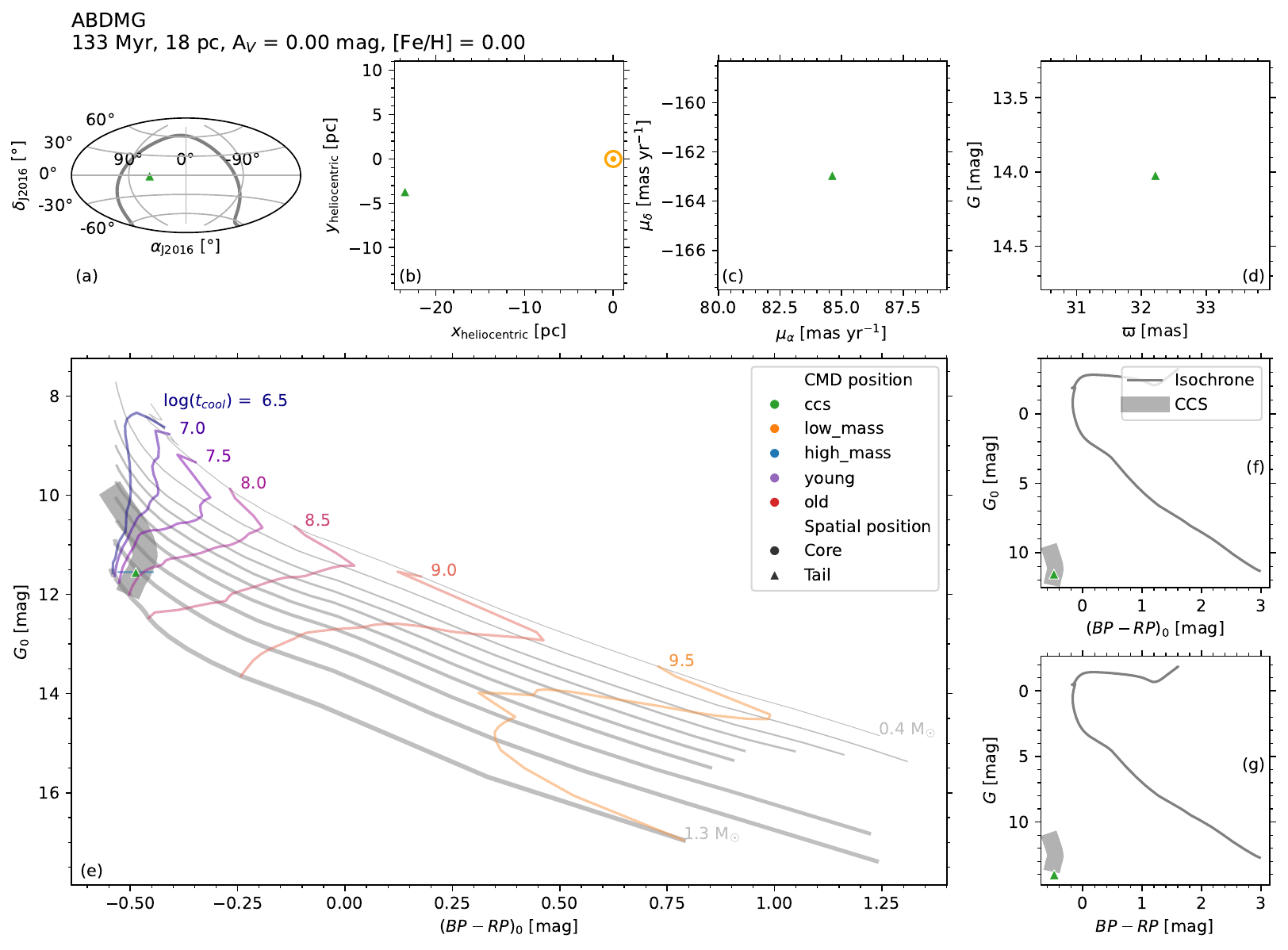}
\includegraphics[width=0.85\linewidth]{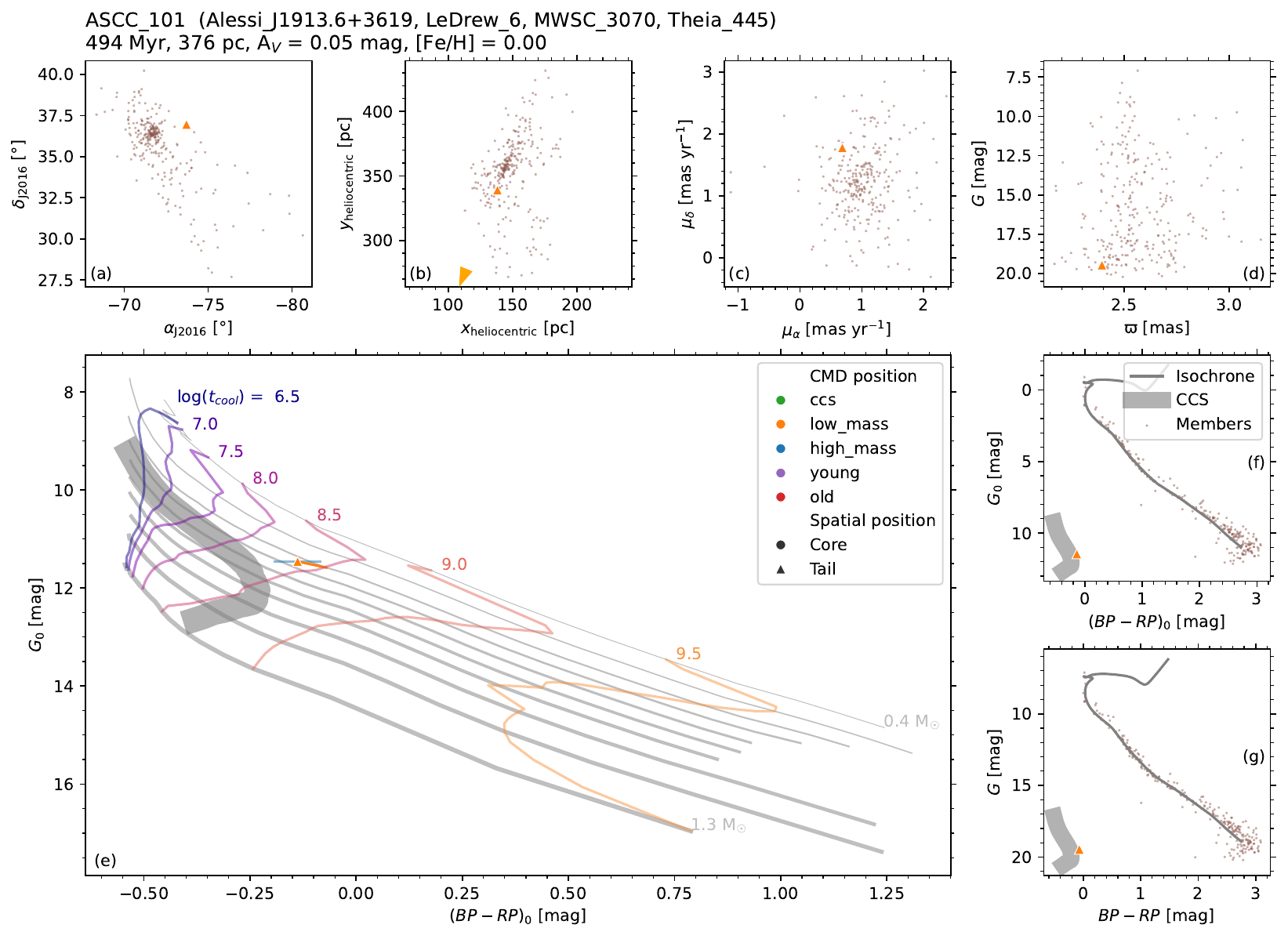}
\caption{Diagnostic plots for ABDMG and ASCC 101. All details are similar to Figure~\ref{fig:combo_Melotte_25}.}
\label{fig:combo_ASCC_101_appendix}
\end{figure}
\begin{figure}
\centering
\includegraphics[width=0.85\linewidth]{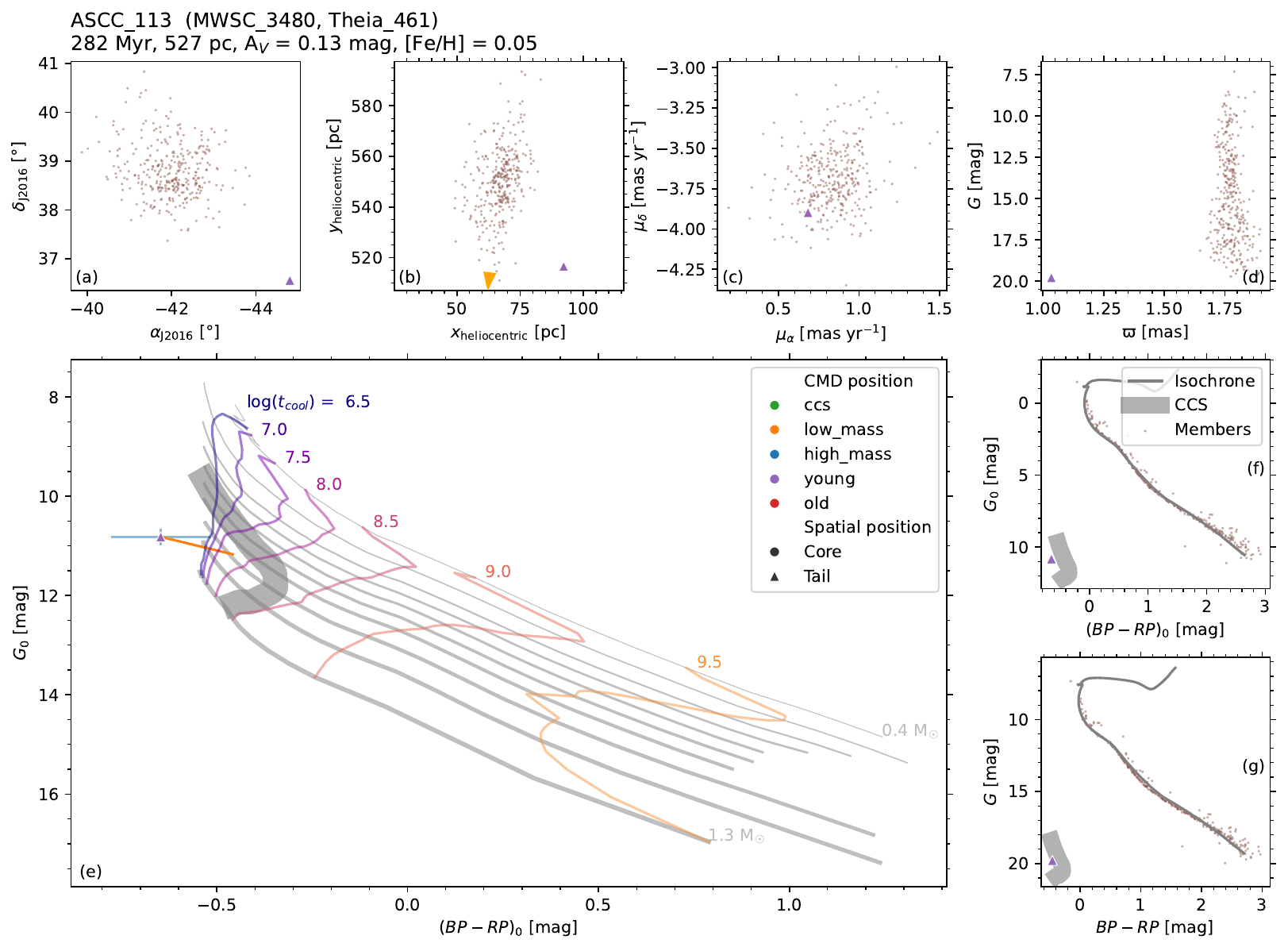}
\includegraphics[width=0.85\linewidth]{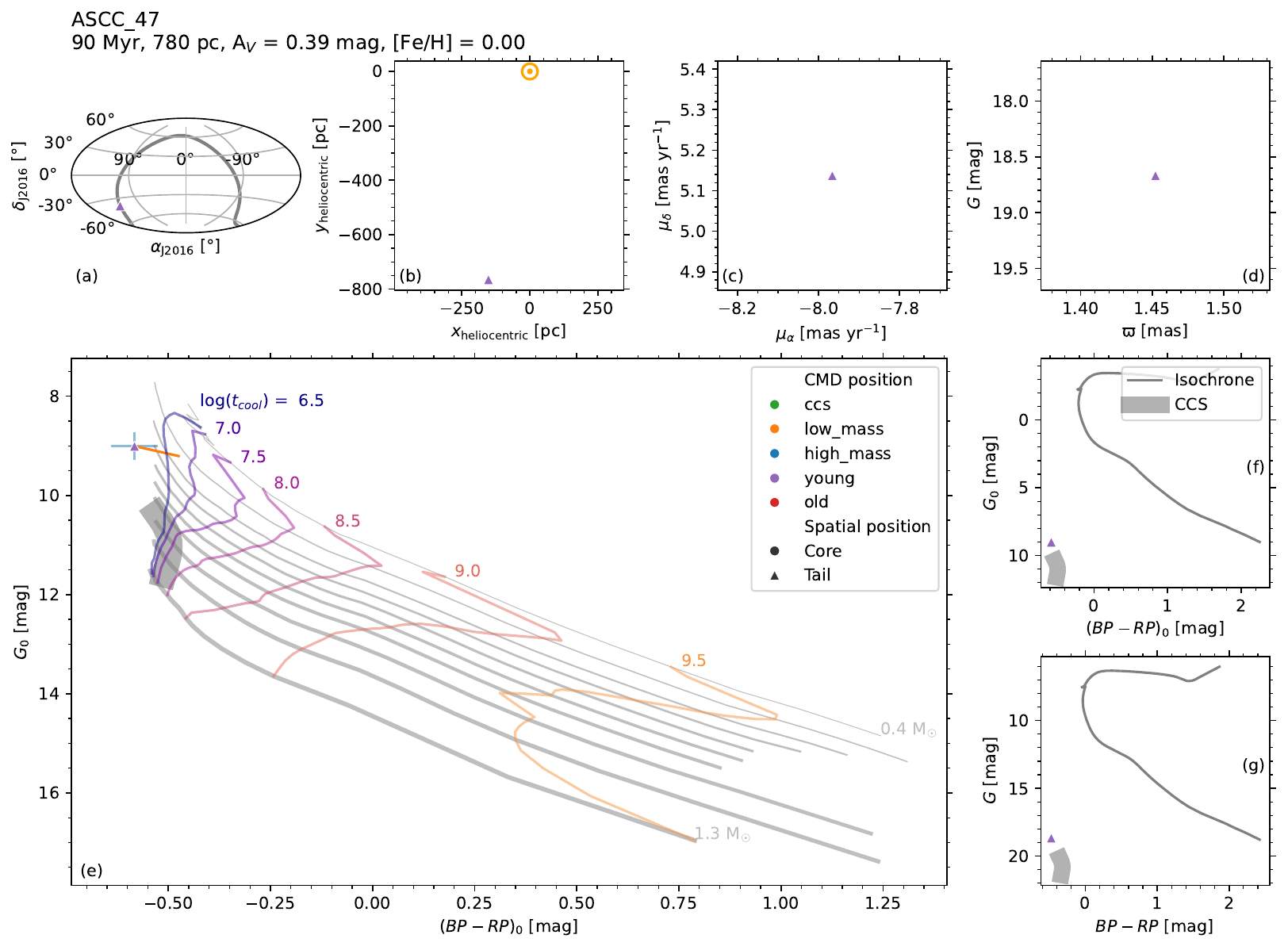}
\caption{Diagnostic plots for ASCC 113 and ASCC 47. All details are similar to Figure~\ref{fig:combo_Melotte_25}.}
\label{fig:combo_ASCC_47_appendix}
\end{figure}
\begin{figure}
\centering
\includegraphics[width=0.85\linewidth]{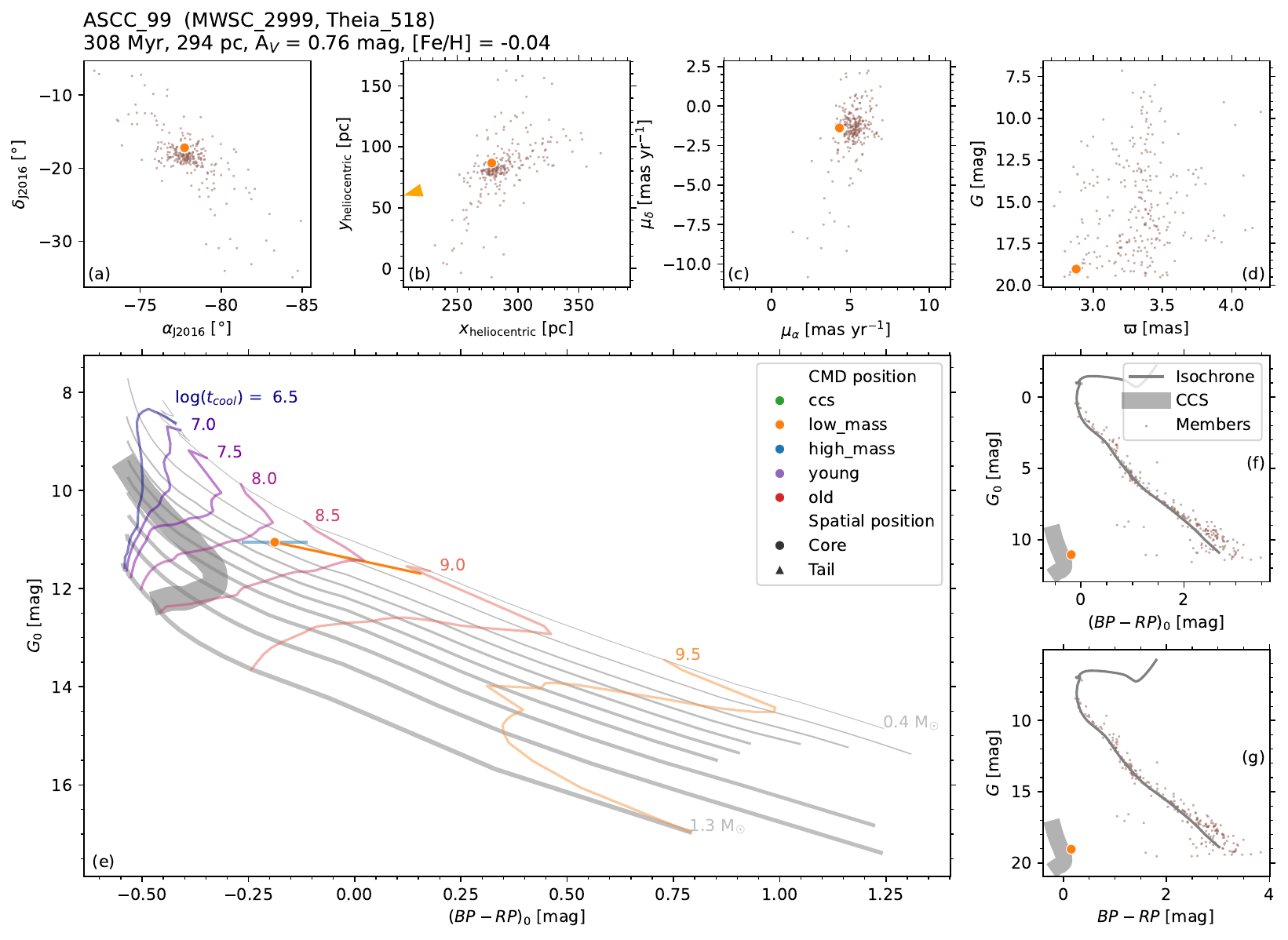}
\includegraphics[width=0.85\linewidth]{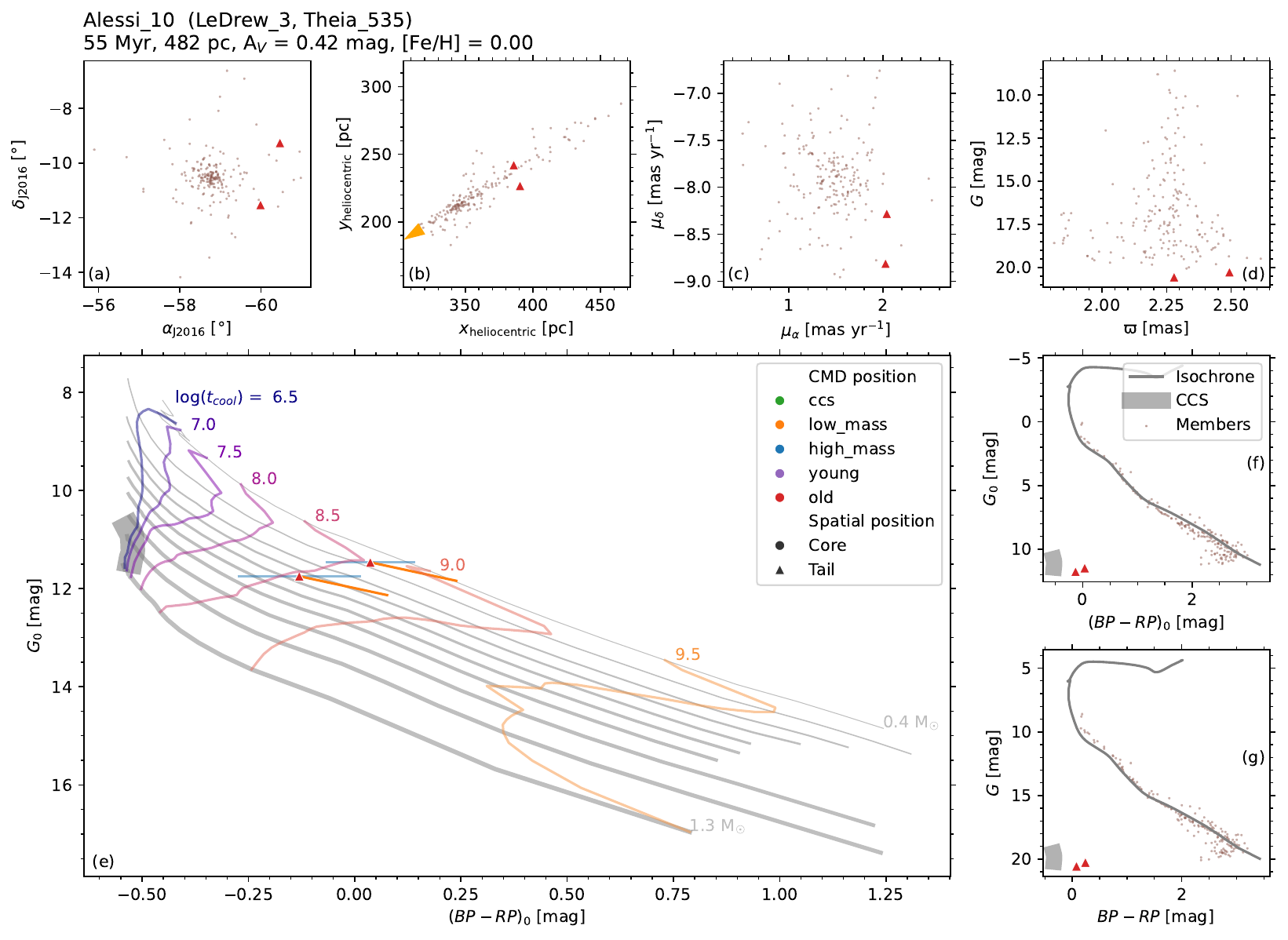}
\caption{Diagnostic plots for ASCC 99 and Alessi 10. All details are similar to Figure~\ref{fig:combo_Melotte_25}.}
\label{fig:combo_Alessi_10_appendix}
\end{figure}
\begin{figure}
\centering
\includegraphics[width=0.85\linewidth]{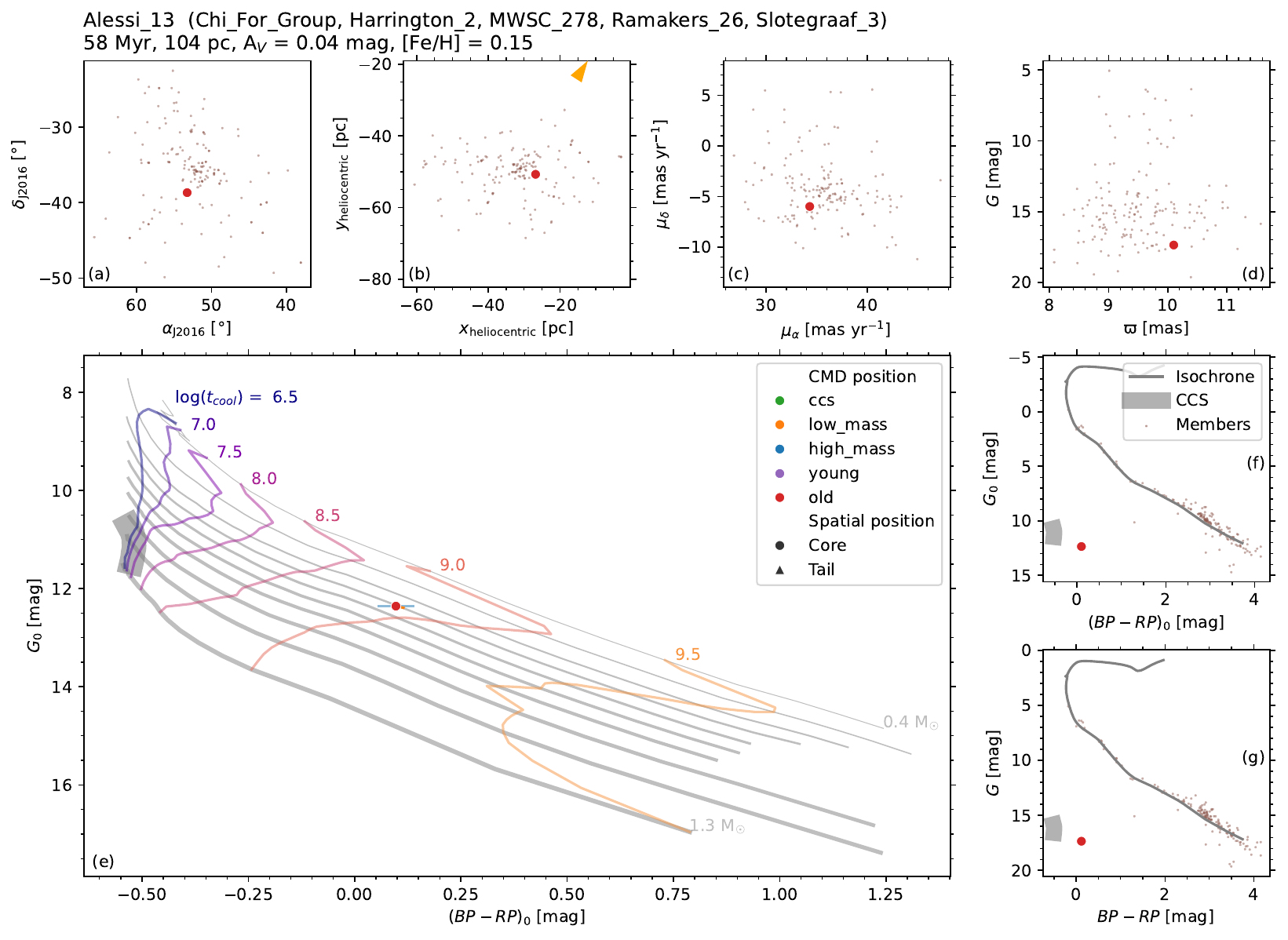}
\includegraphics[width=0.85\linewidth]{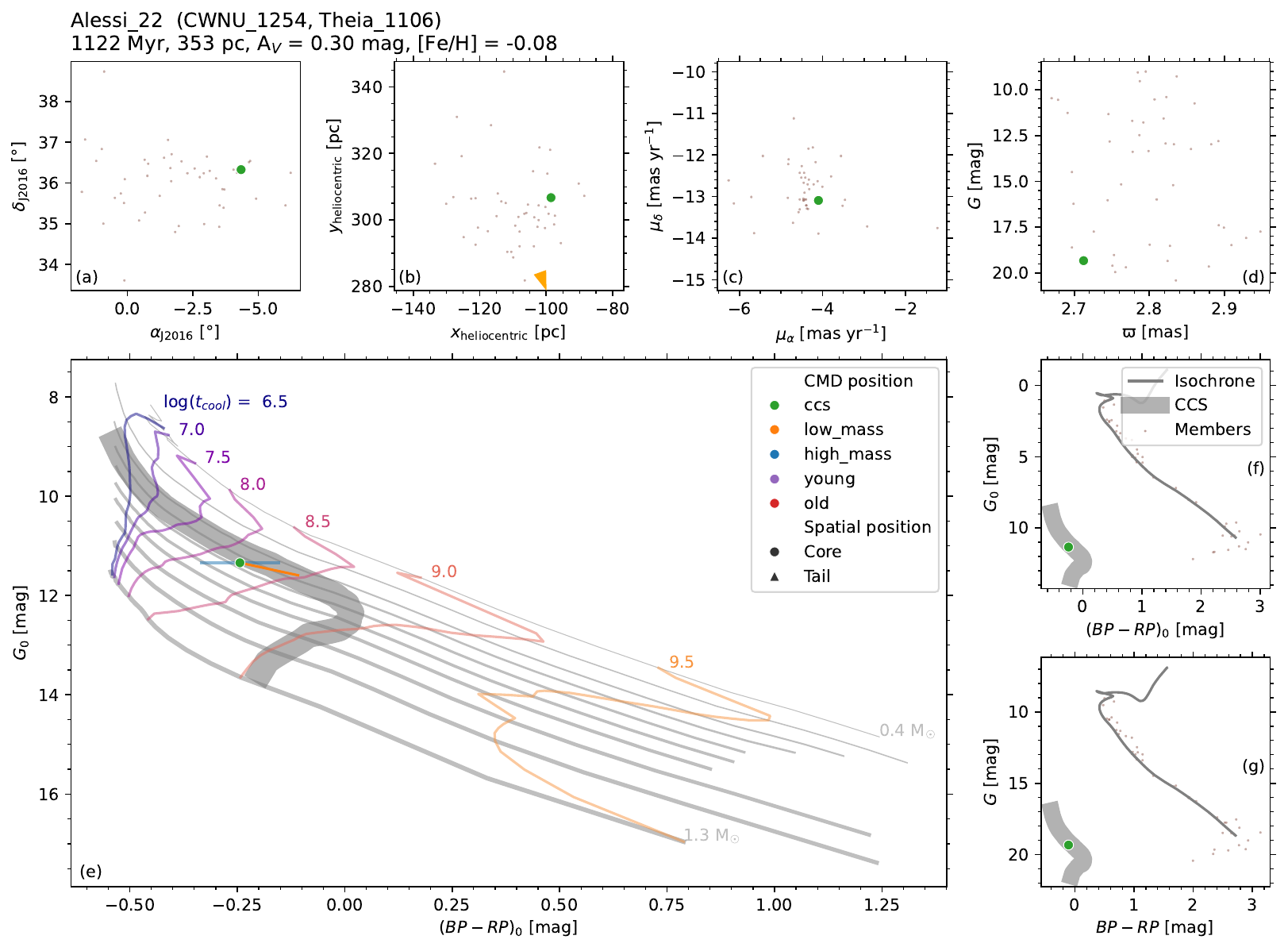}
\caption{Diagnostic plots for Alessi 13 and Alessi 22. All details are similar to Figure~\ref{fig:combo_Melotte_25}.}
\label{fig:combo_Alessi_22_appendix}
\end{figure}
\begin{figure}
\centering
\includegraphics[width=0.85\linewidth]{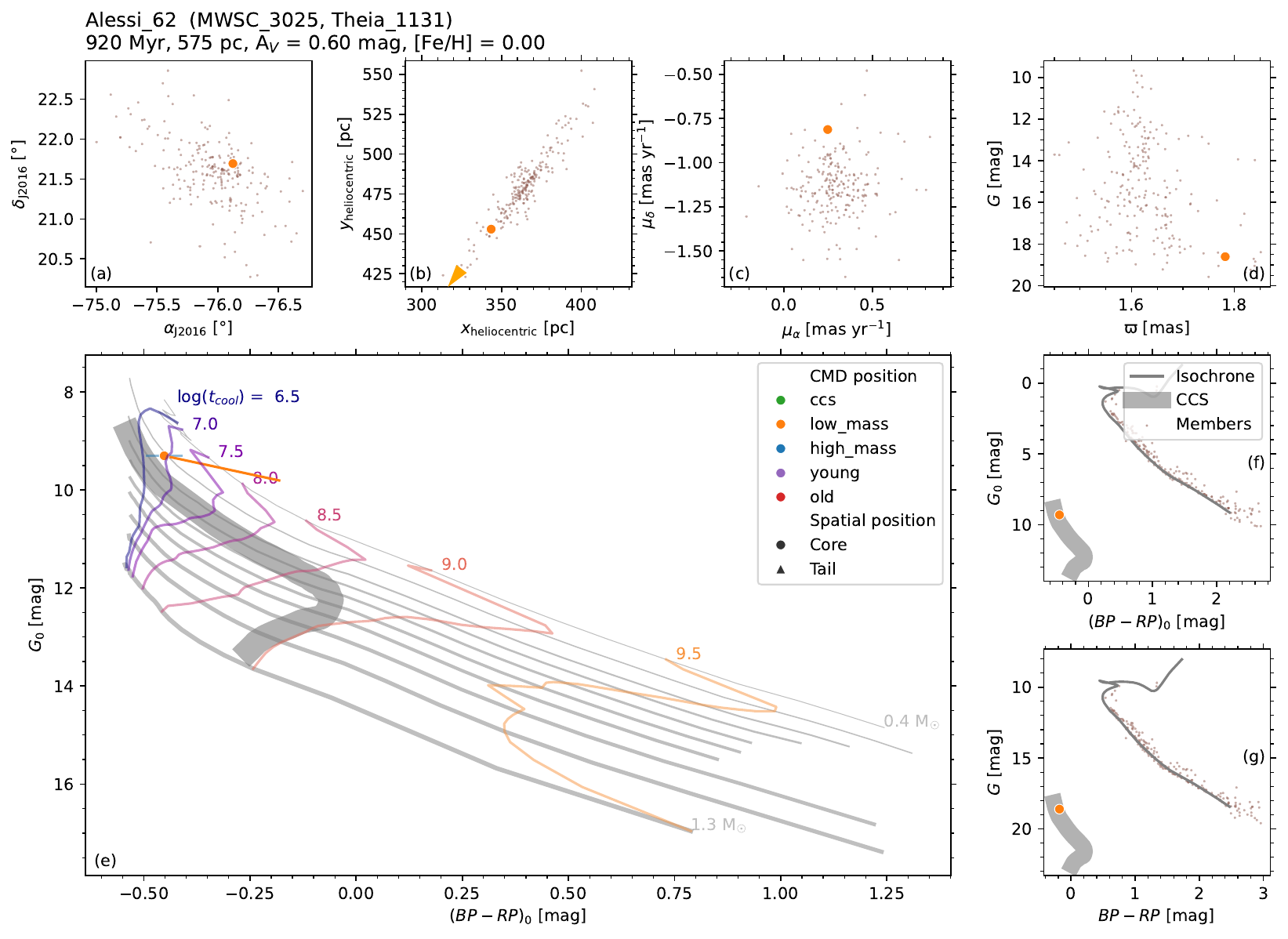}
\includegraphics[width=0.85\linewidth]{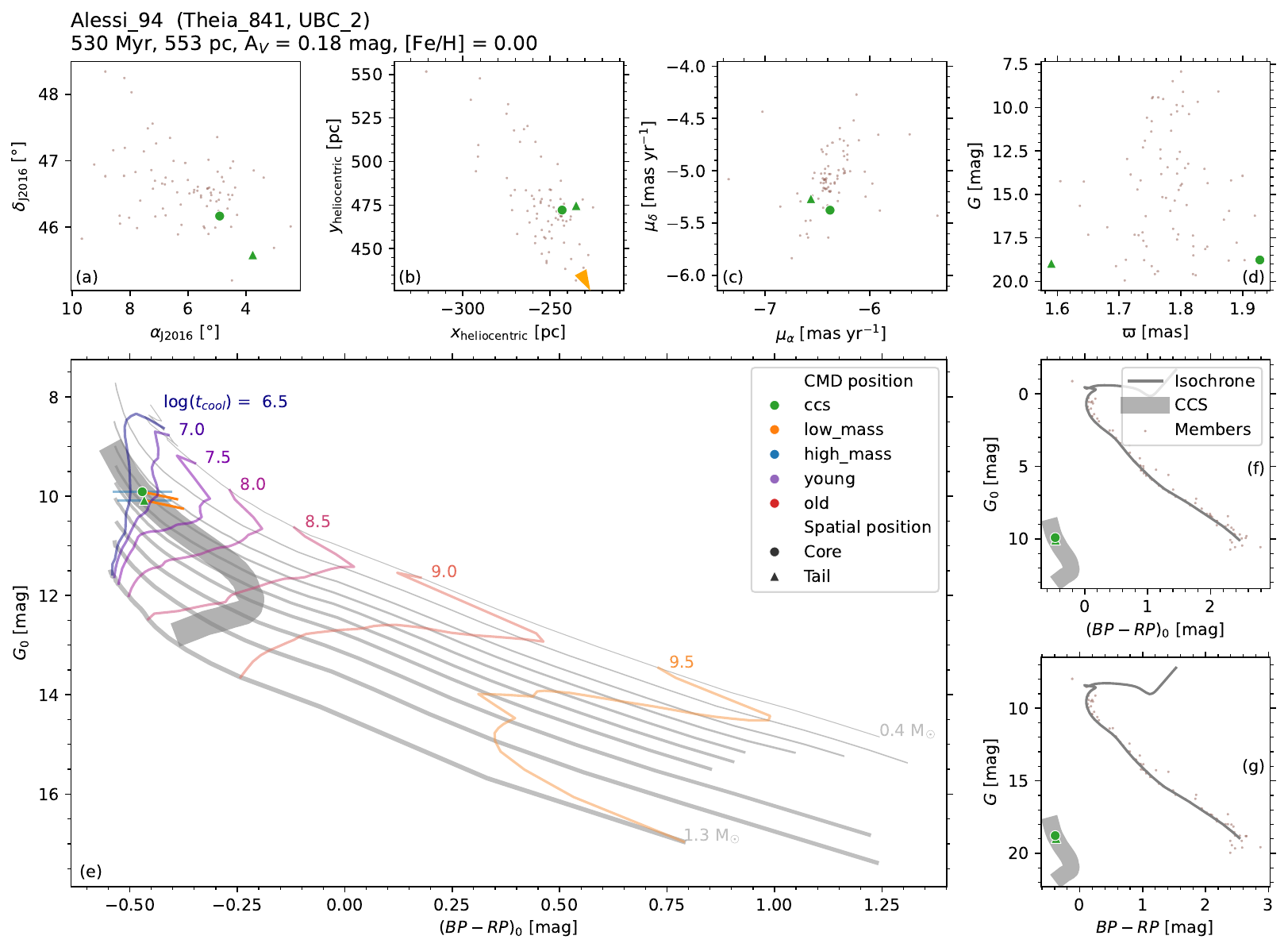}
\caption{Diagnostic plots for Alessi 62 and Alessi 94. All details are similar to Figure~\ref{fig:combo_Melotte_25}.}
\label{fig:combo_Alessi_94_appendix}
\end{figure}
\begin{figure}
\centering
\includegraphics[width=0.85\linewidth]{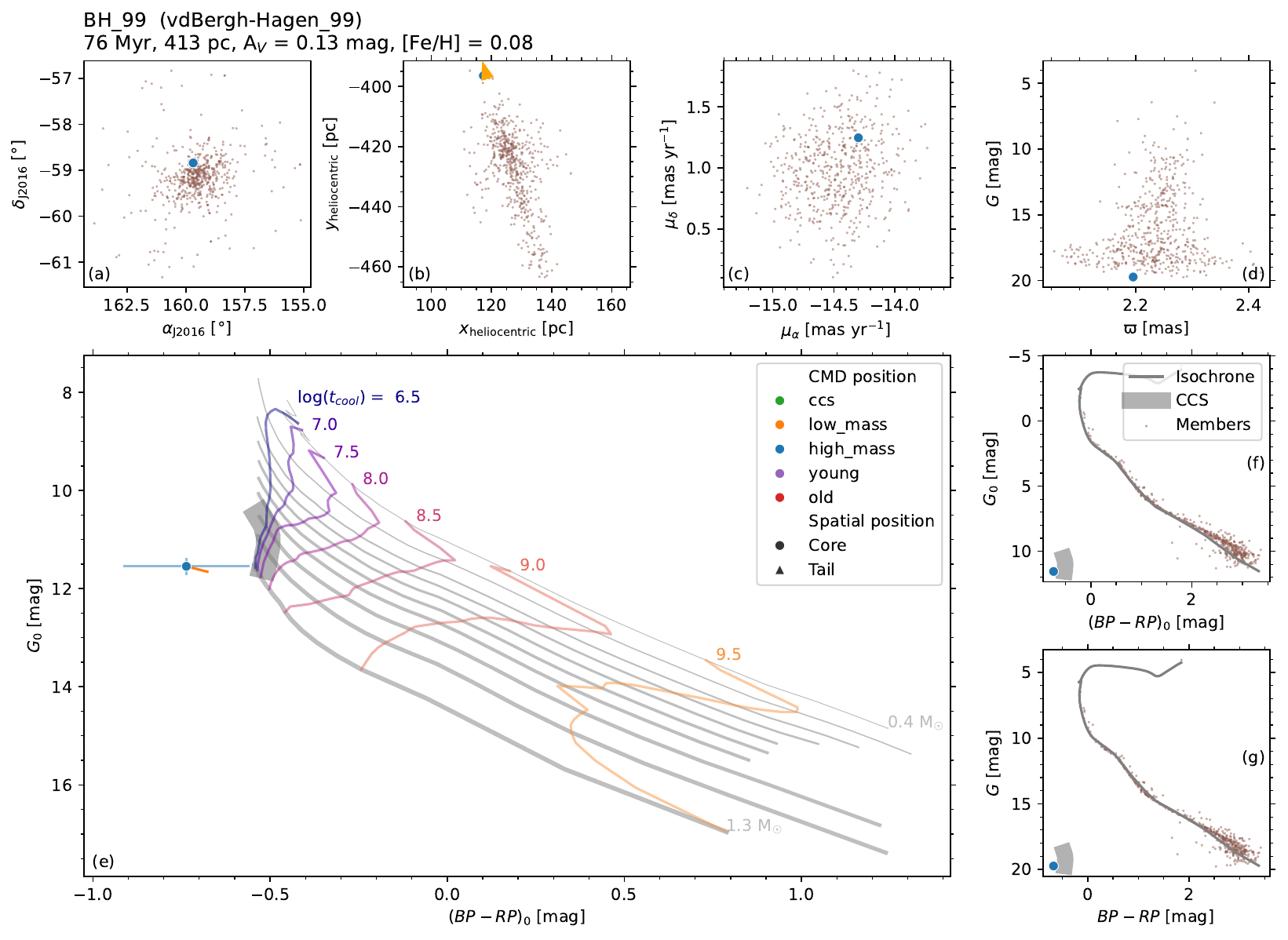}
\includegraphics[width=0.85\linewidth]{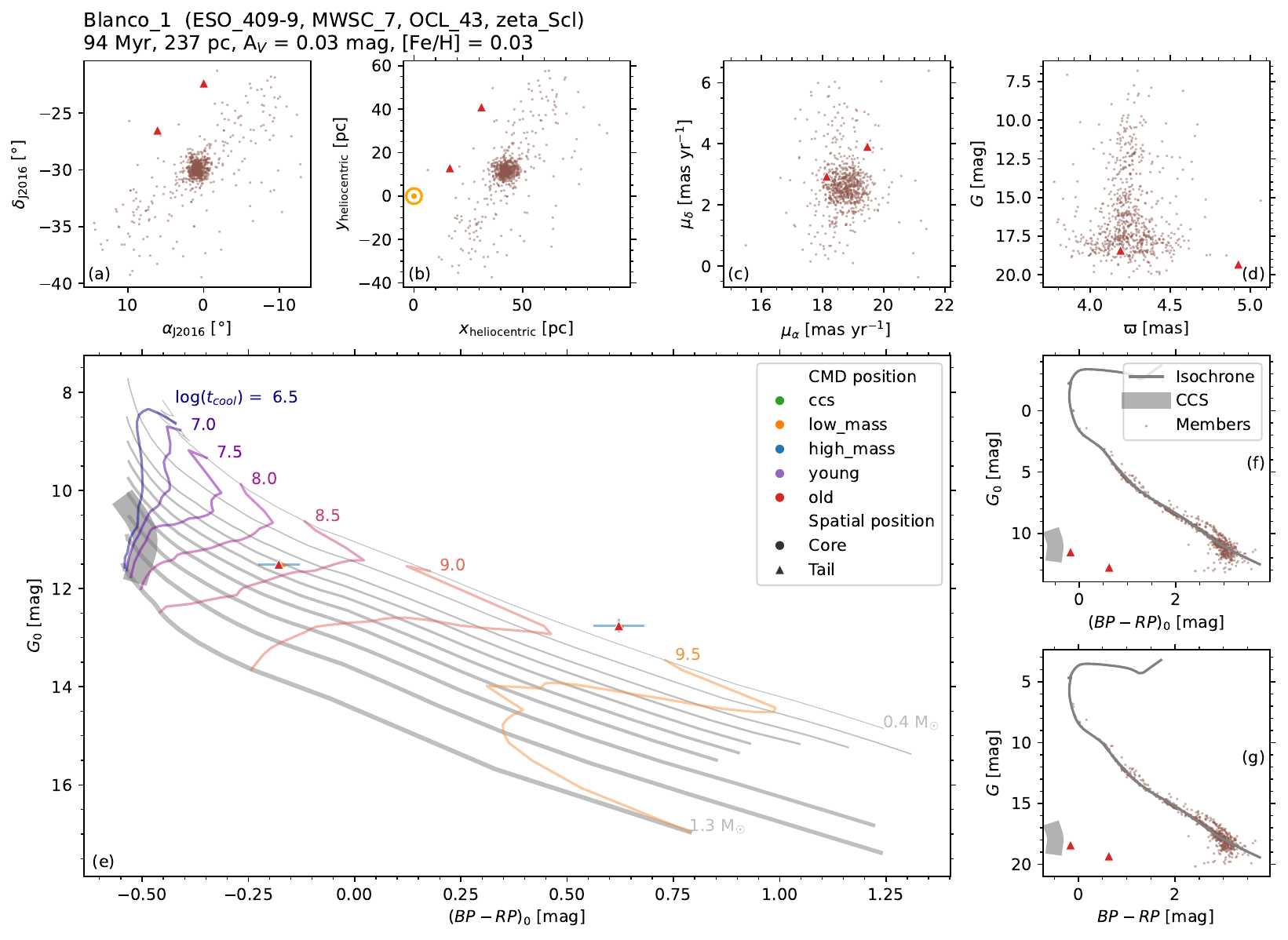}
\caption{Diagnostic plots for BH 99 and Blanco 1. All details are similar to Figure~\ref{fig:combo_Melotte_25}.}
\label{fig:combo_Blanco_1_appendix}
\end{figure}
\begin{figure}
\centering
\includegraphics[width=0.85\linewidth]{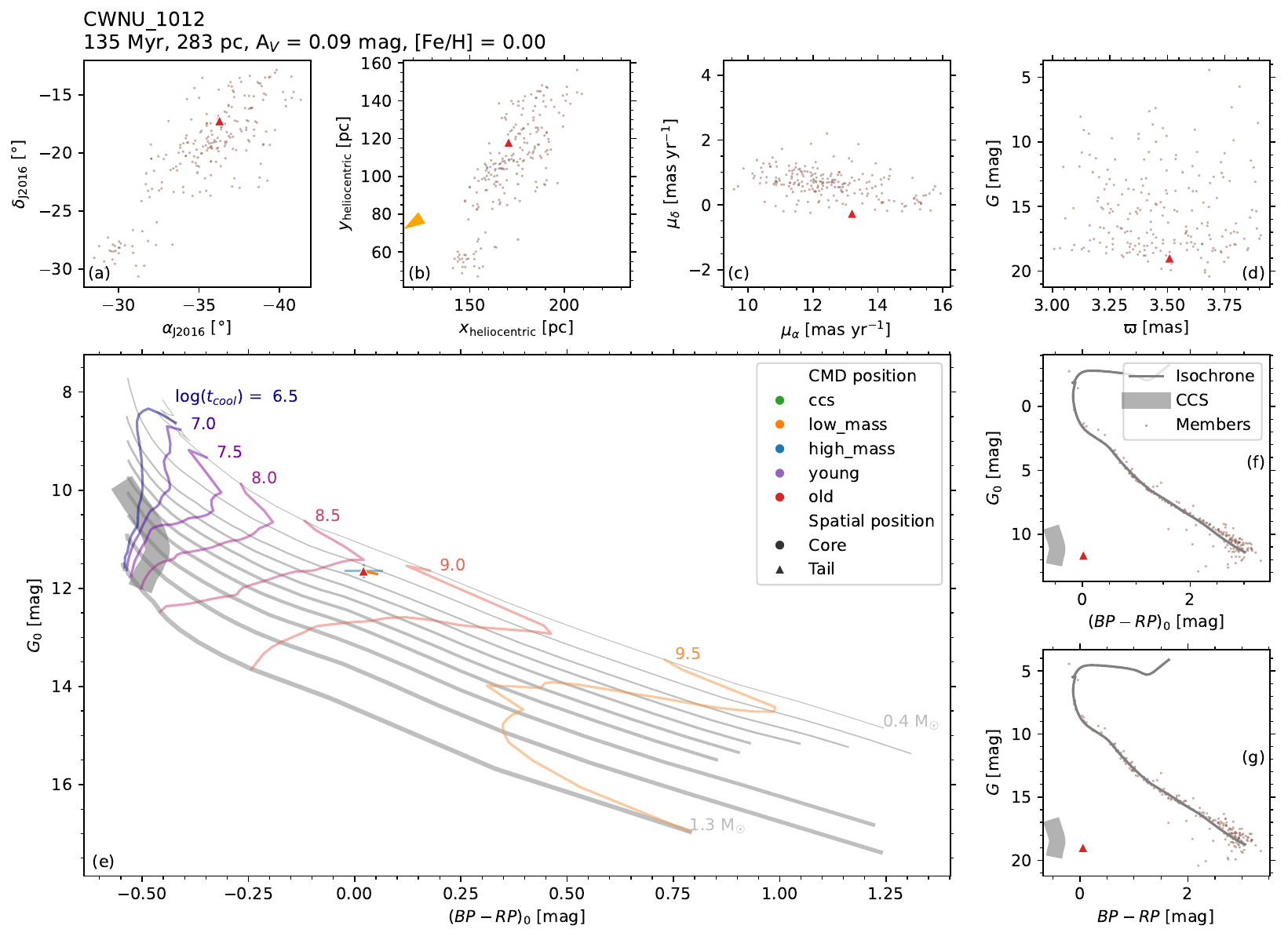}
\includegraphics[width=0.85\linewidth]{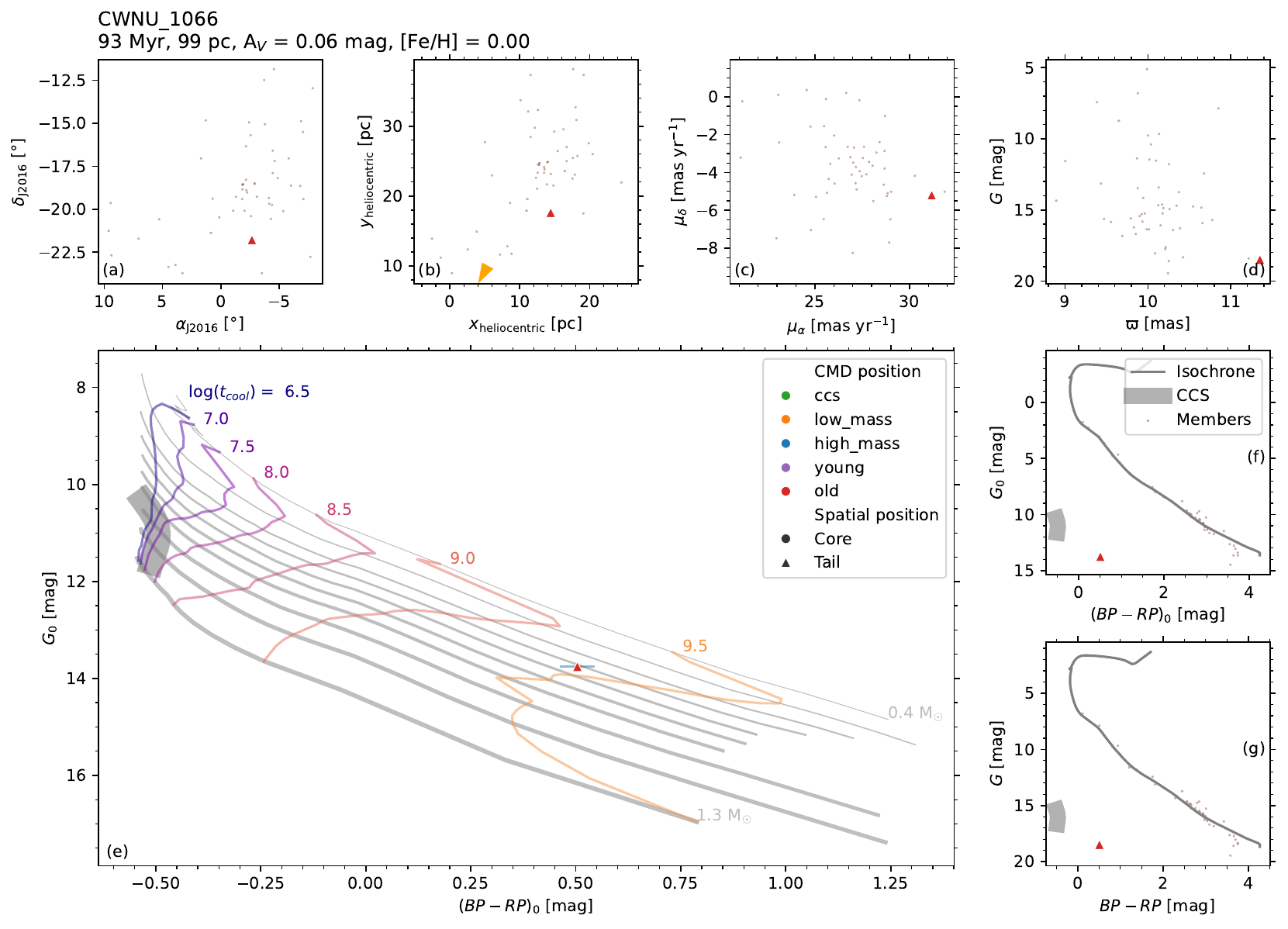}
\caption{Diagnostic plots for CWNU 1012 and CWNU 1066. All details are similar to Figure~\ref{fig:combo_Melotte_25}.}
\label{fig:combo_CWNU_1066_appendix}
\end{figure}
\begin{figure}
\centering
\includegraphics[width=0.85\linewidth]{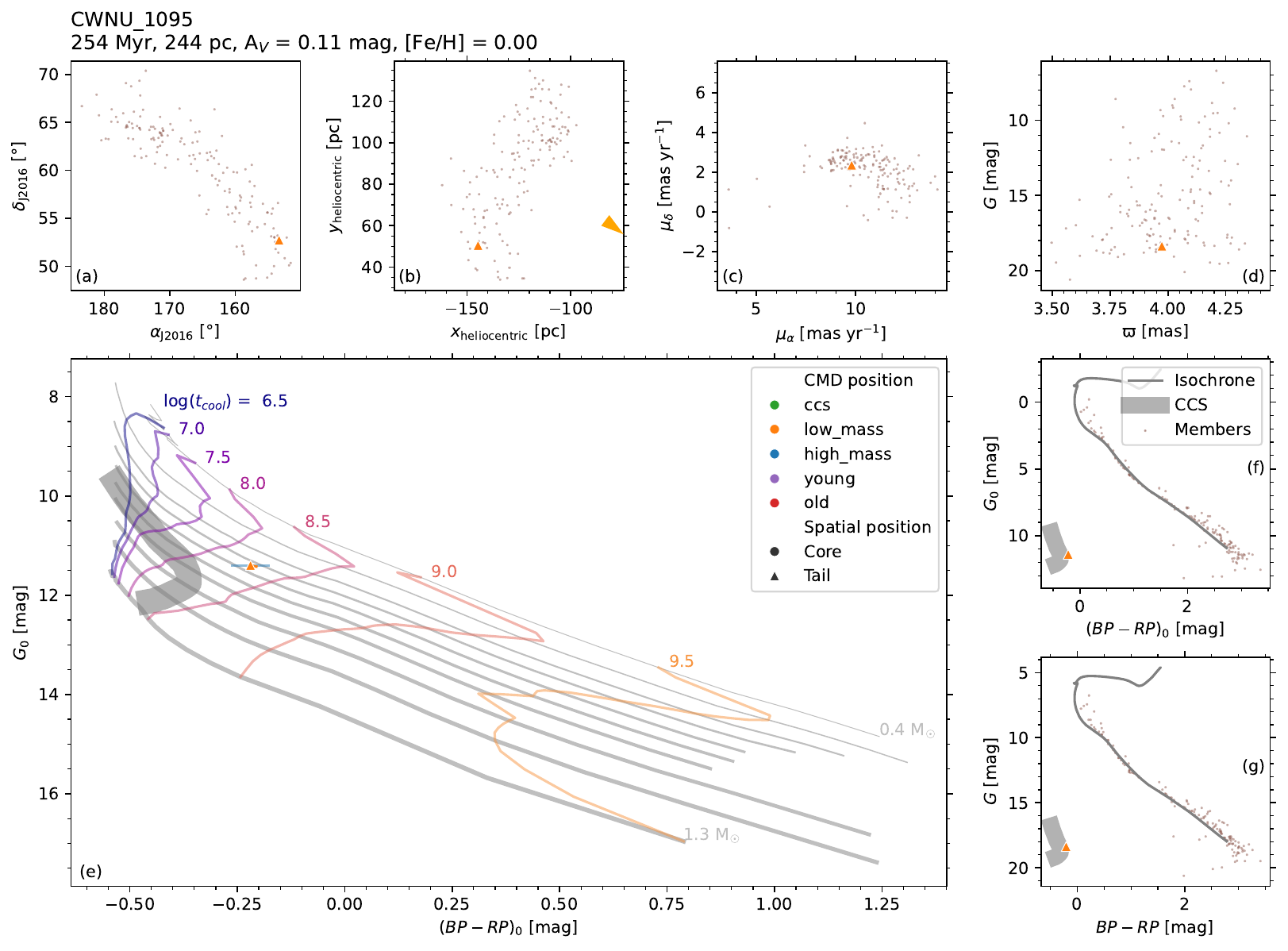}
\includegraphics[width=0.85\linewidth]{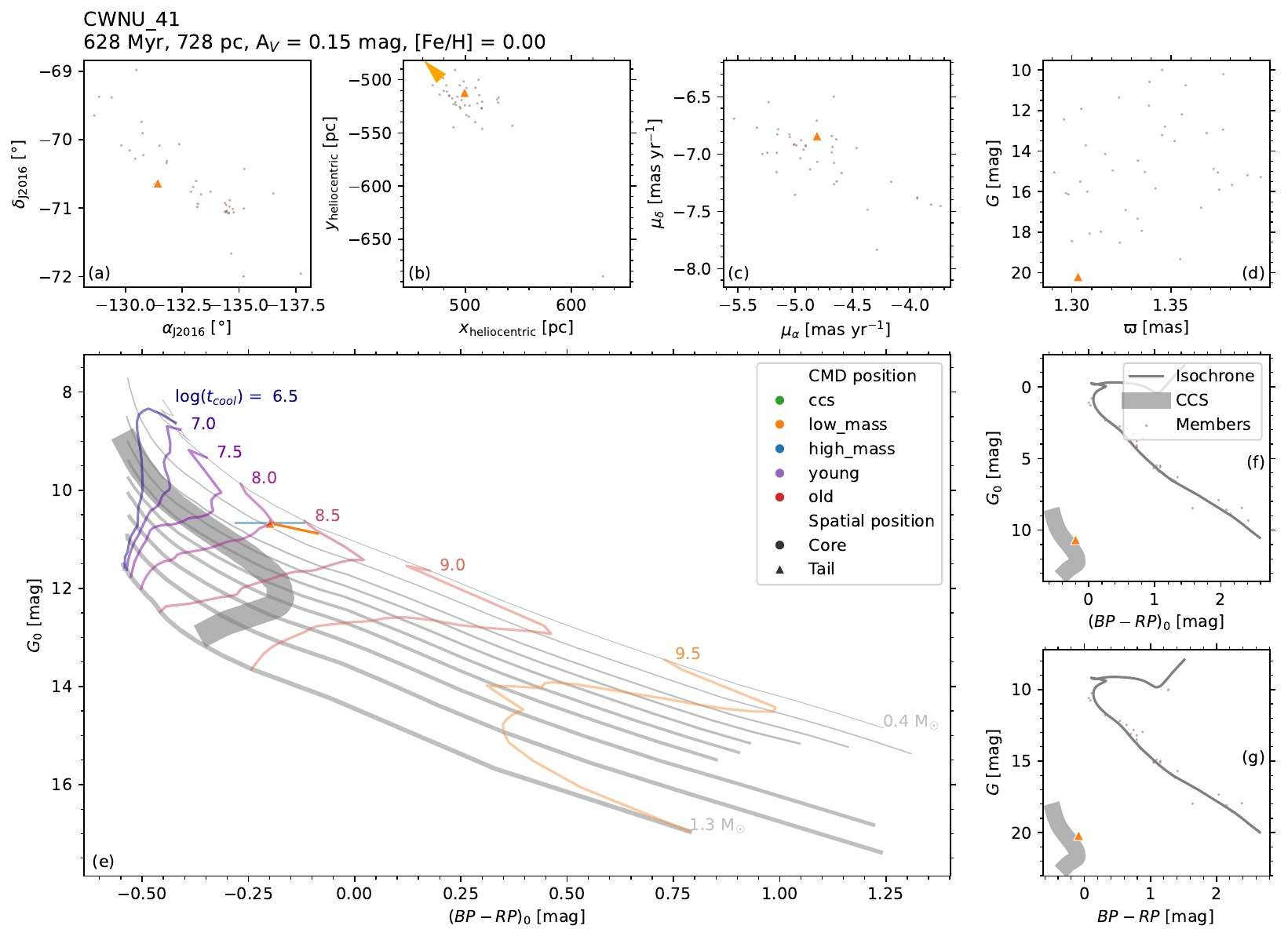}
\caption{Diagnostic plots for CWNU 1095 and CWNU 41. All details are similar to Figure~\ref{fig:combo_Melotte_25}.}
\label{fig:combo_CWNU_41_appendix}
\end{figure}
\begin{figure}
\centering
\includegraphics[width=0.85\linewidth]{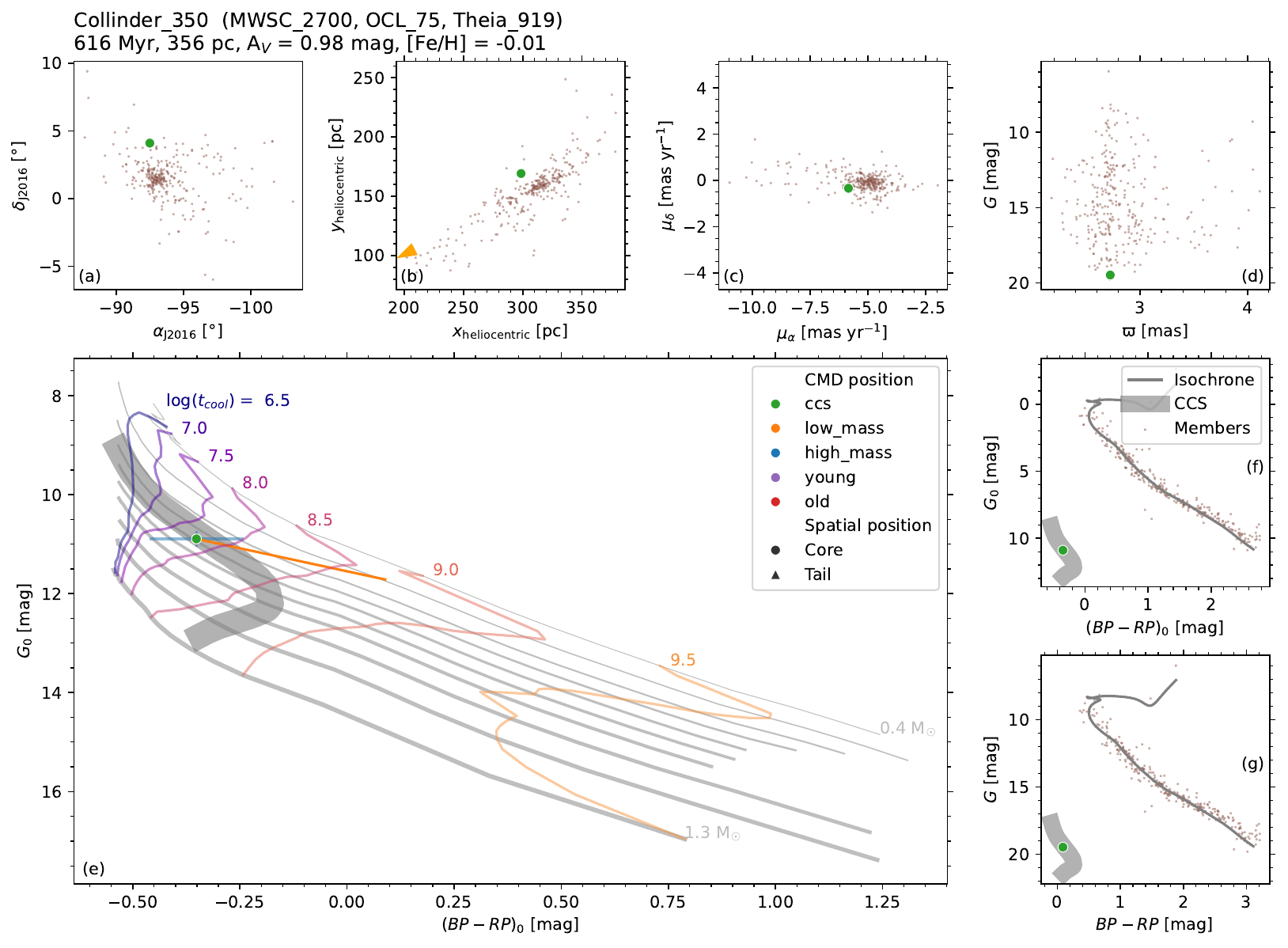}
\includegraphics[width=0.85\linewidth]{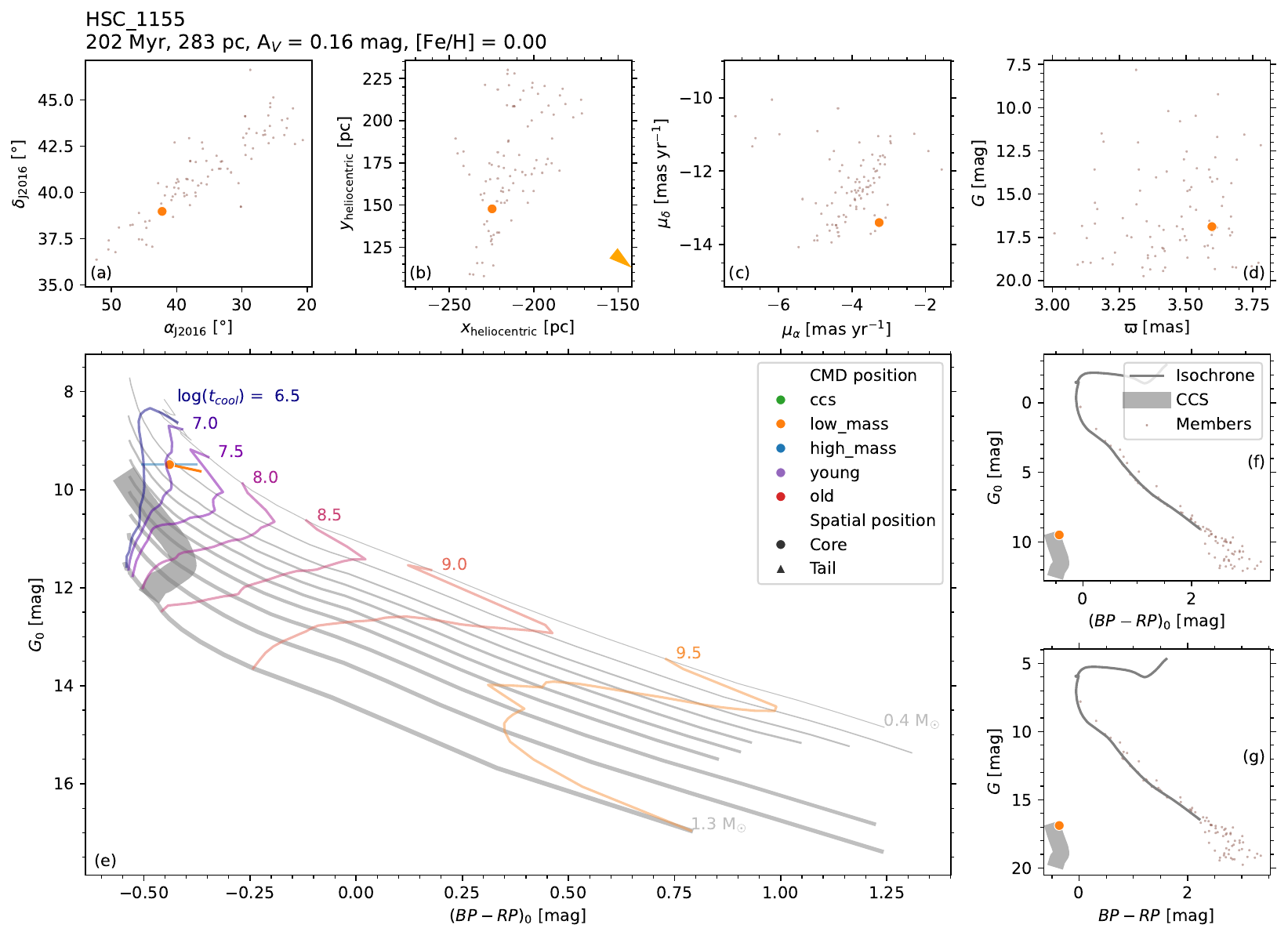}
\caption{Diagnostic plots for Collinder 350 and HSC 1155. All details are similar to Figure~\ref{fig:combo_Melotte_25}.}
\label{fig:combo_HSC_1155_appendix}
\end{figure}
\begin{figure}
\centering
\includegraphics[width=0.85\linewidth]{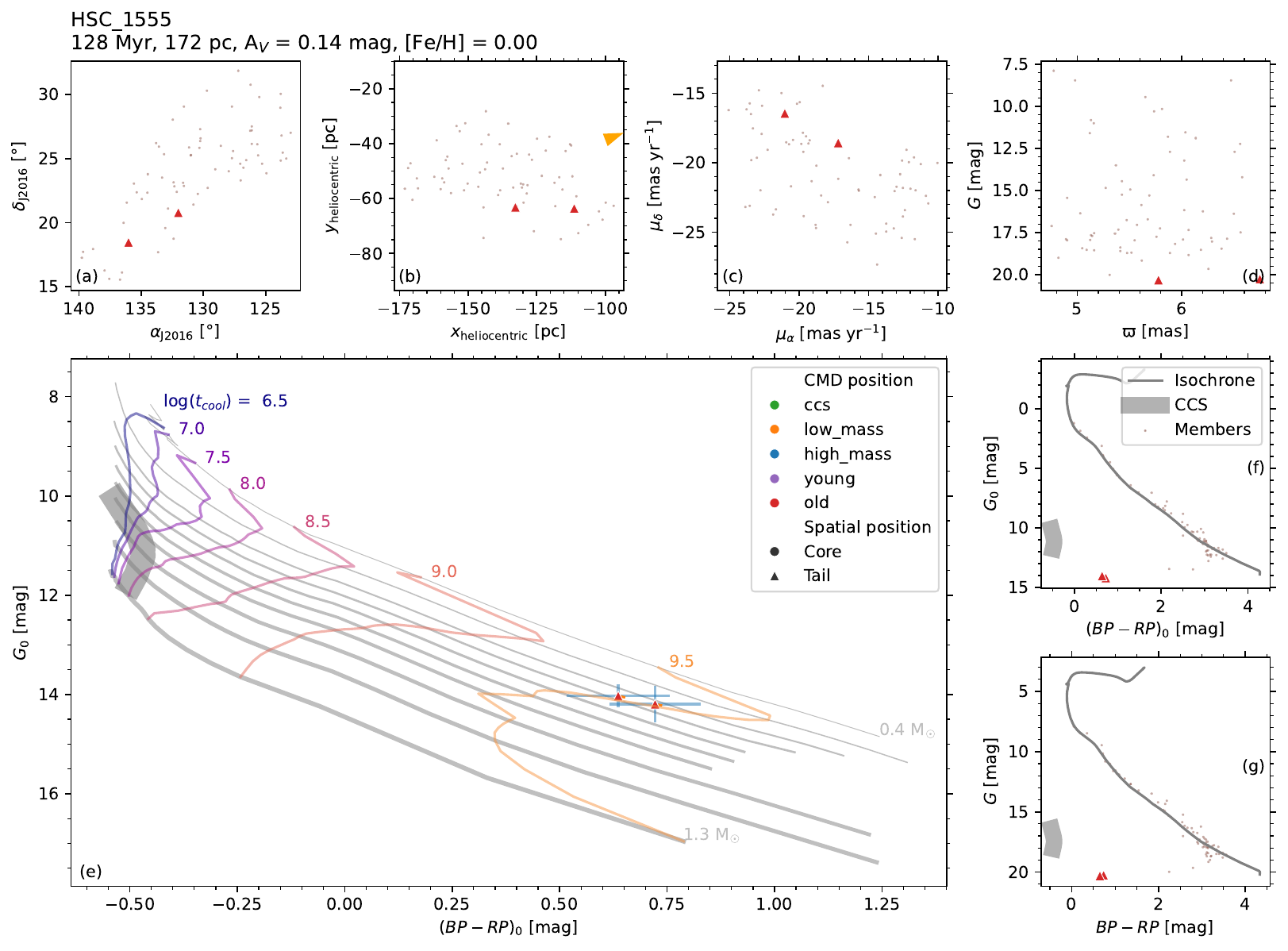}
\includegraphics[width=0.85\linewidth]{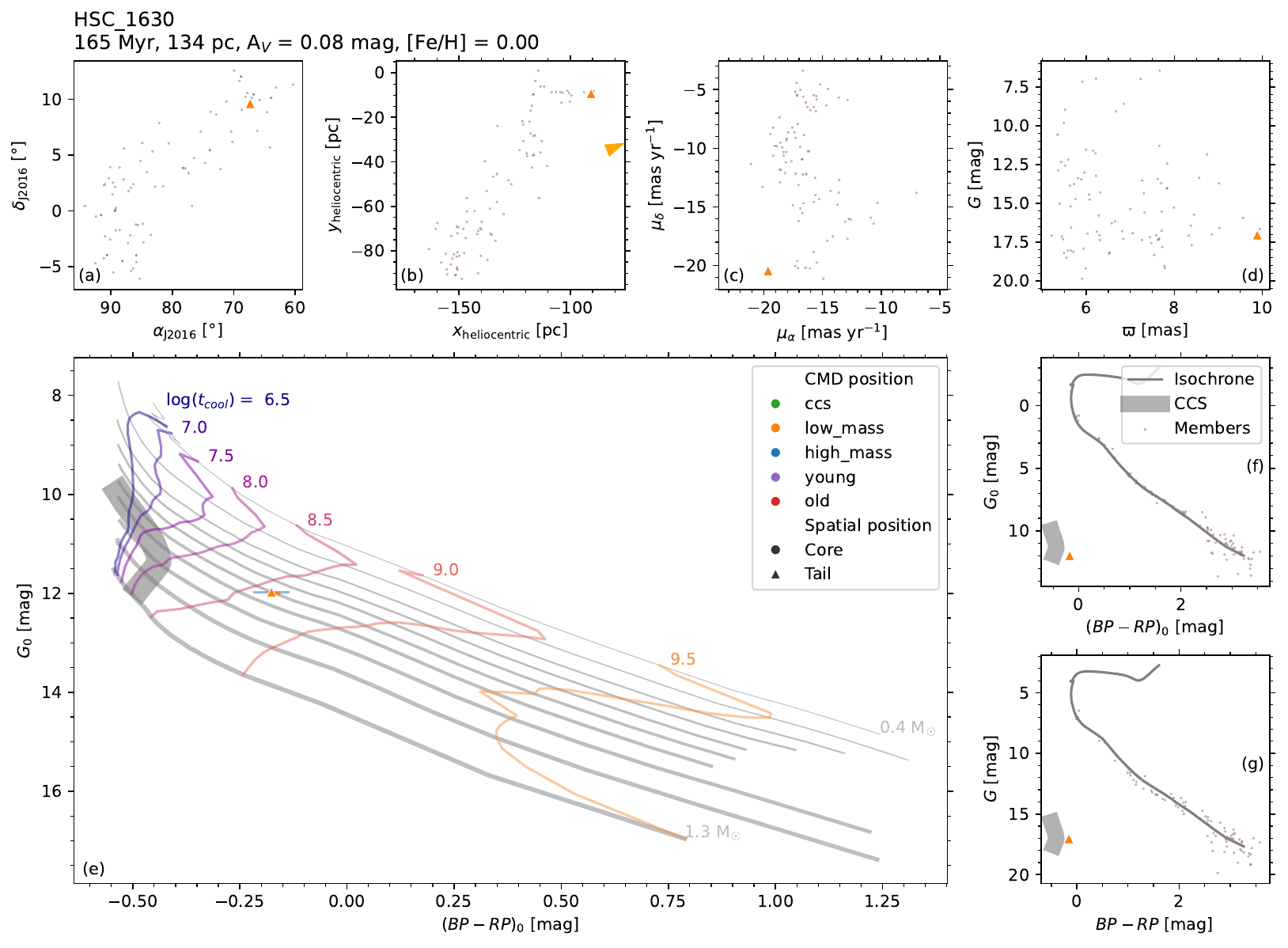}
\caption{Diagnostic plots for HSC 1555 and HSC 1630. All details are similar to Figure~\ref{fig:combo_Melotte_25}.}
\label{fig:combo_HSC_1630_appendix}
\end{figure}
\begin{figure}
\centering
\includegraphics[width=0.85\linewidth]{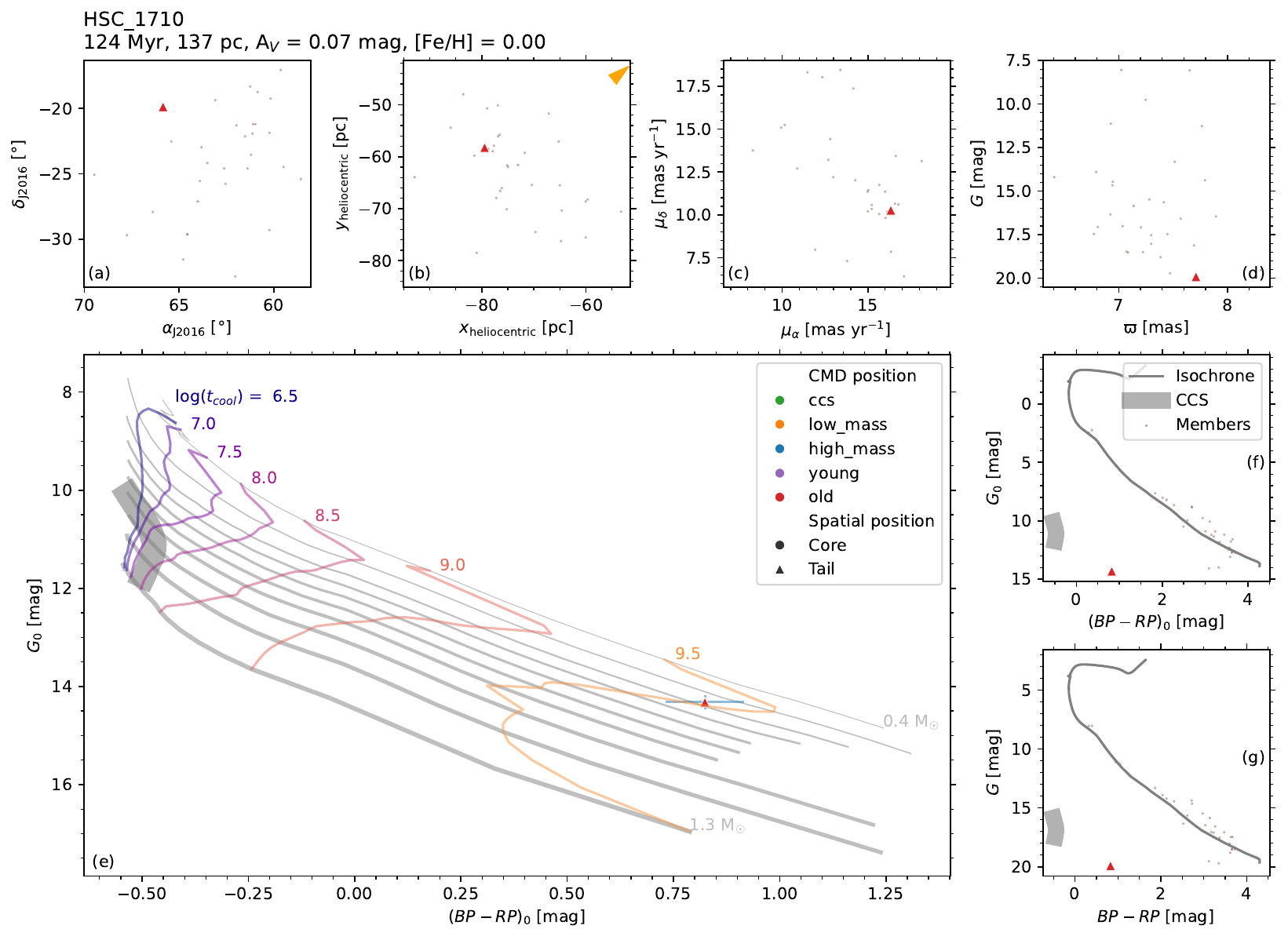}
\includegraphics[width=0.85\linewidth]{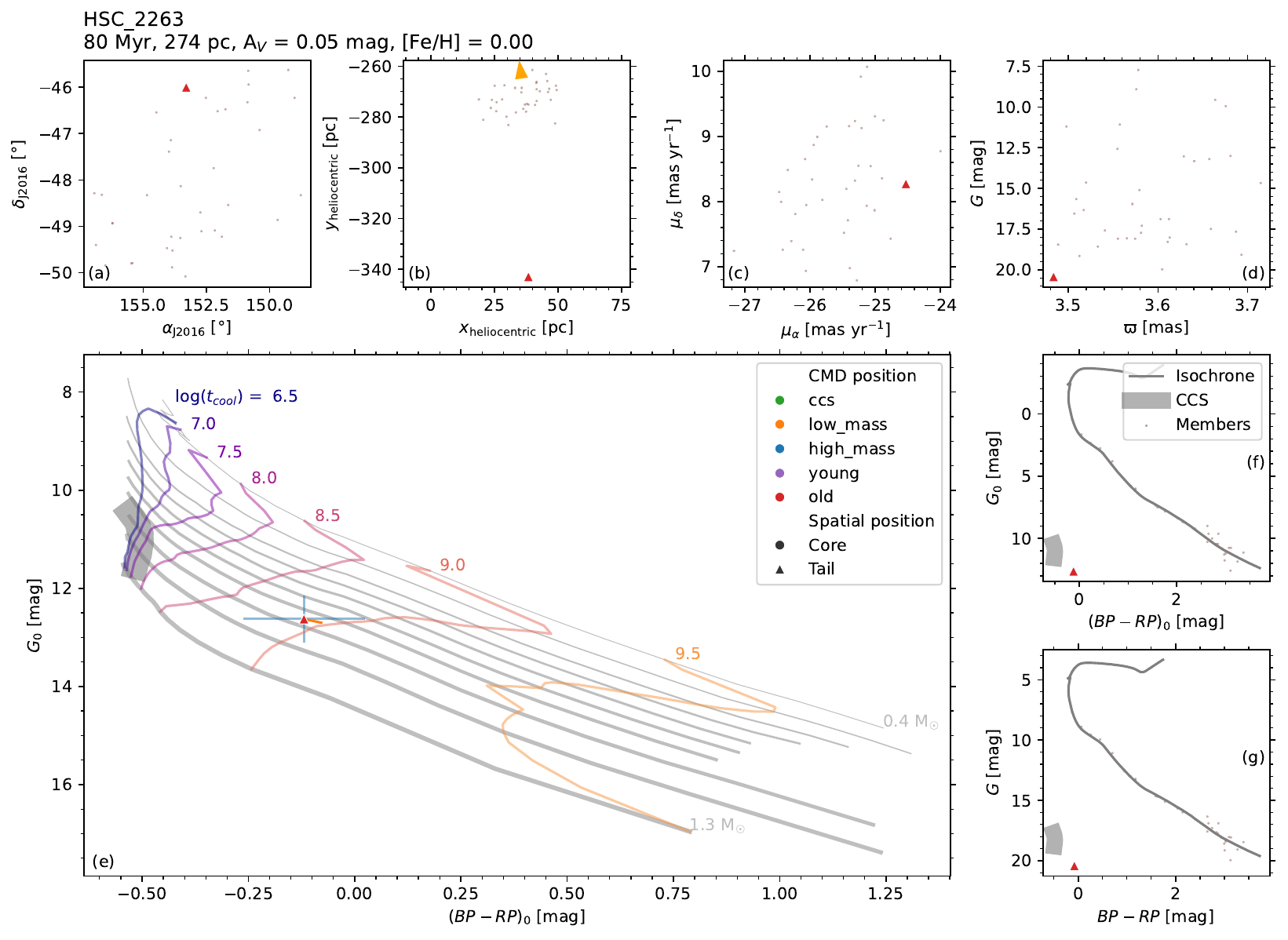}
\caption{Diagnostic plots for HSC 1710 and HSC 2263. All details are similar to Figure~\ref{fig:combo_Melotte_25}.}
\label{fig:combo_HSC_2263_appendix}
\end{figure}
\begin{figure}
\centering
\includegraphics[width=0.85\linewidth]{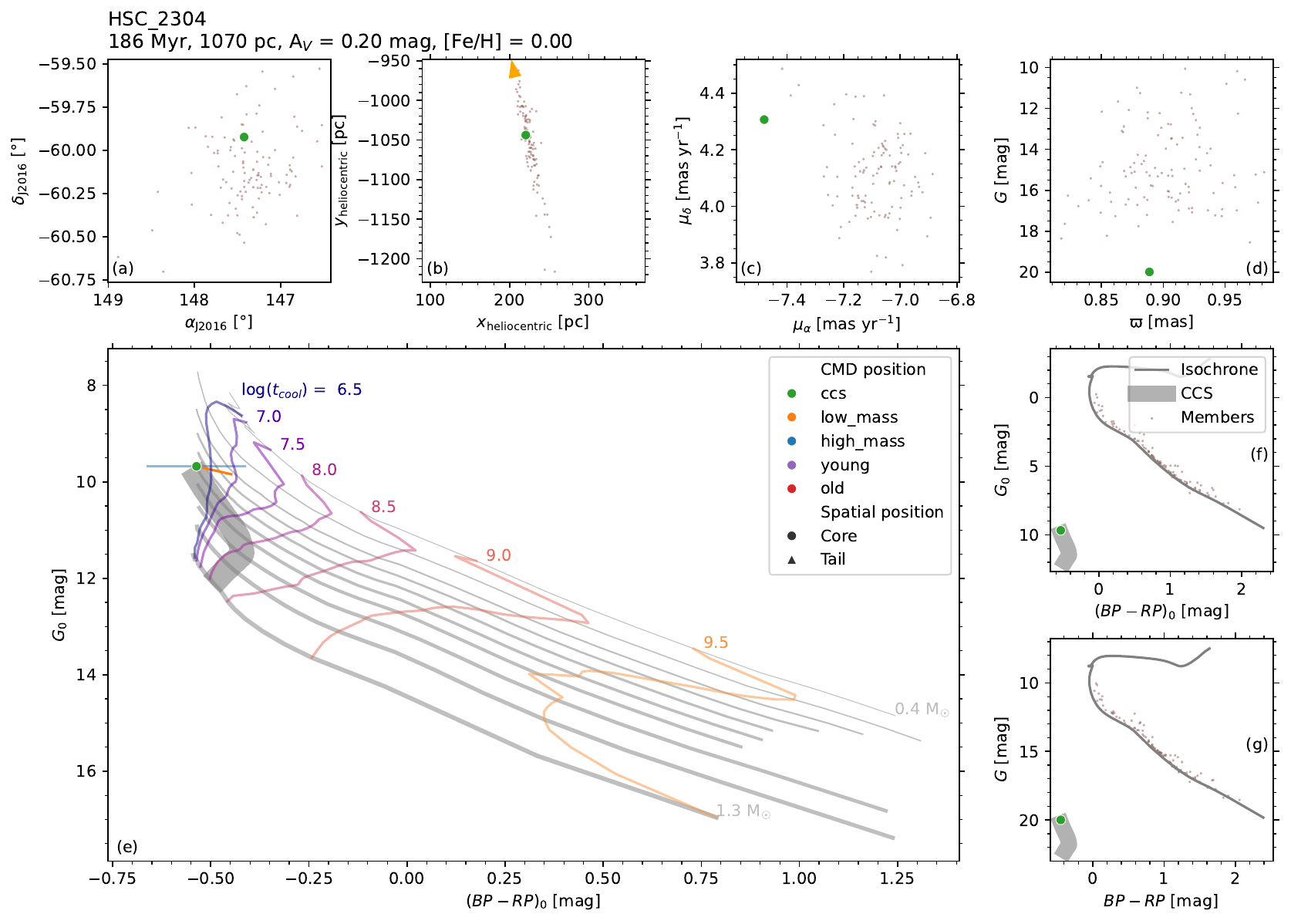}
\includegraphics[width=0.85\linewidth]{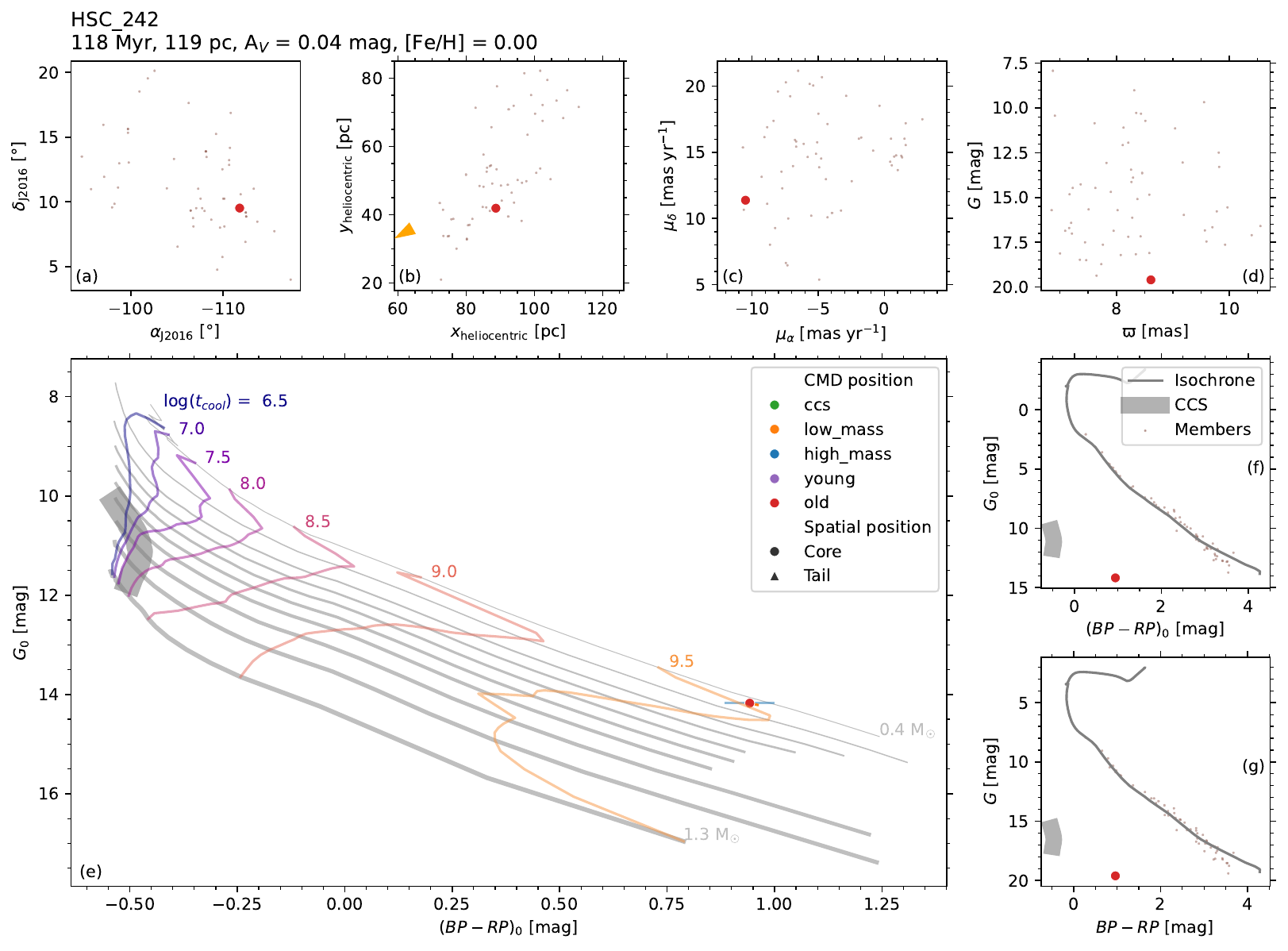}
\caption{Diagnostic plots for HSC 2304 and HSC 242. All details are similar to Figure~\ref{fig:combo_Melotte_25}.}
\label{fig:combo_HSC_242_appendix}
\end{figure}
\begin{figure}
\centering
\includegraphics[width=0.85\linewidth]{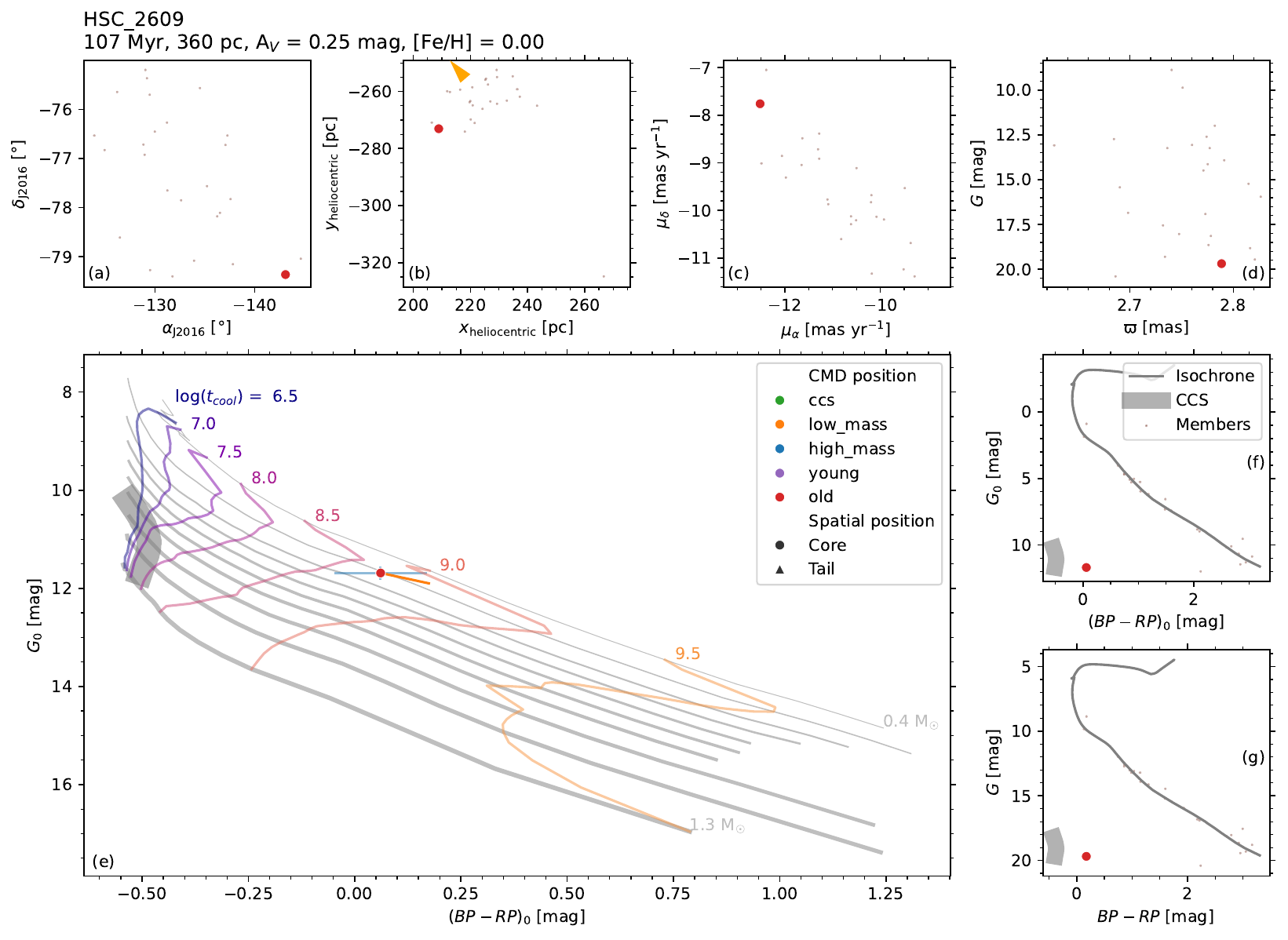}
\includegraphics[width=0.85\linewidth]{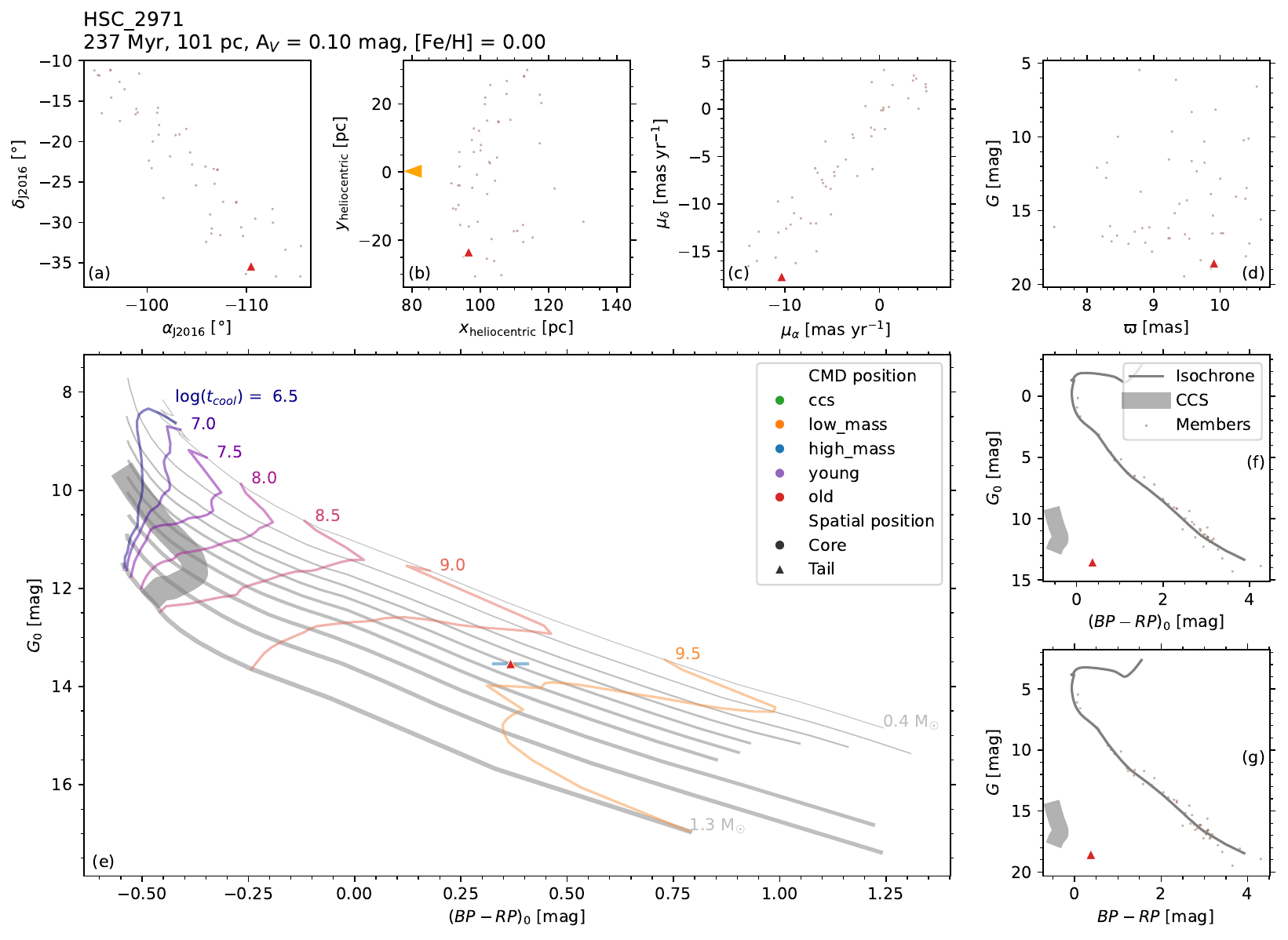}
\caption{Diagnostic plots for HSC 2609 and HSC 2971. All details are similar to Figure~\ref{fig:combo_Melotte_25}.}
\label{fig:combo_HSC_2971_appendix}
\end{figure}
\begin{figure}
\centering
\includegraphics[width=0.85\linewidth]{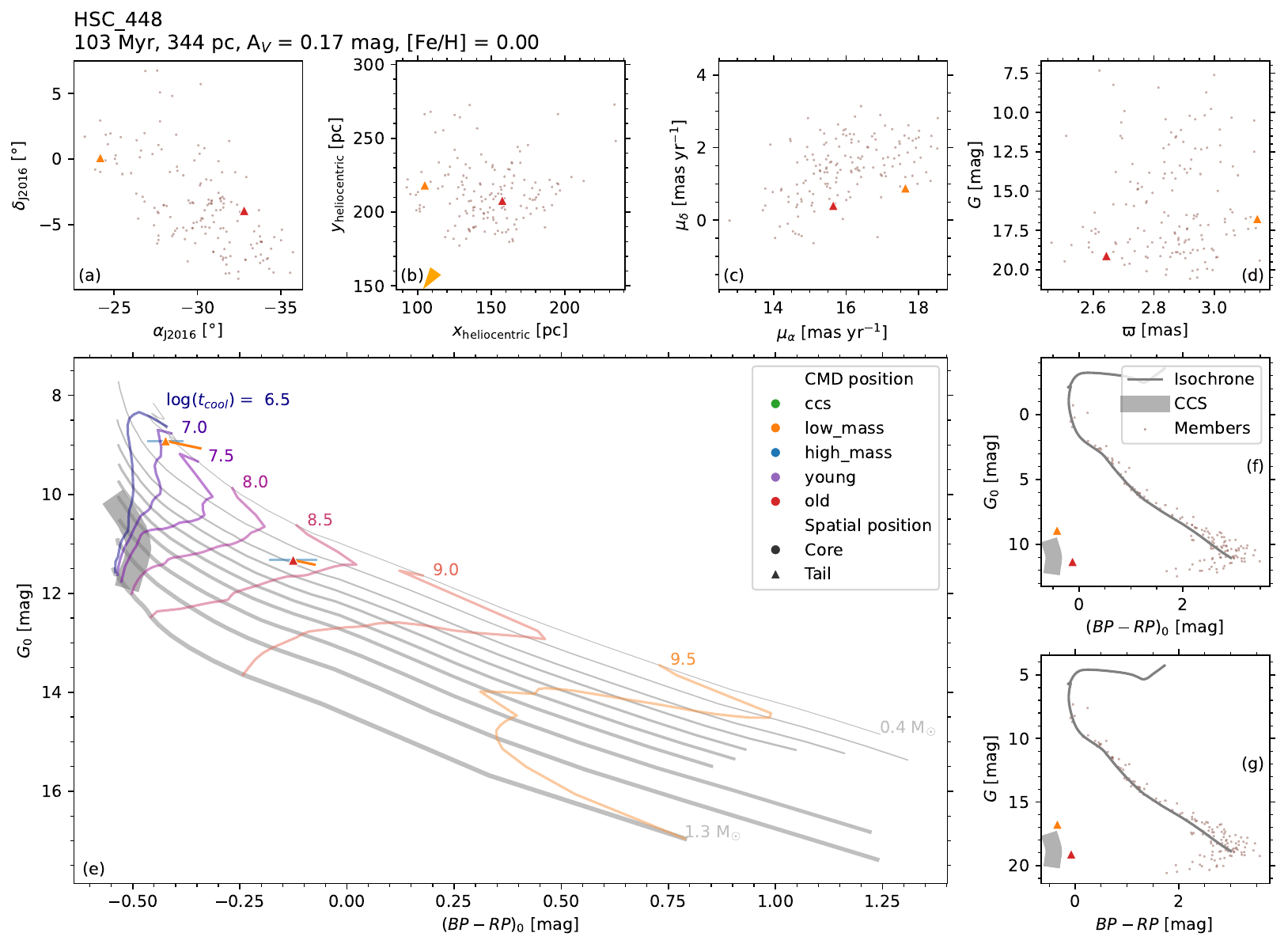}
\includegraphics[width=0.85\linewidth]{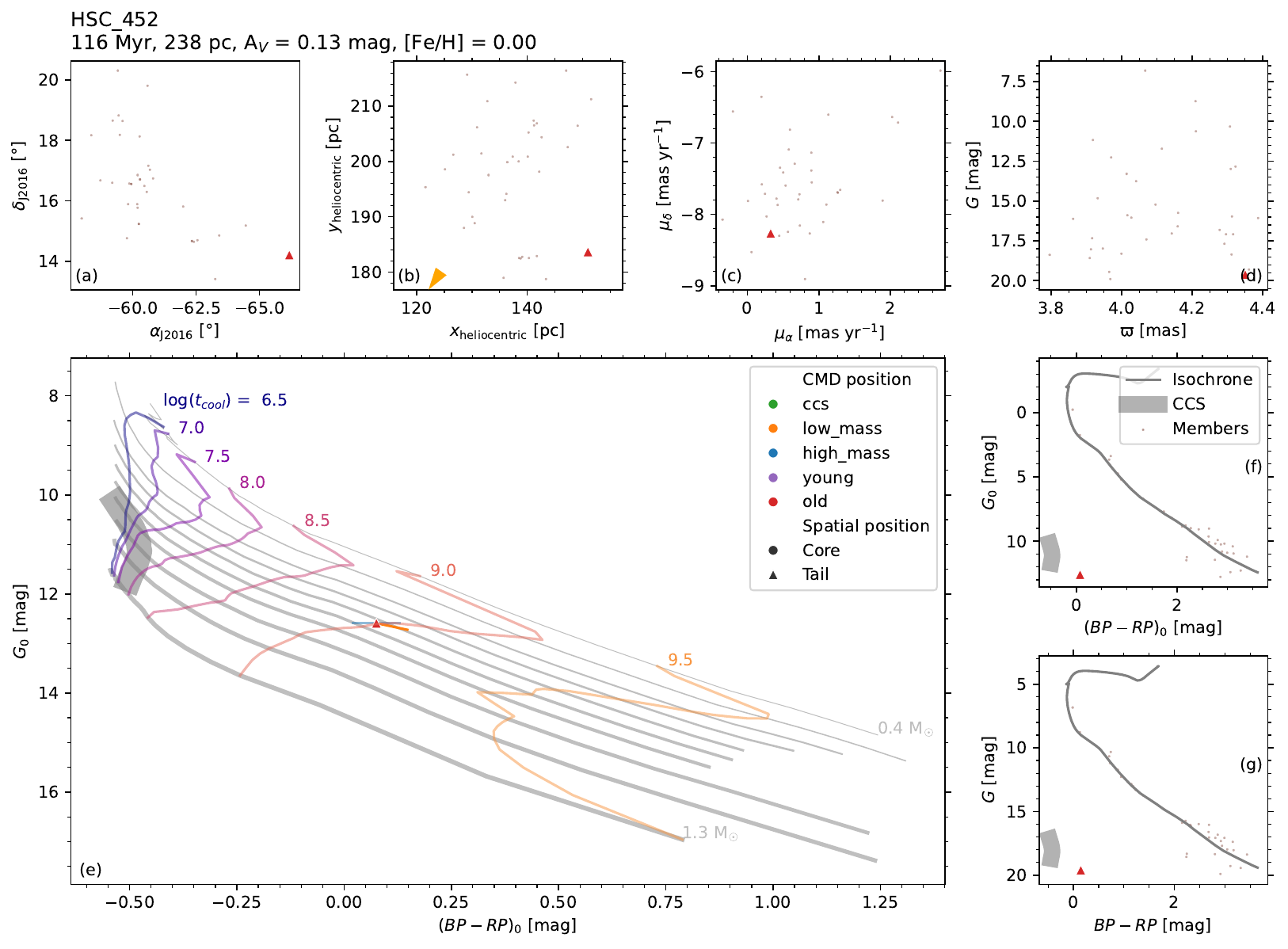}
\caption{Diagnostic plots for HSC 448 and HSC 452. All details are similar to Figure~\ref{fig:combo_Melotte_25}.}
\label{fig:combo_HSC_452_appendix}
\end{figure}
\begin{figure}
\centering
\includegraphics[width=0.85\linewidth]{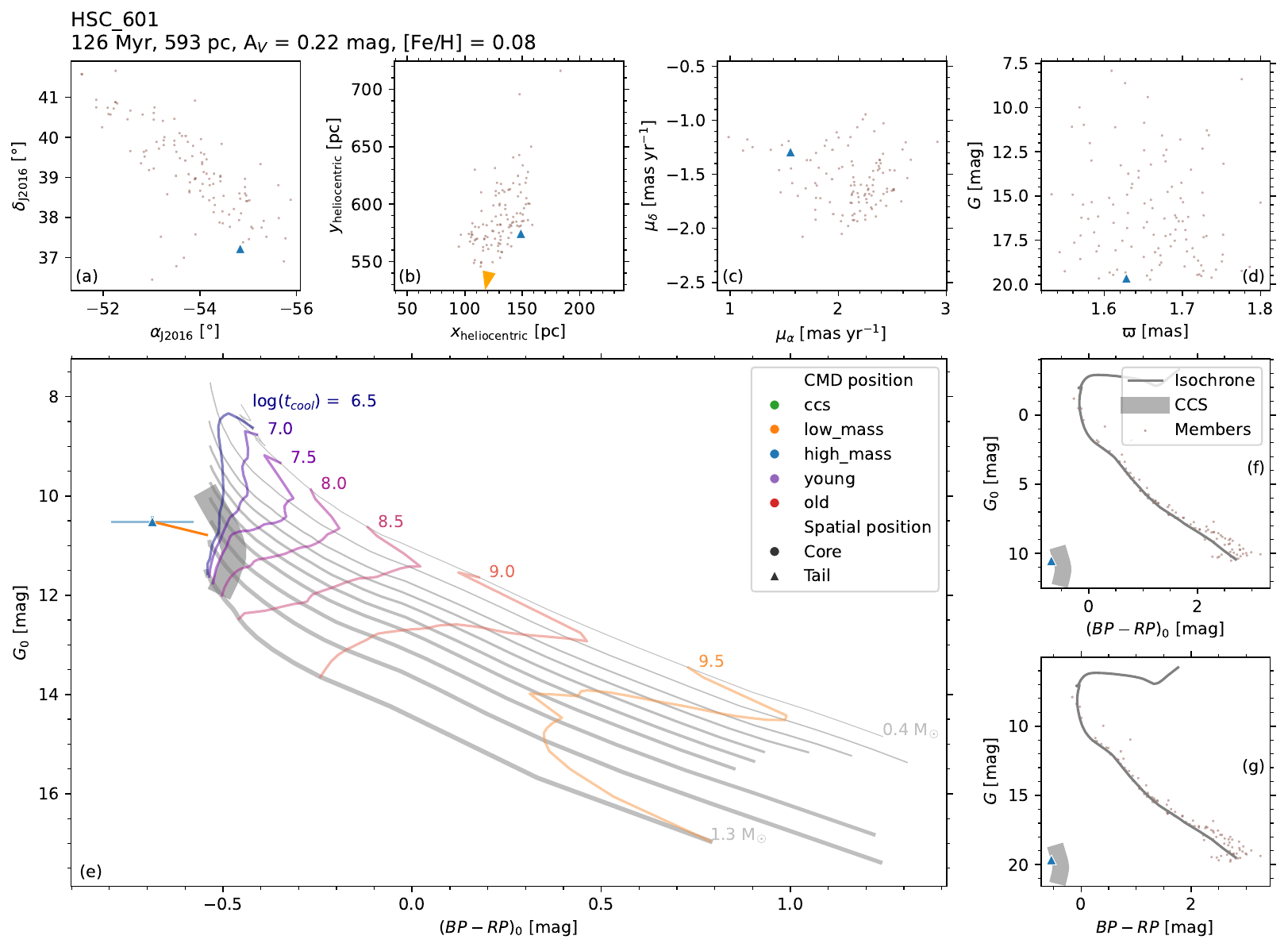}
\includegraphics[width=0.85\linewidth]{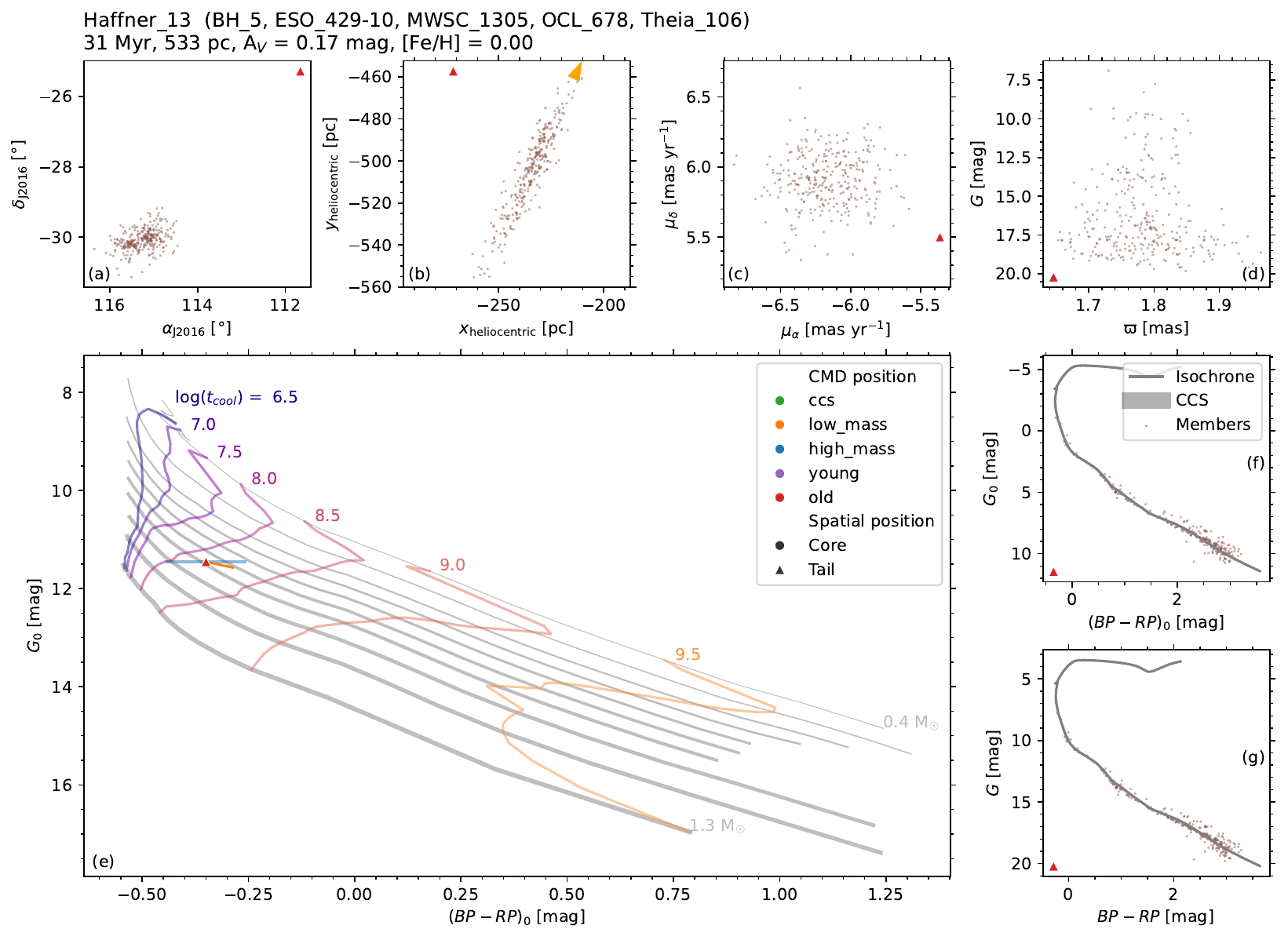}
\caption{Diagnostic plots for HSC 601 and Haffner 13. All details are similar to Figure~\ref{fig:combo_Melotte_25}.}
\label{fig:combo_Haffner_13_appendix}
\end{figure}
\begin{figure}
\centering
\includegraphics[width=0.85\linewidth]{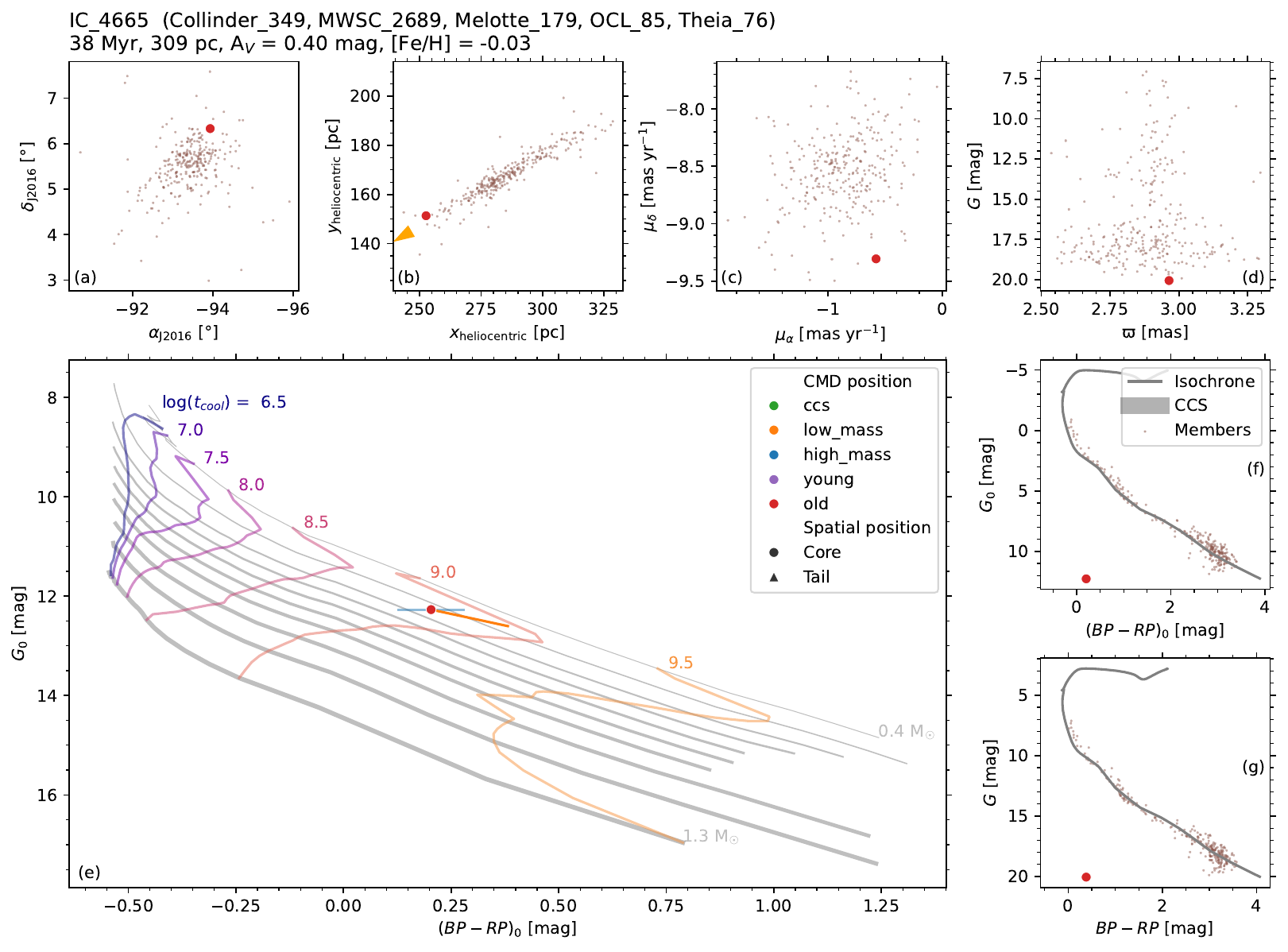}
\includegraphics[width=0.85\linewidth]{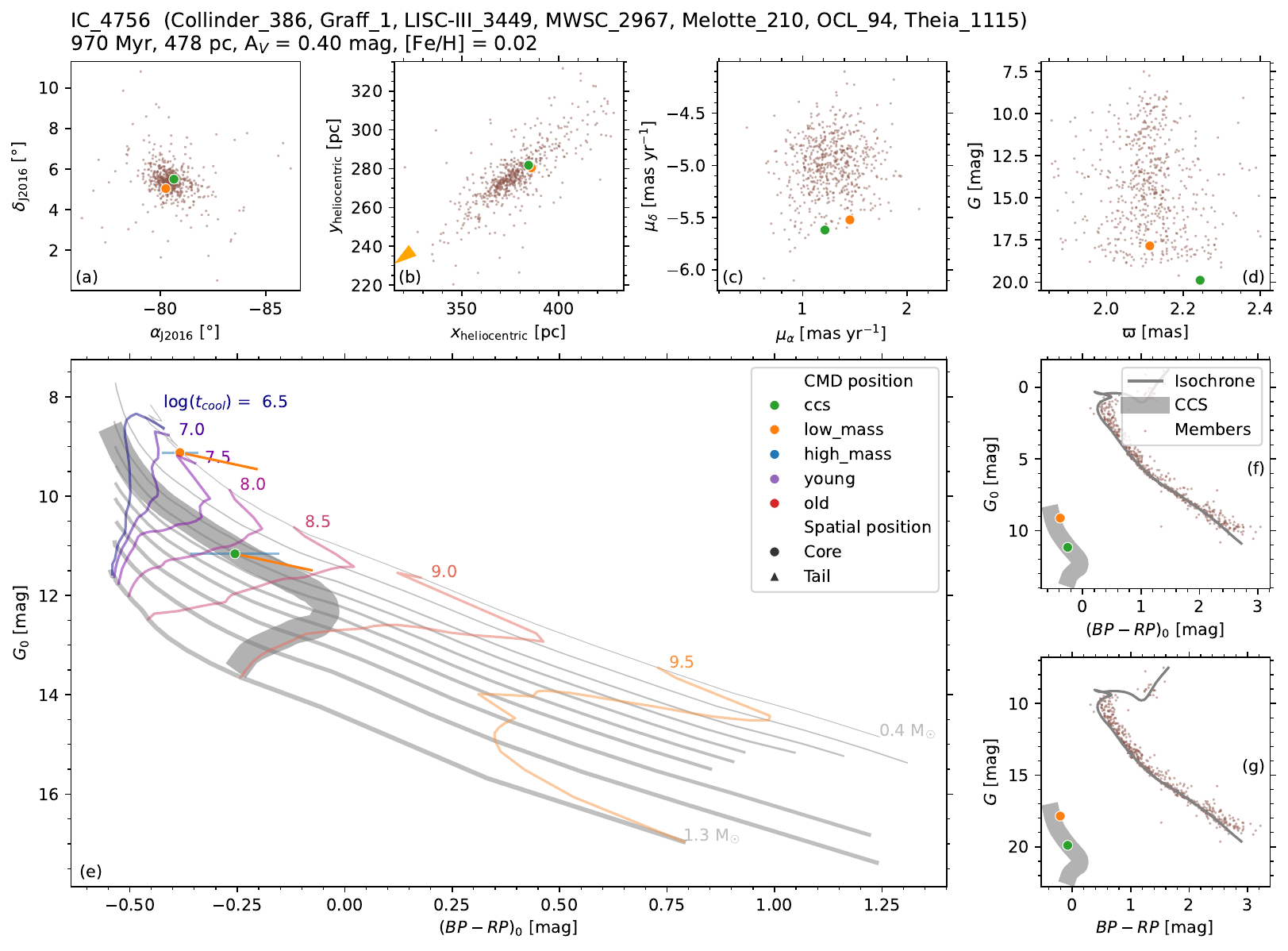}
\caption{Diagnostic plots for IC 4665 and IC 4756. All details are similar to Figure~\ref{fig:combo_Melotte_25}.}
\label{fig:combo_IC_4756_appendix}
\end{figure}
\begin{figure}
\centering
\includegraphics[width=0.85\linewidth]{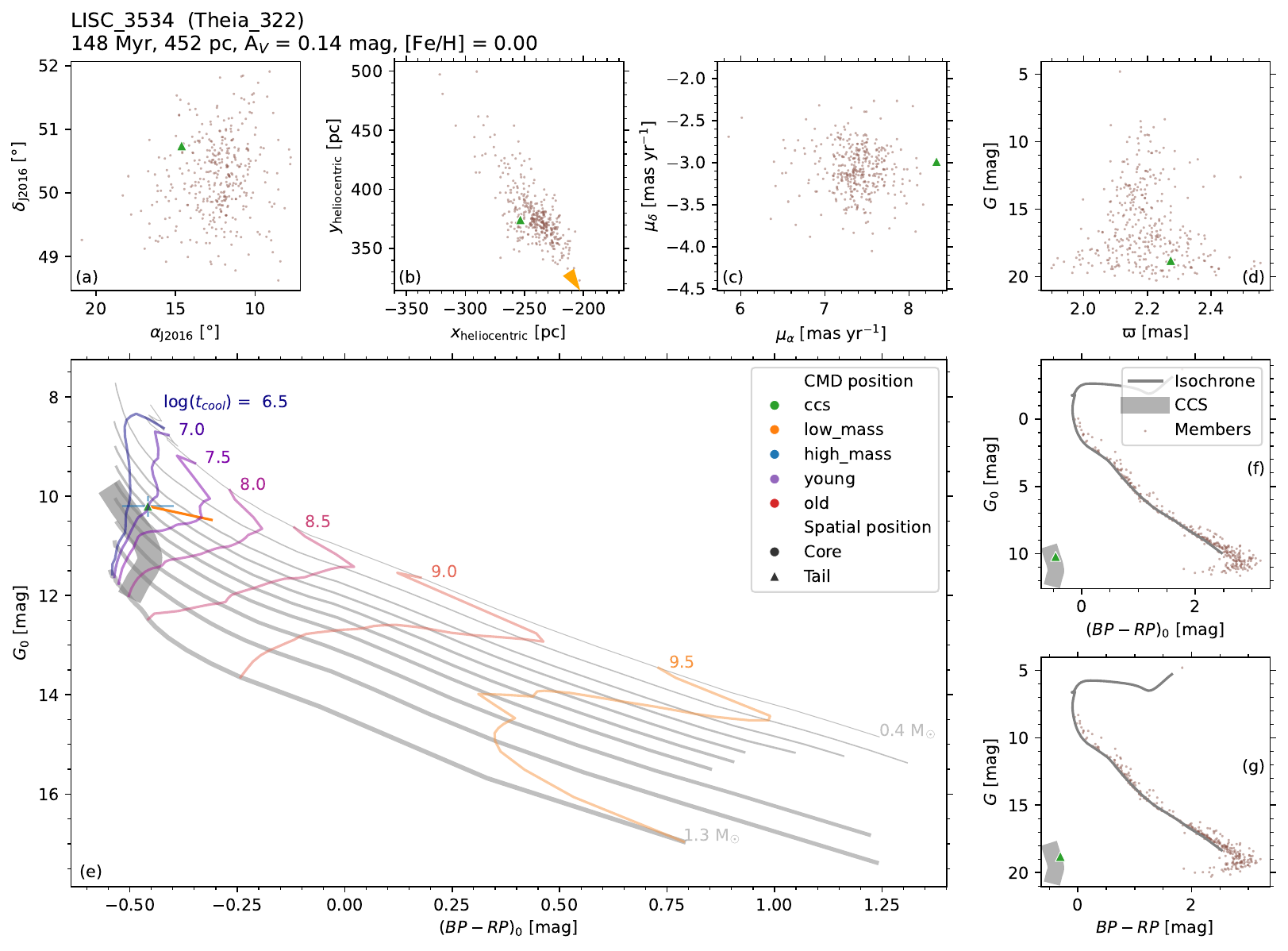}
\includegraphics[width=0.85\linewidth]{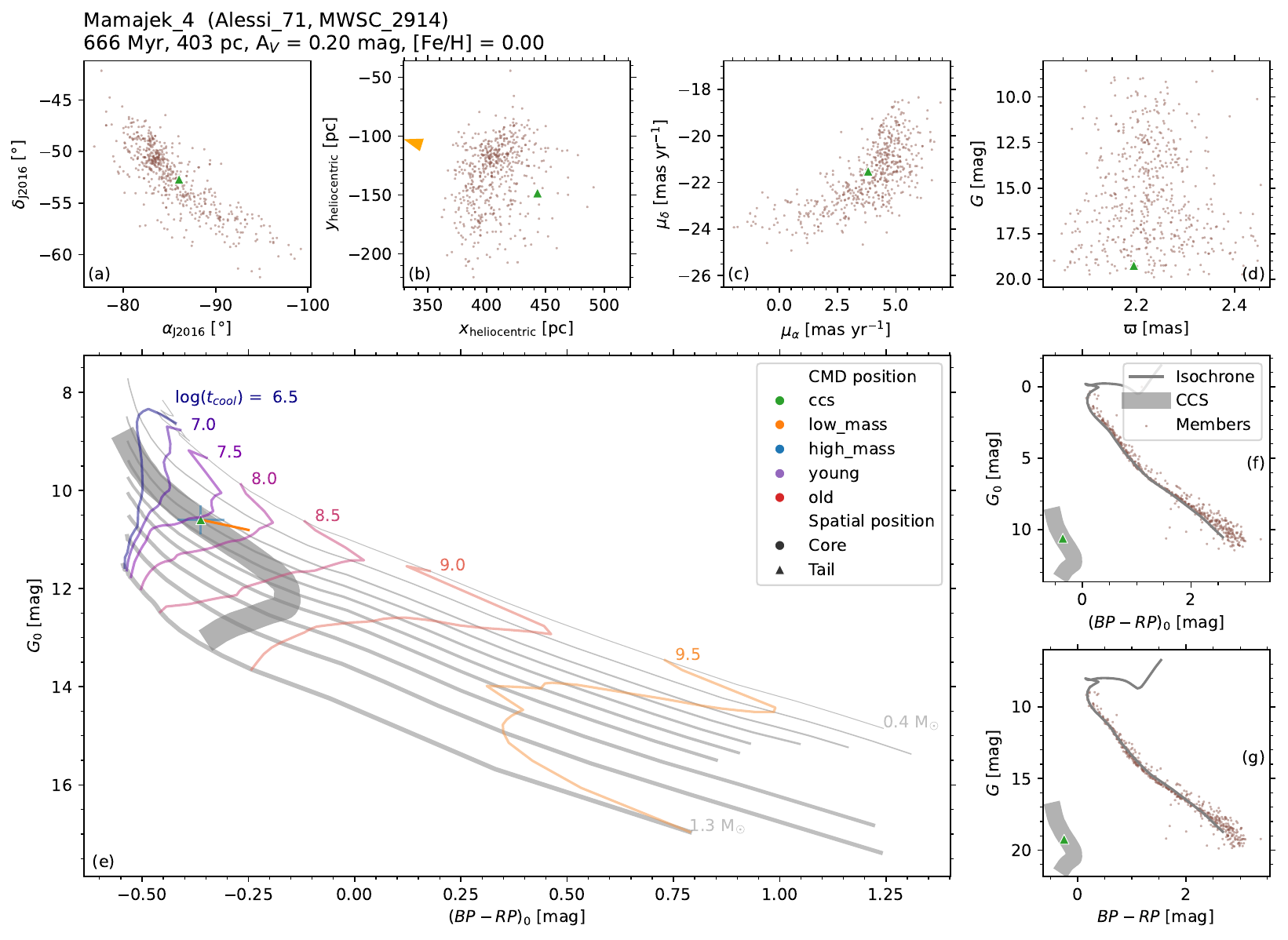}
\caption{Diagnostic plots for LISC 3534 and Mamajek 4. All details are similar to Figure~\ref{fig:combo_Melotte_25}.}
\label{fig:combo_Mamajek_4_appendix}
\end{figure}
\begin{figure}
\centering
\includegraphics[width=0.85\linewidth]{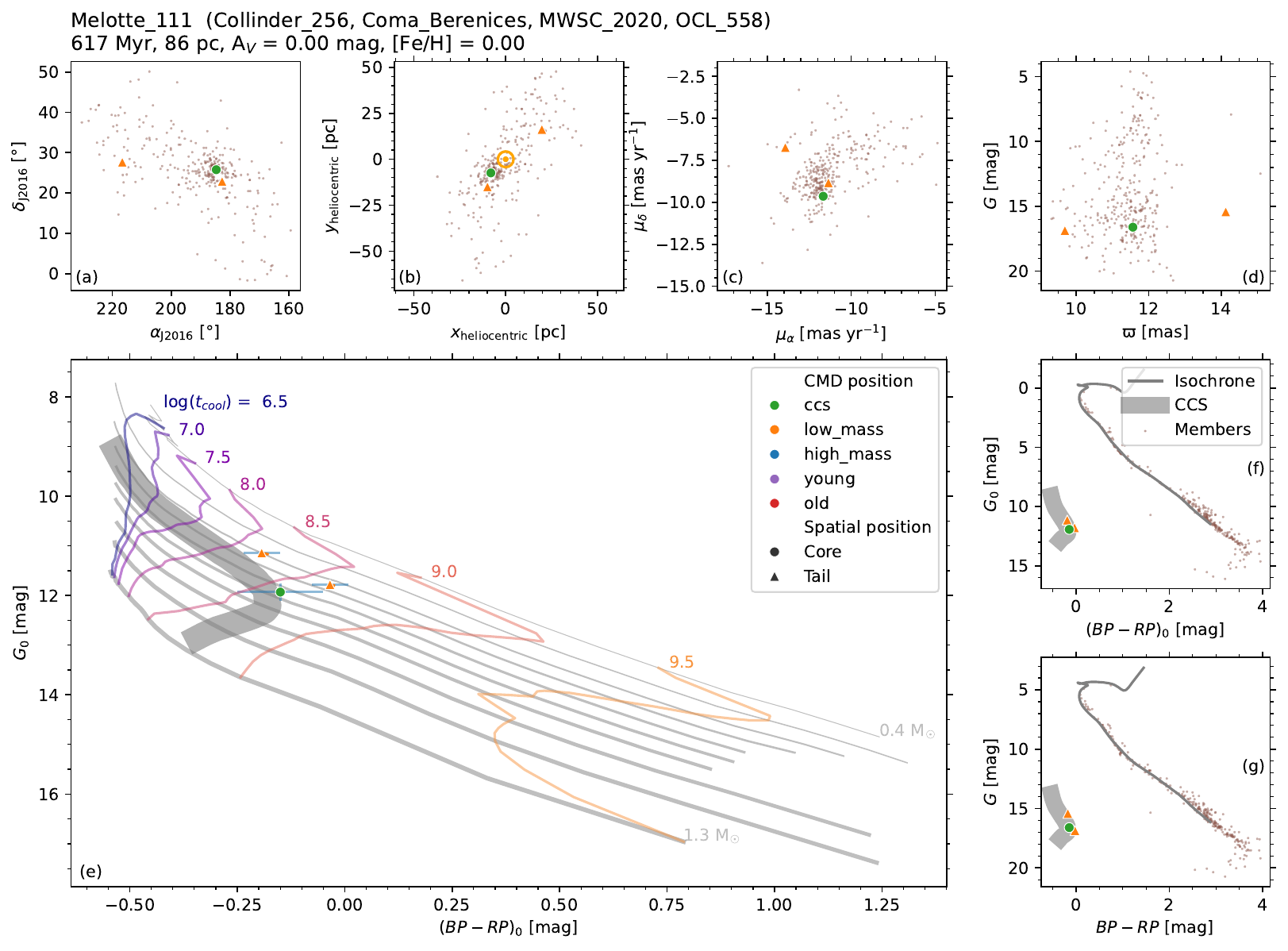}
\includegraphics[width=0.85\linewidth]{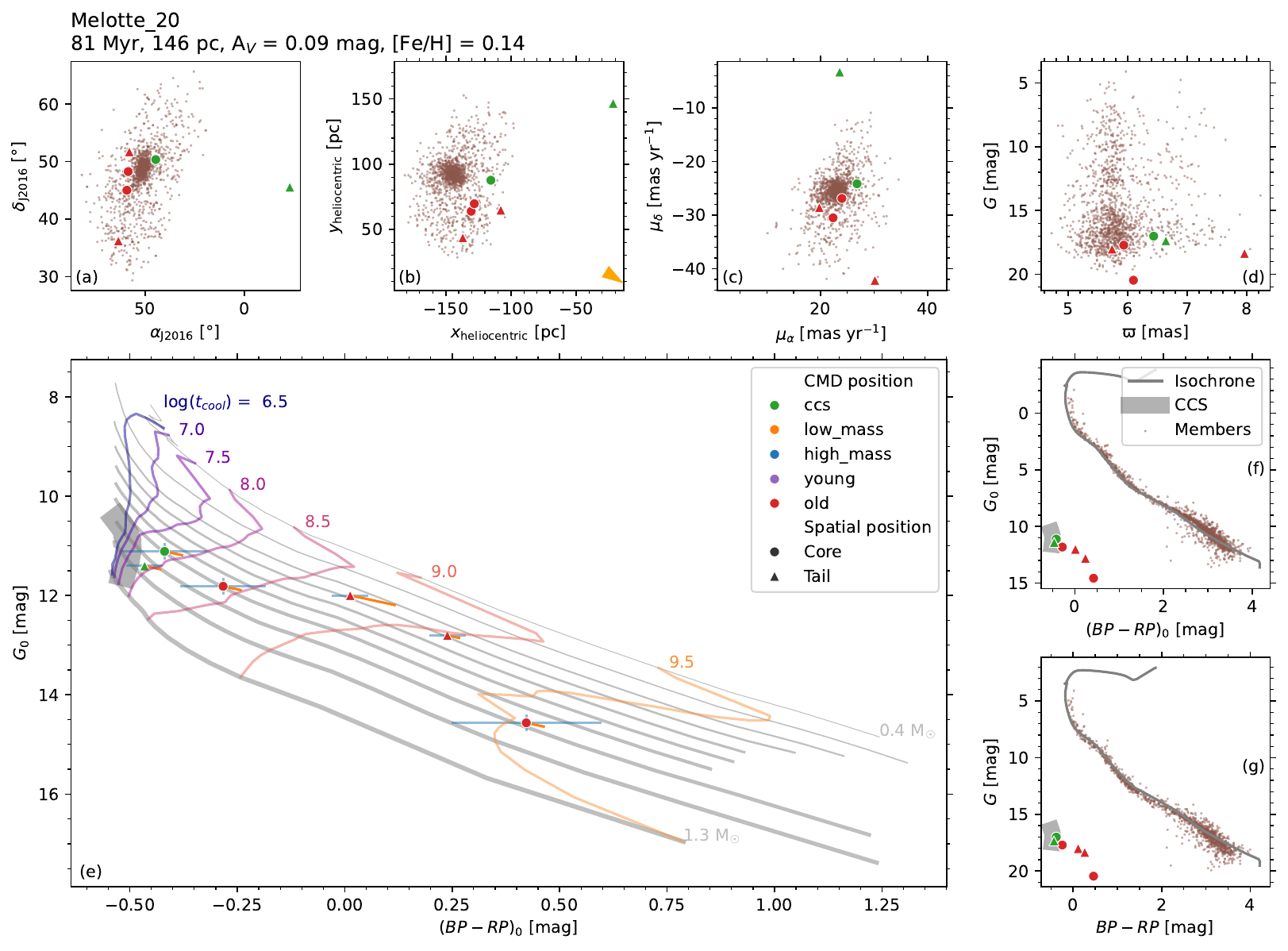}
\caption{Diagnostic plots for Melotte 111 and Melotte 20. All details are similar to Figure~\ref{fig:combo_Melotte_25}.}
\label{fig:combo_Melotte_20_appendix}
\end{figure}
\begin{figure}
\centering
\includegraphics[width=0.85\linewidth]{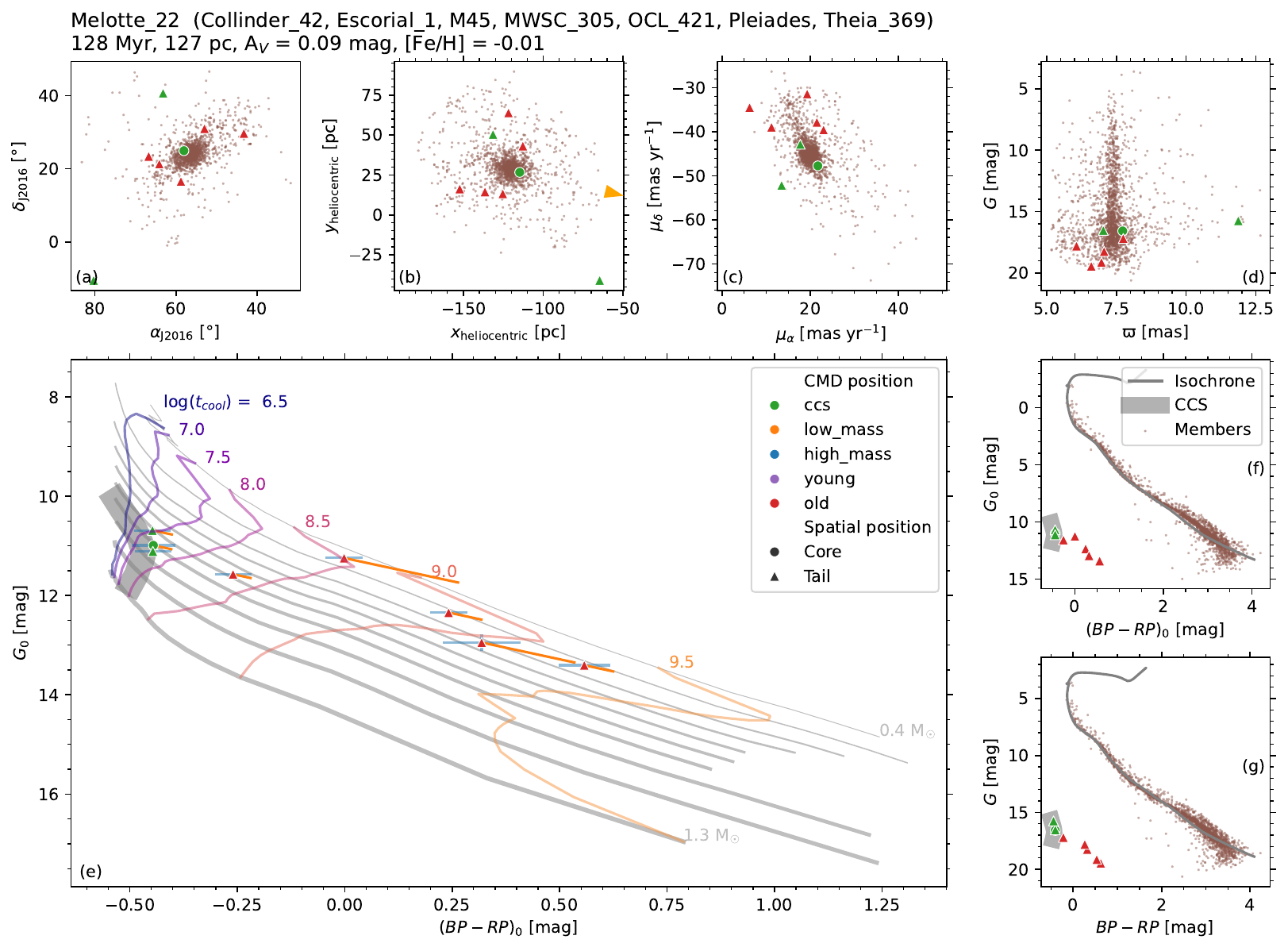}
\includegraphics[width=0.85\linewidth]{combo_figures/Melotte_25.pdf}
\caption{Diagnostic plots for Melotte 22 and Melotte 25. All details are similar to Figure~\ref{fig:combo_Melotte_25}.}
\label{fig:combo_Melotte_25_appendix}
\end{figure}
\begin{figure}
\centering
\includegraphics[width=0.85\linewidth]{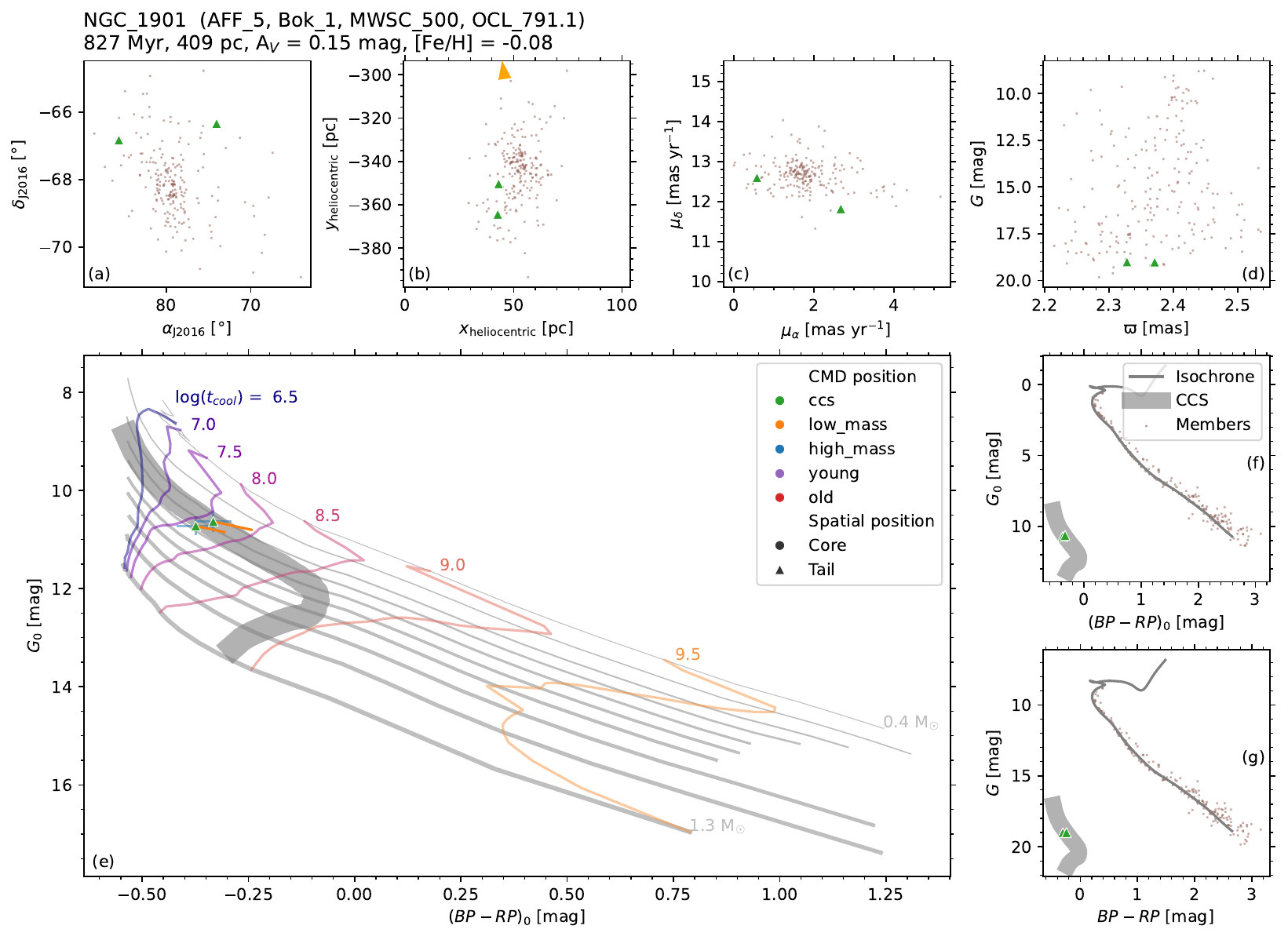}
\includegraphics[width=0.85\linewidth]{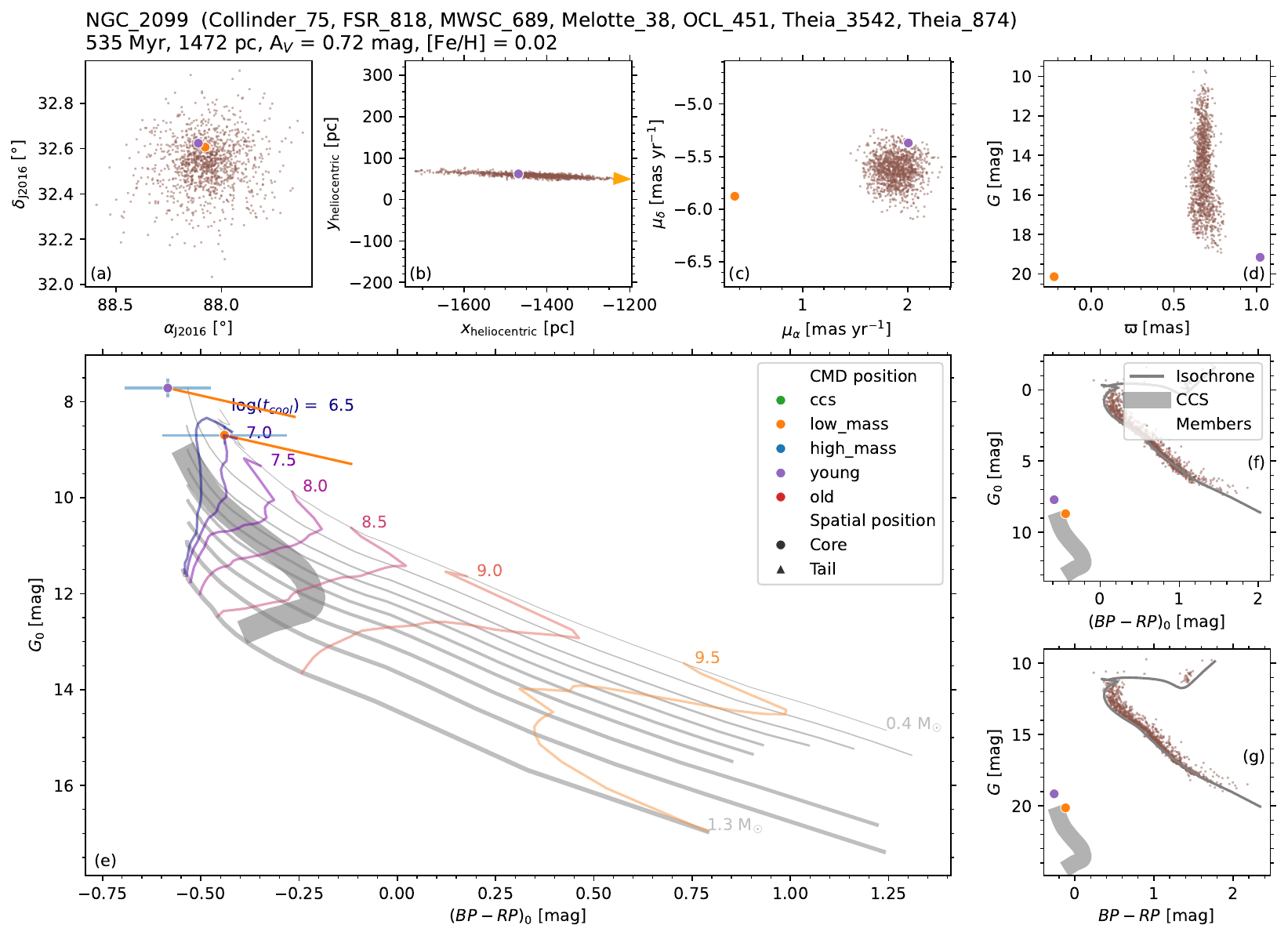}
\caption{Diagnostic plots for NGC 1901 and NGC 2099. All details are similar to Figure~\ref{fig:combo_Melotte_25}.}
\label{fig:combo_NGC_2099_appendix}
\end{figure}
\begin{figure}
\centering
\includegraphics[width=0.85\linewidth]{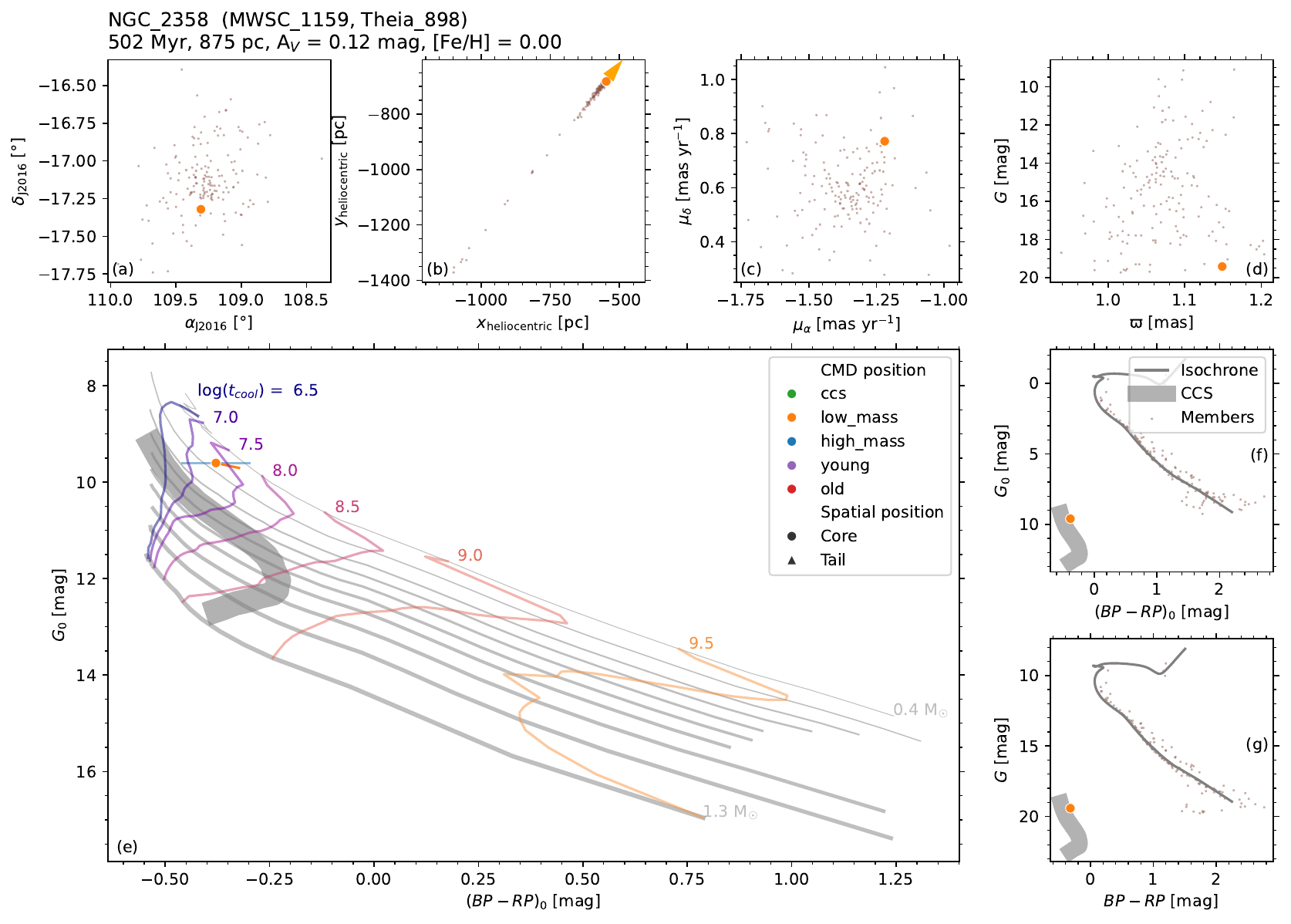}
\includegraphics[width=0.85\linewidth]{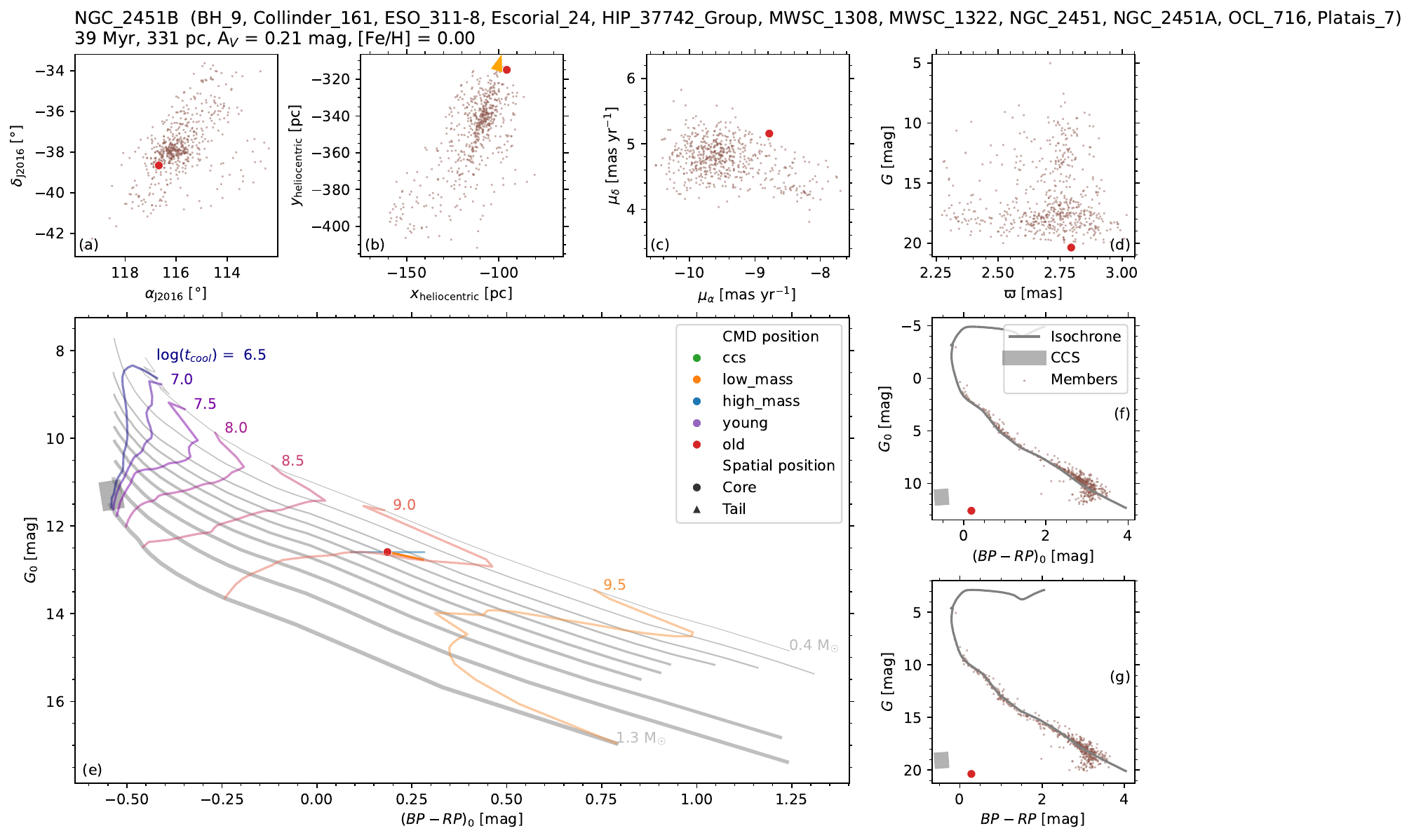}
\caption{Diagnostic plots for NGC 2358 and NGC 2451B. All details are similar to Figure~\ref{fig:combo_Melotte_25}.}
\label{fig:combo_NGC_2451B_appendix}
\end{figure}
\begin{figure}
\centering
\includegraphics[width=0.85\linewidth]{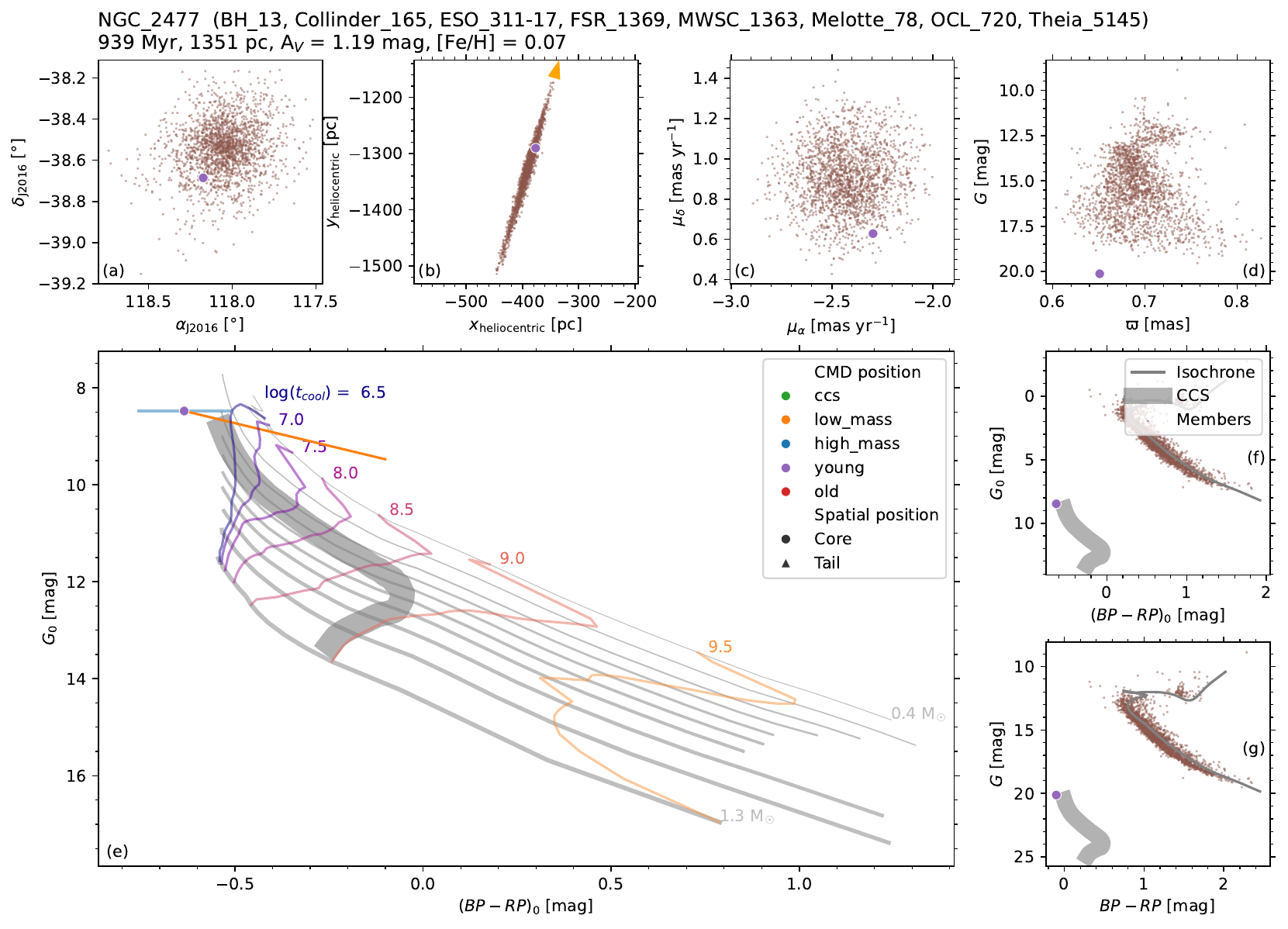}
\includegraphics[width=0.85\linewidth]{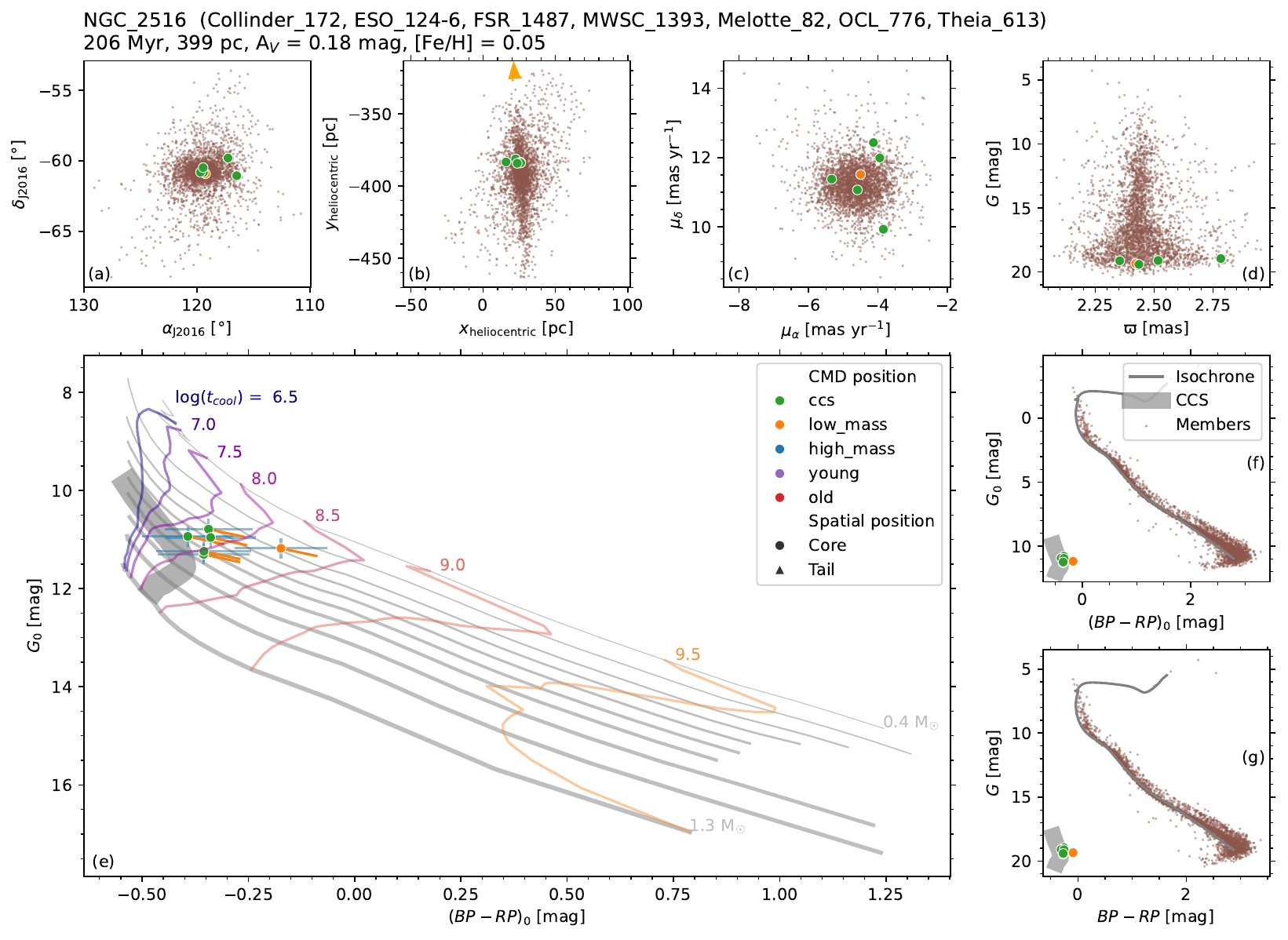}
\caption{Diagnostic plots for NGC 2477 and NGC 2516. All details are similar to Figure~\ref{fig:combo_Melotte_25}.}
\label{fig:combo_NGC_2516_appendix}
\end{figure}
\begin{figure}
\centering
\includegraphics[width=0.85\linewidth]{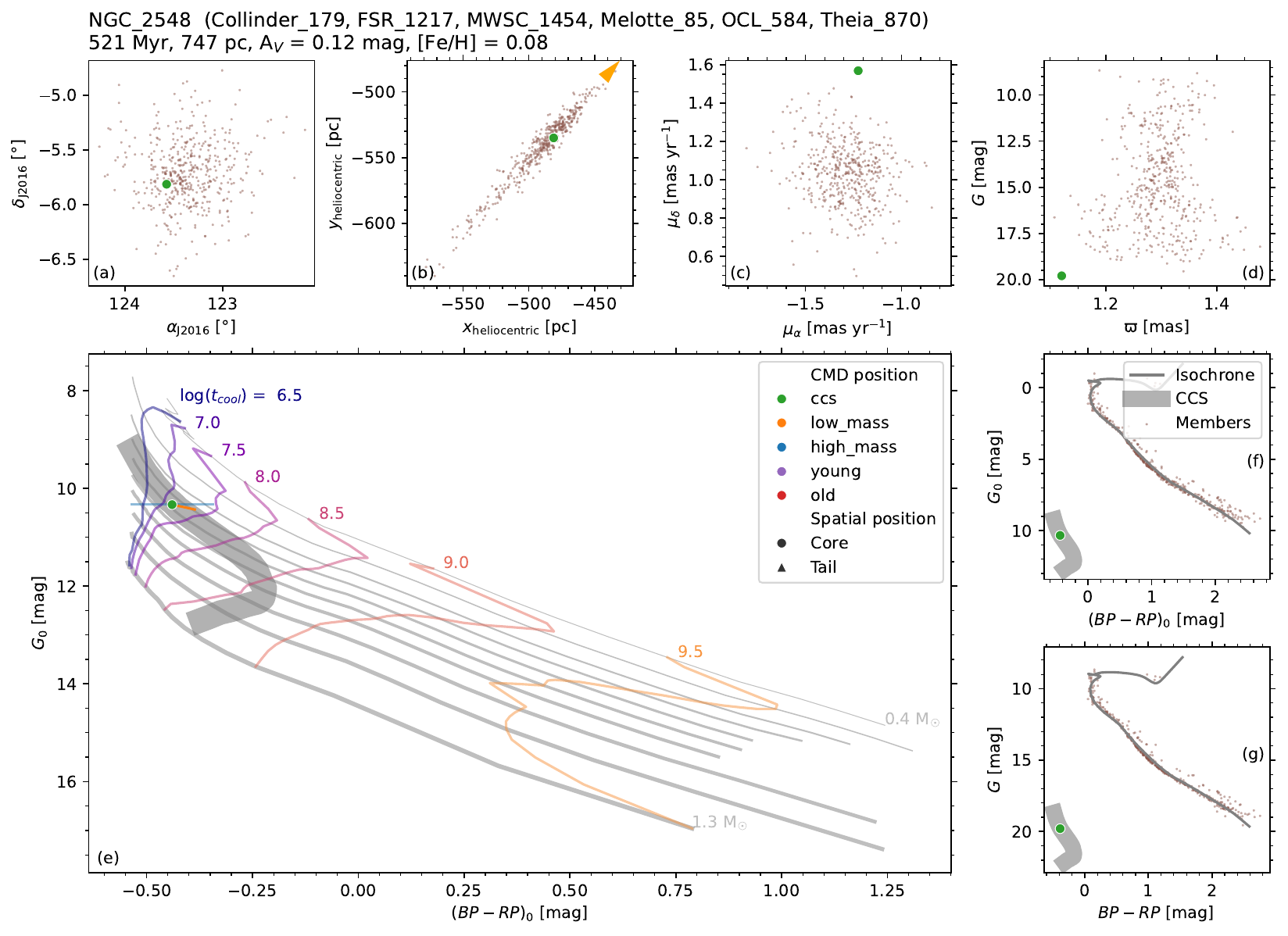}
\includegraphics[width=0.85\linewidth]{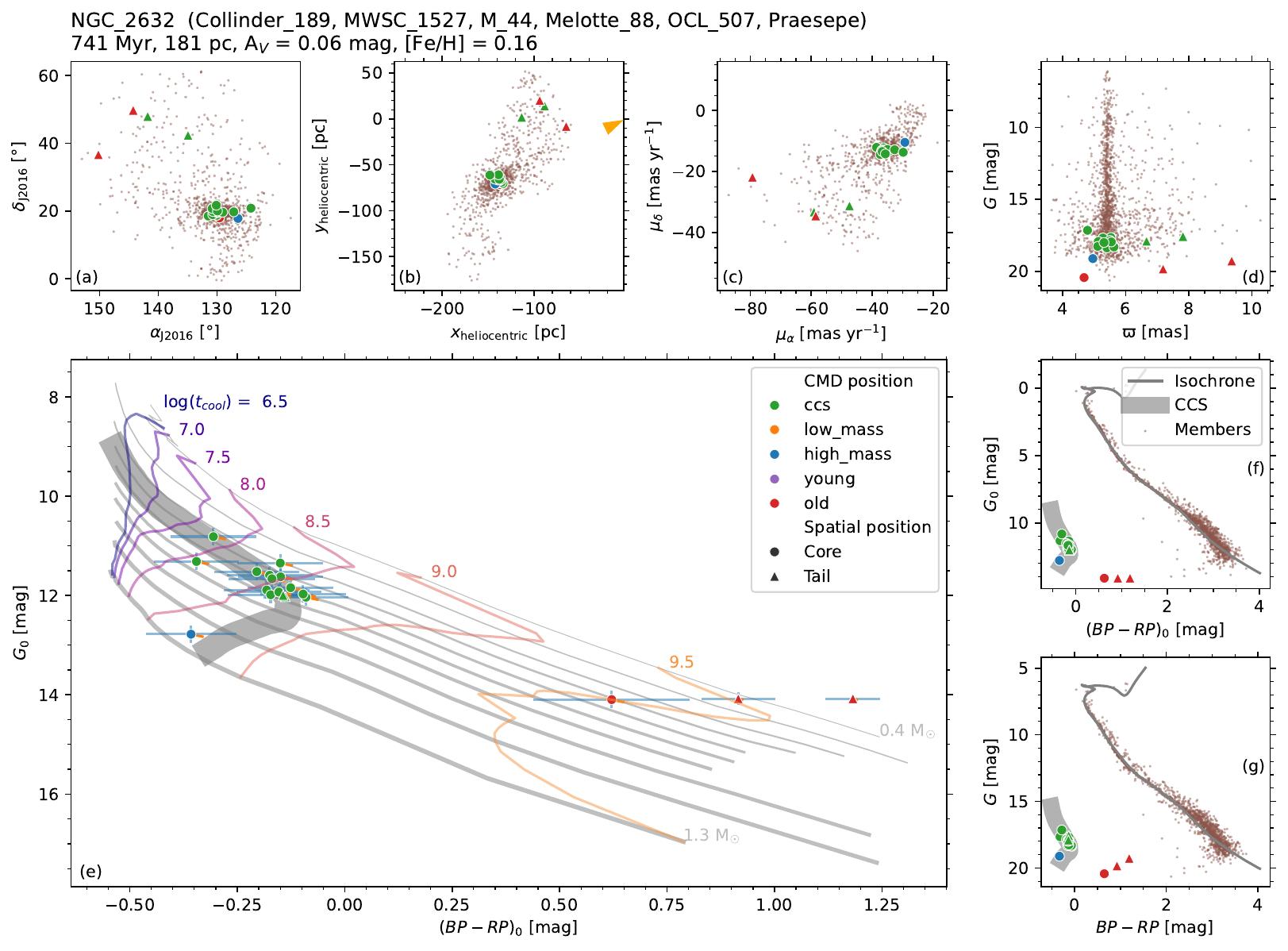}
\caption{Diagnostic plots for NGC 2548 and NGC 2632. All details are similar to Figure~\ref{fig:combo_Melotte_25}.}
\label{fig:combo_NGC_2632_appendix}
\end{figure}
\begin{figure}
\centering
\includegraphics[width=0.85\linewidth]{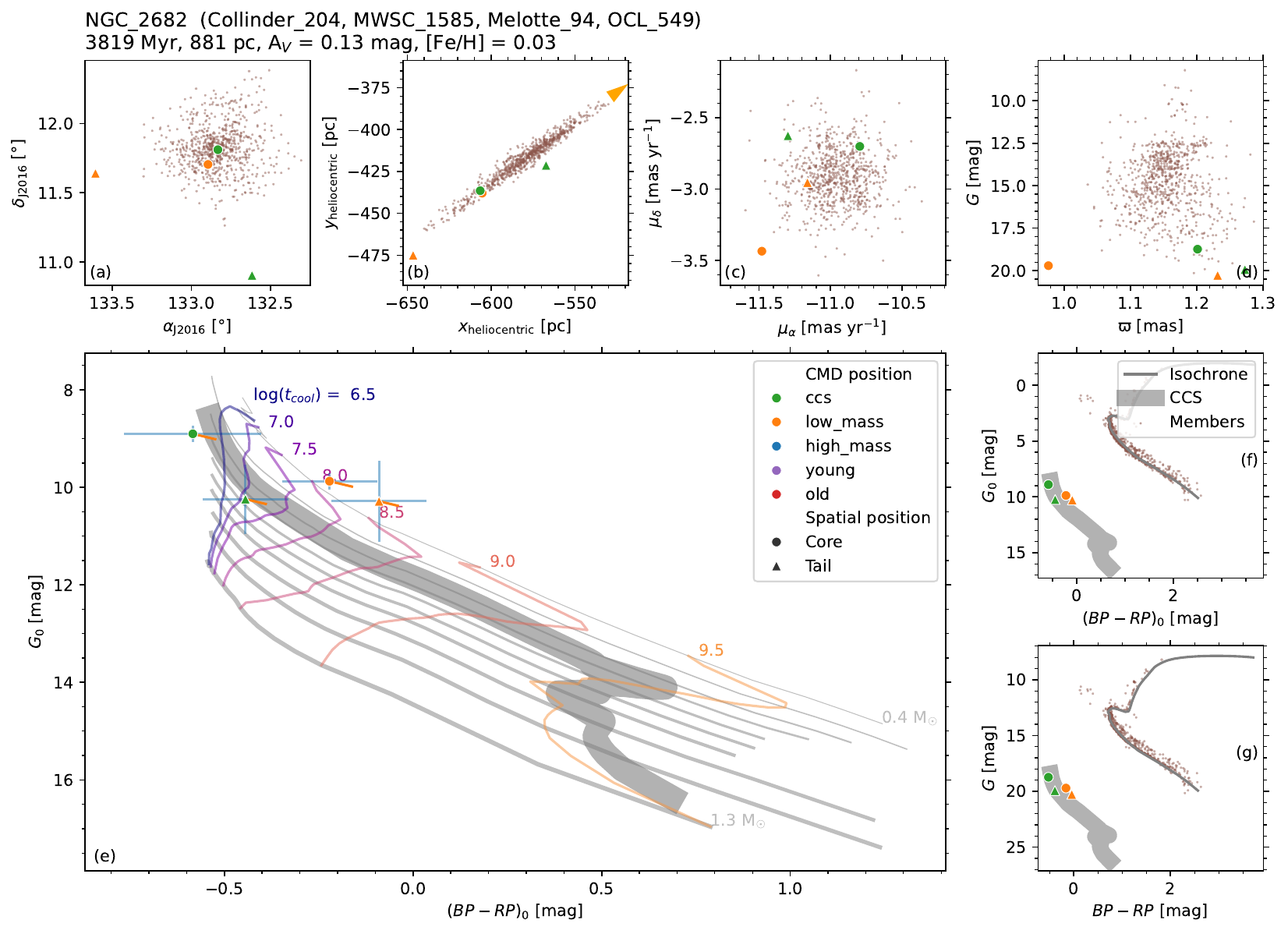}
\includegraphics[width=0.85\linewidth]{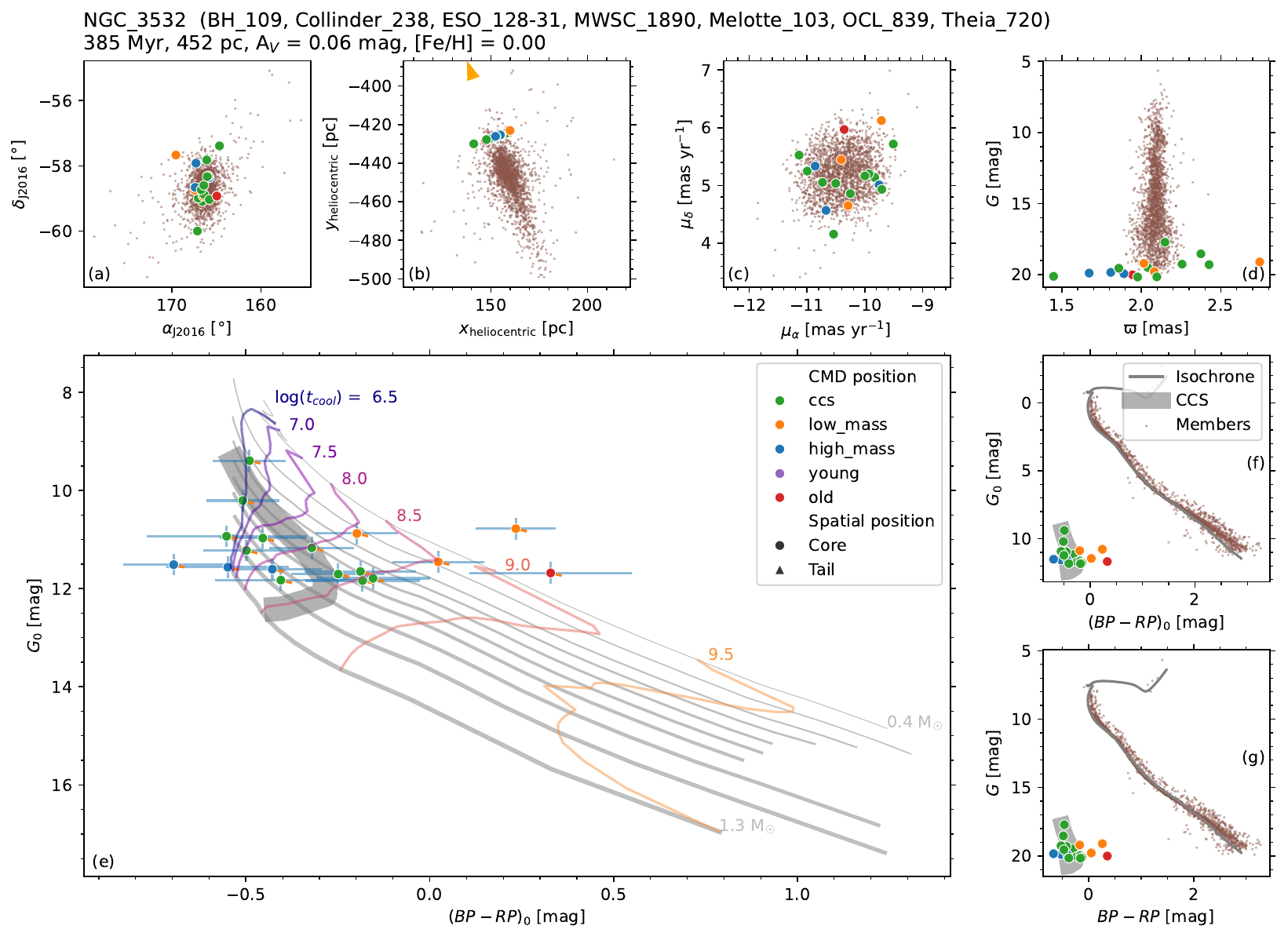}
\caption{Diagnostic plots for NGC 2682 and NGC 3532. All details are similar to Figure~\ref{fig:combo_Melotte_25}.}
\label{fig:combo_NGC_3532_appendix}
\end{figure}
\begin{figure}
\centering
\includegraphics[width=0.85\linewidth]{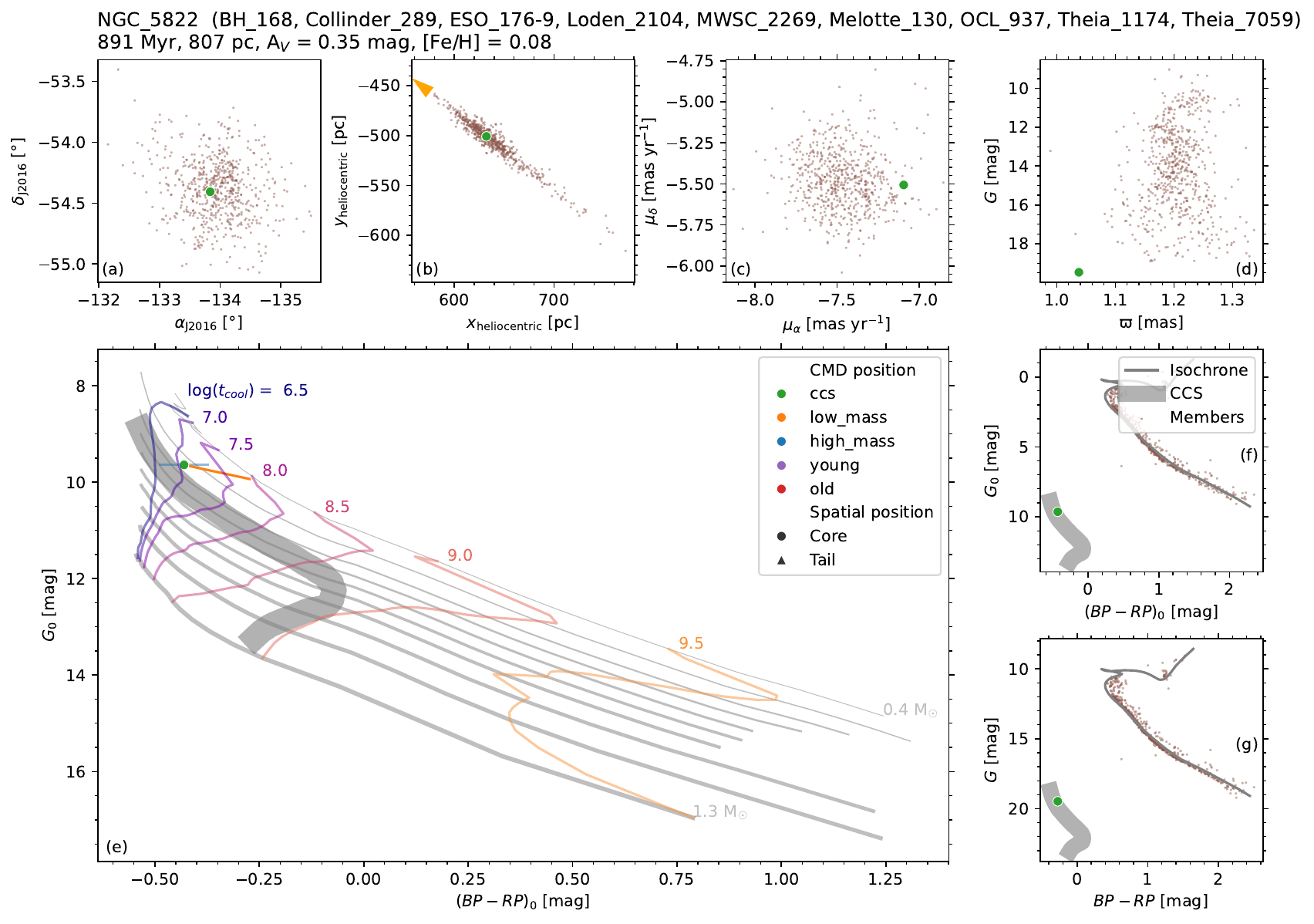}
\includegraphics[width=0.85\linewidth]{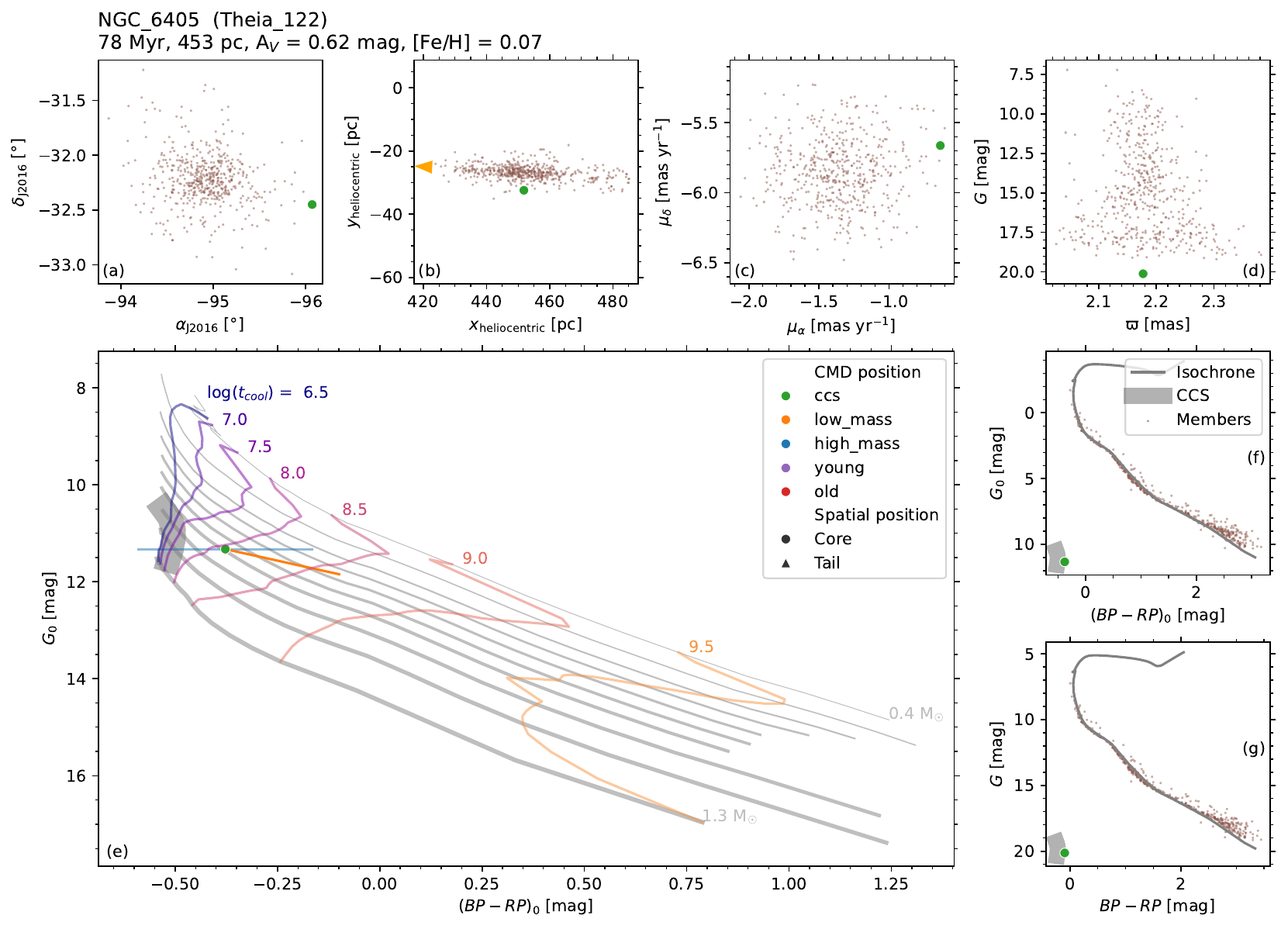}
\caption{Diagnostic plots for NGC 5822 and NGC 6405. All details are similar to Figure~\ref{fig:combo_Melotte_25}.}
\label{fig:combo_NGC_6405_appendix}
\end{figure}
\begin{figure}
\centering
\includegraphics[width=0.85\linewidth]{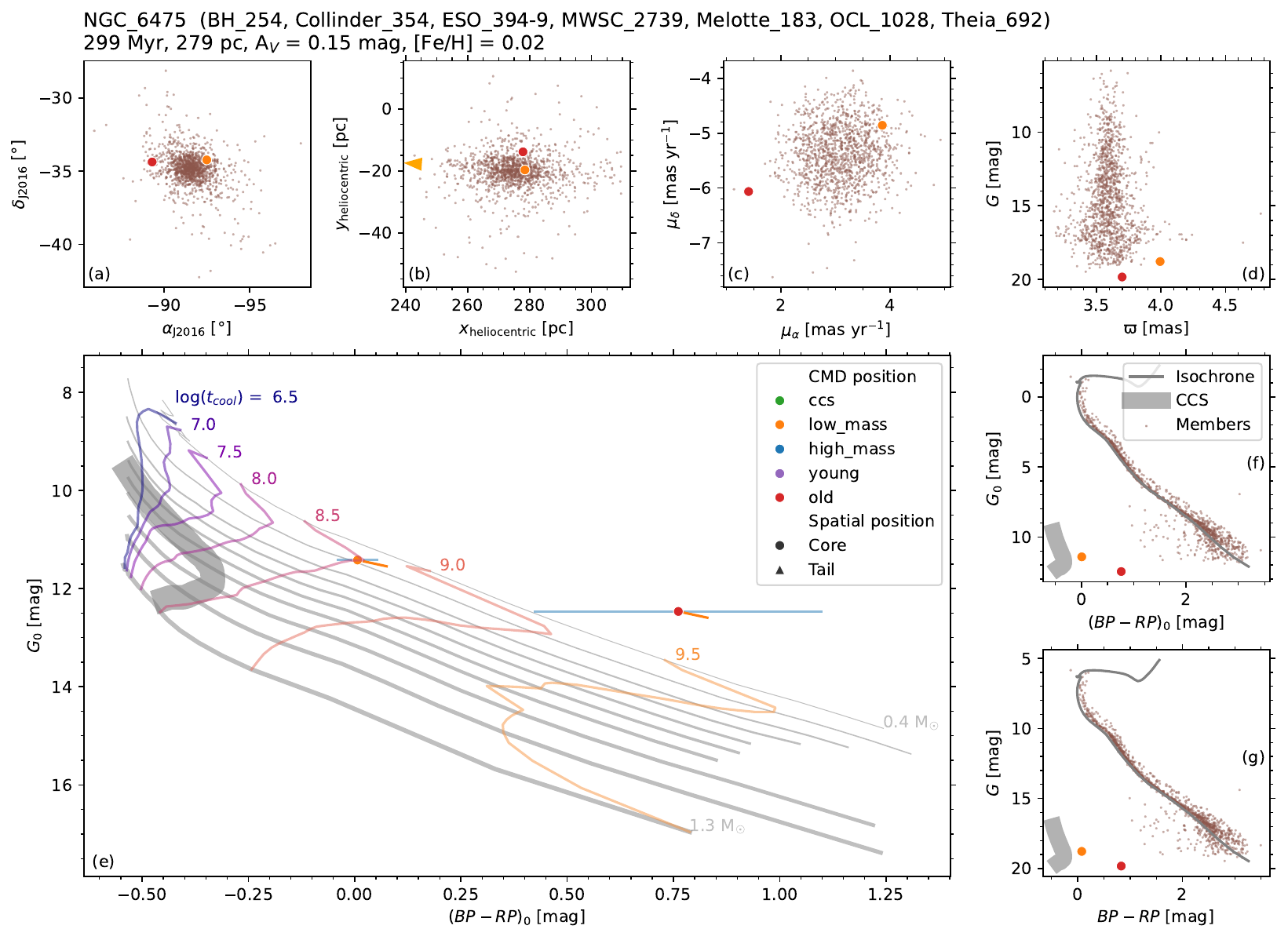}
\includegraphics[width=0.85\linewidth]{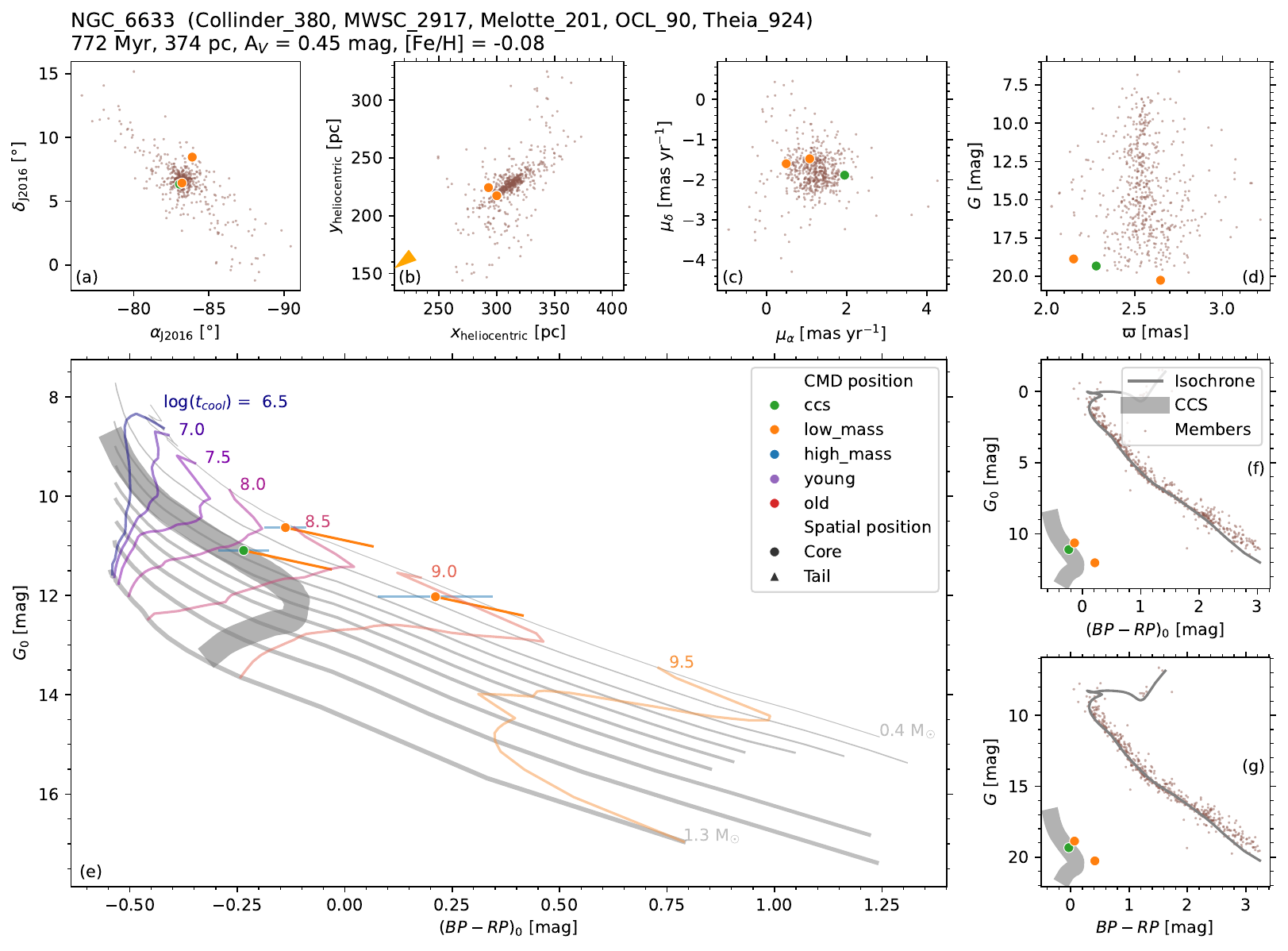}
\caption{Diagnostic plots for NGC 6475 and NGC 6633. All details are similar to Figure~\ref{fig:combo_Melotte_25}.}
\label{fig:combo_NGC_6633_appendix}
\end{figure}
\begin{figure}
\centering
\includegraphics[width=0.85\linewidth]{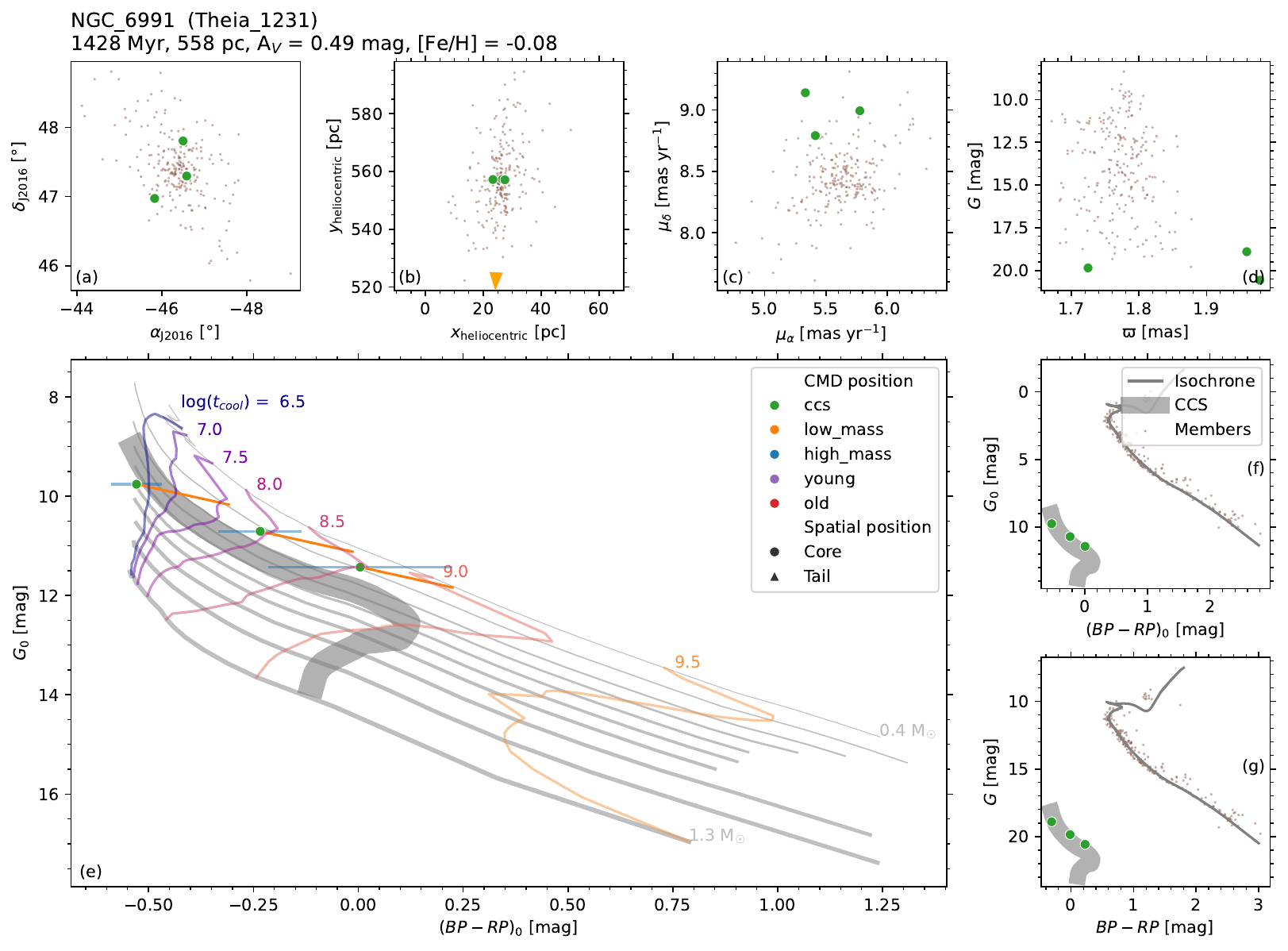}
\includegraphics[width=0.85\linewidth]{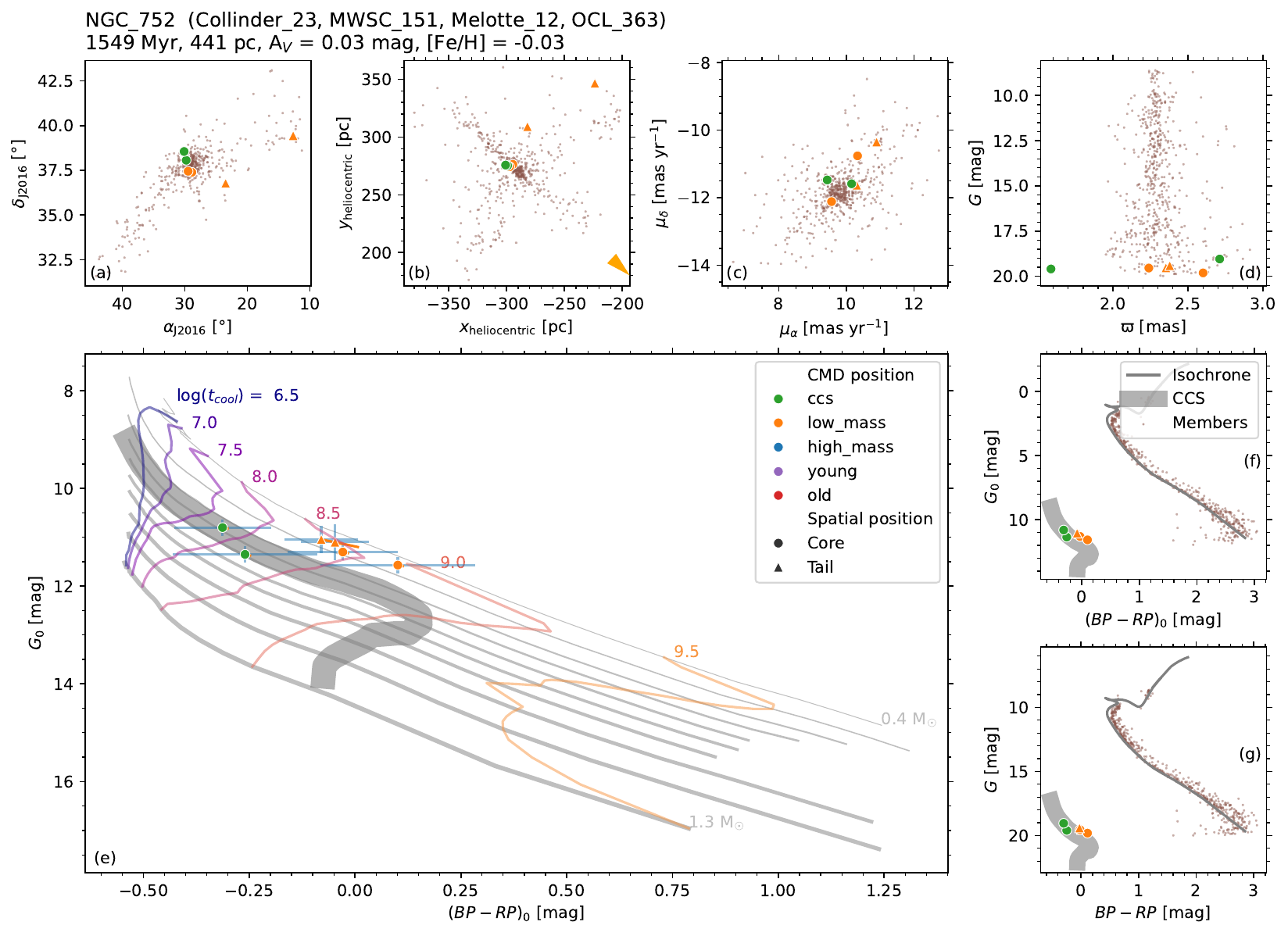}
\caption{Diagnostic plots for NGC 6991 and NGC 752. All details are similar to Figure~\ref{fig:combo_Melotte_25}.}
\label{fig:combo_NGC_752_appendix}
\end{figure}
\begin{figure}
\centering
\includegraphics[width=0.85\linewidth]{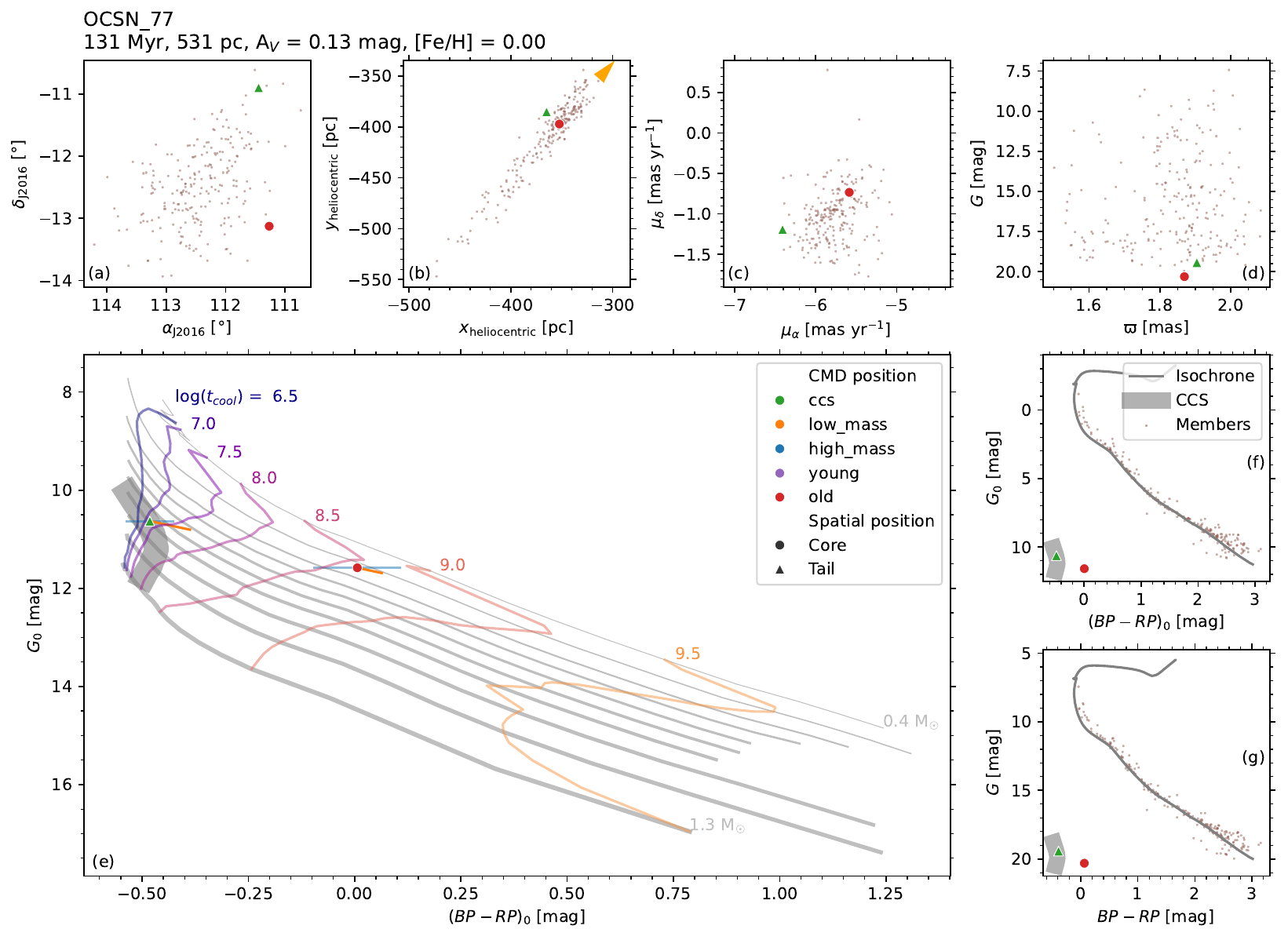}
\includegraphics[width=0.85\linewidth]{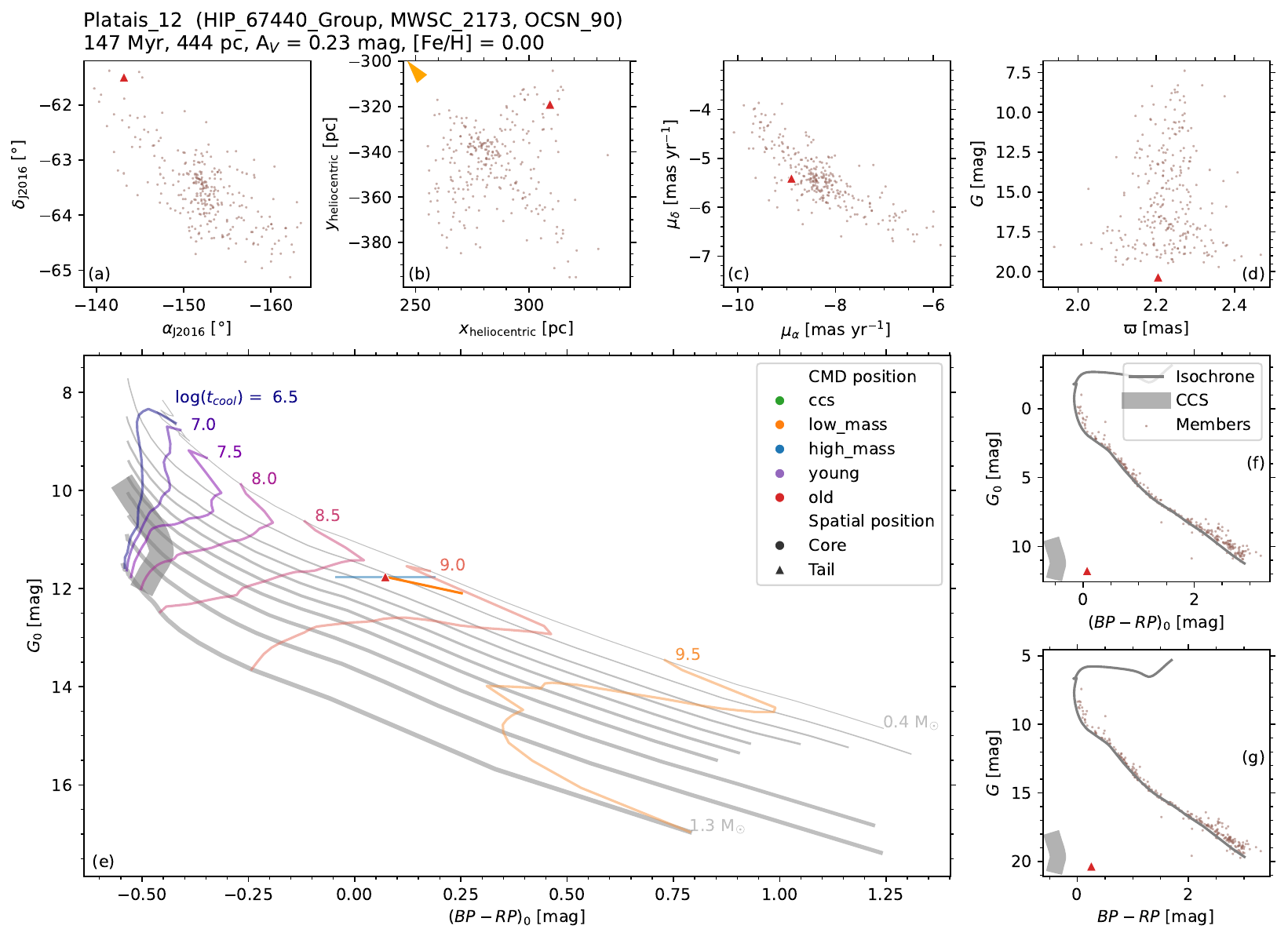}
\caption{Diagnostic plots for OCSN 77 and Platais 12. All details are similar to Figure~\ref{fig:combo_Melotte_25}.}
\label{fig:combo_Platais_12_appendix}
\end{figure}
\begin{figure}
\centering
\includegraphics[width=0.85\linewidth]{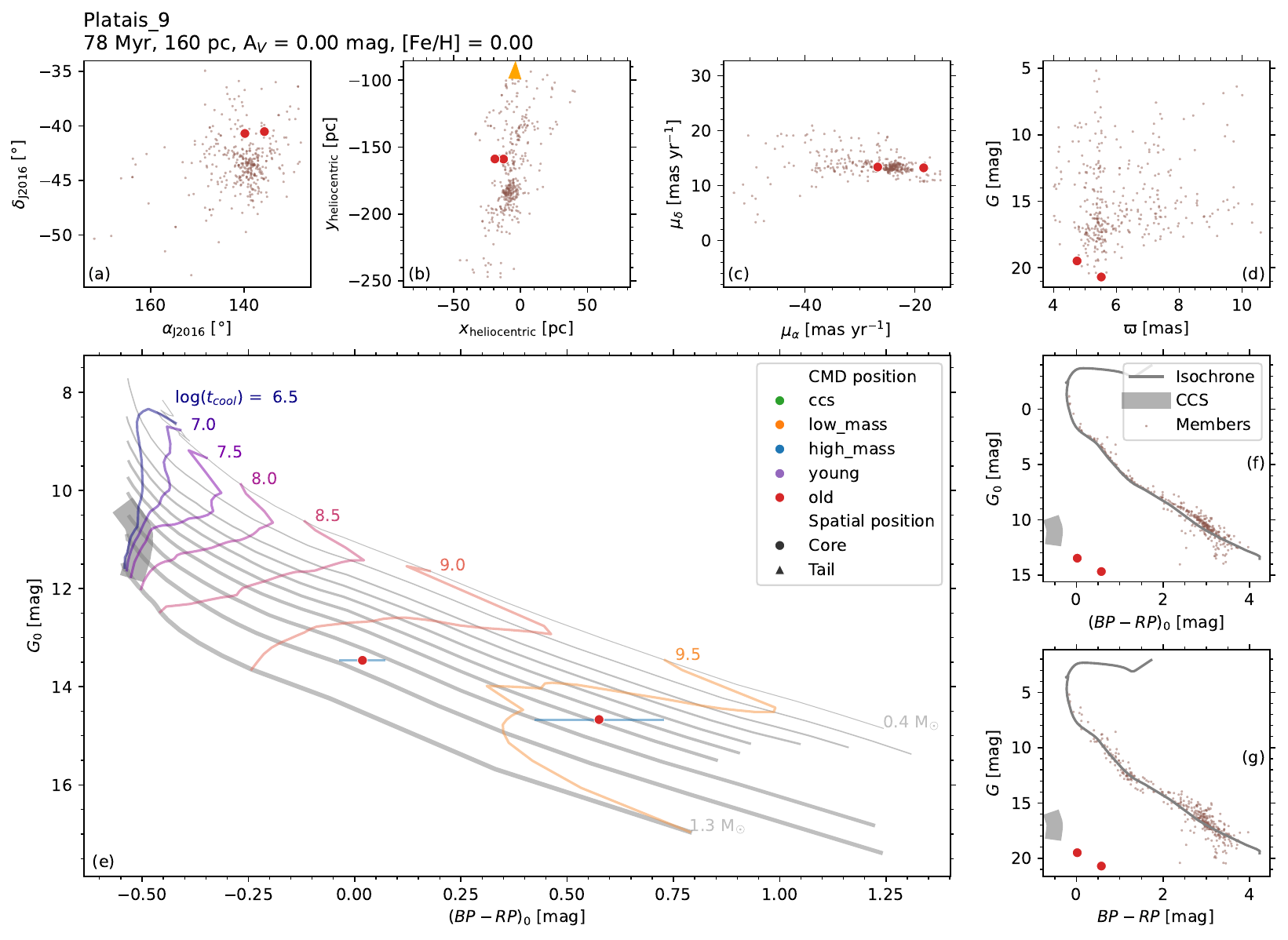}
\includegraphics[width=0.85\linewidth]{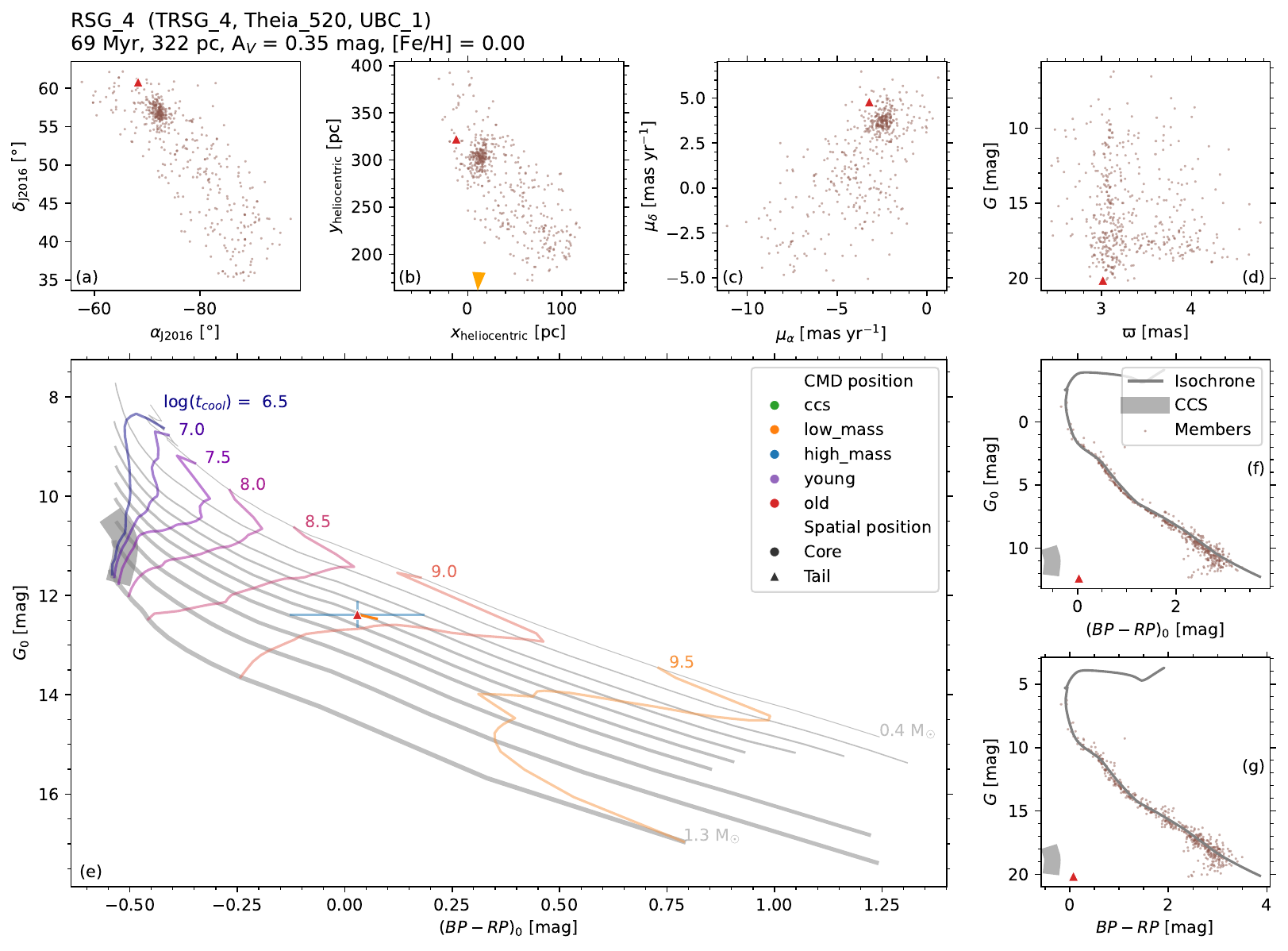}
\caption{Diagnostic plots for Platais 9 and RSG 4. All details are similar to Figure~\ref{fig:combo_Melotte_25}.}
\label{fig:combo_RSG_4_appendix}
\end{figure}
\begin{figure}
\centering
\includegraphics[width=0.85\linewidth]{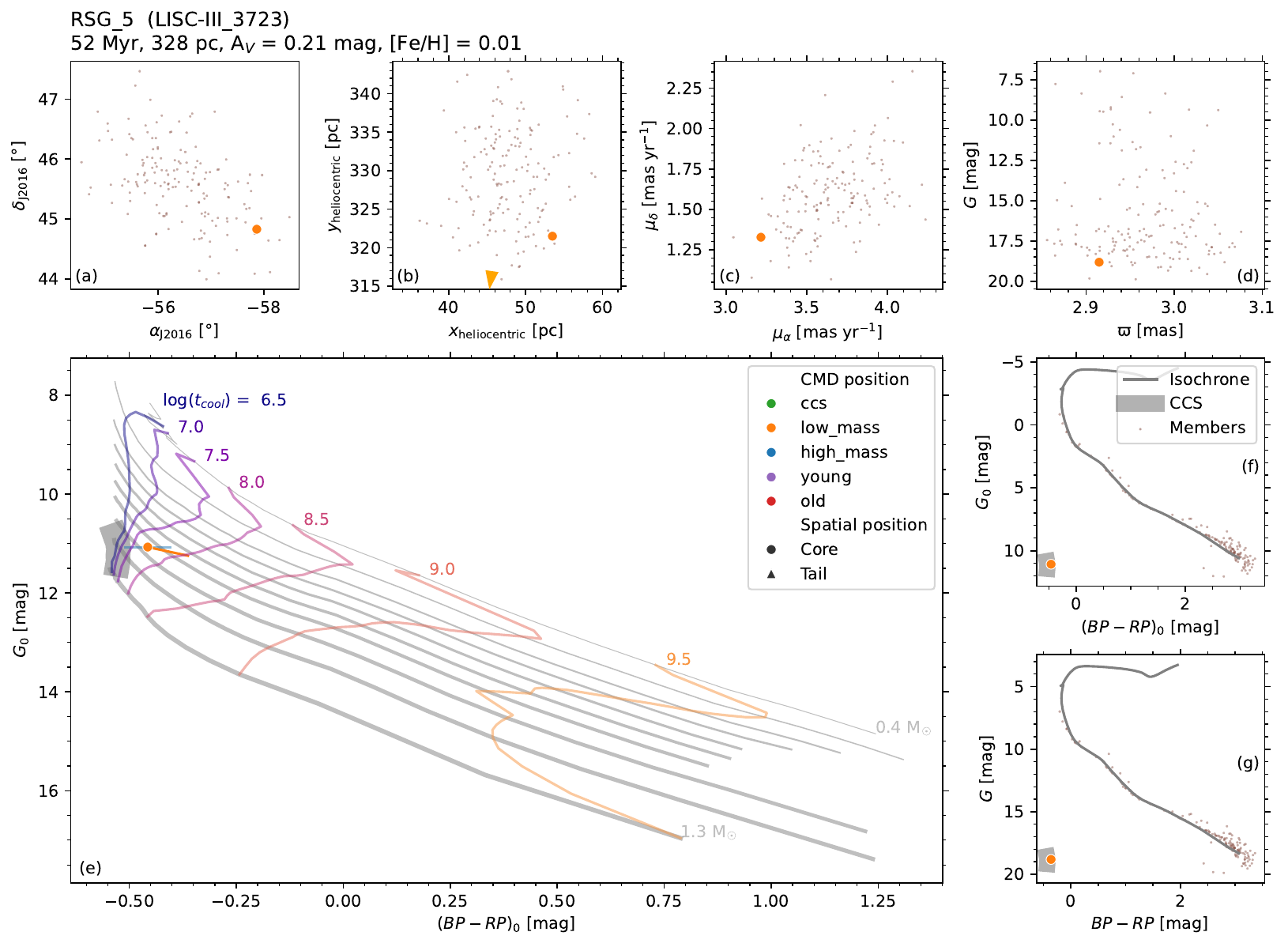}
\includegraphics[width=0.85\linewidth]{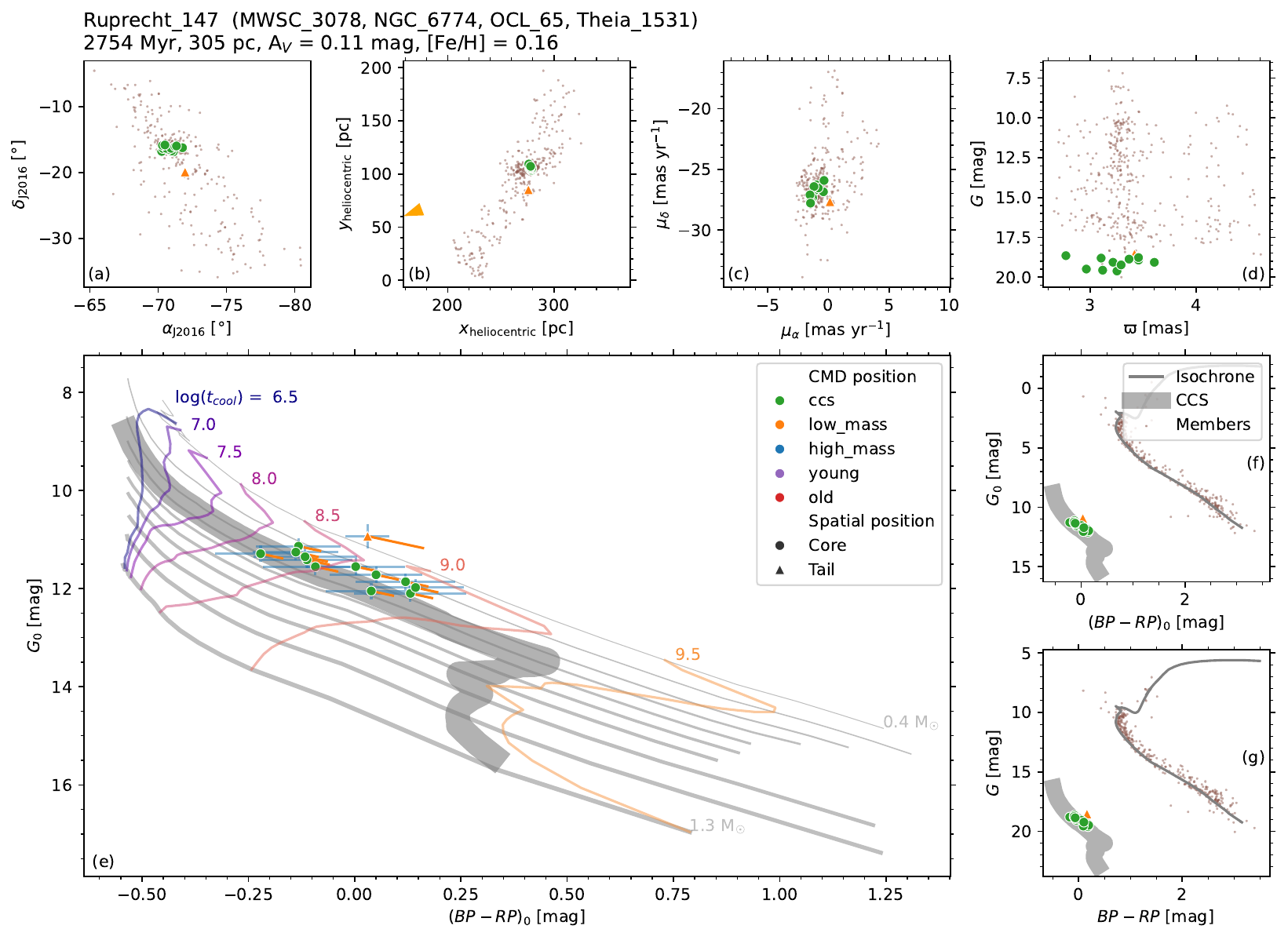}
\caption{Diagnostic plots for RSG 5 and Ruprecht 147. All details are similar to Figure~\ref{fig:combo_Melotte_25}.}
\label{fig:combo_Ruprecht_147_appendix}
\end{figure}
\begin{figure}
\centering
\includegraphics[width=0.85\linewidth]{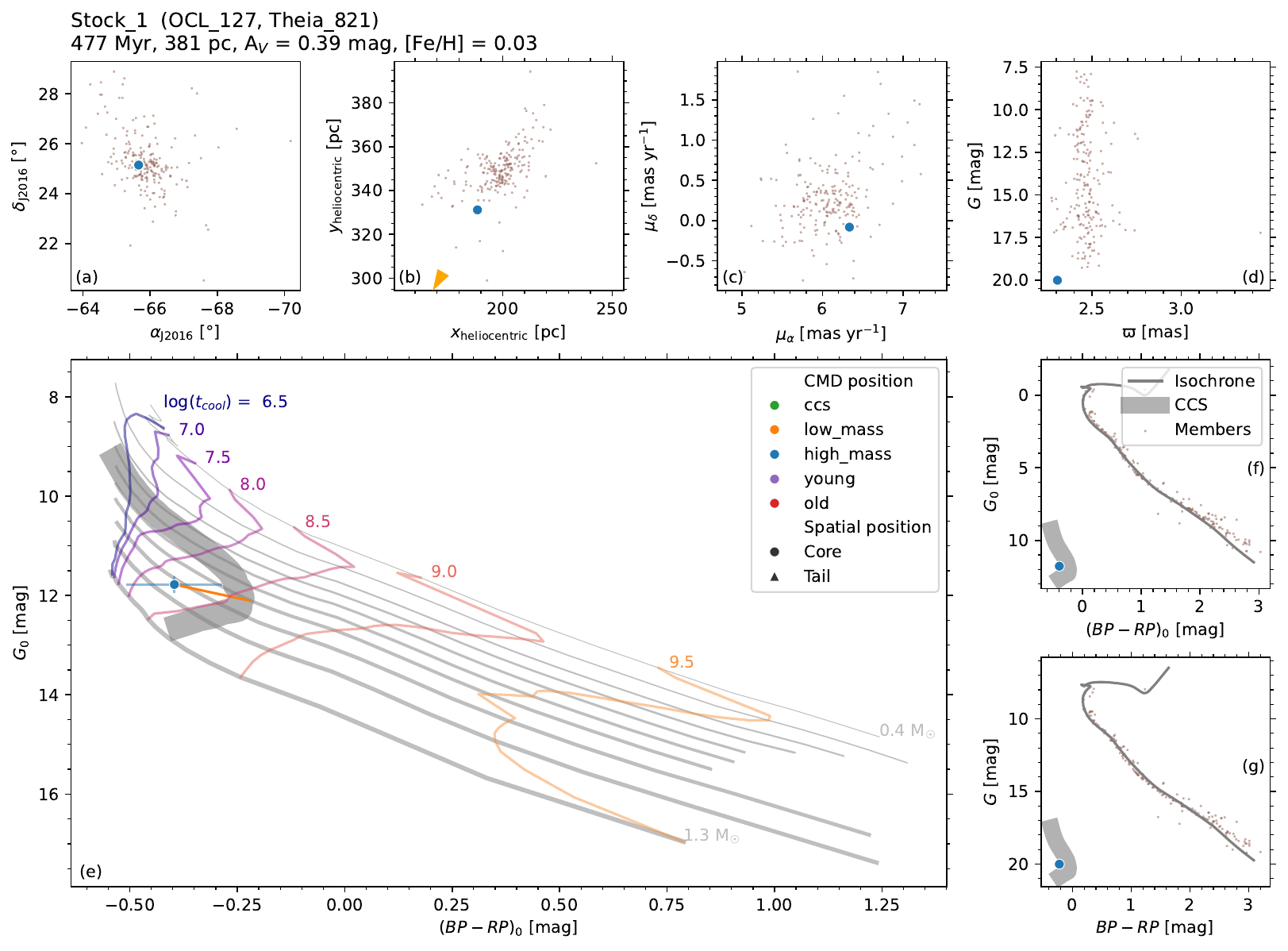}
\includegraphics[width=0.85\linewidth]{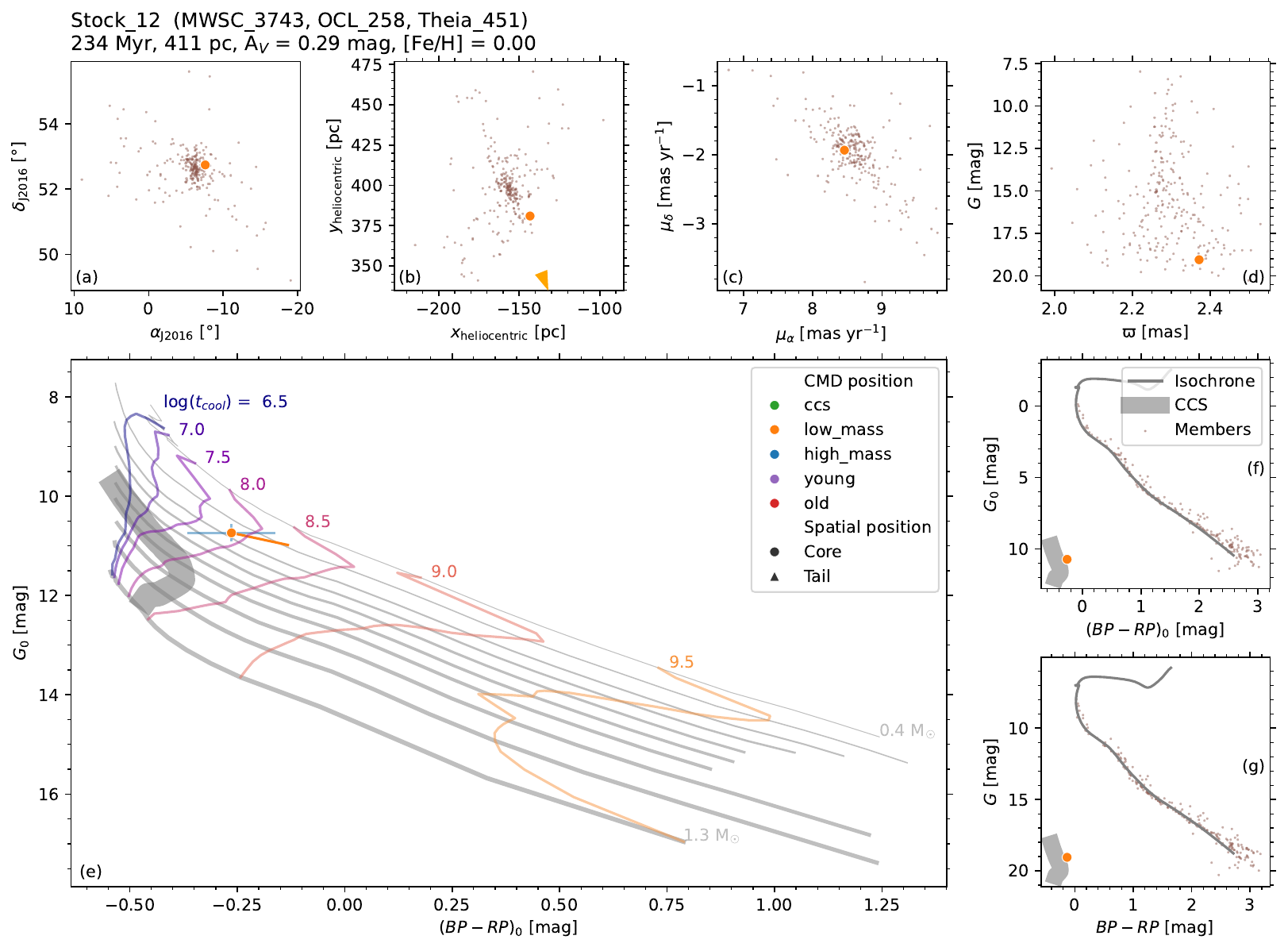}
\caption{Diagnostic plots for Stock 1 and Stock 12. All details are similar to Figure~\ref{fig:combo_Melotte_25}.}
\label{fig:combo_Stock_12_appendix}
\end{figure}
\begin{figure}
\centering
\includegraphics[width=0.85\linewidth]{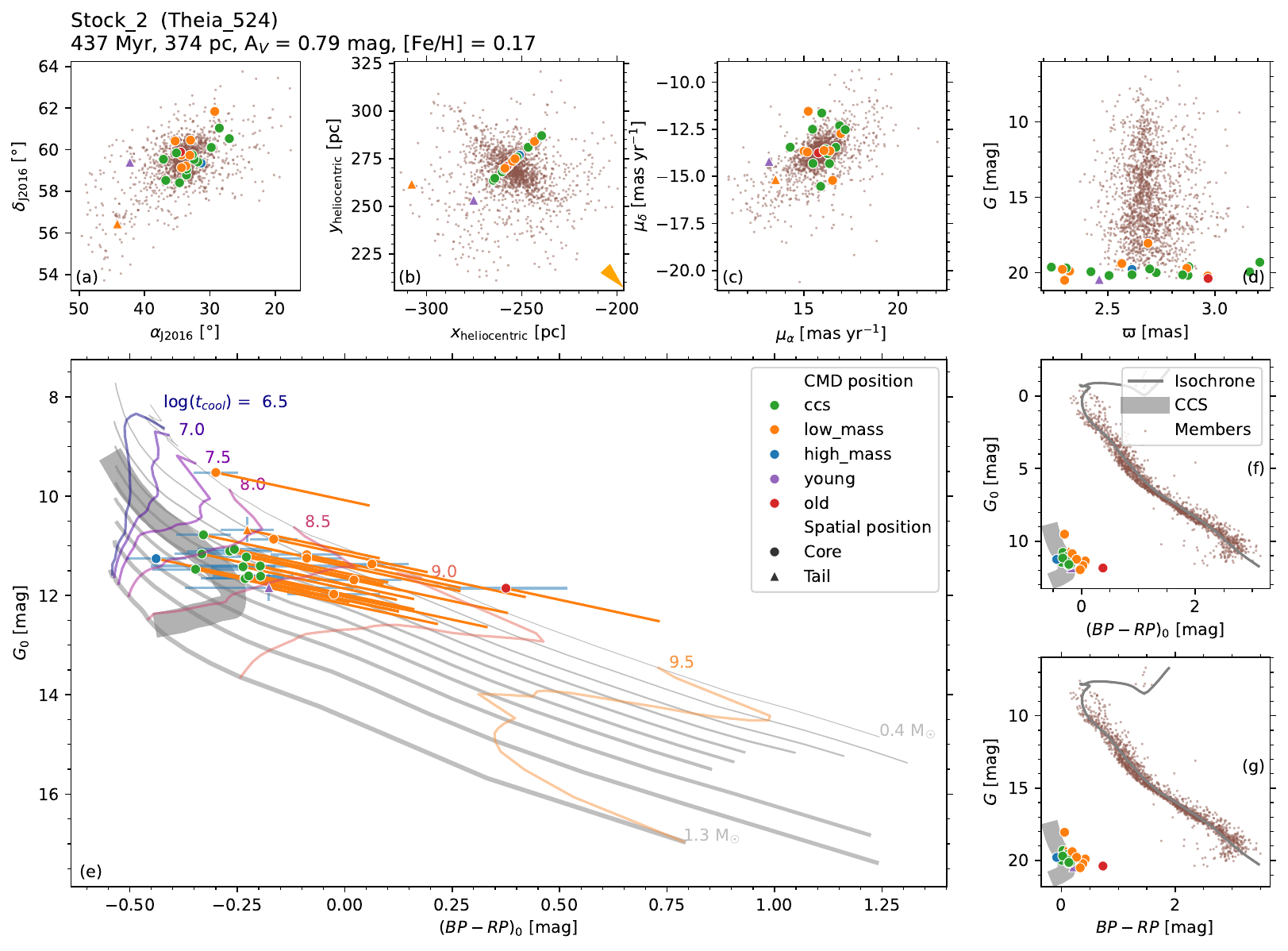}
\includegraphics[width=0.85\linewidth]{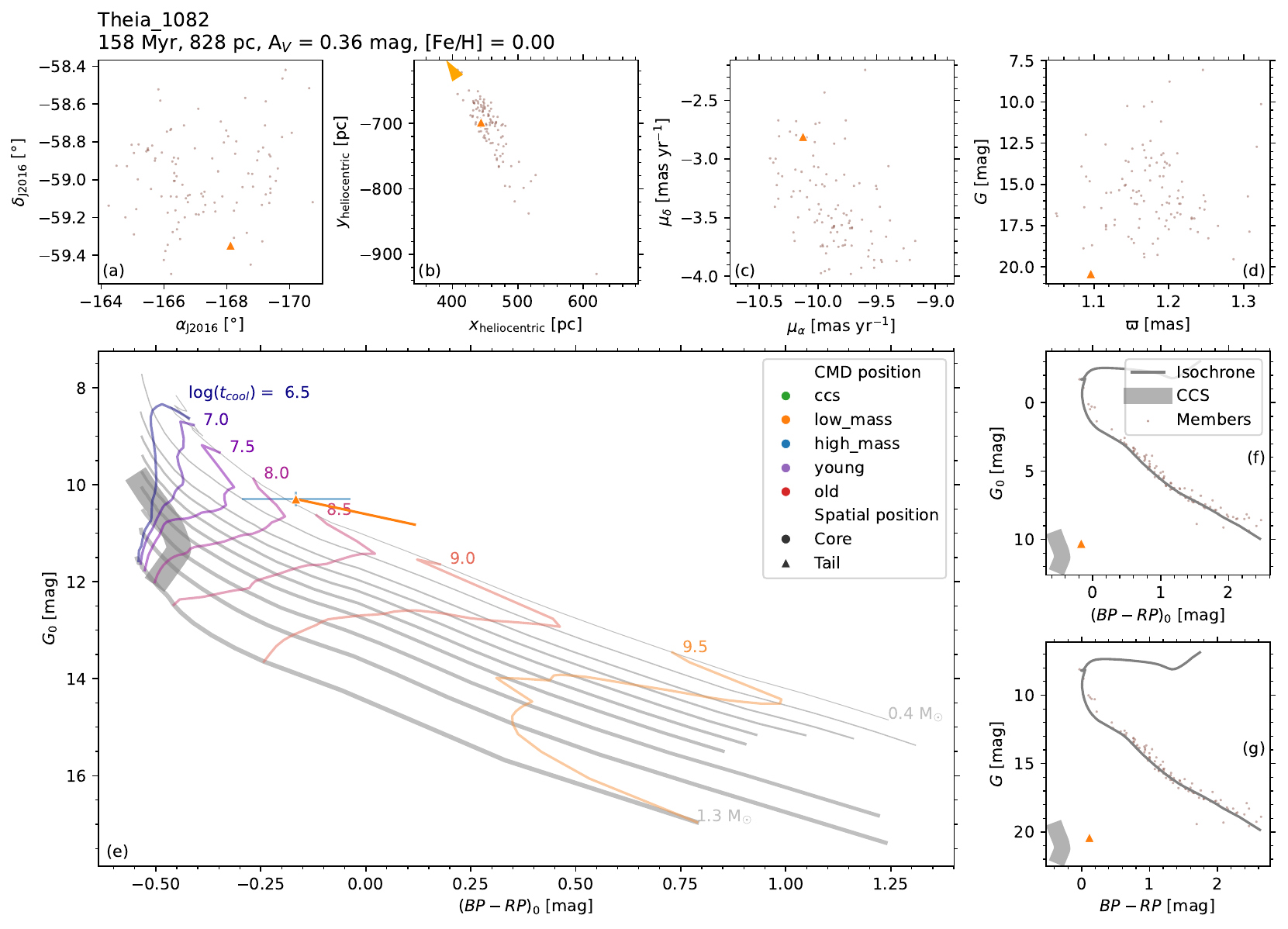}
\caption{Diagnostic plots for Stock 2 and Theia 1082. All details are similar to Figure~\ref{fig:combo_Melotte_25}.}
\label{fig:combo_Theia_1082_appendix}
\end{figure}
\begin{figure}
\centering
\includegraphics[width=0.85\linewidth]{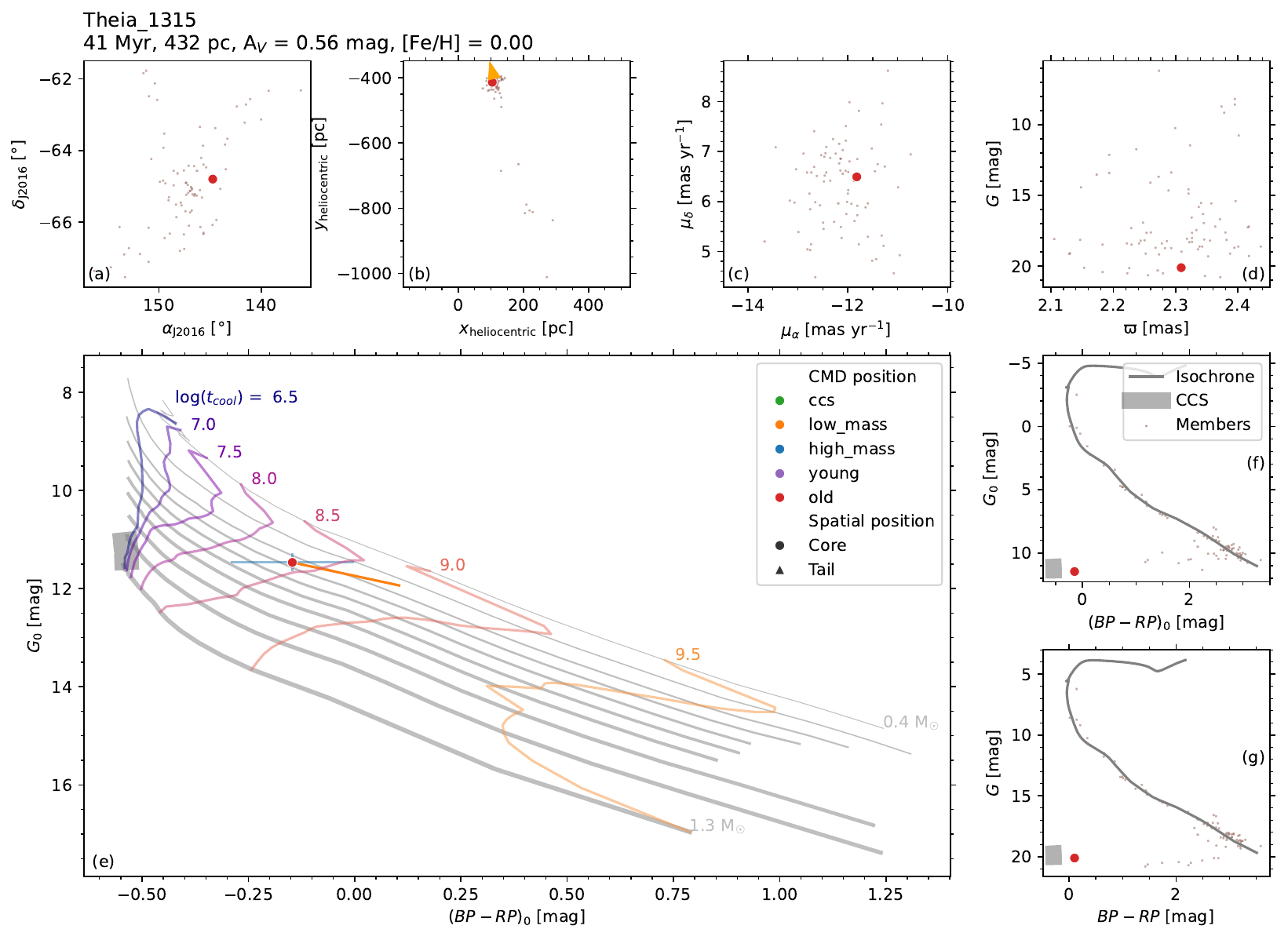}
\includegraphics[width=0.85\linewidth]{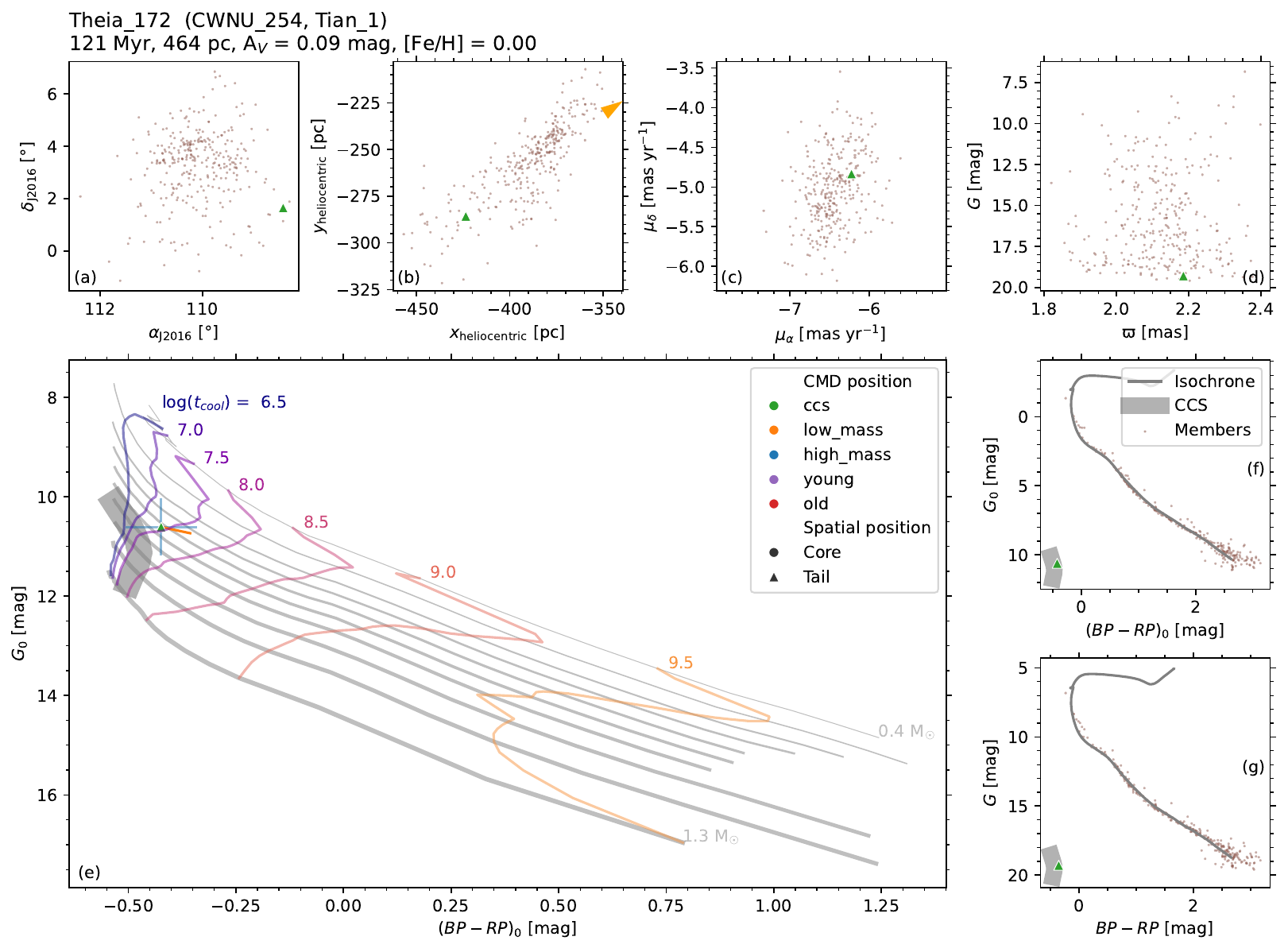}
\caption{Diagnostic plots for Theia 1315 and Theia 172. All details are similar to Figure~\ref{fig:combo_Melotte_25}.}
\label{fig:combo_Theia_172_appendix}
\end{figure}
\begin{figure}
\centering
\includegraphics[width=0.85\linewidth]{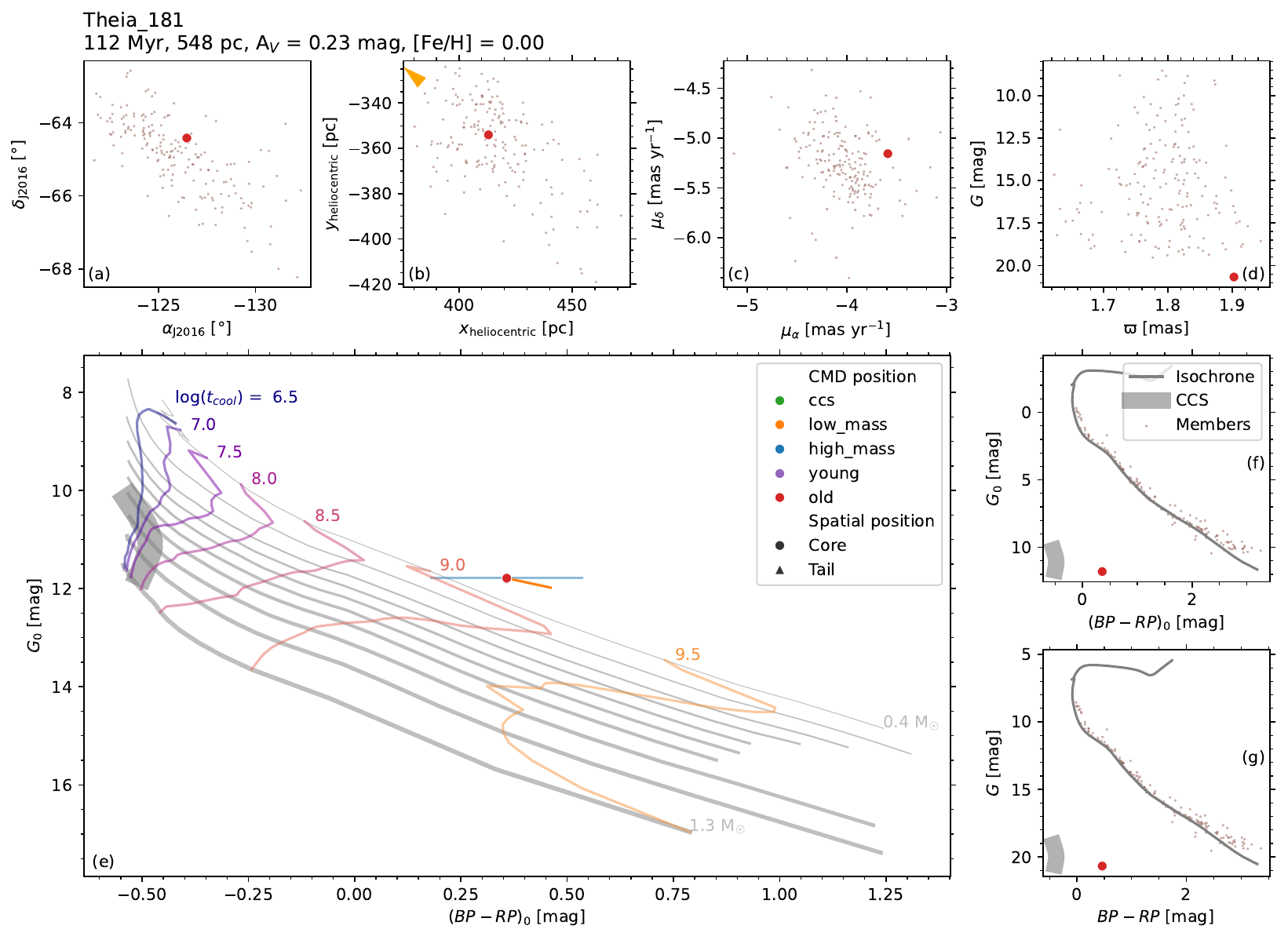}
\includegraphics[width=0.85\linewidth]{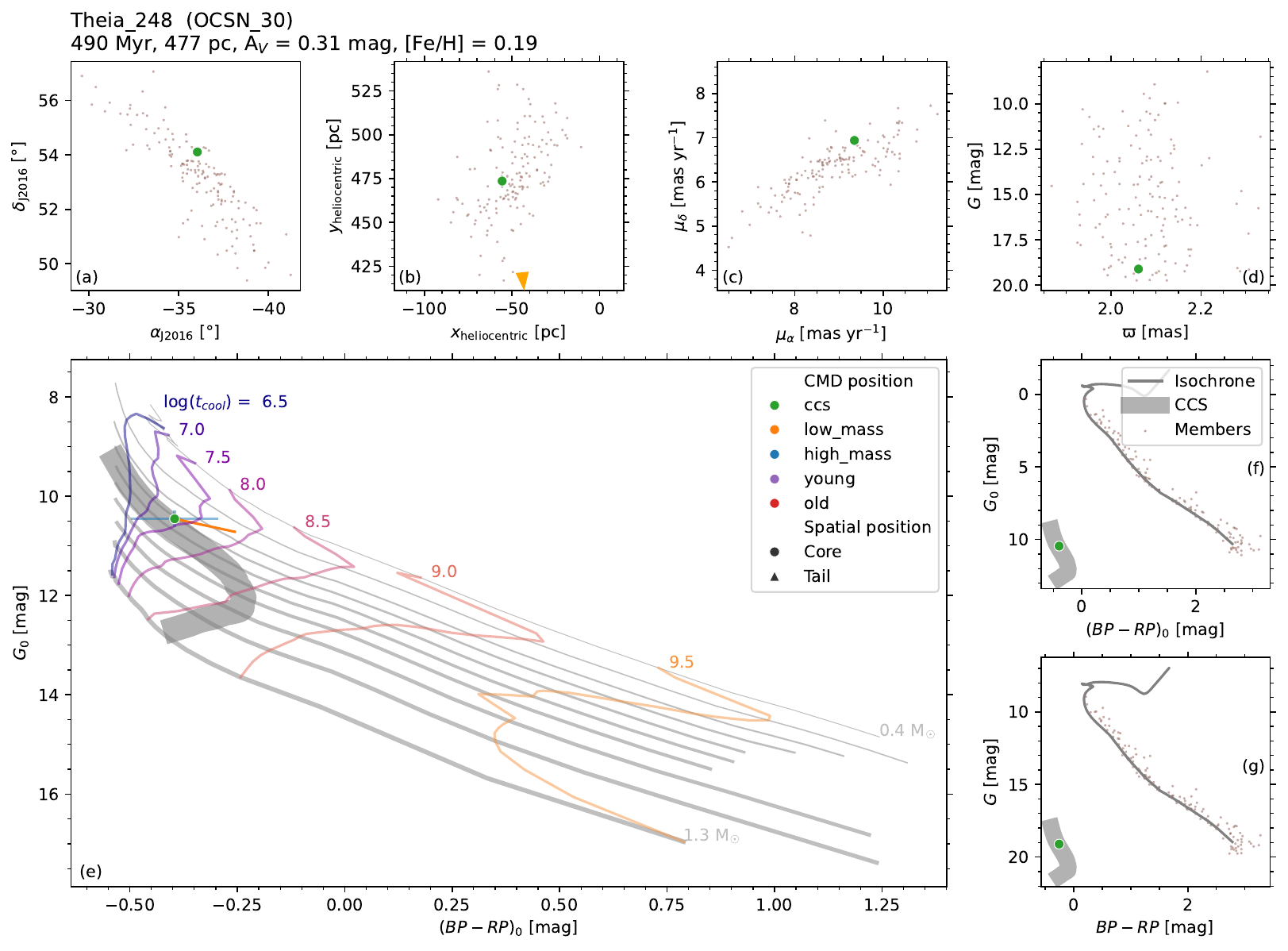}
\caption{Diagnostic plots for Theia 181 and Theia 248. All details are similar to Figure~\ref{fig:combo_Melotte_25}.}
\label{fig:combo_Theia_248_appendix}
\end{figure}
\begin{figure}
\centering
\includegraphics[width=0.85\linewidth]{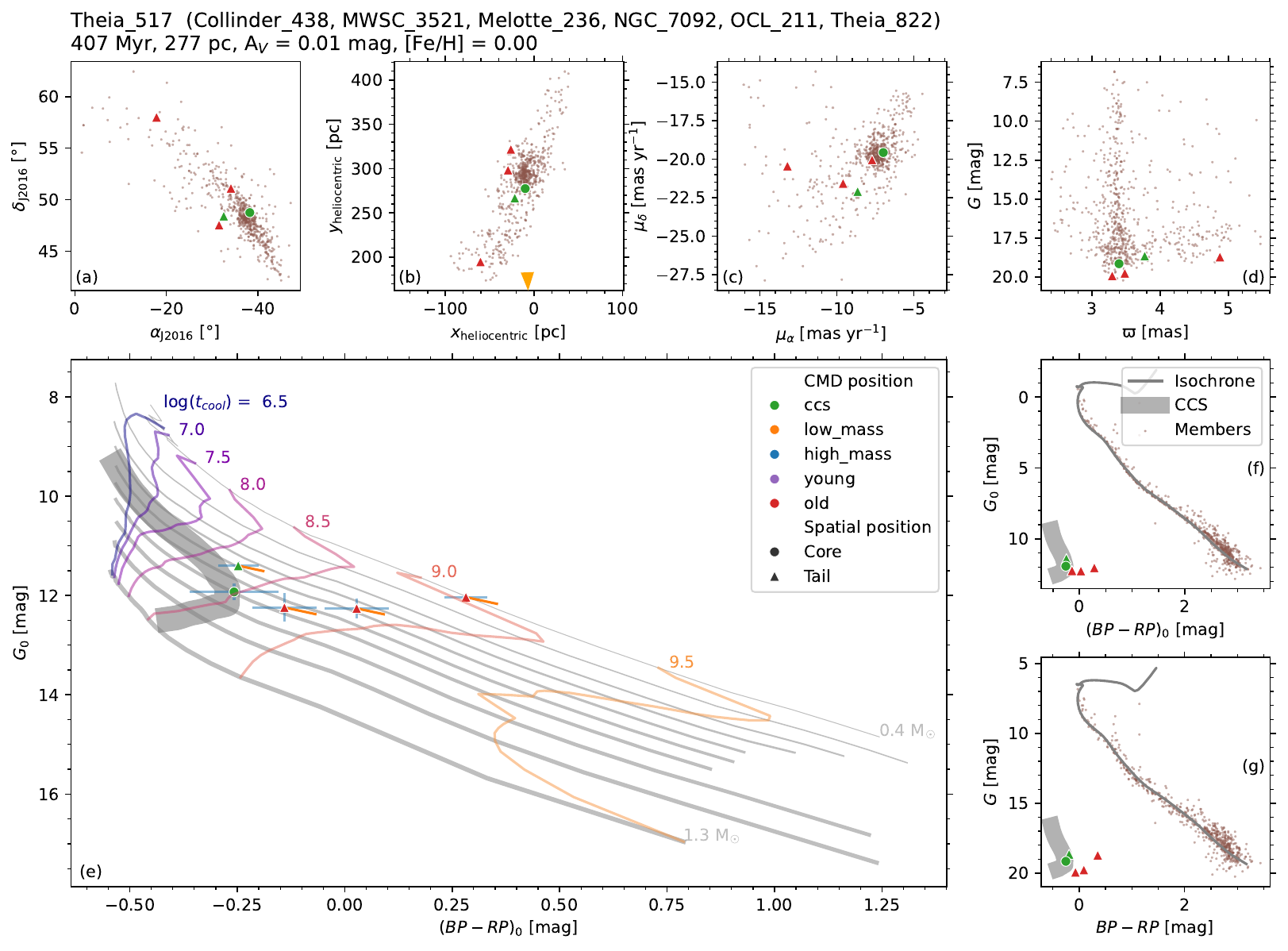}
\includegraphics[width=0.85\linewidth]{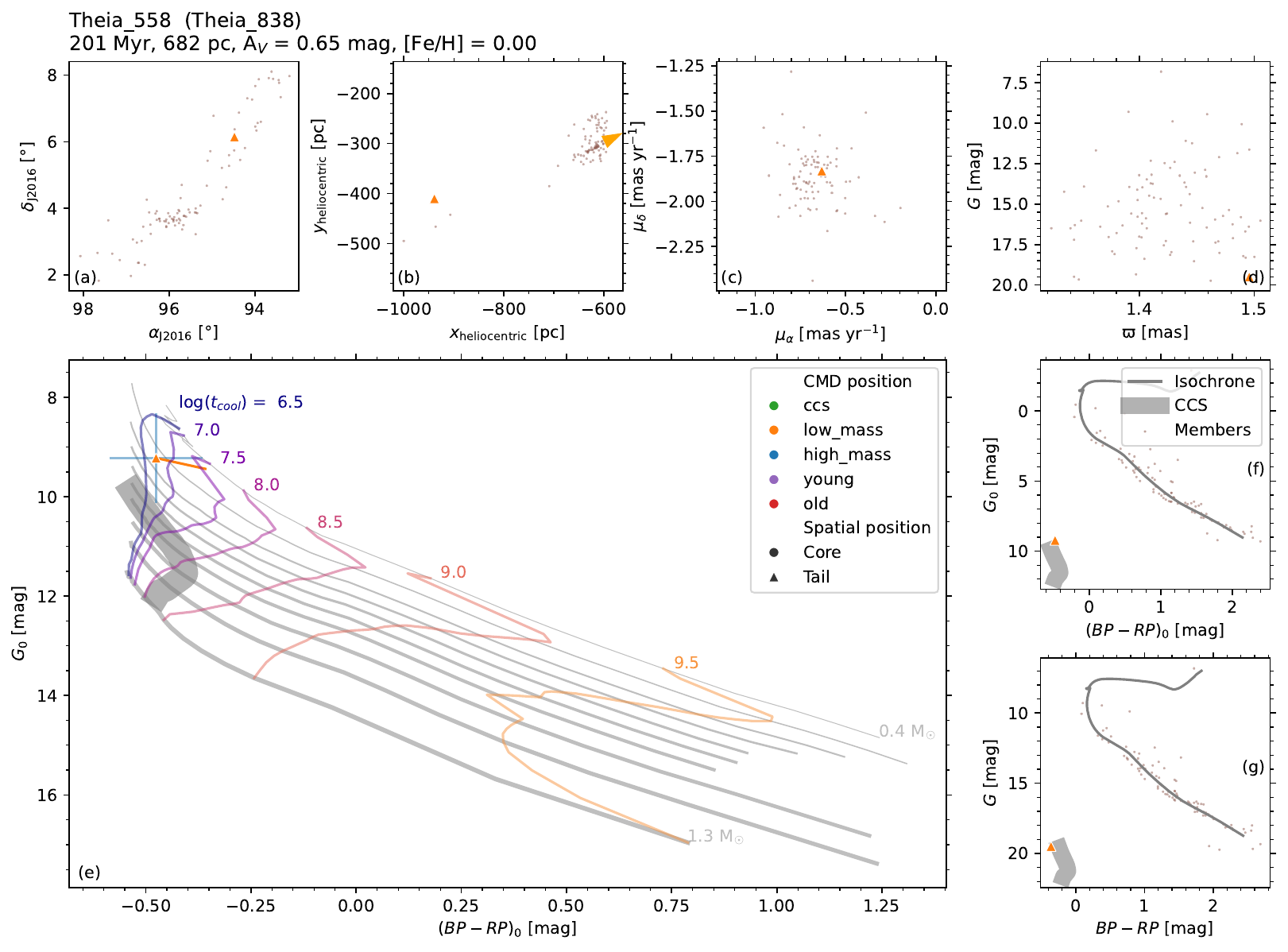}
\caption{Diagnostic plots for Theia 517 and Theia 558. All details are similar to Figure~\ref{fig:combo_Melotte_25}.}
\label{fig:combo_Theia_558_appendix}
\end{figure}
\begin{figure}
\centering
\includegraphics[width=0.85\linewidth]{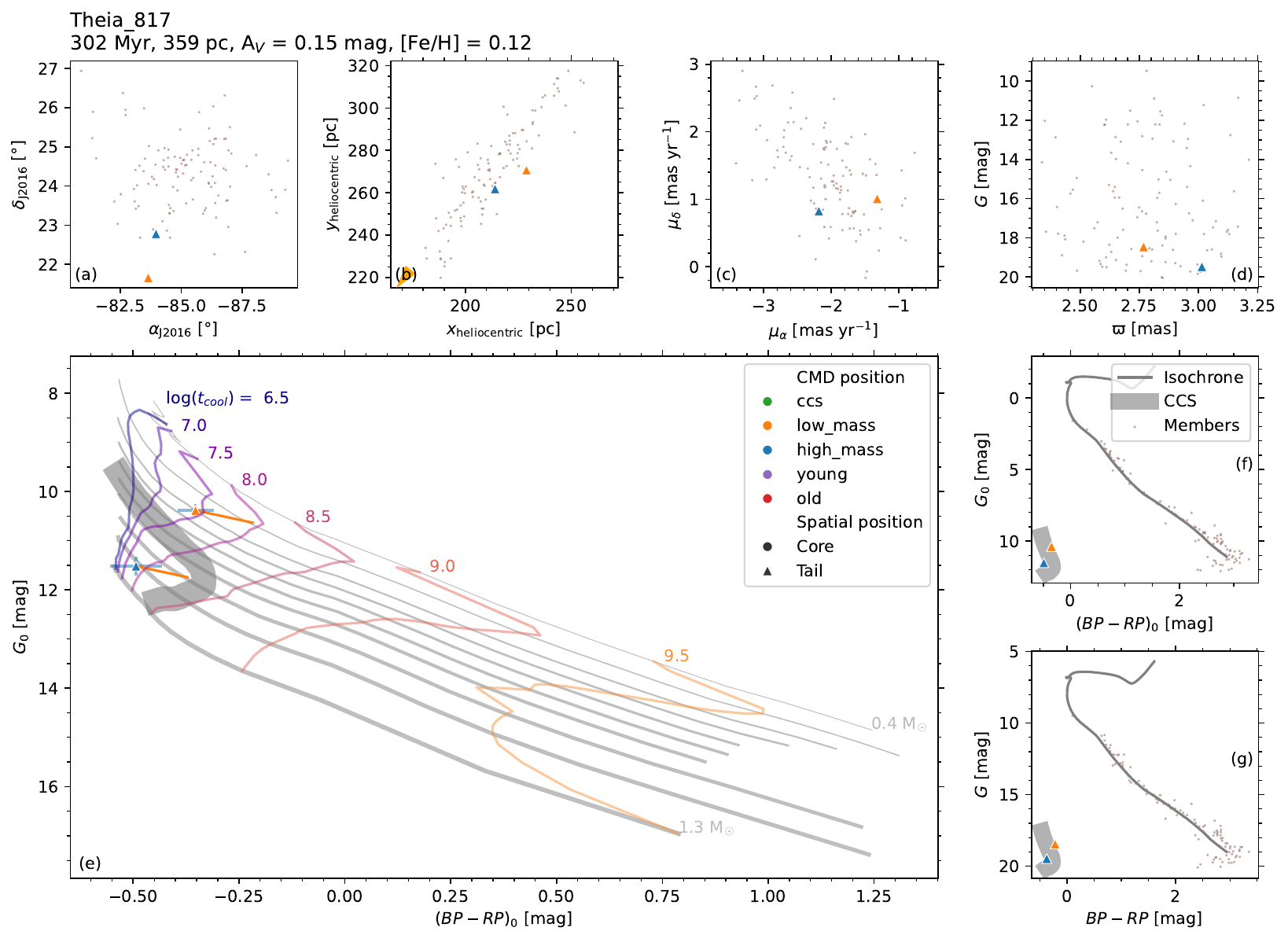}
\includegraphics[width=0.85\linewidth]{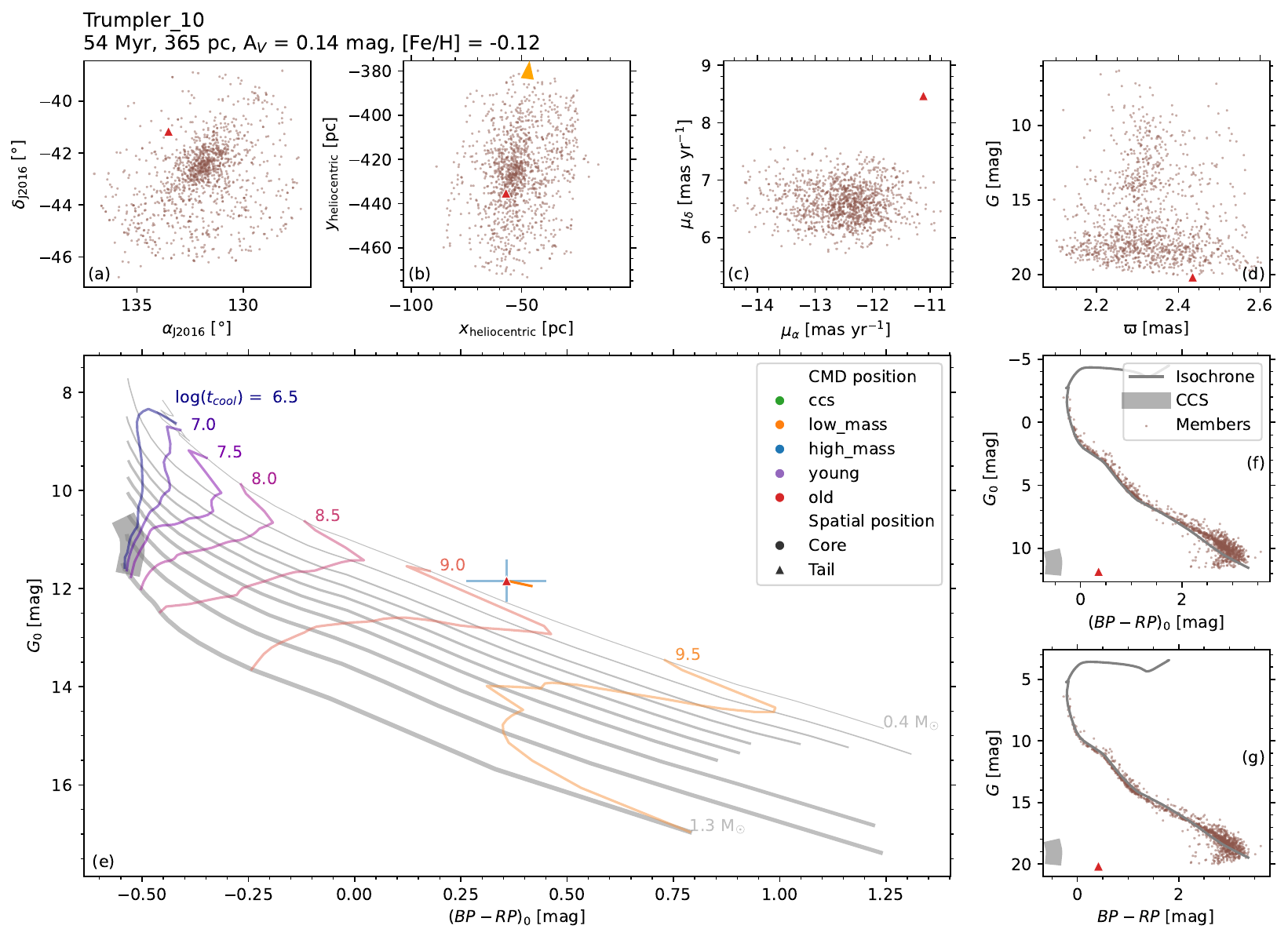}
\caption{Diagnostic plots for Theia 817 and Trumpler 10. All details are similar to Figure~\ref{fig:combo_Melotte_25}.}
\label{fig:combo_Trumpler_10_appendix}
\end{figure}
\begin{figure}
\centering
\includegraphics[width=0.85\linewidth]{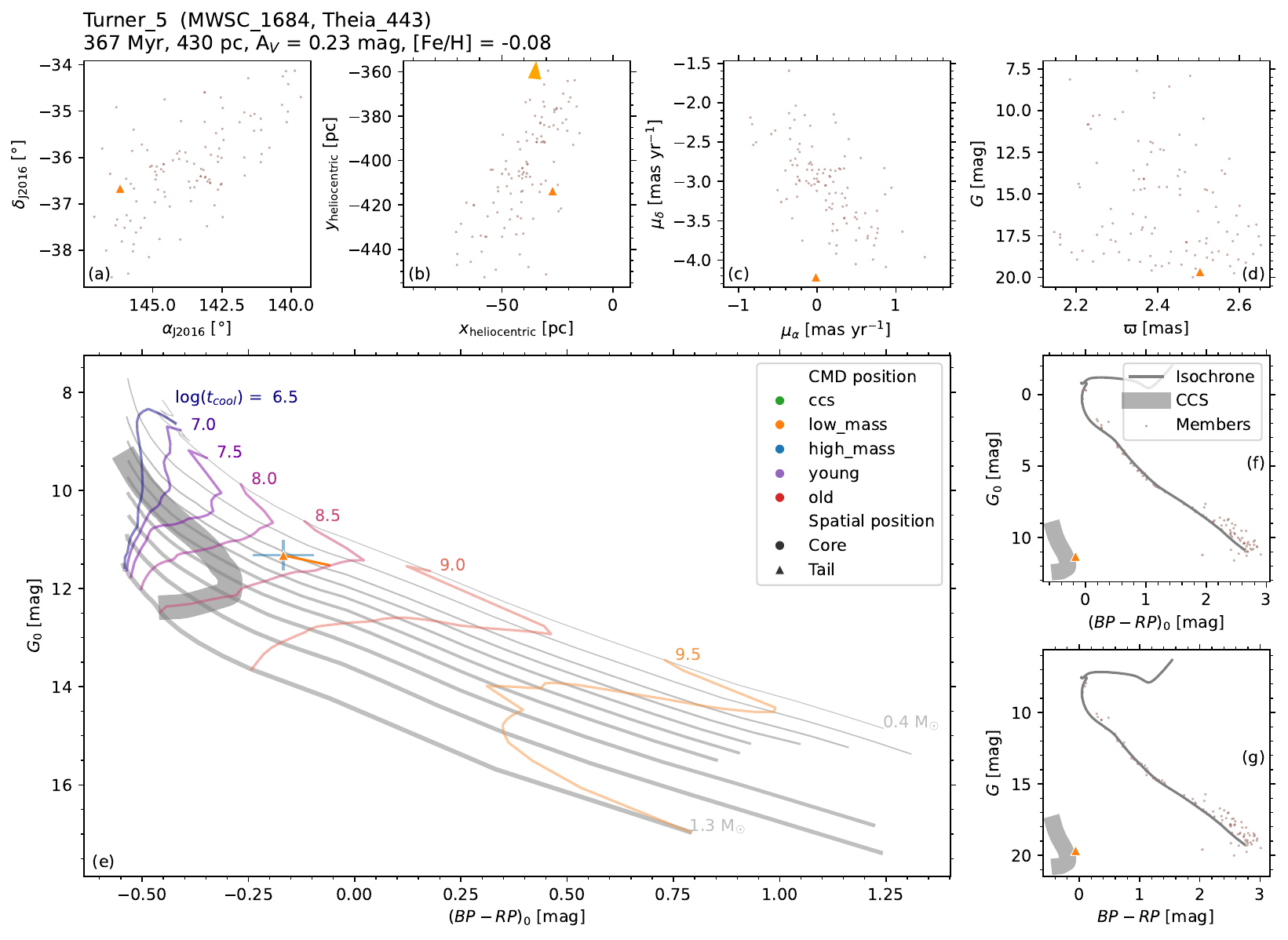}
\includegraphics[width=0.85\linewidth]{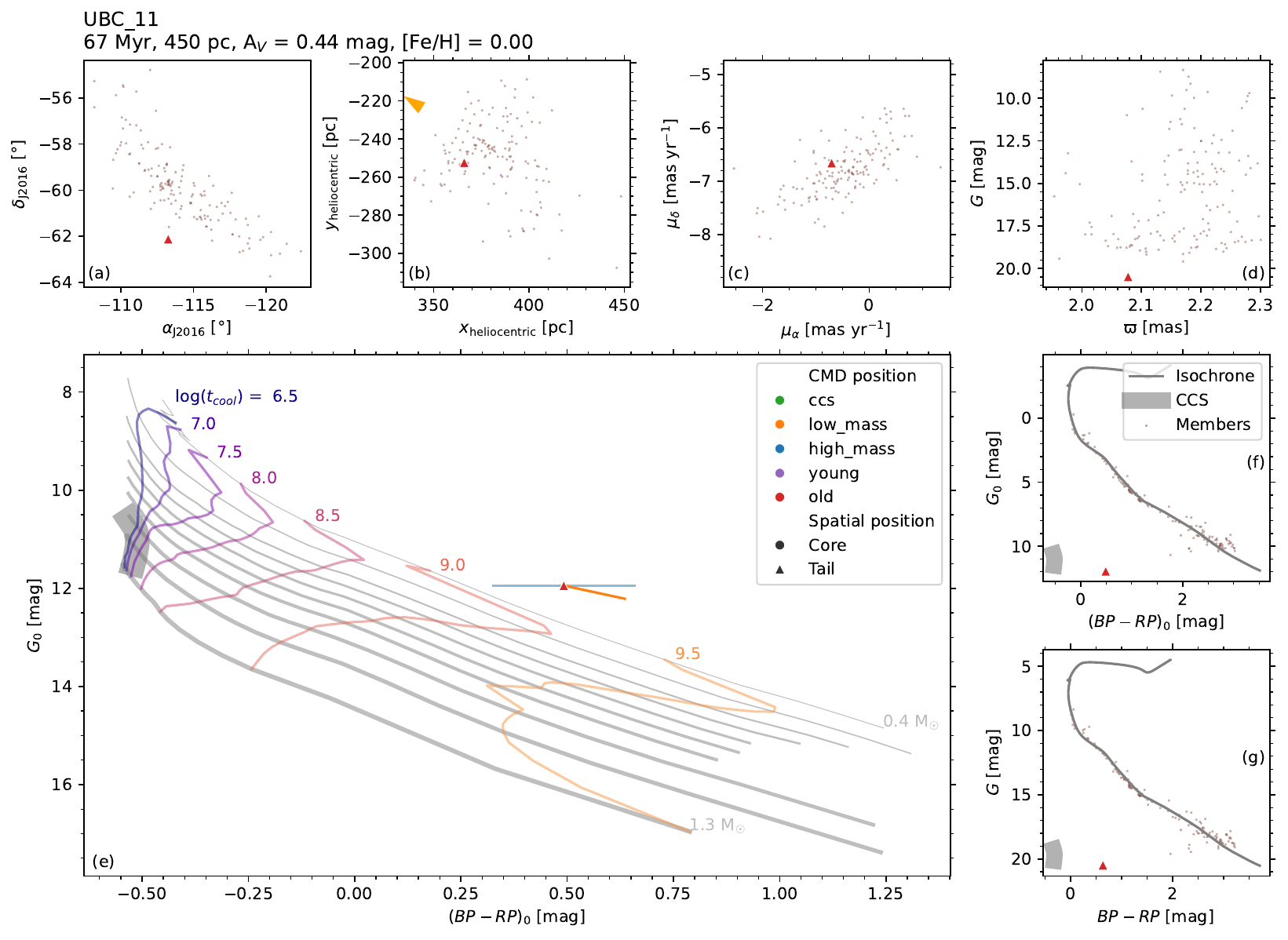}
\caption{Diagnostic plots for Turner 5 and UBC 11. All details are similar to Figure~\ref{fig:combo_Melotte_25}.}
\label{fig:combo_UBC_11_appendix}
\end{figure}
\begin{figure}
\centering
\includegraphics[width=0.85\linewidth]{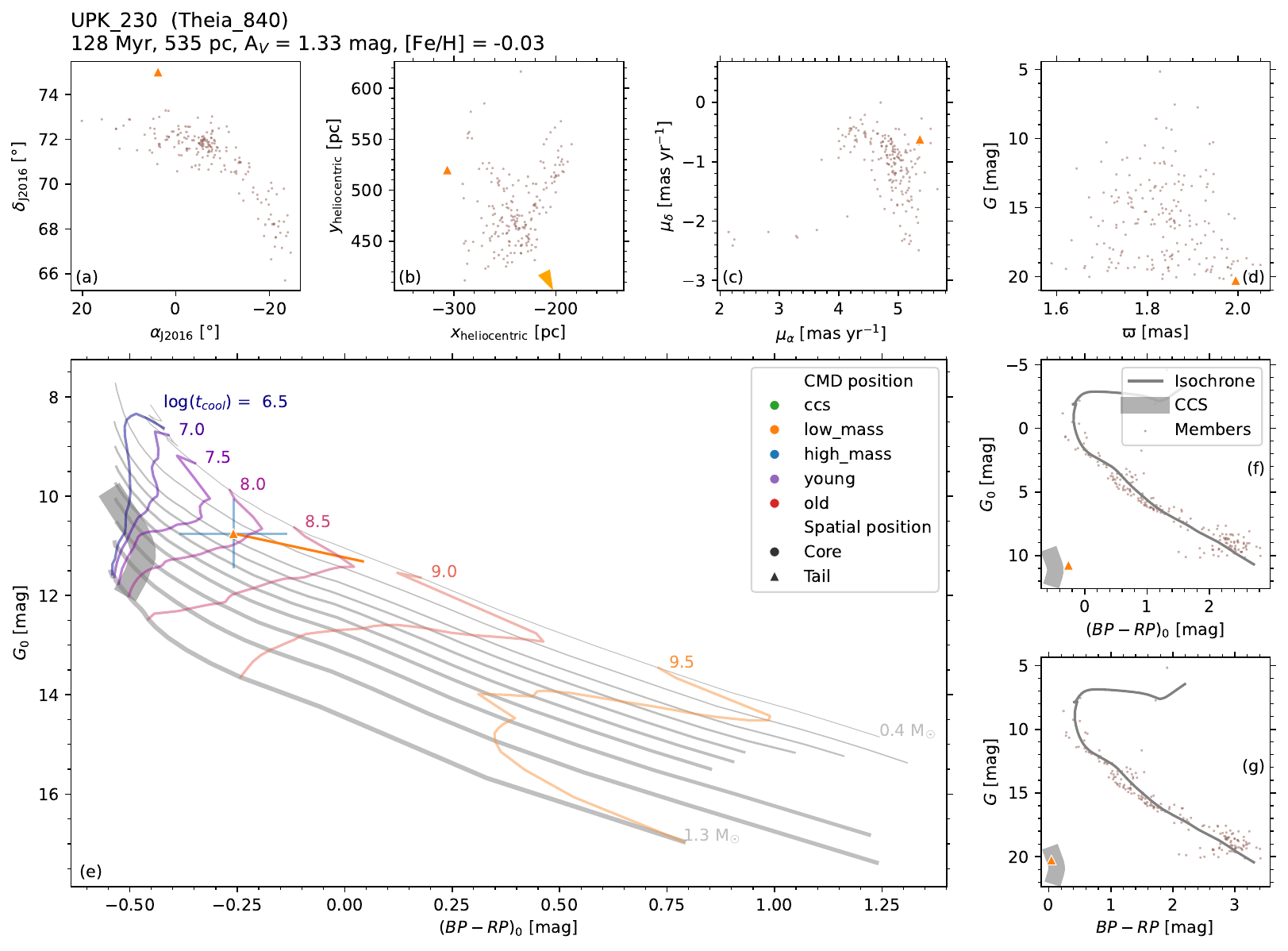}
\includegraphics[width=0.85\linewidth]{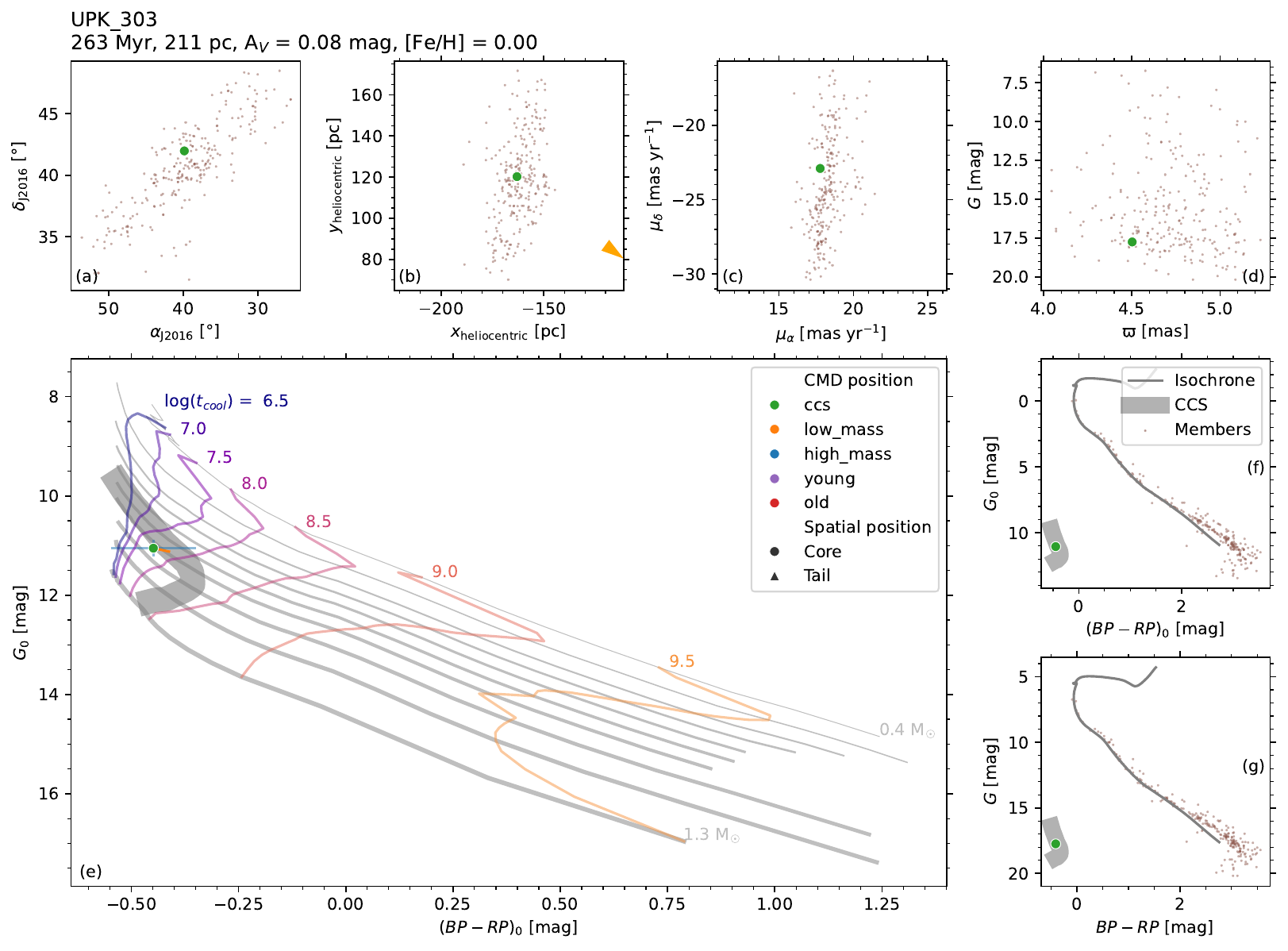}
\caption{Diagnostic plots for UPK 230 and UPK 303. All details are similar to Figure~\ref{fig:combo_Melotte_25}.}
\label{fig:combo_UPK_303_appendix}
\end{figure}
\begin{figure}
\centering
\includegraphics[width=0.85\linewidth]{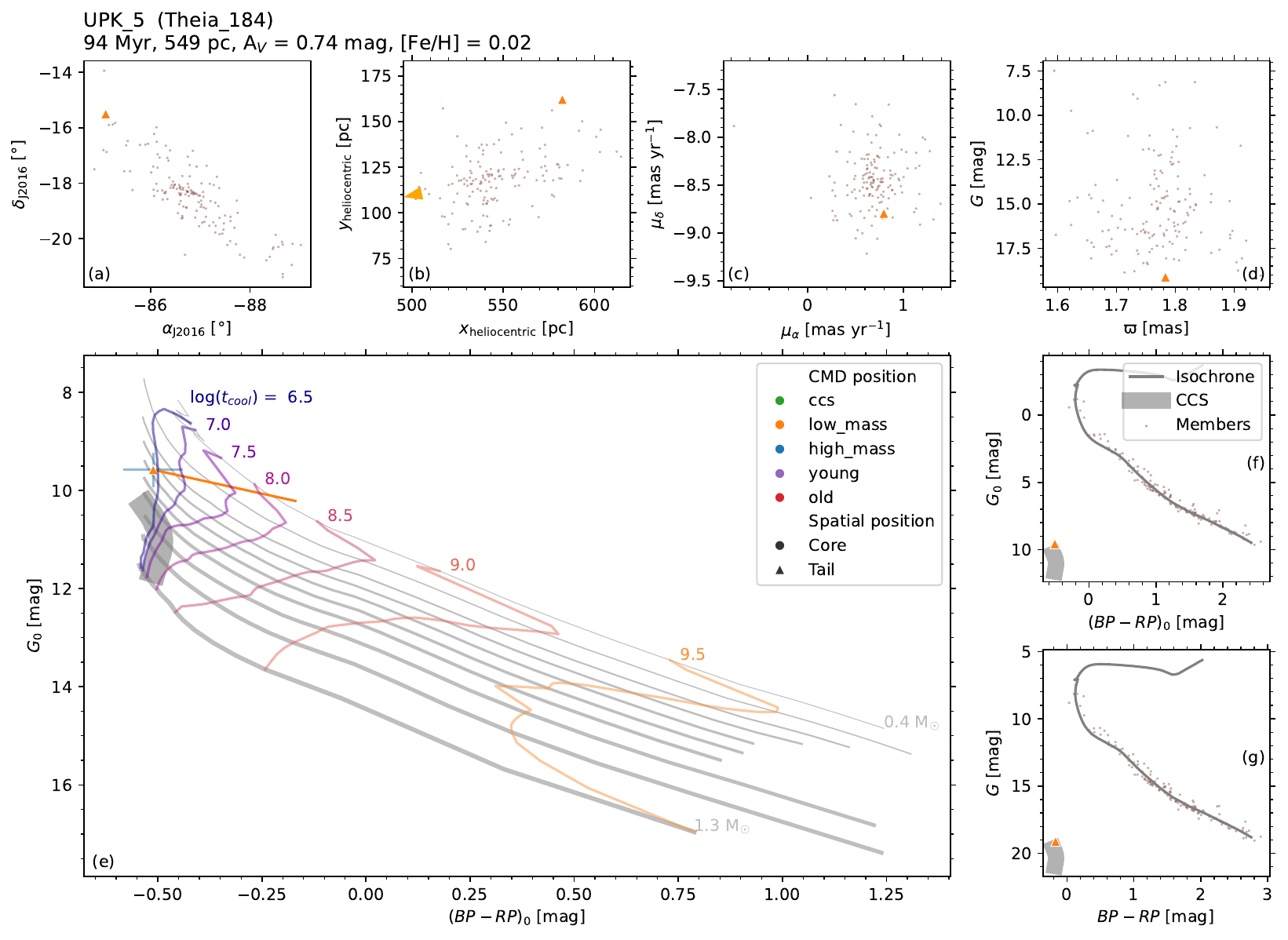}
\includegraphics[width=0.85\linewidth]{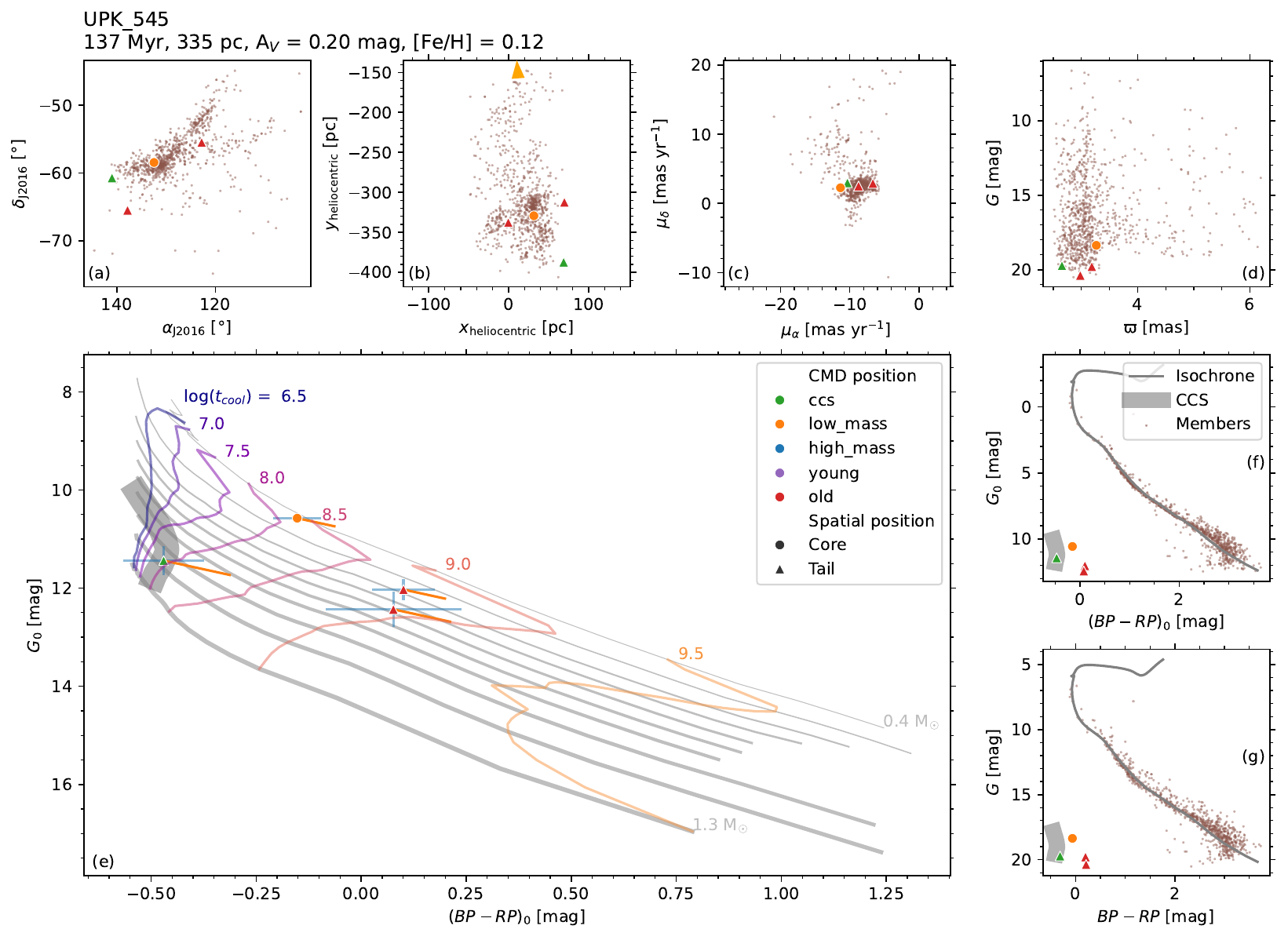}
\caption{Diagnostic plots for UPK 5 and UPK 545. All details are similar to Figure~\ref{fig:combo_Melotte_25}.}
\label{fig:combo_UPK_545_appendix}
\end{figure}
\begin{figure}
\centering
\includegraphics[width=0.85\linewidth]{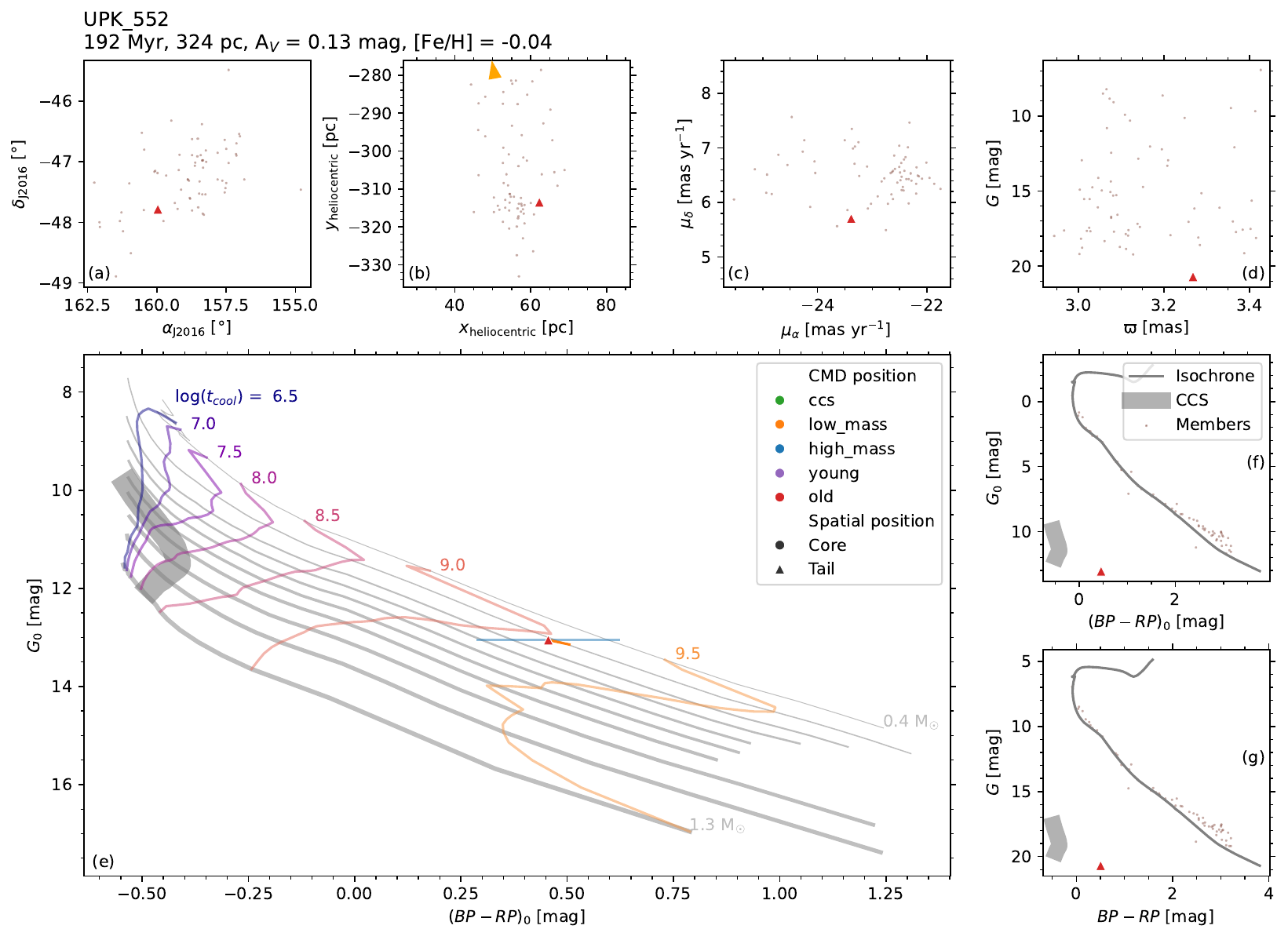}
\includegraphics[width=0.85\linewidth]{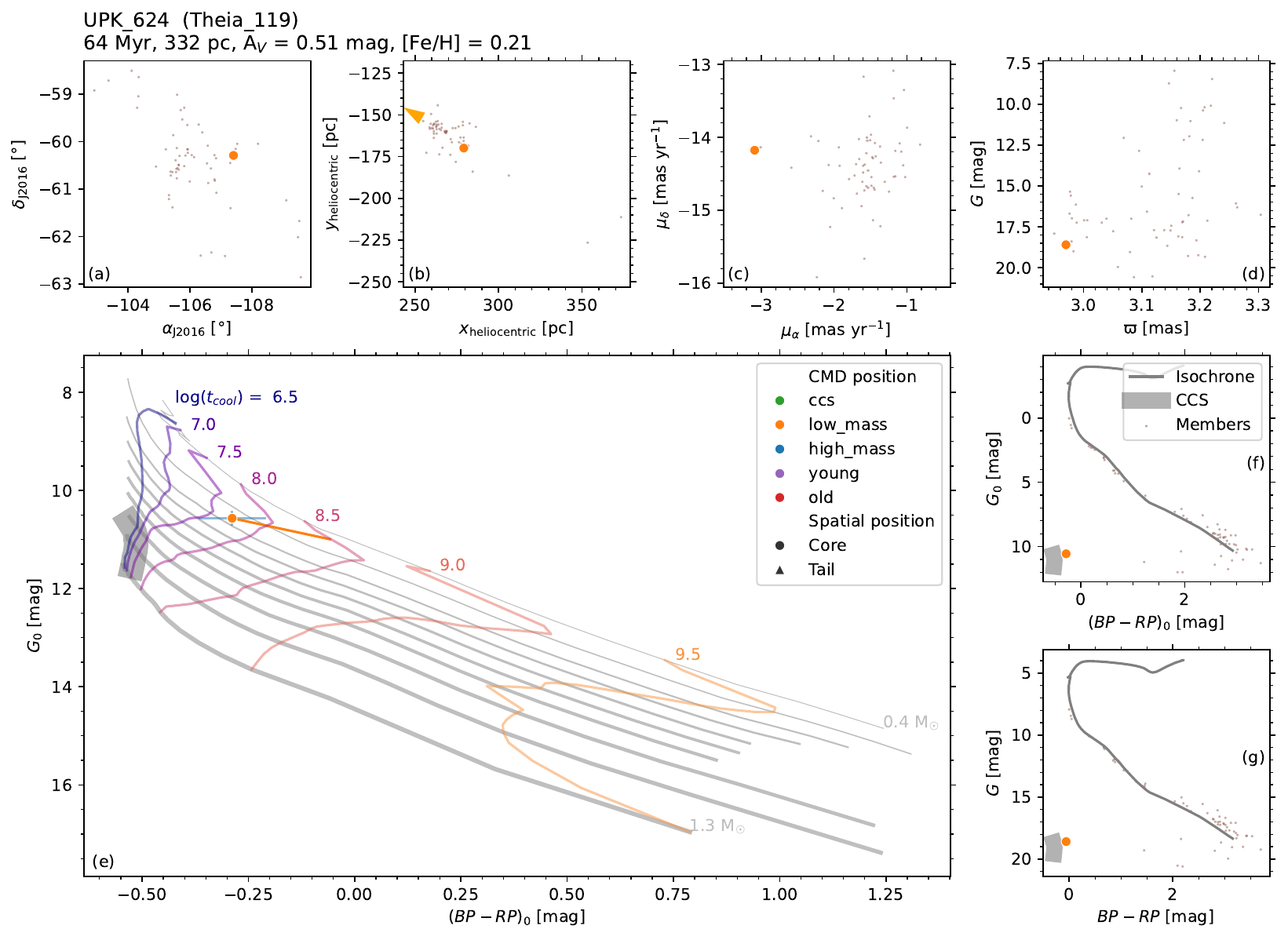}
\caption{Diagnostic plots for UPK 552 and UPK 624. All details are similar to Figure~\ref{fig:combo_Melotte_25}.}
\label{fig:combo_UPK_624_appendix}
\end{figure}